\documentclass[aps,pra,reprint,showkeys,longbibliography]{revtex4-2} 
\usepackage{amsmath}
\usepackage{amssymb}
\usepackage{graphicx} 
\usepackage{bm} 
\usepackage{xcolor}
\usepackage[toc,page]{appendix}
\usepackage{braket}
\usepackage{mathrsfs}

\makeatletter

\@ifclasswith{revtex4-2}{draft}{%
    \usepackage[notref,notcite]{showkeys}
    
}{}

\makeatother

\def\eg{e.g.\ } 
 
\def\etc{etc.\ } 
 
\def\rhs{r.h.s.\ }

\newcommand{\nn}{\nonumber}

\newcommand{\eps}{\varepsilon}
\newcommand{\htilde}{\tilde{h}}
\newcommand{\QCD}{\text{QCD}}
\newcommand{\GUT}{\text{GUT}}
\newcommand{\gra}{(\text{gr})}
\newcommand{\cri}{(\text{cr})}

\begin{document}

\title{Fermi scale from quantum gravity scaling solution}

\author{Christof Wetterich}
\affiliation{Institut f\"ur Theoretische Physik\\
    Universit\"at Heidelberg\\
    Philosophenweg 16, D-69120 Heidelberg}

\begin{abstract}
    We propose that quantum gravity may predict the Fermi scale.
    Fundamental scale invariance implies the scale invariant standard model.
    Both the Fermi scale and the Planck mass are given by fields, and their ratio is dictated by a dimensionless cosmon-Higgs coupling. 
    For an ultraviolet fixed point of quantum gravity this coupling is an irrelevant parameter of the renormalization flow and becomes predictable. 
    An analytic scaling solution for quantum gravity admits no free parameter for the mass term of the Higgs boson. 
    We discuss a new asymptotically safe quantum gravity fixed point for which the scalar potential is not flat.
    If the largest intrinsic mass scale generated by the renormalisation flow away from this fixed point is sufficiently below the Fermi scale, the couplings of the scale invariant standard model are determined by the scaling solution. 
    For a given short distance model remaining valid to infinitely small distances the ratio Fermi scale over Planck mass can then be predicted. 
    With reasonable assumptions for the ultraviolet fixed point a numerical solution finds a tiny value for the ratio between the Fermi and Planck scales, very close to a second order quantum electroweak phase transition. 
    This could explain the observed gauge hierarchy.
\end{abstract}

\maketitle


\section{Introduction}
In the standard model of particle physics the Fermi scale $\varphi_0 = 175\;\mathrm{GeV}$ denotes the vacuum expectation value of the Higgs scalar field.
It sets the masses of the $W$- and $Z$-boson and the charged fermions.
The characteristic scale for gravity is the (reduced) Planck mass $M_p = 2.4\times 10^{18}\;\mathrm{GeV}$.
A consistent description of physics needs a theory of quantum gravity.
Any realistic theory of quantum gravity has to accommodate the observed tiny ratio of $\varepsilon_\varphi = \varphi_0^2 / M_p^2\approx10^{-32}$.
From the point of view of the renormalization group for the standard model the Fermi scale is associated to a relevant parameter, as typically given by the quadratic term (mass term) $\mu_H^2 H^\dagger H$ in the effective potential for the Higgs field $H$.
It is a free parameter which cannot be predicted.
The question arises if this situation extends to quantum gravity.

\indent There is actually no problem of (technical) naturalness associated to the small ratio $\varepsilon_\varphi$ (or a similar small ratio in grand unified theories --- the gauge hierarchy \cite{GIL, SWGH}).
The value $\varphi_0 = 0$ or $\varepsilon_\varphi = 0$ is associated to a symmetry, namely scale symmetry or dilatation symmetry \cite{FTP}, see also ref~\cite{WBAR, HEM, PSW, MENI, MENI2, FKV, SHZEN, CWWTL, SHSH, KSZ, BOLI}.
This "particle physics scale symmetry" is linked to the critical behaviour of a quantum second order phase transition.
In the absence of strong interaction effects the vacuum (zero temperature) electroweak phase transition is a quantum phase transition associated to the variation of the mass parameter $\mu_H^2$.
It is a second order phase transition.
As for any second order phase transition there exists a critical surface in the space of running couplings for which the flow ends in the infrared at criticality, \eg $\mu_H^2 = 0$.
The critical surface cannot be crossed by any trajectory of running couplings.
If the couplings are precisely on the critical surface in the ultraviolet, they remain on the critical surface for the whole flow to the infrared.
A relevant parameter is associated to the deviation from the critical surface, resulting in nonzero $\mu_H^2$.
Problems of ``fine tuning'' in certain perturbative expansions only reflect that such expansions do not take into account properly this basic property of a second order phase transition \cite{FTP}.
From the point of view of physics at the Planck scale the question of why $\varepsilon_\varphi$ is tiny amounts to the question why our world is so close to the second order quantum phase transition.

\indent Functional renormalization (FRG) based on the effective average action \cite{EEE, REUW, TETW, RGREV, ELL, MOR} is a suitable tool for a quantitative investigation of these issues.
Already the first FRG computation of the effective potential for the Higgs field in the standard model \cite{CWQR} has clearly established the approximate second order quantum phase transition and the flow of the relevant parameter for a deviation from the critical surface.
This basic feature persists in the subsequent development of rather powerful and elaborate techniques \cite{GGS, GISO, KRALA1, KRALA2, EGJP, GISO2, REG, SON, HESO, PASPA, GISZ, GGL}.

\indent For the present purpose it is a crucial advantage of FRG that it can deal with quantum gravity formulated as a quantum field theory for the metric (or similarly for the vierbein).
Quantum gravity based on the Einstein--Hilbert action or a suitable generalization is not perturbatively renormalizable.
Quantum gravity can, however, be asymptotically safe \cite{WEI, REU} or asymptotically free \cite{STE, FRT, AVBAR, SWY}.
The FRG approach allows one to cope with the non-perturbative issues for both possibilities.
In this paper we address the issue of the Fermi scale within functional renormalization for the standard model of particle physics coupled to gravity.
The main new point beyond the FRG investigations of the effective potential for the Higgs scalar in refs.~\cite{SHAW, EHLY, PRWY, CWQS, CWMY, CWEP} is the focus on the scaling solution which is needed in order to define an ultraviolet complete theory.

\indent In particular, we ask if a Fermi scale compatible with observations can be realized for fundamental scale invariance \cite{CWFSI} for which no relevant parameters are available.
This requires a new asymptotically safe fixed point of quantum gravity which differs from the extended Reuter fixed point by a field-dependent effective Planck mass, a non-flat scalar potential and field dependent couplings.
This fixed point generalises the dilaton-gravity fixed point \cite{DQG1, DQG2, A2last} to the inclusion of the particles of the standard model.
For the scaling solution characterizing this fixed point both the Planck mass and the Fermi scale are proportional to the scalar fields. 
Their ratio is given by a dimensionless cosmon-Higgs coupling $\lambda_m$. 
For fundamental scale invariance there is no flow away from the fixed point.
There exists therefore no relevant parameter for the Fermi scale. 
This situation remains essentially valid if the largest intrinsic mass scale for a possible flow away from the scaling solution is much smaller than the Fermi scale.

\indent The central question of this paper asks if quantum gravity can predict the ratio Fermi scale over Planck mass in a setting where the Fermi scale is not determined by a relevant parameter.
This amounts to the question if quantum gravity can predict the value of the cosmon-Higgs coupling.

\section{Overview}

\indent Within FRG, an ultraviolet complete theory needs the existence of a scaling solution for the effective average action.
This scaling solution generalizes the notion of an ultraviolet fixed point for a finite number of flowing couplings.
One has now to deal with functions of fields or momenta.
For our purpose we can focus on quantities at zero or small momentum.
The central function we need is the effective scalar potential which determines the Fermi scale as a (partial) minimum with respect to variations of the Higgs field.
In addition to the Higgs field $H$ we consider a singlet scalar field $\chi$, needed for cosmology for the epoch of inflation or for dynamical dark energy.
The "cosmon" $\chi$ plays the role of the dilaton for the scale invariant standard model. We assume a discrete symmetry $\chi\xrightarrow{}-\chi$. 
In our setting the effective potential $U(H, \chi; k)$ will depend on the Higgs field $H$ through the invariant $H^\dagger H$, and on the scalar singlet field $\chi$ through the invariant $\rho = \chi^2 / 2$.

\indent The dependence of $U$ on the renormalization scale $k$ obeys a functional flow equation which obtains as an approximation to the exact flow equation for the effective average action \cite{EEE}.
For the scaling solution the dimensionless potential $u = U/k^4$ only depends on the dimensionless fields $\tilde{h} = H^\dagger H / k^2$ and $\tilde{\rho} = \rho / k^2$, without any explicit dependence on $k$.
This permits the extension of the theory to arbitrary short distances, as encoded in the ultraviolet limit $k \to \infty$.
The existence of a scaling solution is not trivial since $u$ has to exist for all values of fields, $0 \le \tilde{h} < \infty$, $0 \le \tilde{\rho} < \infty$.
We will see that the scaling solution has to obey a differential equation for its dependence on $\tilde{\rho}$ or $\tilde{h}$ at fixed $h = \tilde{h} / \tilde{\rho}$.
While local solutions for a finite range of fields are found without major problems, the extension to a global solution in the limits $\tilde{h} \to 0$, $\tilde{\rho} \to 0$ or $\tilde{h} \to \infty$, $\tilde{\rho} \to \infty$ typically poses constraints.
These constraints make quantum gravity predictive, a first example being the prediction of the mass of the Higgs boson \cite{SHAW} before discovery.

\indent The issue of the Fermi scale depends on the role of the Planck mass.
For a first possibility, $M_p$ corresponds to a relevant parameter characterizing the flow away from the scaling solution.
The Fermi scale appears then typically as a second relevant parameter, and one has to understand the tiny ratio $\varepsilon_\varphi$ as a ratio between two relevant parameters.
In this case a rather simple candidate for the scaling solution is a completely flat scalar potential, and a generalization to an effective action which is invariant under shifts $H \to H + c$.
Gauge and Yukawa interactions have to vanish for this type of scaling solution, and there is no non-minimal coupling of the type $H^\dagger H R$, with $R$ the curvature scalar.
Such a scaling solution is the generalization of the Reuter fixed point \cite{REU} in the presence of matter \cite{DP, WSOU, LAUREU, SAUREU, NP, PV, DELP, EP, LPF}.
In this scenario the field-dependence of the effective potential for the Higgs scalar is generated completely by the flow away from the scaling solution due to relevant (including marginal) couplings. The quartic scalar coupling, and therefore the mass of the Higgs boson, remains predictable for a given value of the top-quark Yukawa coupling.
This is due to the fact that the metric fluctuations render quartic scalar couplings irrelevant in the ultraviolet region.
In contrast, the Fermi scale could become predictable only if in the ultraviolet region a phenomenon of ``self organized criticality'' renders the mass parameter irrelevant \cite{CWMYGH, BOW, GIS, GIRS, CWMY, GIPI}.
Otherwise it remains as a free parameter.

\indent As a second alternative, the observed Planck mass is not associated to a relevant parameter, but rather given by the present value of the scalar field $\chi$.
This requires a different type of scaling solution without the shift symmetry $\chi\xrightarrow{}\chi+c$. 
The effective potential for this scaling solution is no longer completely flat.
For this setting the effective action contains a term $- \xi \chi^2 R/ 2$ which dominates the gravitational interactions for the present cosmological epoch.
The scaling solution is compatible with a term $-f(\chi^2/k^2)k^2 R/2$, which includes the non-minimal coupling for $f = \xi \chi^2 / k^2$.
If no relevant parameter is responsible for the Planck mass, the issue of the Fermi scale changes profoundly.
The largest intrinsic mass scale (LIMS) generated by the flow away from the fixed point can be much smaller than the Planck mass.
Such models can solve the cosmological constant problem dynamically, similar to the first proposal for dynamical dark energy or quintessence \cite{CWQ}.
If the effective potential approaches a constant for $\chi^2 \to \infty$ (or increases less fast than $\sim \chi^4$), and the effective Planck mass increases $M_p \sim \chi$, the observable ratio $U/M_p^4$ decreases to zero for $\chi \to \infty$.
For such models the characteristic cosmological solutions are ``runaway cosmologies'' for  which $\chi$ increases towards infinity as time increases towards the infinite future.
Thus the ratio $U/M_p^4$, which corresponds to the observable cosmological ``constant'', or the cosmological ``constant'' in the Einstein frame, vanishes in the infinite future.
An observable dark energy component reflects the fact that for the finite age of the Universe the ratio $U/M_p^4$ still differs from zero.
If the LIMS is of the order of the neutrino masses, say $\sim 10^{-3}\;\mathrm{eV}$, a rather interesting cosmology with dynamical dark energy emerges \cite{CWIQ, CWQGSS, DEQG}.

\indent Cosmology provides a strong motivation to investigate the possibility of a field-dependent Planck mass more in detail.
The scaling solution for this scenario has to be different from the Reuter fixed point or its generalization to matter with a shift-invariant potential.
Since the term $\sim \chi^2 R$ plays a crucial role for the scaling solution, this scaling solution cannot be shift-invariant under $\chi \to \chi + c$.
As a consequence, for $\xi \neq 0$ the scaling solution for the effective scalar potential cannot be shift invariant either \cite{CWEP}.
The scaling solution for the standard model coupled to gravity has rather to be a generalization of the scaling solution for dilaton quantum gravity \cite{DQG1, DQG2, A2last}.

\indent One may ask if in such a scenario the Fermi scale $\varphi_0$ could be associated to a relevant parameter for the flow away from the scaling solution, thus constituting the largest intrinsic mass scale (LIMS).
If the Fermi scale is intrinsic and the Planck mass field dependent, the ratio $\varepsilon_\varphi$ would be field-dependent.
This could realize the ``Dirac hypothesis'' of explaining the small ratio of electron mass over Planck mass by the huge age of the universe, since $\chi$ can have grown to very large values in units of the electron mass.
Observation seems to indicate that this possibility may not be realized.
For the runaway cosmologies envisaged here there are strong bounds on the time variation of the ratio Fermi scale over Planck scale \cite{CWTVC, primeiro, segundo, terceiro, quarto, quinto}.
A possible way out would need a cosmological scenario where the decrease of $\varepsilon_\varphi$ would be substantial in the very early universe, and essentially stop for the present cosmological epoch.

\indent A more natural alternative seems to be that the Fermi scale is not an intrinsic mass scale.
In this case the mass parameter in the Higgs potential is field-dependent, associated to a quartic scalar coupling $\sim \lambda_m H^\dagger H\chi^2$ between the Higgs field $H$ and the cosmon $\chi$.
This is compatible with a scaling solution for the effective potential for which the term quadratic in $H$ takes the form $u \sim \lambda_m \tilde{h} \tilde{\rho}$.
If $\chi$ increases during the cosmological evolution, both $M_p$ and $\varphi_0$ increase at the same rate $\sim\chi$.
The ratio $\varepsilon_\varphi$ remains constant, in accordance with observation.
With particle masses proportional to $\chi$ this extends to other mass ratios.

\indent If the largest intrinsic mass scale (LIMS) associated to a relevant parameter is much smaller than the confinement scale $\Lambda_{\text{QCD}}$ of quantum chromodynamics (QCD) quantum gravity realizes the scale invariant standard model \cite{CWQ, SHAZEN1, SHAZEN2, SHT, BSZ, KASH, GBRSZ, FHR, CPRU, FHNR, RUB, CKPR}.
For the scaling solution all mass scales (except possibly the neutrino masses) are proportional to $\chi$.
With constant $\varepsilon_\varphi$ the Fermi scale is proportional to $\chi$, $\varphi_0 = \sqrt{\varepsilon_\varphi} \chi$.
The dimensionless gauge and Yukawa couplings can depend on the ratios $H/\chi$, or $k/\chi$ but not on $H$, $\chi$ or $k$ separately.
(For non-zero momenta $q$ they can also depend on $q^2/ H^\dagger H$ or $q^2/\chi^2$.)
In consequence, the confinement scale $\Lambda_\QCD$ is proportional to $\chi$, such that the ratio of nucleon mass over Planck mass is independent of $\chi$.
The assumption that the LIMS is much smaller than $\Lambda_{\text{QCD}}$ entails that the standard model is determined precisely by the scaling solution.
This setting \textit{predicts} the scale invariant standard model.

\indent The scale invariant standard model is compatible with the strong observational bounds on the time variation of couplings even for cosmologies where $\chi$ changes with time.
Due to this simple consequence of quantum scale symmetry \cite{CWQS} no screening mechanism is needed for dynamical dark energy.
This extends to possible apparent deviations from general relativity due to forces induced by an almost massless scalar field, which take the form of violations of the equivalence principle.
Quantum scale symmetry implies that all those effects are tiny.
The scale invariance of the quantum effective action, rather than scale invariance for the classical action \cite{FUJ}, is crucial for these properties. 

\indent The scaling solution for a scenario where the LIMS is much smaller than the Fermi scale cannot be shift invariant for the Higgs field, $H \to H+c$.
The scaling form of the effective potential $u(\tilde{h}, \tilde{\rho})$ necessarily shows a non-trivial dependence on $\tilde{h}$.
A term linear in $\tilde{h}$, $U=\mu_h \tilde{h}$, results in a mass term for the Higgs field $U = \mu_h k^2 H^\dagger H$ which vanishes for $k \to 0$.
A non-zero LIMS may replace $k^2$ by an intrinsic scale $\bar{k}_0^2$.
Still, a mass term $\sim \bar{k}_0^2 H^\dagger H$ is negligible if the LIMS is many orders of magnitude smaller than the Fermi scale.
A term quadratic in $\tilde{h}$ in the scale invariant potential, $u=\lambda_h \tilde{h}^2 / 2$, results in a non-zero quartic coupling, $U = \lambda_h (H^\dagger H)^2 / 2$.
A field-dependent mass term for the Higgs scalar is generated by the cosmon-Higgs coupling $u = \lambda_m \tilde{h} \tilde{\rho}$ or $U = \lambda_m H^\dagger H \chi^2 / 2$.
For a realistic Fermi scale the coupling $\lambda_m$ has to be tiny, since $\varepsilon_\varphi \sim \lambda_m$.

\indent The present paper computes the scaling potential $u(\tilde{h}, \tilde{\rho})$ in an approximation to the exact flow equation for the effective potential \cite{EEE}.
A key question will be the size of $\lambda_m$.
We find a family of local scaling solutions which are parametrized by an ``integration constant'' or ``boundary value'' $\bar{\lambda}_m$, which corresponds to $\lambda_m (\tilde{\rho} \rightarrow \infty)$.
Here ``local'' refers to a region in field space, with $\tilde{\rho}$ sufficiently large so that quantum gravity effects are small, and $\tilde{h} \ll \tilde{\rho}$.
For these local solutions $\bar{\lambda}_m$ is a free parameter.
If these local solutions can be extended to global solutions valid for arbitrary $\tilde{\rho}$ and $\tilde{h}$, the Fermi scale $\varphi_0$ cannot be predicted.
For this case it would not be associated to a relevant parameter for the flow away from the scaling solution.
Rather it would be related to the existence of a continuous family of scaling solutions.

\indent On the other hand, if the extension of the local solution to a global solution is possible only for a certain value of $\bar{\lambda}_m$ (or a discrete set), our scenario can compute the Fermi scale, or more precisely the ratio of Fermi scale over Planck mass.
In this case observation can falsify (of confirm) a scenario of quantum gravity with a largest intrinsic mass scale much smaller than the Fermi scale.
We will find that this scenario is realized for an analytic global scaling solution. 
The Fermi scale becomes predictable in principle if the LIMS is much smaller than $\Lambda_{\mathrm{QCD}}$.

\indent The predictivity for the ratio of the Fermi scale over the Planck mass does not depend on details of the computation or additional assumptions. 
The logic leading to predictivity may be summarized as follows
\begin{widetext}
\begin{center}
\textbf{Chain of predictivity for the ratio of the Fermi scale over the
Planck mass}
\end{center}

\begin{center}
\begin{tabular}{l@{\qquad}c@{\qquad}l}
local scaling solution
&
:
&
has free parameters as integration constants
\\[0.5em]
\multicolumn{1}{c}{$\Downarrow$}
& &
\\[-0.2em]
global scaling solution
&
:
&
fixes integration constants
\\[0.5em]
\multicolumn{1}{c}{$\Downarrow$}
& &
\\[-0.2em]
analyticity of the global scaling solution
&
:
&
no free parameters in the scaling solution
\\[0.5em]
\multicolumn{1}{c}{$\Downarrow$}
& &
\\[-0.2em]
$\mathrm{LIMS}\ll\Lambda_{\mathrm{QCD}}$
&
:
&
no relevant parameter for the Fermi scale
\\[0.5em]
& $\Downarrow$ &
\\[-0.2em]
\multicolumn{3}{c}{no free parameter for the Fermi scale:
predictivity}
\end{tabular}
\end{center}
\end{widetext}

\indent More in detail, the scaling solution obeys a differential equation for the $\tilde{\rho}$-dependence of $u$ which is rather similar to the flow equation for the dependence on $k$.
The flow of couplings with $k$ at fixed $\rho$ corresponds for the scaling solution to a flow with $\tilde{\rho} = \rho / k^2$.
For fixed $\rho$ the relevant parameters reflect for the flow with decreasing $k$ a dependence of the corresponding couplings on the initial conditions at large $k$.
For the local scaling solution this maps to a dependence of field-dependent couplings on ``initial values'' for some fixed small $\tilde{\rho}$.
One finds a trace of the relevant parameters in the existence of continuous families of local scaling solutions

\indent There are, however, two new ingredients which are specific for the scaling solution.
The first involves the necessity of a global solution which imposes restrictions from the existence of the solution for $\tilde{\rho} \to 0$ or $\tilde{\rho} \to \infty$.
As well known for non-linear differential equations the existence of a global solution imposes constraints on the "extendable" local solutions.
The second is related to the fact that for a scaling solution the dimensionless couplings become functions of the field.
An example is a field-dependent coefficient of the kinetic term for the gauge bosons $Z_F(\tilde{\rho}, \tilde{h}) F^{\mu\nu} F_{\mu\nu}$, which translates to a field-dependent gauge coupling by $g^2 = Z_F^{-1}$.
For a field-dependent $Z_F$ one has new vertices for the interactions between gauge bosons and the scalar fields $H$ and $\chi$.
Typically this field-dependence becomes negligible in the range of large $\tilde{h}$ corresponding to $k$ much smaller than $\Lambda_\QCD$.
Nevertheless, it affects the scaling solution in the range of smaller $\tilde{h}$.
For our purpose we will infer the logarithmic field-dependence of the gauge and Yukawa couplings from perturbation theory.
For $\tilde{\rho}$ large enough, such that the gravitational fluctuations decouple, the perturbative flow with $k$ translates then to the flow with $\tilde{\rho}$.
This will permit a reasonable control of the local scaling solutions.

\indent We find that generically the requirement of a global scaling solution for all values of $\tilde{\rho}$ and $\tilde{h}$ conflicts with the presence of a continuous family of scaling solutions. There exists, at best, a discrete set of global scaling solutions.
Therefore an ultraviolet complete model for the particles of the standard model coupled to quantum gravity renders the ratio $\varphi_0/M_p$ predictable if the LIMS is sufficiently below the Fermi scale.
This generalizes to extensions of the standard model as, for example, grand unified theories.
The predicted value of $\varphi_0/M_p$ may depend, however, on the details of the model for the physics around and beyond the Planck scale.
While "predictivity in principle" seems to be generic, "predictivity in practice" requires an assumption about the particles relevant near the Planck scale,as well as sufficient computational control of the scaling solution in this energy range.

\indent A key ingredient for the elimination of free parameters for the global scaling solution is analyticity for $\tilde{\rho}\to0$. 
As an example, consider the $\tilde{\rho}$-dependence of the mass term $\mu_h (\tilde \rho)$ for the Higgs boson. 
The cosmon--Higgs coupling $\lambda_m(\tilde{\rho})$ corresponds to the $\tilde{\rho}$-derivative of $\mu_h$,
\begin{equation}
\lambda_m(\tilde{\rho})
=
\partial_{\tilde{\rho}}\,\mu_h(\tilde{\rho})\,.
\label{eq:AL1}
\end{equation}
For a global scaling solution we require that $\lambda_m$ remains finite for $\tilde{\rho}\to0$. 
This implies that the first derivative $\partial_{\tilde{\rho}}\mu_h$ remains finite for $\tilde{\rho}\to0$. 
This continues to higher derivatives, since $\partial_{\tilde{\rho}}\lambda_m(\tilde{\rho})$ is again a coupling which should remain finite for $\tilde{\rho}\to0$. 
Thus, the finiteness of couplings in the ultraviolet limit $\tilde{\rho}\to0$ is equivalent to the analyticity of the function $\mu_h(\tilde{\rho})$ for $\tilde{\rho}\to0$.

\indent The qualitative implications can be understood in a simplified setting. 
Near a UV fixed point $\mu_h^*=\mu_h(0)$, the $\tilde{\rho}$-dependence of $\mu_h(\tilde{\rho})$ has to obey a differential ``scaling equation'' of the form
\begin{equation}
\tilde{\rho}\,
\partial_{\tilde{\rho}}\,\mu_h
=
\frac{1}{2}(2-A)
\left(\mu_h-\mu_h^*\right)
+
d\tilde{\rho}
+\cdots\,,
\label{eq:AL2}
\end{equation}
where $(2-A)$ reflects the dimension two and $A$ is the anomalous dimension.
This is the analogue of the flow equation for the dependence on $k$.
The general solution for $\tilde{\rho}\to 0$ involves a free integration constant $c$,
\begin{equation}
\mu_h(\tilde{\rho})
=
\mu_h^*
+
c\tilde{\rho}^{\,1-\frac{A}{2}}
+
\frac{2d}{A}\tilde{\rho}
+\cdots\,.
\label{eq:AL3}
\end{equation}
For $d=0$ and $A<2$, the function $\mu_h(\tilde{\rho})$ departs from
the fixed-point value as $\tilde{\rho}$ increases -- this is the analogue of a relevant parameter.

\indent Analyticity requires $c=0$ whenever $A$ differs from an even integer, which is the case in general. 
For $c\neq0$, the cosmon-Higgs coupling diverges for $\tilde{\rho}\to0$ if $A>0$
\begin{equation}
\lambda_m(\tilde{\rho})
=
c
\left(1-\frac{A}{2}\right)
\tilde{\rho}^{-\frac{A}{2}}
+
\frac{2d}{A}
+\cdots\,.
\label{eq:AL4}
\end{equation}
For $A<0$, the cosmon-Higgs coupling remains finite, but $\partial_{\tilde{\rho}}\lambda_m$ or some higher coupling diverges.
In consequence, one has no free integration constant $c$ independently of the sign of $A$.
This general feature carries over to the coupled system for several (or many) coupling functions as we will discuss in detail later. 
This discussion also covers the effective appearance of the term $d\tilde{\rho}$ in \eqref{eq:AL2}, which may lead to a nonzero
fixed-point value $\lambda_{m*} = \lambda_m(\tilde{\rho}=0)$.

\indent Having established general predictivity for the ratio $\varphi_0/M_p$ the question arises what a given model actually predicts.
We will find that the prediction yields a very small value for $\epsilon_\varphi$ --- how small may depend on the details of the model around and beyond the Planck scale. The small predicted value of $\epsilon_\varphi$ is closely linked to the presence of an (approximate) second order quantum electroweak phase transition.
For $A>0$ the trajectories of dimensionless couplings as functions of $\tilde{\rho}$ are attracted towards the critical surface of this phase transition for increasing $\tilde{\rho}$.
If the critical surface is compatible with analyticity the prediction amounts to a Fermi scale on the critical surface. 
Then the scaling solution of quantum gravity realizes a new form of self-organized criticality.

\indent For a LIMS sufficiently below the Fermi scale the other couplings of the standard model become predictable as well.
The scaling solution admits no free parameters for the gauge and Yukawa couplings.
This renders our scenario very restrictive.
A given short-distance model can be falsified by many observations.
It seems very unlikely that the pure standard model coupled to quantum gravity can survive these cosntraints.
Additional particles with mass in the vicinity of the Planck mass, as for grand unified theories, are likely to be needed.
In this sense the assumption of the LIMS below the Fermi scale becomes a strong constraint on viable short distance modelds.
The benefit is a very high degree of predictivity.

\indent Since the issue of the scaling solution for the effective potential $u(\tilde{\rho}, \tilde{h})$ is rather complex, we proceed by a stepwise extension of our truncation.
This will reveal crucial features one by one.
This will also help to disentangle which features depend on the truncation, and which are robust with respect to changes of truncation and the precise setting of the flow equation. 
The general feature of predictivity of $\varphi_0/M_p$ is truncation independent, involving only analyticity of the scaling solution and the LIMS being much smaller than the Fermi scale.
The presence of a second order quantum electroweak phase transition and the associated critical surface is truncation independent. 
Only QCD effects turn this transition into a crossover.
The attraction towards the critical surface involves only the sign of $A$. 
It is robust to the extent that this sign is robust. 
In contrast, the precise value of the prediction for $\varphi_0/M_p$ may be affected by the details of the truncation and other approximations. 
An assertion of the strength of this dependence will need further systematic investigations.
In any case, the prediction typically depends on physics near and beyond the Planck scale.

\indent In sect.~\ref{sec:II} we present the flow equation for $k\partial_k u (\tilde{\rho},\tilde{h},k)$ and the corresponding scaling equation for the scaling potential $u(\tilde{\rho},\tilde{h})$.
In our truncation the flow equation has a very simple structure.
Nevertheless, it covers all relevant features for our discussion. More extended truncations are not expected to change the general outcome. In sec.~\ref{sec:III} we discuss the local scaling solution in the limit of field-independent gauge and Yukawa couplings.
In this limit the standard model exhibits a second order quantum phase transition. We also employ a simple approximation for the fluctuations of the scalar field.
The family of local scaling solutions is then parametrized by a cosmon-Higgs coupling $\lambda_m$ which is field independent. Sect.~\ref{sec:IV} discusses the infrared limit of very large $\tilde{\rho}$. This is is needed for a continuation of the scaling solution to $\tilde{\rho}\rightarrow \infty$.
This limit is also relevant for the question of time varying fundamental couplings for cosmologies where $\tilde{\rho}$ increases with time.
We find that these effects are tiny, such that the class of models discussed here does not enter into conflict with observation from this side.

\indent Sect.~\ref{sec:V} addresses the field-dependence of gauge and Yukawa couplings which is predicted for the scaling solution in our setting. This field dependence renders the properties of the critical surface for the electroweak transition more complex.
It also implies that the transition becomes a crossover rather than a second order transition. The characteristic scale for the crossover turns out, however, to be three orders of magnitude below the observed Fermi scale. For realistic models a second order phase transition remains a very good approximation.

\indent In sect.~\ref{sec:VI} we include the effects of the metric fluctuations on the scaling equations for the gauge and Yukawa couplings. 
For the ultraviolet limit we are open to possible extensions of the standard model by particles with masses in the vicinity of the Planck mass. We discuss possible fixed points with non-zero values for the gauge and Yukawa couplings. 
This issue is important for understanding initial values for the flow and field-dependence of gauge and Yukawa couplings for $k$ smaller than the Planck mass. 
On this basis we discuss in sect.~\ref{sec:VII} the local scaling solution for $u(\tilde{\rho},\tilde{h})$ in the presence of field-dependent gauge and Yukawa couplings.
The main emphasis concerns here the dependence on the Higgs field in the region of small values of the ratio $h=\tilde{h}/\tilde{\rho}$.

\indent In sect.~\ref{sec:VIII} we turn to the flow or field-dependence of the cosmon-Higgs coupling $\lambda_m$.
This coupling is directly related to the ratio $\varphi_0/M_p$, constituting a key quantity for our investigation.
Since $\lambda_m$ is dimensionless, the question if it is relevant or irrelevant differs strongly from a similar question for the mass term of the Higgs boson.
We find that even for $k$ below $M_p$ the cosmon-Higgs coupling has the features of an irrelevant parameter.
This property becomes much more pronounced once we include in sect.~\ref{sec:IX} the effects of the metric fluctuations for $k$ near or above $M_p$.
We discuss a new form of self-organized criticality which renders the ratio $\varphi_0/M_p$ predictable if the flow away from the scaling solution plays no role.
For the scaling solution all couplings have to assume for $\tilde{\rho}\xrightarrow{}0$ precisely their fixed point values.
This holds both for irrelevant and relevant parameters.
There are no free parameters for the flow away from the fixed point.
The crossover from the UV-fixed point for $\tilde{\rho}\xrightarrow{}0$ for the IR fixed point for $\tilde{\rho}\xrightarrow{}\infty$ is uniquely due to the decoupling of the gravitational fluctuations once $\tilde{\rho}$ grows sufficiently larger than one.

\indent Sect.~\ref{sec:X} complements out analytic discussion by a numerical solution for a restriction of the flow to a finite number of couplings.
One can see explicitly all features of the analytic discussion.
The numerical solution confirms the predictivity for the Fermi scale.
For reasonable assumptions without much tuning of the UV-properties we find a scaling solution with the observed tiny value of $\varphi_0/M_p$.
In sect.~\ref{sec:XI} we summarise our results.

\indent For convenience we summarize our conventions for fields and the scalar potential in Appendix~\ref{app:notations}. 
Appendix~\ref{app:A} provides for a detailed and coherent exposition of the modifications due to strong interactions. 
Since the confinement scale $\Lambda_{\mathrm{QCD}}$ is three orders of magnitude below the Fermi scale these effects turn out to be only smallcorrections. 
Further appendices are devoted to more technical details.

\section{Flow equation and scaling equation for the effective potential}
\label{sec:II}

In this section we focus on the functional flow equation for the effective potential for a region of field variables and renormalization scale for which the effects of the gravitational degrees of freedom can (almost) be neglected.
This concerns values of $k$ sufficiently below the Planck mass, where the latter may be given by the scalar field $\chi$.
The flow equation for this region is rather well established.
It can be based directly on the exact flow equation of ref.~\cite{EEE}, with extensions to gauge theories \cite{REUW} and fermions \cite{ELLW, BOW2, CWQR}.
We employ here the gauge invariant flow equation \cite{CWGIF} in the form of the simplified flow equation \cite{SFE}.
The flow equation is one-loop exact.
The dominant effects beyond one-loop perturbation theory are the threshold functions which describe the decoupling of particles with masses larger than $k$.
Higher loop perturbative effects relevant for our question are accounted for qualitatively.
Two-loop exactness would require, however, an extension of the approximation (truncation) with additional resolution for the kinetic terms \cite{PAPW}. 
The quantum gravity effects will be included in sect.~\ref{sec:IX}.
This extends the range to $k>\chi$.

\subsection*{Effective scalar potential}
\label{sec:ESP}
We consider the standard model with an additional singlet scalar field $\chi$.
We employ a truncation of the effective average action
\begin{equation}
    \begin{aligned}
    \label{eq:ESP1}
	\Gamma_k = \int_x \sqrt{g'} \bigg\{ &-\frac{F}{2} R' + U + \frac{K}{2} D^\mu \chi D_\mu \chi \\
					    & + Z_H D^\mu H^\dagger D_\mu H + \dots \bigg\}\,.
    \end{aligned}
\end{equation}
Here $H$ denotes the Higgs doublet and the dots stand for the pieces involving the fermions and gauge bosons of the standard model.
The effective scalar potential $U$ depends on $H$ and $\chi$, as well as on the ``renormalization scale'' $k$, and similar for $F$, $K$ and $Z_H$.
Here the covariant derivatives $D_\mu$ are formed with the metric $g_{\mu\nu}'$, where the prime recalls that we are typically not in the Einstein frame.
Due to the gauge symmetry of the standard model the functions $U$, $F$, $K$ and $Z_H$ depend on $H$ only via the invariant $H^\dagger H$.
For simplicity we first neglect the anomalous dimension of the Higgs scalar by setting $Z_H = 1$.
The renormalized Higgs field which includes effects of the anomalous dimension will be discussed at the end of this section.

\indent We are interested in the range of $\chi$ much larger than the Fermi scale or the renormalization scale $k$.
In this range we approximate the scaling solution by 
\begin{equation}
    \label{eq:A1}
    F = \xi \chi^2 + \xi_H H^\dagger H + f_\infty k^2\,.
\end{equation}
The effective potential in the Einstein frame reads
\begin{equation}
    \label{eq:A2}
    U_E = \frac{M^4}{F^2}U\,.
\end{equation}
The fixed Planck mass $M$ is introduced only by the variable transformation (Weyl scaling)
\begin{equation}
    \label{eq:3A}
    g_{\mu\nu} = \frac{F}{M^2}g'_{\mu\nu}\,.
\end{equation}
It is not a parameter of our model.
The Fermi scale $\varphi_0$ is given by the value of $H$ at the partial minimum of $U_E$ with respect to variation of $H$
\begin{equation}
    \label{eq:A3}
\frac{\partial U_E}{\partial H} (H_0) = 0\,,\quad
\varphi_0 = |{H_0}|\,.
\end{equation}
It may depend on $\chi$ and $k$.
The effective action $\Gamma_k$ and the associated effective potential $U$ only include quantum fluctuations with (Euclidean) momentum larger than $k$.
The observable Fermi scale should therefore be evaluated at $k\ll \varphi_0$.

\indent We may expand $F$ in the small quantities $H^\dagger H / \chi^2$ and $k^2/\chi^2$, leading to
\begin{equation}
    \label{eq:A4}
    U_E = \frac{M^4 U}{\xi^2 \chi^4} \left(1-\frac{2\xi_H H^\dagger H}{\xi \chi^2} - \frac{2 f_\infty k^2}{\xi \chi^2}\right)\,.
\end{equation}
A non-zero Fermi scale $\varphi_0$ indicates spontaneous symmetry breaking of the $SU(2)\times U(1)$ gauge symmetry, requiring $\partial U_E / \partial (H^\dagger H) = 0$ at $H^\dagger H=\varphi_0^2$ and $k=0$.
This condition reads
\begin{equation}
\label{eq:A5}
    \frac{\partial U}{\partial H^\dagger H} \left(\chi,H^\dagger H =\varphi_0^2\right) = \frac{2\xi_H}{\xi\chi^2} U\left(\chi,H^\dagger H=\varphi_0^2\right)\,.
\end{equation}
For the region of interest of this note the \rhs of eq.~\eqref{eq:A5} is tiny as compared to $\varphi_0^2$ and will be neglected.

\indent One may be interested in cosmologies for which $\chi$ varies in time.
We choose units such that the present value $\chi_0$ is related to the observed Planck mass, $\xi \chi_0^2 = M^2$.
The present value of $U_E$ coincides then with $U$.
We may associate $U(\chi,H^\dagger H =\varphi_0^2)$ with a type of $\chi$- and $k$-dependent ``cosmological constant''.
Observation tells us that the time-variation of the ratio Fermi scale over Planck mass must be tiny in the recent cosmological epoch.
If $\chi$ varies with time, and $\xi$ can be taken constant, a realistic cosmology requires $\varphi_0^2 = \eps_\varphi \chi^2$ with constant $\eps_\varphi$.
The value of $\xi$ depends on the normalization of the cosmon field $\chi$.
A useful normalization is specified by $\xi = 1$, such that $\chi$ corresponds to the Planck mass for the region $\chi/k \gg 1$.
For this choice the kinetic term of $\chi$ does not have the standard normalization, with $K_\infty = K(\chi \to \infty)$ a free parameter.

\indent Experiments measure the Fermi scale in the Einstein frame $\varphi_{0E}$, or more precisely the ratio $\varphi_{0E}/M$.
Translating to the frame with varying $F$ yields
\begin{equation}
	\label{eq:A6}
	\frac{\varphi_{0E}^2}{M^2} = \frac{\varphi_0^2}{F} = \frac{\varphi_0^2}{\xi \chi^2}\,.
\end{equation}
We will therefore be interested in the ratio
\begin{equation}
\label{eq:A7}
h_0 = \frac{2\varphi_0^2}{\chi^2} = 2\xi \frac{\varphi_{0E}^2}{M^2}\,.
\end{equation}
The quartic scalar coupling $\lambda_h$, which determines the mass of the Higgs boson by
\begin{equation}
\label{eq:A8}
\frac{m_H^2}{M^2} = \frac{2\lambda_h\varphi_0^2}{\xi\chi^2} = \frac{\lambda_{h,0}}{\xi}h_0\,,
\end{equation}
is defined as 
\begin{equation}
	\label{eq:A9}
	\lambda_{h,0} = \frac{\partial^2 U}{\partial(H^\dagger H)^2} (\chi,H^\dagger H =\varphi^2_0)\,.
\end{equation}

\subsection*{Flow equation}
\label{sec:FESS}
The dependence of $U$ on $k$ at fixed $\chi$ and $H$ obeys an exact flow equation \cite{EEE}.
We focus first on scales $k$ sufficiently below $M_p = \sqrt{F}$ such that the metric fluctuations only contribute  a field-independent term. 
Our truncation first assumes field-independent $Z_H$ and $K$ (local potential approximation).
The simplified flow equation maintains gauge invariance, leading to a simple form of the contribution from gauge boson fluctuations.
With these approximations the flow of $U$ takes the simplified form
\begin{equation}
    \label{eq:AA1}
    k \partial_k U = \frac{\mathcal{N}k^4}{32\pi^2} = 4c_U k^4\,.
\end{equation}
Here $\mathcal{N}$ denotes the effective number of degrees of freedom, with an appropriate minus sign for fermions.
If all particles are massless, $\mathcal{N}$ is a constant and eq.~\eqref{eq:AA1} accounts for the contribution to the cosmological constant from fluctuations between $k$ and $k+\mathrm{d}k$.
The field-dependence of the effective potential arises from the field-dependence of $\mathcal{N}$ due to mass thresholds.
Only particles with mass smaller than $k$ contribute effectively to $\mathcal{N}$, while particles with mass much larger than $k$ decouple.
The mass thresholds arise naturally from functional renormalization.

\indent We introduce dimensionless functions of dimensionless fields
\begin{equation}
	\label{eq:FESS2}
	u=\frac{U}{k^4}\,,\quad
	\tilde\rho = \frac{\chi^2}{2 k^2}\,,\quad
	\tilde h = \frac{H^\dagger H}{k^2}\,,\quad
	h = \frac{\tilde h}{\tilde \rho} = \frac{2 H^\dagger H}{\chi^2}\,.
\end{equation}
Besides the dependence on $k$ the dimensionless potential $u$ depends on two field variables, that we may choose as $(\tilde\rho, h)$ or $(\tilde h, h)$ or $(\tilde\rho, \tilde h)$.
Different pairs of field variables will be optimal for different issues, and we will indicate the field variables explicitly when needed.

\indent If $u(\tilde\rho, h)$ has no partial minimum at $h=0$, i.e.\ $\partial_h u(\tilde \rho, 0) <0$, the partial minimum of $U$ occurs for $H^\dagger H > 0$.
In this case the $h$-derivative of $u(\tilde\rho, h)$ at fixed $\tilde\rho$ determines the Fermi scale by eq.~\eqref{eq:A7} and 
\begin{equation}
	\label{eq:FESSA2}
	\frac{\partial u}{\partial h} (\tilde\rho, h_0) = 0\,.
\end{equation}
The quartic Higgs scalar coupling \eqref{eq:A9} is given by
\begin{equation}
	\label{eq:FESSB2}
	\hat{\lambda}_h = \frac{\partial^2 u}{\partial h^2} = \lambda_h \tilde{\rho}^2\,,
\end{equation}
or
\begin{equation}
    \label{eq:AA2}
\lambda_{h,0} = \frac{1}{\tilde{\rho}^2} \partial_h^2 u(\tilde\rho, h_0)\,.
\end{equation}
This determines the mass of the Higgs boson by eq.~\eqref{eq:A8}.

\indent Our central equation is the flow equation for $u(\tilde\rho, h)$ which is evaluated for fixed $\tilde\rho$ and $h$,
\begin{equation}
	\label{eq:FESS3}
	k \partial_k u = -4u+2\tilde\rho \partial_{\tilde\rho} u + 4c_U\,.
\end{equation}
The term $-4u$ arises from $u\sim k^{-4}$, the term $2\tilde\rho\partial_{\tilde\rho} u$ reflects the switch from fixed $\chi$ to fixed $\tilde{\rho}$, and $c_U = k^{-3} \partial_k U$ accounts for the fluctuation contributions to the effective potential $U$ at fixed $\chi$ and $H$.
For the range of large $\tilde\rho \gg 1$, the scaling solution for $F$ can be approximated by
\begin{equation}
\label{eq:FESS4}
F = \xi \chi^2\,,
\end{equation}
with $\xi$ and $K$ constants.
For $\xi \tilde\rho \gg 1$ the metric fluctuations are effectively decoupled from the flow, being suppressed by inverse powers of $\xi \tilde\rho$.
An exception is a field independent contribution to $c_U$.
We recall that we neglect first the anomalous dimension of the Higgs doublet.

\indent In the range $\tilde\rho \gg 1$, $h < 1$ the simplified flow equation \cite{SFE} yields a simple expression for the mass thresholds and therefore for $c_U$ \cite{DEQG,PRWY,CWQGSS,CWEP},
\begin{equation}
	\label{eq:FESS5}
	c_U = \frac{1}{128\pi^2} \left( \bar{N}_S + 2\bar{N}_V - 2\bar{N}_F + 2\right)\,.
\end{equation}
Here $\bar{N}_S$, $\bar{N}_V$ and $\bar{N}_F$ are the effective numbers of scalars, gauge bosons and Weyl fermions, respectively.
They depend on the fields via the mass thresholds that reflect the decoupling of particles with masses larger than $k$.
The last term is the contribution of the two physical graviton degrees of freedom.
For the fermions we approximate the neutrinos as massless, while for the charged fermions the squared masses are
\begin{equation}
\label{eq:FESS6}
m_f^2 = y_f^2 H^\dagger H\,,
\end{equation}
where the Yukawa couplings $y_f$ can depend on $\tilde\rho$, $h$ and $k$.
One obtains
\begin{equation}
	\label{eq:FESS7}
	\begin{aligned}
		\bar{N}_F &= \sum_f \left(1+\frac{m_f^2}{k^2}\right)^{-1} + 3 \\
					&= \sum_f \left(1+y_f^2 h \tilde\rho\right)^{-1} + 3\,,
\end{aligned}
\end{equation}
with $\sum_f$ extending over quarks and charged leptons.

\indent For the gauge bosons one has
\begin{equation}
\label{eq:FESS8}
2\bar{N}_V = \frac{6}{1+\tilde{m}_W^2} + \frac{3}{1+\tilde{m}_Z^2} - 3 + 18\,,
\end{equation}
where
\begin{equation}
\label{eq:FESS9}
    \tilde{m}_W^2 = \frac{g_2^2 h \tilde\rho}{2}\,,\quad
    \tilde{m}_Z^2 = \left(\frac{g_2^2}{2} + \frac{3g_1^2}{10}\right) h \tilde\rho\,,
\end{equation}
with $g_2$ and $g_1$ the gauge couplings of the $SU(2)$ and $U(1)$ groups of the standard model.
The measure term subtracts one for $W^\pm$ and $Z$.
Eight massless gluons and the massless photon account for the last term +18. 
Finally, the effective number of scalars amounts to
\begin{equation}
\label{eq:FESS10}
    \bar{N}_s = \sum_S\left(1+\frac{m_s^2}{k^2}\right)^{-1}\,,
\end{equation}
with $m_s^2$ the eigenvalues of the mass matrix
\begin{equation}
    \label{eq:FESS11}
    M_{ij}^2 = \frac{\partial^2 U}{\partial \varphi_i \partial \varphi_j}\,,
\end{equation}
where $\varphi_i = (\chi,\tilde{H}_\gamma)$ with $\tilde{H}_\gamma$ the four real fields in $H$, \eg $H_1=\frac{1}{\sqrt{2}} \left(\tilde{H}_1 + i \tilde{H}_2 \right)$, $H_2 = \frac{1}{\sqrt{2}} \left(\tilde{H}_3 + i \tilde{H}_4 \right)$, $H^\dagger H = \frac{1}{2}\sum_{\gamma} \tilde{H}_\gamma^2$.
For the simplified flow equation the mass threshold function for scalars coincides with the standard approach using the Litim cutoff \cite{LIT}. 

\subsection*{Scaling equation for the scaling solution}
\indent The scaling solution obtains from the flow equation \eqref{eq:FESS3} by setting $k\partial_k u = 0$.
This results in a non-linear differential equation for $u(\tilde\rho,h)$,
\begin{equation}
    \label{eq:FESS12}
	\tilde\rho \partial_{\tilde\rho} u = 2\left(u - c_U\right)\,.
\end{equation}
In our approximation it has to be supplemented by an ansatz for the field dependence of the gauge and Yukawa couplings.
We will see below that for the range $\tilde\rho \gg 1$, $h < 1$ this field dependence can be well approximated by the one-loop perturbative flow equations for the running dimensionless couplings of the standard model.
The $\tilde\rho$-derivative in eq.~\eqref{eq:FESS12} is evaluated at fixed $h$.
For fixed $h$ one has $\tilde h\partial_{\tilde h} = \tilde\rho \partial_{\tilde\rho}$.
We can therefore equivalently use for $u(\tilde h, h)$ the flow equation
\begin{equation}
    \label{eq:24A}
    \tilde h \partial_{\tilde h} u = 2(u-c_U)\,,
\end{equation}
which $\tilde h$-derivative taken at fixed $h$.
The differential equation \eqref{eq:FESS12} or \eqref{eq:24A} for the scaling solution of the flow equation will be called the ``scaling equation''.

\indent The structure of the flow equation \eqref{eq:AA1} and the associated scaling equation is very simple.
As long as the particle masses are much smaller than $k$ each degree of freedom contributes to $k \partial_k U$ a constant $k^4/(32\pi^2)$, with a minus sign for fermions.
This is precisely the result of a naive perturbative computation, which treats free fields as a family of uncoupled harmonic oscillators for the different momentum modes.
The usual result diverges or involves some ultraviolet cutoff.
The flow equation focuses on the change if a cutoff is varied.
This change only includes a finite range of momentum modes around $k$, and therefore yields a finite result.
The power $k^4$ simply reflects the dimension of $U$.
This perturbative result is only modified by the threshold functions for the decoupling of particles with mass larger than $k$.
At first sight the threshold functions may appear as a purely technical device.
They contain, however, key physical features.
Without the threshold functions a field-independent potential would remain field-independent during the flow.
The field dependence of $u$ is a direct consequence of the threshold functions.
Expanding the threshold functions in powers of fields we will recover the standard perturbative one-loop result for the running of the quartic Higgs coupling.
It is the strength of functional renormalization that a simple equation for $u$ covers at once all couplings associated to field-derivatives of $u$.

\indent The scaling equation \eqref{eq:FESS12} for $u$ is rather similar to the flow equation for the $k$-dependence of $u$ at fixed $\chi$ (instead for fixed $\tilde{\rho}$), which obtains from eq.~\eqref{eq:AA1}.
One simply replaces $k \partial_k$ by $-2\tilde{\rho} \partial_{\tilde{\rho}}$.
It is therefore not surprising that features of the flow at fixed $\chi$ (which may be associated with a fixed Planck mass) are found again for the local scaling solution.
This concerns, in particular, the relevant parameters for the flow with $k$.
They appear as free parameters labeling the members of a family of local scaling solutions.

\indent Despite the formal analogy with the $k$-flow at fixed Planck mass there are important differences for the physical setting.
We deal now with an effective action depending on two fields $H$ and $\chi$.
Cosmology is based on the solution of the field equations for both fields.
The $\chi$-dependence of gauge couplings and Yukawa couplings induces new interactions.
And most importantly for our purpose, the local scaling solution needs to be extended to a global solution covering all field values.
From the mathematical point of view the scaling equation is a non-linear differential equation for a function of two variables $\tilde{\rho}$ and $\tilde{h}$ or $\tilde{\rho}$ and $h$.
This complex problem gets further enhanced by the presence of field-dependent functions for the various gauge and Yukawa couplings.

\indent The ranges of the field variables we have to cover are huge.
For $\tilde{h} \ll 1$ the masses of the standard model particles are much smaller than $k$.
Perturbation theory is valid.
For $\tilde{h}$ around one and smaller the threshold functions lead to a subsequent decoupling of particles.
For $\tilde{h} \gg \varphi_0 / m_e$ the electron has decoupled and the particle physics of the standard model is settled except for the neutrinos.
We focus on $h\ll 1$, as appropriate for a small ratio of Fermi scale over Planck mass.
In this case the value of $\tilde{\rho} = \tilde{h} / h$ is already very large for values of $\tilde{h}$ near one.
Finally, as $\tilde{\rho}$ approaches one the gravitational fluctuations become important and we have to extend the flow equation \eqref{eq:AA1}, and correspondingly the scaling equations \eqref{eq:AA2}, \eqref{eq:FESS3}.
The two quantities $\tilde{h}$ and $h$ measure the Higgs field in very different units.
They therefore differ by many orders of magnitude, as given by $\tilde{\rho}$.

\subsection*{Anomalous dimension for the Higgs field}
\indent Fluctuation effects also lead to a scale dependence of the kinetic term for the Higgs scalar.
For fixed fields $\chi$ and $H$ one has
\begin{equation}
    \label{eq:AD1}
    k \partial_k Z_{H\;|\chi,H} = - \eta_H Z_H\,.
\end{equation}
For large enough $\tilde{\rho}$, such that the gravitational fluctuations can be neglected, functional renormalization amounts in leading order to the one-loop perturbative result
\begin{equation}
    \label{eq:AD2}
    \eta_H = -k \partial_k \ln Z_H = \frac{3}{8\pi^2}\left(y_t^2 - \frac{3}{4}g_2^2 - \frac{3}{20} g_1^2\right)\,.
\end{equation}
Here $y_t$ is the Yukawa coupling of the top quark and $g_2$, $g_1$ are the gauge couplings of the gauge groups $SU(2)$ and $U(1)$ of the standard model, respectively.
For realistic values of these couplings $\eta_H$ is a small quantity.

\indent Taking the $k$-dependence of $Z_H$ into account one defines $\tilde{h}$ and $h$ in terms of a renormalized field
\begin{equation}
    \label{eq:AD3}
    \tilde{h} = \frac{H_R^\dagger H_R}{k^2} = \frac{Z_H H^\dagger H}{k^2}\,,\quad
    h = \frac{\tilde{h}}{\tilde{\rho}}\,.
\end{equation}
With
\begin{equation}
    \label{eq:AD4}
    k \partial_{k\;|\chi,H} = k \partial_{k\;|\chi,H_R} - \frac{\eta_H}{2}H_R \frac{\partial}{\partial H_R}_{\;|\chi,k},
\end{equation}
one has
\begin{equation}
    \label{eq:AD5}
    k \partial_{k\;| \tilde{\rho},h} = k \partial_{k\;|\chi,H} + \eta_H h \partial_{h\;|\tilde{\rho}} + 2 \tilde{\rho} \partial_{\tilde{\rho}\;|h}\,.
\end{equation}
This replaces in the flow equation for fixed $\tilde{\rho}$ and $h$, as well as for the scaling solution,
\begin{align}
    \label{eq:AD6}
    2 \tilde{\rho} \partial_{\tilde{\rho}\;|h} \to 2 \tilde{\rho} \partial_{\tilde{\rho}\;|h} + \eta_H h \partial_{h\;|\tilde{\rho}}\,, \nn\\
    2 \tilde{h} \partial_{\tilde{h}\;|h} \to (2+\eta_H) \tilde{h} \partial_{\tilde{h}\;|h} + \eta_H h \partial_{h\;|\tilde{h}}\,.
\end{align}

\indent For the scaling solution $y_t$, $g_2$, $g_1$ will be functions of $\tilde{\rho}$ and $h$.
Then $\eta_H$ only depends on $\tilde{\rho}$ and $h$ without any explicit dependence on $k$.
The anomalous dimension $\eta_H$ only induces small quantitative corrections without changing the qualitative features.
For the sake of simplicity of the discussion we will omit it for most of our discussion.
We add its effect at the appropriate places, as for the numerical solution. The field-dependence of $\eta_H$ and $Z_H$ implies that the field transformation \eqref{eq:AD3} to renormalized fields is non-linear. In principle, this introduces corrections to the flow equation $\sim\tilde{\rho}\partial_{\tilde{\rho}}\eta_H$ \cite{CWFT}, which are further suppressed as compared to the corrections $\sim \eta_H$ by the $\beta$-functions of the Yukawa and gauge couplings. They  will be omitted.

\section{Local scaling solution for constant Yukawa and gauge couplings}
\label{sec:III}
\indent In this section we discuss the local scaling solution in the limit where $c_U$ is a function of $\tilde{\rho}$ and $h$.
For this purpose the scalar mass terms are approximated by zero for the Goldstone modes and the cosmon, and by $\tilde{m_r}^2 = 2 \lambda_h \tilde{h} $ for the radial mode. Then $c_U$ does not involve $u$ or its derivatives.
We may investigate the contributions of the different particles of the standard model, starting with the top quark which has the largest coupling to the Higgs boson.
Subsequently we include the couplings of the scalars and the gauge bosons, as well as other fermions.
In the approximation of this section the field-dependence of the Yukawa and gauge couplings is neglected.

\subsection*{Top quark fluctuations}
In order to get a glance on qualitative features of the scaling solution of eq.~\eqref{eq:FESS12} we first set all gauge couplings and Yukawa couplings except for the top quark to zero.
The top quark Yukawa coupling is approximated by a constant $y_t$.
At this stage we approximate the contribution from scalar fluctuations by setting $\tilde{m}_i^2 = \frac{m_s^2}{k^2} = 0$, or $\bar{N}_s = 5$.
In this approximation one has
\begin{equation}
    \label{eq:TQF13}
	\bar{N}_S = 5\,,\quad
	\bar{N}_V = 12\,,\quad
	\bar{N}_F = 39 + \frac{6}{1 + y_t^2 h \tilde\rho}\,,
\end{equation}
or
\begin{equation}
    \label{eq:TQF14}
c_U = -\frac{1}{128\pi^2} \left(47 + \frac{12}{1 + y_t^2 h \tilde\rho} \right)\,.
\end{equation}

\indent Let us first discuss the overall effect of the top quark fluctuations.
For the region $y_t^2 h \tilde\rho \ll 1$ the general local solution of eq.~\eqref{eq:FESS12} reads
\begin{equation}
	\label{eq:TQF15}
	u = L_{-}(h) \tilde{\rho}^2 - \frac{59}{128\pi^2} + \dots
\end{equation}
Similarly, for $y_t^2 h \tilde\rho \gg 1$ one finds approximately 
\begin{equation}
	\label{eq:TQF16}
    u=L_{+}(h) \tilde{\rho}^2 - \frac{47}{128\pi^2} + \dots
\end{equation}
We will see below that the ``integration constants'' $L_+$ and $L_-$ coincide.
Thus the top quark fluctuations shift $u$ by a constant value $3/(32\pi^2)$ between the regions in $\tilde{\rho}$ for which the top quark is effectively massless, and the region where its mass is much larger than $k$.

\indent For an investigation of the dependence of $u$ on the Higgs doublet we take a $h$-derivative of eq.~\eqref{eq:FESS12}
\begin{equation}
    \label{eq:TQF17}
	\tilde{\rho} \partial_{\tilde\rho} (\partial_h u) = 2 \partial_h u - \frac{3 y_t^2 \tilde\rho}{16 \pi^2 (1 + y_t^2 h \tilde\rho)^2}\left(1 + \frac{\partial \ln y_t^2}{\partial \ln h}\right)\,.
\end{equation}
In the present approximation we neglect the small term $\left|\partial \ln y_t^2 / \partial \ln h\right| \ll 1$.
This yields for the approximate mixed second derivative
\begin{equation}
	\label{eq:TQF18}
	\partial_{\tilde{\rho}} \partial_h u = \frac{2}{\tilde\rho}\partial_h u - \frac{3y_t^2}{16\pi^2 (1+y_t^2 h \tilde\rho)^2}\,.
\end{equation}
For the second $h$-derivative this approximation implies
\begin{equation}
    \label{eq:TQF19}
	\hat{\lambda}_h = \partial_h^2 u\,,\quad
	\partial_{\tilde{\rho}} \hat{\lambda}_h = \frac{2}{\tilde\rho} \hat{\lambda}_h + \frac{3y_t^4 \tilde\rho}{8\pi^2 (1+y_t^2 h \tilde\rho)^3}\,.
\end{equation}

\indent For $y_t^2 h \tilde\rho \ll 1$ the approximative solution of eqs.~\eqref{eq:TQF18} \eqref{eq:TQF19} is found by an integration over $\tilde{\rho}$,
\begin{align}
	\label{eq:TQF20}
	\partial_h u &= \partial_h L_{-}(h) \tilde{\rho}^2 + \frac{3y_t^2\tilde{\rho}}{16\pi^2} + \frac{3y_t^4 h \tilde{\rho}^2}{8\pi^2}\ln(h\tilde\rho) + \dots\,,\nn\\
	\partial_h^2 u &= \partial_h^2 L_{-}(h) \tilde{\rho}^2 + \frac{3y_t^4 \tilde{\rho}^2}{8\pi^2}\left(\ln(h\tilde{\rho}) + 1\right) + \dots\,.
\end{align}
or 
\begin{equation}
	\label{eq:TQF21}
	u = L_-(h) \tilde{\rho}^2 
	- \frac{59}{128\pi^2} 
	+ \frac{3y_t^2\htilde}{16\pi^2}
	+ \frac{3y_t^4\htilde^2}{16\pi^2} \left(\ln(\htilde) - \frac{1}{2}\right) + \dots
\end{equation}
Eq.~\eqref{eq:TQF21} involves the first three terms in an expansion of $c_U$ in powers of $y_t^2 h \tilde\rho$, accounting for the two lowest correction terms to eq.~\eqref{eq:TQF15}.
We observe that the fluctuation contribution only depends on $\tilde{h} = H^\dagger H / k^2$.

\indent Restoring dimensions yields
\begin{equation}
    \begin{aligned}
    \label{eq:TQF22}
    U =&\; \frac{1}{4}L_-(h) \chi^4
	- \frac{59 k^4}{128 \pi^2}
	+ \frac{3y_t^2 k^2 H^\dagger H}{16\pi^2} \\
       &+ \frac{3 y_t^4 (H^\dagger H)^2}{16\pi^2} \left(\ln\left(\frac{H^\dagger H}{k^2}\right) - \frac{1}{2}\right)\,.
    \end{aligned}
\end{equation}
Evaluating the flow with $k$ at fixed $\chi$ and $H$,
\begin{equation}
	\label{eq:TQF23}
	k \partial_k U_{\chi,H} = -\frac{59}{32\pi^2} k^4
	+ \frac{3 y_t^2 k^2 H^\dagger H}{8 \pi^2} - \frac{3 y_t^4 (H^\dagger H)^2}{8\pi^2}\,,
\end{equation}
one finds the one-loop contribution from top quarks for the mass term and quartic coupling in the Higgs potential,
\begin{equation}
    \label{eq:44A1}
    k\partial_k \lambda_h = -\dfrac{3y^2_t}{4\pi^2}\,.
\end{equation}
Adding the part from the anomalous dimension $\eta_H$ this is the standard perturbative result.

\indent For $y_t^2 h \tilde\rho \gg 1$ the top quark decouples from the flow.
The contribution of the top fluctuations are suppressed by powers of $(y_t^2 h \tilde\rho)^{-1}$.
Only other effectively massless particles contribute to the flow of $u$.

\subsection*{Mass thresholds for the scaling potential}
\indent At fixed $h$ one has $\tilde{\rho} \partial_{\tilde{\rho}} = \htilde \partial_{\htilde}$, and we can write the scaling equation \eqref{eq:FESS12} as
\begin{equation}
	\label{eq:TQF24}
	\htilde \partial_{\htilde} u = 2u + \frac{1}{64\pi^2}\left(47 + \frac{12}{1 + y_t^2 \htilde}\right)\,.
\end{equation}
For constant $y_t$ one can find the explicit solution 
\begin{equation}
    \label{eq:TQF25}
	u = \tilde{L}(h) \htilde^2 - \frac{47}{128\pi^2} - \frac{3}{32\pi^2} t_u(\tilde{m}_t^2),\quad
	\tilde{m}_t^2 = y_t^2\htilde\,.
\end{equation}
The threshold function \cite{DEQG,CWQGSS}
\begin{equation}
	\label{eq:TQF26}
	t_u(\tilde{m}^2) = 1 - 2\tilde{m}^2 - 2\tilde{m}^4\ln \frac{\tilde{m}^2}{1 + \tilde{m}^2}\,,
\end{equation}
obeys
\begin{equation}
    \label{eq:TQF27}
	\tilde{m}^2 \partial_{\tilde{m}^2} t_u
	= 2 t_u - \frac{2}{1 + \tilde{m}^2}\,.
\end{equation}
It interpolates between $t_u = 1$ for $\tilde{m}^2 \to 0$ and $t_u = 0$ for $\tilde{m}^2 \to \infty$. With $t_u$ shifted by one as the top-quark mass threshold is crossed the potential $u$ is shifted by the constant value $-3/(32\pi^2)$. This relates the two asymptotic expressions \eqref{eq:TQF15}, \eqref{eq:TQF16} with $L_+(h) = L_-(h)$.

\indent The threshold functions $t_u(\tilde{m}^2)$ contain important information about the flow of couplings. This can be seen by evaluating the derivatives of the threshold function \eqref{eq:TQF23}, 
\begin{equation}
    \label{eq:QC1}
    \frac{\partial t_u}{\partial \tilde{m}^2} = - \left(2 + \frac{2\tilde{m}^2}{1+\tilde{m}^2} + 4 \tilde{m}^2 \ln \left(\frac{\tilde{m}^2}{1+\tilde{m}^2}\right)\right)\,,
\end{equation}
and
\begin{equation}
    \label{eq:QC2}
    \frac{\partial^2 t_u}{(\partial \tilde{m}^2)^2} = - \!\left(\! 4 \ln \!\left(\! \frac{\tilde{m}^2}{1+\tilde{m}^2} \!\right) + \frac{4}{1 + \tilde{m}^2} + \frac{2}{(1+\tilde{m}^2)^2} \!\right)\!.
\end{equation}
Derivatives of the potential with respect to $\tilde{h}$ are directly related to derivatives of $t_u$ with respect to $\tilde{m}^2$.

\indent We may define a type of quartic scalar coupling by
\begin{equation}
    \label{eq:QC2A}
    \lambda'_h(\htilde) = \frac{\partial^2 u}{\partial \htilde^2}_{|h} 
	= \bar{\lambda}_h(h) - \frac{3y_t^4}{32\pi^2} \frac{\partial^2 t_u}{(\partial \tilde{m}_t^2)^2}\,.
\end{equation}
It has a simple relation (see below) to the quartic scaling coupling $\lambda_h$ by switching from constant $h$ to constant $\tilde{\rho}$. Eq.~\eqref{eq:QC2} implies
\begin{align}
    \label{eq:QC3}
    \lambda'_h = \,\bar{\lambda}'_h + \frac{3y_t^4}{8\pi^2} \bigg[\ln \left(y_t^2\frac{H^\dagger H}{k^2}\right) - \ln \left(1 + y_t^2 \frac{H^\dagger H}{k^2}\right) \nn\\ 
    + \frac{k^2}{y_t^2 H^\dagger H + k^2} + \frac{k^4}{2(y_t^2 H^\dagger H + k^2)^2}\bigg]\,.
\end{align}
For $k^2 \gg y_t^2 H^\dagger H$ we observe the logarithmic increase of $\lambda'_h$ with decreasing $k$ due to the top quark fluctuations 
\begin{equation}
    \label{eq:QC4}
    \lambda'_h = \bar{\lambda}'_h + \frac{3y_t^4}{8\pi^2} \left[\ln \left(y_t^2 \frac{H^\dagger H}{k^2}\right) + \frac{3}{2}\right]\,.
\end{equation}
This accounts again for the top-contribution to the perturbative running at fixed $H^\dagger H$,
\begin{equation}
    \label{eq:QC5}
    k \partial_k \lambda'_h = -\frac{3 y_t^4}{4\pi^2}\,.
\end{equation}

\indent For arbitrary values of $\tilde{h}$ we may also directly evaluate the flow of $\lambda'_h(\htilde)$ from eq.~\eqref{eq:TQF24}
\begin{align}
    \label{eq:QC6}
    \htilde \partial_{\htilde} \lambda'_h &= \htilde \partial_{\htilde} (\partial_{\htilde}^2 u) = \tilde{\partial}_h^2 \left(\htilde \partial_{\htilde} u\right) - 2\lambda_h' \nn\\
					      &= \frac{3}{16\pi^2} \partial_{\htilde}^2 \left(1 + y_t^2\htilde\right)^{-1} = \frac{3y_t^4}{8\pi^2 (1+y_t^2\htilde)^3}\,.
\end{align}
This shows that $\lambda'_h$ is monotonically increasing for increasing $\htilde$.
The solution of eq.~\eqref{eq:QC6},
\begin{equation}
    \label{eq:QC7}
    \lambda'_h = \bar{\lambda}'_h + \frac{3y_t^4}{8\pi^2} \!\left(\!\ln \!\left(\!\frac{y_t^2 \htilde}{1+y_t^2\htilde}\!\right)\! + \frac{1}{1 + y_t^2 \htilde} + \frac{1}{2(1+y_t^2\htilde)^2}\!\right),
\end{equation}
agrees with eq.~\eqref{eq:QC3}.
This demonstrates that the threshold function \eqref{eq:TQF23} encodes both the logarithmic increase of the quartic scalar coupling for $k^2 \gg m_t^2$, and the stop of this increase for $k^2 \ll m_t^2$.
The integration constant $\bar{\lambda}'_h$ corresponds to the IR-value of $\lambda'_h$ for $k \to 0$.

\indent In a similar way we may define
\begin{equation}
    \label{eq:QC8}
    \mu'_h(\htilde) = \partial_{\htilde} u(\htilde,h)_{|h}\,.
\end{equation}
The flow of $\mu_h'$ with $\htilde$,
\begin{align}
    \label{eq:QC9}
    \htilde \partial_{\tilde{h}} \mu'_{h\;|h} &= \partial_{\htilde} \left(\htilde \partial_{\htilde} u\right) - \mu'_h \nn\\
				  &= \mu'_h - \frac{3y_t^2}{16\pi^2 (1+y_t^2\tilde{h})^2}\,,
\end{align}
has the solution
\begin{equation}
    \label{eq:QC10}
    \mu'_h = c_\mu \tilde{h} + \frac{3y_t^2}{16\pi^2} \left(1 + \frac{y_t^2 \tilde{h}}{1 + y_t^2 \tilde{h}} + 2 y_t^2 \tilde{h} \ln \left(\frac{y_t^2 \htilde}{1 + y_t^2 \htilde}\right)\right)\,.
\end{equation}
The integration constant $c_\mu$ multiplies $\htilde$, such that there is no free integration constant for $\mu_h'(\tilde{h}=0)$.
Comparison with eq.~\eqref{eq:TQF22} implies
\begin{equation}
    \label{eq:11}
    c_\mu = \bar{\lambda}_h' = 2\tilde{L}\,.
\end{equation}
The only free integration constant for the scaling solution is the coefficient $\tilde{L}$ of the term quadratic in $\htilde$.
From the simple relation
\begin{equation}
    \label{eq:QC12}
    \mu_h'(\htilde = 0) = \frac{3y_t^2}{16\pi^2}\,,
\end{equation}
one infers that the minimum of $U$ occurs for $H=0$ if $\tilde{L}$ is independent of $h$.
The mass term for the Higgs scalar decreases $\sim k^2$ for this scaling solution,
\begin{equation}
    \label{eq:QC13}
    m_H^2 = \frac{\partial U}{\partial(H^\dagger H)}_{|H=0} = \frac{3y_t^2 k^2}{16\pi^2}\,.
\end{equation}
It vanishes for $k\to 0$.

\indent Comparison with eqs.~\eqref{eq:TQF15},\eqref{eq:TQF16},\eqref{eq:TQF21} yields
\begin{equation}
    \label{eq:TQF28}
	\tilde{L}(h) = \frac{L_-(h)}{h^2}
	= \frac{L_+(h)}{h^2}
	= \frac{L(h)}{h^2}\,.
\end{equation}
Except for a possible $h$-dependence of the boundary term $\tilde L(h)$ the scaling potential $u(\tilde h,h)$ is only a function of $\tilde h$.
We will see that this is characteristic for a second order quantum phase transition.

\indent The quantities that are directly related to the $H$-derivatives of $U$ at fixed $\chi$ and $k$ correspond for the scaling solution to $\tilde{h}$-derivatives at fixed $\tilde{\rho}$
\begin{equation}
    \label{eq:PP1}
    \mu_h = \partial_{\tilde{h}} u_{|\tilde{\rho}} = k^{-2} \frac{\partial U}{\partial H^\dagger H}_{|\chi}\,,
\end{equation}
and
\begin{equation}
    \lambda_h = \partial_{\tilde{h}}^2 u_{|\tilde{\rho}} = \frac{\partial^2 U}{(\partial H^\dagger H)^2}_{|\chi}\,.
\end{equation}
We can take these as functions of the fields or for certain fixed fields, say at $\tilde{h} = 0$.
For the observables related to the Higgs potential the function $u(\tilde{\rho}, \tilde{h})$ is most appropriate.
On the other hand, the scaling equations \eqref{eq:FESS12} and \eqref{eq:24A} assume their simple form for derivatives at fixed $h$.
One may switch between derivatives at fixed $h$ or $\tilde{\rho}$ by the identity 
\begin{equation}
    \label{eq:PP3}
    \partial_{\tilde{h}\;|\tilde{\rho}} = \partial_{\tilde{h}\;|h} + \frac{h}{\tilde{h}} \partial_{h\;| \tilde{h}}\,.
\end{equation}
The difference between $\mu'_h$ and $\mu_h$, or $\lambda_h'$ and $\lambda_h$, therefore concerns only the boundary term $\tilde{L}(h)$.

\subsection*{Fermi scale}
\label{sec:FS}
The local solution involves the free function $L(h)$ which sets the ``boundary values'' or ``initial values'' for the local solution of the differential equation \eqref{eq:FESS12}.
These boundary values encode the behaviour for $\tilde{h}\xrightarrow{}\infty$, or for $k\xrightarrow{}0$ at fixed $H^\dagger H$.
They are therefore directly related to the Fermi scale.
For small $h$ we may approximate
\begin{equation}
    \label{eq:FS29}
    L(h) = \frac{1}{2} \bar{\lambda}_\chi + \bar{\lambda}_m h + \frac{1}{2} \bar{\lambda}_h h^2 (1+\Delta(h))\,.
\end{equation}
The normalisation of $\Delta(h)$ affects the precise definition of $\bar{\lambda}_h$.
For a minimum of $L(h)$ at $h=0$ we take $\Delta(0)=0$.
In case of a minimum of $L(h)$ for $h_0\neq0$ we choose the normalisation such that the relation between $h_0$ and $\bar{\lambda}_h$ is simple.
Inserting the approximation \eqref{eq:PP3} the effective potential reads
\begin{align}
    \label{eq:FS30}
    U =&\; \frac{1}{8} \bar{\lambda}_\chi \chi^4 + \frac{1}{2}\bar{\lambda}_m \chi^2 H^\dagger H \nn\\
       &+ \frac{1}{2} \bar{\lambda}_h (H^\dagger H)^2 \left(1+\Delta \left(\frac{2H^\dagger H}{\chi^2}\right)\right)\\
	&- \frac{k^4}{128\pi^2} \left[47 + 12 t_u \left(\frac{y_t^2 H^\dagger H}{k^2}\right)\right]\,.\nn
\end{align}
For $k\to 0$ one employs
\begin{equation}
    \label{eq:FS31}
    \lim_{\tilde{m}^2 \to \infty} t_u(\tilde{m}^2) = \frac{2}{3\tilde{m}^2}\,.
\end{equation}
The term $\sim k^4$ vanishes and we conclude that the couplings $\bar{\lambda}_\chi$, $\bar{\lambda}_m$ and $\bar{\lambda}_h$ correspond in the approximation \eqref{eq:FS31} to the boundary values for $k\to 0$ or $\tilde{\rho} \to \infty$.

\indent For $\bar{\lambda}_m < 0$ one observes spontaneous breaking of the $SU(2)\times U(1)$-symmetry according to 
\begin{equation}
    \label{eq:FS32}
    U_{k\to 0} = \frac{1}{2} \bar{\lambda}_h (H^\dagger H - \varphi_0^2)^2 + \frac{\delta}{8}\chi^4 + \Delta U\,,
\end{equation}
where $\bar{\lambda}_h$ is specified by
\begin{equation}
    \label{eq:44A}
    \partial_h L(h)_{|h_0} - \partial_h L(h)_{|0} = \bar{\lambda}_h h_0\,,
\end{equation}
and $\bar{\lambda}_m = \partial_h L(h)_{|0}$.
This specifies the constant part in $\Delta$ by the relation
\begin{equation}
    \label{eq:44B}
    \Delta (h_0) + \frac{1}{2}h \partial_h \Delta(h_0) = 0\,.
\end{equation}
With this normalization the term
\begin{equation}
    \label{eq:44C}
    \Delta U = \frac{1}{2} \bar{\lambda}_h (H^\dagger H)^2\Delta 
\end{equation}
does not contribute to $\partial U / \partial(H^\dagger H)$ at $\varphi_0^2$.
The Fermi scale $\varphi_0$ is proportional to $\chi$,
\begin{equation}
    \label{eq:FS33}
    \varphi_0^2 = -\frac{\bar{\lambda}_m}{2\bar{\lambda}_h} \chi^2\,,\quad
    \delta = \bar{\lambda}_\chi - \frac{\bar{\lambda}_m^2}{\bar{\lambda}_h}\,.
\end{equation}

\indent The observable quartic Higgs coupling obeys 
\begin{equation}
    \label{eq:45A}
    \lambda_h = \frac{\partial^2 U}{\partial(H^\dagger H)^2}_{|\varphi_0} = \bar{\lambda}_h \left(1 + \frac{3}{2}h \partial_h \Delta + \frac{1}{2} h^2 \partial_{h^2} \Delta\right)_{|h_0}\,.
\end{equation}
For the scaling solution a possible logarithmic dependence of the field-dependent quartic scalar coupling $\partial^2 U/\partial(H^\dagger H)^2$ only arises through $\Delta(h)$ in the boundary term.
This may be understood from the observation that for $k = 0$ and zero momentum any logarithm needs a second scale, as present in the ratio $H^\dagger H/\chi^2$.
We will often omit $\Delta$ since a quadratic approximation to $L(h)$ is sufficient for many purposes.

\indent With
\begin{equation}
    \label{eq:FS33A}
    h_0 = \frac{2\varphi_0^2}{\chi^2} = -\frac{\bar{\lambda}_m}{\bar{\lambda}_h}
\end{equation}
a small value $h_0$ requires small $\left|\bar{\lambda}_m / \bar{\lambda}_h\right|$.
For the local solution the couplings $\bar{\lambda}_h$ and $\bar{\lambda}_\chi$, as well as the ratio $\varphi_0^2 / \chi^2$, are free parameters.
On this level there is no prediction for the Fermi scale, the mass of the Higgs boson or a possible cosmological constant.
Restrictions on the form of $L(h)$ or the parameters $\bar{\lambda}_\chi$, $\bar{\lambda}_m$ and $\bar{\lambda}_h$ may arise from the requirement of a global scaling solution that exists for all values of $\tilde{\rho}$ and $h$.
These global requirements will increase the predictive power of the scaling solution.

\subsection*{Scalar fluctuations}
\label{sec:SF}
We next evaluate the effective number of scalars $\bar{N}_S$.
For this purpose we have to diagonalize the scalar mass matrix \eqref{eq:FESS11}.
By virtue of the $SU(2) \times U(1)$ symmetry it is sufficient to evaluate the second derivatives for a configuration $\tilde{H}_1 = \sqrt{2\htilde k^2}$, $\tilde{H}_2 = \tilde{H}_3 = \tilde{H}_4 = 0$.
The mass matrix becomes block diagonal, with three ``Goldstone directions''
\begin{equation}
    \label{eq:SF34}
    \tilde{m}_g^2 = \partial_{\htilde} u\,.
\end{equation}
The radial mode of the Higgs scalar and the cosmon $\chi$ are mixed, with $2\times 2$ matrix
\begin{equation}
    \label{eq:SF35}
    \tilde{m}^2_s = \begin{pmatrix} \partial_{\htilde} u + 2 \htilde \tilde{\partial}_h^2 u & 2\sqrt{\htilde\tilde{\rho}}\, \partial_{\tilde{\rho}} \partial_{\htilde} u \\
    2 \sqrt{\htilde\tilde{\rho}}\, \partial_{\tilde{\rho}} \partial_{\htilde} u & \partial_{\tilde{\rho}} u + 2\tilde{\rho} \partial_{\tilde{\rho}}^2 u
    \end{pmatrix}
    = \begin{pmatrix}
    \tilde{m}_r^2 & \tilde{m}_{rc}^2 \\ \tilde{m}_{rc}^2 & \tilde{m}_c^2
    \end{pmatrix}
    \,.
\end{equation}
The $\htilde$-derivatives are taken at fixed $\tilde{\rho}$, such that $\partial_{\htilde} = \tilde{\rho}^{-1} \partial_h$, and the $\tilde{\rho}$-derivatives are taken at fixed $\tilde{h}$.

\indent The eigenvalues of the mass matrix \eqref{eq:SF35} are 
\begin{equation}
    \label{eq:SF36}
    \tilde{m}_\pm^2 = \frac{1}{2}\left(\tilde{m}_r^2 + \tilde{m}_c^2 \pm \sqrt{ (\tilde{m}_r^2 - \tilde{m}_c^2)^2 + 4\tilde{m}_{rc}^4}\right)\,.
\end{equation}
If needed, the entries in the mass matrix \eqref{eq:SF35} can also be expressed in terms of $u(\tilde{\rho},h)$ with $\tilde{\rho}$-derivatives taken at fixed $h$,
\begin{align}
    \label{eq:SF37}
    \tilde{m}_r^2 &= \frac{1}{\tilde{\rho}} \Big(\partial_h u + 2\hat{\lambda}_h h\Big)\,\nn\\
    \tilde{m}_{rc}^2 &= 2\sqrt{h} \partial_{h\;|\tilde{\rho}\vphantom{\tilde{h}}} \partial_{\tilde{\rho}\;|\tilde{h}} u \nn\\
		     &= 2\sqrt{h} \left(\partial_{\tilde{\rho}} \partial_h u(\tilde\rho, h) - \frac{1}{\tilde{\rho}} \partial_h u - \lambda_hh \tilde{\rho}\right)\,.
\end{align}
For the expressions involving $\tilde{\rho}$-derivatives at fixed $h$ one uses
\begin{equation}
    \label{eq:59A}
    \partial_{\tilde{\rho}\;|\tilde{h}} = \partial_{\tilde{\rho}\;|h} - \frac{h}{\tilde{\rho}} \partial_{h\;|\tilde{\rho}}\,.
\end{equation}
In terms of the eigenvalues \eqref{eq:SF36} the effective number of scalars is given by
\begin{equation}
    \label{eq:SF38}
    \bar{N}_s = \frac{3}{1 + \tilde{\rho}^{-1}\partial_h u} + \frac{1}{1+\tilde{m}_+^2} + \frac{1}{1+\tilde{m}_-^2} \,.
\end{equation}

\indent Consider the $\tilde{\rho}$-dependent partial minimum for $h$ at $h_0(\tilde{\rho})$,
\begin{equation}
    \label{eq:SF39}
    \partial_h u (\tilde{\rho}, h_0(\tilde\rho)) = 0\,.
\end{equation}
At this partial minimum the Goldstone boson mass vanishes, and $\tilde{m}_r^2 \sim h_0\tilde{\rho}$,
\begin{equation}
    \label{eq:SF40}
    \tilde{m}_g^2 = 0\,,\quad
    \tilde{m}_r^2 = 2 \lambda_h h_0 \tilde{\rho}\,.
\end{equation}

\indent For the off-diagonal part one finds
\begin{align}
    \tilde{m}_{rc}^2 &= - \sqrt{h_0} \bigg(\tilde{m}_r^2 - 2\partial_{\tilde{\rho}} \partial_h u(\tilde{\rho},h_0)\bigg)     \nn\\
		     &= - \sqrt{h_0} \bigg(\tilde{m}_r^2 + \frac{3y_t^2}{8\pi^2(1+y_t^2h\tilde{\rho})^2}\bigg)\,.
\end{align}
The ratio $\tilde{m}_{rc}^2/\tilde{m}_r^2$ is proportional to $\sqrt{h_0}$ and therefore small for small $h_0$.
To a good approximation we can replace $\tilde{m}_+^2$ by $\tilde{m}_r^2$, with
\begin{equation}
    \label{eq:SF46}
    \tilde{m}_r^2 = 2\lambda_h \htilde_0 = 2\lambda_h h_0 \tilde{\rho}\,.
\end{equation}
For $h$ near $h_0$ the effect of the radial Higgs boson fluctuation is similar to the top fluctuations, with $y_t^2$ replaced by $2\lambda_h$ and a prefactor one instead of $-12$.

\indent Away from the partial minimum the effect of the scalar fluctuations is more complex.
This concerns the flow of $h$-derivatives of $u$.
For example, the Goldstone boson fluctuations contribute to the flow of the quartic scalar coupling $\lambda_h$, or scalar fluctuations induce an anomalous dimension for the running of $\lambda_m$.
We will come back to these issues below.

\subsection*{Integration constants}
The local solution of the scaling equation \eqref{eq:FESS12} has free ``integration constants'' or boundary values.
We may view eq.~\eqref{eq:FESS12} as a family of first order non-linear differential equations, one for each value of $h$.
There are then infinitely many integration constants, one for each value of $h$.
This is reflected by the boundary function $L(h)$.
There will be additional requirements on $L(h)$, as continuity and differentiability.
If the first few derivatives of $L(h)$ exist at $h=0$, we may use a Taylor expansion in the small quantity $h$.
If $L(h)$ is analytic in a neighborhood of $h_0$, the coefficients of a Taylor expansion around $h_0$ may be associated to the infinitely many integration constants for the local solution of eq.~\eqref{eq:FESS12}.
For very small $h$ and $h_0$ only the few lowest expansion coefficients will matter in practice.

\indent The simple form of the boundary term $L(h)\tilde{\rho}^2$ holds  globally for all $\tilde{\rho}$ only in the approximation that $c_U$ is a function of $\tilde{\rho}$ and $h$ which does not involve $u$.
The contribution of the scalar fluctuations violates this property since the field-dependent scalar masses involve $\partial_{\tilde{\rho}} u$ and $\partial_{\tilde{\rho}}^2 u$.
Nevertheless, in the limit where the scalar masses can be approximated by zero or by functions of $\tilde{\rho}$ and $h$ which do not involve $u$ explicitly, the global form of the boundary term remains preserved.
In this case a family of local solutions can be found by starting from a particular solution $\bar{u}(\tilde{\rho}, h)$ by constructing the general solution as 
\begin{equation}
    \label{eq:55A}
    u(\tilde{\rho}, h) = \bar{u}(\tilde{\rho}, h) + L(h)\tilde{\rho}^2\,.
\end{equation}
We will first work with this approximation and discuss the modifications due to the scalar fluctuations at a later stage.

\indent More generally, the local scaling solutions are always characterized by an "initial value function" or "boundary function" $L(h)$.
Only the global form \eqref{eq:55A} does no longer hold.
For general field-dependent couplings we may define the threshold functions such that they vanish in the limit $\tilde{m}_i^2 \to \infty$.
There is a region in field space for $\tilde{h} \gg 1$ where all massive particles of the standard model except the neutrinos have masses $\tilde{m}_i^2 \gg 1$, while for the others one can approximate $\tilde{m}_i^2 \approx 0$.
In this region the form \eqref{eq:55A} becomes a good approximation for $h$ near the partial minimum $h_0$.
We could specify $L(h)$ as the "initial value function" defined by some $\tilde{h}_{in}$ in this region.
We will actually take $\tilde{h}_{in}$ to infinity and define $L(h)$ by the boundary value for $\tilde{\rho}\xrightarrow{}\infty$, $\tilde{h}\xrightarrow{}\infty$. 

\subsection*{Scaling solution for constant gauge and Yukawa couplings}
Many features of our discussion of the top-quark fluctuations features persist if we include the effects of fluctuations of other particles of the standard model.
Neglecting the field-dependence of Yukawa and gauge couplings the scaling solution for $u$ becomes
\begin{align}
    \label{eq:FS43B}
	u = L(h) \tilde{\rho}^2 \,+\, &\frac{1}{128\pi^2} \bigg\{17 - 2\sum_f n_f t_u (\tilde{m}_f^2) \nn\\
							   & +3\sum_v t_u(\tilde{m}_v^2) + \sum_s t_u(\tilde{m}_s^2) \bigg\}\,.
\end{align}
Here $n_f = 1$ for neutrinos, $n_f=2$ for charged leptons, $n_f=6$ for quarks, and $\tilde{m}_f = m_f/k$ the dimensionless fermion mass.
The sum $\sum_v$ is over the three massive vector bosons $W^\pm$, $Z$ and $\sum_s$ sums the five scalars with mass eigenvalues $\tilde{m}_s^2 = m_s^2/k^2$ in an approximation where they can be taken to be proportional to $\tilde{h}$ or zero.

\indent For $k\to 0$ or $\tilde{\rho} \to \infty$ the potential is dominated by the boundary function $L(h)$, which encodes the boundary information for $\tilde{\rho} \to \infty$.
Possible restrictions on the behavior for $\tilde{\rho} \to 0$ will have to be translated to restrictions on $L(h)$.
The scaling solution implies the important property of the scale invariant standard model \cite{CWQ,SHAZEN1,SHAZEN2} that the Fermi scale is proportional to the field $\chi$, given by the location of the minimum of $L(h)$ at $h_0$.

\indent Let us consider the scaling solution for a range of $k$ larger than the mass of the charm quark.
In this range the flow equations \eqref{eq:FESS12}, \eqref{eq:24A} may be a reasonable approximation, since the effects of the strong interactions are not yet very large.
We will discuss the continuation to lower $k$ later. We approximate 

\begin{equation}
    \label{eq:78}
    L(h)\tilde{\rho}^2 = \bar{\lambda}_m\tilde{\rho}\tilde{h}+\frac{1}{2}\bar{\lambda}_m\tilde{h}^2 +\frac{1}{2}\bar{\lambda}_\chi\tilde{\rho}^2\,.
\end{equation}

\indent In the limit of constant gauge and Yukawa couplings and vanishing neutrino masses the scaling solution for the effective potential is given by eq.~\eqref{eq:FS43B},
\begin{align}
    \label{eq:N2}
    u =\bar{\lambda}_m\tilde{\rho}\tilde{h}&+\frac{1}{2}\bar{\lambda}_h\tilde{h}^2 +\frac{1}{2}\bar{\lambda}_\chi\tilde{\rho}^2  + \frac{1}{128\pi^2} \Bigg\{3 t_u (\tilde{m}_Z^2)  \nn\\ 
				  &+ 6t_u(\tilde{m}_W^2)+ 3t_u(\tilde{m}_g^2) + t_u(\tilde{m}_r^2) - 12t_u(\tilde{m}_t^2)  \nn\\
				  &- 12t_u(\tilde{m}_b^2)- 12 t_u(\tilde{m}_{ch}^2) - 4 t_u(\tilde{m}_\tau^2) - 32\Bigg\}\,.
\end{align}
Here we have neglected the Yukawa couplings of the $s$, $u$ and $d$ quark and the electrons and muons, and we approximate the cosmon mass by zero.
For the non-zero dimensionless vector boson mass terms one has
\begin{equation}
    \label{eq:N3}
    \tilde{m}_W^2 = \frac{1}{2} g_2^2 \htilde\,,\quad
    \tilde{m}_Z^2 = \left(\frac{1}{2}g_2^2 + \frac{3}{10}g_1^2\right)\htilde\,,
\end{equation}
with $g_2$ and $g_1$ the gauge couplings of the $SU(2)$ and $U(1)$-groups of the standard model.
For the top-, bottom- and charm quark we take
\begin{equation}
    \label{eq:N4}
    \tilde{m}_t^2 = y_t^2 \htilde\,,\quad
    \tilde{m}_b^2  = y_b^2 \htilde\,,\quad
    \tilde{m}_{ch}^2 = y_{ch}^2 \tilde{h}\,,
\end{equation}
and for the scalars
\begin{equation}
    \label{eq:N5}
    \tilde{m}_g^2 = \partial_{\tilde h} u\,,\quad
    \tilde{m}_r^2 = \tilde{\partial}_h u + 2\htilde \partial_{\tilde h}^2 u\,.
\end{equation}

\indent The derivative of eq.~\eqref{eq:N2} with respect to $\tilde h$ reads
\begin{align}
    \label{eq:108A}
	\partial_{\tilde h} u =&\; \bar{\lambda}_m+\bar{\lambda}_h \htilde - \frac{1}{64\pi^2} \Bigg\{
	3g_2^2 s_u(\tilde{m}_W^2) \\
									    &+ \left(\frac{3}{2} g_2^2 + \frac{9}{10}g_1^2 \right) s_u(\tilde{m}_Z^2) -12 y_t^2 s_u(\tilde{m}_t^2) \nn\\
									    &- 12 y_b^2 s_u(\tilde{m}_b^2) -12s_u(\tilde{m}_{ch}^2) - 4s_u(\tilde{m}_\tau^2) \nn\\
									    &+ 3\partial_{\tilde h}^2 u \, s_u (\tilde{m}_g^2) + \left(3\partial_{\tilde h}^2 + 2 \htilde \partial_{\htilde}^3 \right)\! u \, s_u (\tilde{m}_r^2)\Bigg\}\,.\nn
\end{align}
The function
\begin{align}
    \label{eq:N7}
    s_u(\tilde{m}^2)
    &= 1 + 2\tilde{m}^2 \ln \left(\frac{\tilde{m}^2}{1 + \tilde{m}^2}\right) + \frac{\tilde{m}^2}{1 + \tilde{m}^2}\\ \nn
    &= - \frac{1}{\tilde{m}^2} \left(t_u(\tilde{m}^2) - \frac{1}{1 + \tilde{m}^2}\right)
\end{align}
interpolates between the limits
\begin{equation}
    \label{eq:N8}
    s_u(0) = 1\quad \text{and}\quad s_u(\tilde{m}^2 \to \infty) \implies \frac{1}{3\tilde{m}^4}\,.
\end{equation}

\indent With $\lambda_h = \partial_{\tilde{h}}^2 u$ depending only on $\tilde{h}$ we can recover the perturbative one-loop flow equation for the quartic scalar coupling.
Taking the $\tilde{h}$-derivative of eq.~\eqref{eq:108A} yields a logarithmic dependence of $\lambda_h$ on $\tilde{h}$.
Adding the scalar anomalous dimension $\eta_H$ in eq.~\eqref{eq:AD2} one has
\begin{equation}
	k \partial_k \lambda_{h\;|H} = -2 \tilde{h} \partial_{\tilde{h}} \lambda_h(\tilde{h}) - 2 \eta_H \lambda_h (\tilde{h})\,.
\end{equation}
The expansion in powers of the dimensionless couplings yields in the limit where the particle masses are small ($\tilde{h} \ll 1$) the perturbative result.
This demonstrates that essential physics is encoded in the threshold functions.

\indent We recall, however, that the solution \eqref{eq:N2} with threshold function \eqref{eq:TQF26} is only an approximation. For gauge and Yukawa couplings depending on fields the threshold function will be somewhat modified. The appearance of derivatives of $u$ in the scalar mass terms will induce a deviation from the structure \eqref{eq:55A}. 
For a rough picture of the scaling solution these are small effects. In this rough picture a tiny value of $h_0$ is, however, also a rather small feature. 
For the determination of the Fermi scale we will later have to take these modification into account.

\subsection*{Scaling potential for vanishing cosmon-Higgs coupling}

\indent For the scaling potential \eqref{eq:N2} the cosmon-Higgs coupling $\lambda_m$ is approximated by $\bar{\lambda}_m$, which does not depend on fields.
In the limit $\bar{\lambda}_m=0$ the scaling potential is particularly simple.
The partial minimum of $u$ is located for all $\tilde{\rho}$ at $h_0=0$. Indeed, a partial minimum at $\htilde_0 > 0$ can only occur for a range of couplings for which the curly bracket in eq.~\eqref{eq:108A} is positive.
For the region of $\tilde{m}^2 \ll 1$ this requires
\begin{equation}
    \label{eq:N9}
    2 \tilde{m}_W^2 + \tilde{m}_Z^2 + \tilde{m}_h^2 > 4\left(\tilde{m}_t^2 + \tilde{m}_b^2\right)\,,
\end{equation}
where $\tilde{m}_h^2 = 2\lambda_h \tilde{h}=2\partial^2_{\htilde} u$, and $\htilde\partial^3_{\htilde}u$ is neglected.
For gauge and Yukawa couplings in a range compatible with observation this condition is not met.
In the limit $\htilde \ll 1$ the partial minimum for the Higgs potential is located at $H=0$ with positive mass term
\begin{equation}
    \label{eq:N10}
U = m_h^2 H^\dagger H +\cdots= \tilde{m}_h^2 k^2 H^\dagger H+\cdots\,,
\end{equation}
with
\begin{align}
    \label{eq:N11}
    \tilde{m}_h^2 = \left.{\partial_{\tilde h} u}\right|_{\tilde h=0}
	=\, \frac{3}{64\pi^2} \bigg\{&4 y_t^2 + 4y_b^2 + 4y_{ch}^2 + y_{\tau}^2 \nn\\
	 &- \left(\frac{3}{2} g_2^2 + \frac{3}{10}g_1^2 + 2\partial_{\tilde h}^2 u(0)\right)\bigg\}.
\end{align}
As $k$ decreases the different terms in eq.~\eqref{eq:N11} are suppressed by $s_u(\tilde{m}^2) < 1$ as $\htilde$ grows large.
As long as the fermion fluctuations dominate $\tilde{m}_h^2$ remains positive.

\subsection*{Second order phase transition}
If one neglects the running of the gauge and Yukawa couplings, the electroweak theory exhibits a second order quantum phase transition (at zero temperature).
There exists a critical surface in parameter space for which both the expectation values of the Higgs field and the mass of the Higgs particle vanish.
The symmetric phase with $H_0=0$ and $m_H^2 \ge 0$ joins at this phase transition the phase with spontaneous symmetry breaking (SSB-phase) where $H_0 \ge 0$.

\indent The approximate scaling solution \eqref{eq:FS43B} reflects this phase transition in a particularly simple way.
The critical surface is realized by a potential of the form
\begin{equation}
    \label{eq:PTO1}
    u = u_1(\tilde h) + u_2(\tilde \rho)\,,
\end{equation}
where both $u_1$ and $u_2$ depend only on one dimensionless field variable without further dependence on the ratio $h$.
This is realized for $L(h) = (\bar{\lambda}_\chi + \bar{\lambda}_h h^2)/2$, i.e. $\bar{\lambda}_m=0$.
For $k>0$ the (partial) minimum of this potential may occur for $\tilde h_0 > 0$ (spontaneously broken regime), or for $\tilde h_0 = 0$ with $\tilde \mu_h^2 = \partial_{\tilde h} u(\tilde h=0) > 0$ (symmetric regime).
For $k\to 0$ one finds $\varphi_0^2 = \tilde h_0 k^2 \to 0$ or $m_H^2 = \tilde{\mu}_h^2 k^2 \to 0$, respectively.
This constitutes the critical surface.

\indent Moving away from the critical surface to the symmetric or SSB-phase requires deviations from the form \eqref{eq:PTO1}.
The potential \eqref{eq:TQF25} is of the form \eqref{eq:PTO1} for $\bar{\lambda}_m = 0$.
For $\bar{\lambda}_m \neq 0$ the contribution $\sim L$,
\begin{equation}
    \label{eq:PTO2}
    L(h) \tilde{\rho}^2 = \tilde{L}(h) \htilde^2 = \frac{1}{2} \bar{\lambda}_\chi \tilde{\rho}^2 + \frac{1}{2}\bar{\lambda}_h \htilde^2 + \bar{\lambda}_m \htilde \tilde{\rho}\,,
\end{equation}
is the only source of a deviation from the form \eqref{eq:PTO1}.
For $\bar{\lambda}_m > 0$ one ends in the symmetric phase, where eq.~\eqref{eq:FS30} yields
\begin{equation}
    \label{eq:PTO3}
    m_H^2 = \frac{1}{2} \bar{\lambda}_m \chi^2\,,
\end{equation}
while for $\bar{\lambda}_m < 0$ one realizes the SSB-phase with $\varphi_0^2$ given by eq.~\eqref{eq:FS33}.
This generalizes to the potential \eqref{eq:N2}.

\section{Infrared limit}
\label{sec:IV}
In this section we investigate the infrared limit for $k\to 0$ at fixed $\chi$ and $H$, or $\tilde{\rho} \to \infty$ for the scaling solution.
This limit is needed for the initial value function $L(h)$.
The infrared region is relevant for the understanding of dynamical dark energy or quintessence in the recent cosmological epochs \cite{CWIQ, CWQGSS, DEQG, CWQ}, \cite{RP2, CWCMAV, FTSW, FEJO, VL, CLW, CDS, LA1, LACQ, LIN, CWVG, CWGN, ABW}.
We will see that the scaling potential $u$ evaluated at the partial minimum denoted by $h_0(\tilde{\rho})$ approaches a constant for $\tilde{\rho}\xrightarrow{}\infty$.
The present ratio $U_E/M^4\sim u \tilde{\rho}^{-2}$ corresponds to huge values $\tilde{\rho}\approx10^{60}$. For realistic values $h_0\approx10^{-32}$ also $\tilde{h}\approx h_0\tilde{\rho}\approx 10^{28}$ is huge. In this range all particles of the standard model with mass equal or larger than the electron mass have decoupled from the flow. For $\htilde$ away from the partial minimum at $\htilde_0 = h_0\tilde{\rho}$ the potential increases steeply, with $\tilde{m}^2_r = 2\lambda_h \htilde_0 \gg 1$. One expects that $u(\tilde{\rho},\htilde)$ can be reduced effectively to a "cosmon potential" depending only on $\tilde{\rho}$.

\subsection*{Cosmological evolution of Fermi scale}
Independently of the boundary function $L(h)$ the scaling solution predicts the $\tilde{\rho}$-dependence of $h_0$.
The minimum condition \eqref{eq:A2} holds for all $\tilde{\rho}$, implying
\begin{equation}
    \label{eq:FS41}
    \partial_{\tilde{\rho}} \partial_h u(h_0) + \hat{\lambda}_h \partial_{\tilde{\rho}} h_0 = 0\,.
\end{equation}
If we include only the contribution from the top quark fluctuations \eqref{eq:TQF19} one has
\begin{equation}
    \label{eq:FS42}
    \partial_{\tilde{\rho}} \partial_h u(\tilde{\rho}, h_0) = -\frac{3y_t^2}{16\pi^2(1 + y_t^2 h_0 \tilde{\rho})^2}\,.
\end{equation}
The location of the partial minimum slowly increases with $\tilde{\rho}$ for $y_t^2 h_0 \tilde{\rho} \lesssim 1$, and settles to a constant value for $y_t^2 h_0 \tilde{\rho} \gg 1$,
\begin{equation}
    \label{eq:FS43}
    \partial_{\tilde{\rho}} h_0(\tilde{\rho}) = \frac{3y_t^2}{16\pi^2 \hat{\lambda}_h (1+y_t^2 h_0 \tilde{\rho})^2} \,.
\end{equation}

\indent Inserting the quartic Higgs coupling \eqref{eq:FESSB2},
\begin{equation}
    \label{eq:FS44}
    \lambda_h = \partial_{\htilde^2} u = \frac{\hat{\lambda}_h}{\tilde{\rho}^2}\,,
\end{equation}
one obtains
\begin{equation}
    \label{eq:FS43A}
    \tilde{\rho} \partial_{\tilde{\rho}} h_0 = \frac{3y_t^2}{16\pi^2 \lambda_h \tilde{\rho}(1+y_t^2 h_0 \tilde{\rho})^2}\,.
\end{equation}
For large $\tilde{m}^2_t=y^2_th_0\tilde{\rho}$ this change is tiny,
\begin{equation}
    \label{eq:M8AA}
    \frac{\partial \ln h_0}{\partial \ln \tilde{\rho}} = \frac{3}{16\pi^2\lambda_h \tilde{m}^2_t \tilde{\rho}^2}\,.
\end{equation}

\indent For cosmologies with a time-varying $\chi$ the scaling solution predicts in this approximation the absence of a time variation of the Fermi scale in units of the Planck mass.
Additional small contributions to the $\tilde{\rho}$-dependence of $h_0$ for a Yukawa coupling $y_t$ depending on $\tilde{\rho}$ or $h$ will be discussed later.
The situation remains similar once the other dimensionless couplings of the standard model are included. 
The conclusion that cosmologies with a time variation of $\chi$ encounter no problem from this side is rather robust.
It does not depend on details of the scaling potential.

\subsection*{Reduced cosmon potential}
\label{sec:RCP}
The reduced cosmon potential evaluates the effective potential for $h$ at the location of the partial minimum $h_0(\tilde{\rho})$,
\begin{equation}
    \label{eq:RCP47}
    \bar{u}(\tilde{\rho}) = u (\tilde{\rho},h_0(\tilde{\rho}))\,.
\end{equation}
We want to derive an effective scaling equation for the $\tilde{\rho}$-dependence of $\bar{u}(\tilde{\rho})$. For this purpose we show that for the IR-region the scalar mass matrix becomes block diagonal to a very good approximation, with one of the eigenvalues computed in terms of derivatives of $\bar{u}(\tilde{\rho})$. All particles except neutrinos, photon, graviton and cosmon decouple effectively for the scaling equation for $\bar{u}(\tilde{\rho})$.

\indent The minimum condition,
\begin{equation}
    \label{eq:RCP48}
    \partial_h u(\tilde{\rho}, h_0(\tilde{\rho}) = 0\,,
\end{equation}
holds for all $\tilde{\rho}$.
For the $\tilde{\rho}$-derivatives of $\bar{u}(\tilde{\rho})$ one finds
\begin{equation}
    \label{eq:RCP49}
    \partial_{\tilde{\rho}} \bar{u}(\tilde{\rho}) = \partial_{\tilde{\rho}} u(\tilde{\rho}, h_0) + \partial_h u(\tilde{\rho}, h_0) \partial_{\tilde{\rho}} h_0 = \partial_{\tilde{\rho}} u(\tilde{\rho},h_0)\,,
\end{equation}
and
\begin{align}
    \partial_{\tilde{\rho}}^2 \bar{u}(\tilde{\rho}) &= \partial_{\tilde{\rho}}^2 u(\tilde{\rho}, h_0) 
    + 2\partial_{\tilde{\rho}} \partial_h u(\tilde{\rho}, h_0) \,\partial_{\tilde{\rho}} h_0 \notag\\
    &\quad + \partial_h^2 u(\tilde{\rho}, h_0) (\partial_{\tilde{\rho}} h_0)^2\,.
\end{align}
Insertion of eq.~\eqref{eq:FS41} yields
\begin{equation}
    \label{eq:RCP50b}
    \partial^2_{\tilde{\rho}}\bar{u}(\tilde{\rho}) = \partial_{\tilde{\rho}}^2 u(\tilde{\rho}, h_0) - \frac{1}{\hat{\lambda}_h}\Big(\partial_{\tilde{\rho}} \partial_h u(\tilde{\rho},h_0)\Big)^2 \,.
\end{equation}
The term $\sim \hat{\lambda}_h^{-1}$ can be shown to reflect the effective diagonalisation of the scalar mass matrix. 

\indent One may compute the cosmon mass term from the reduced cosmon potential
\begin{align}
    \label{eq:RCP51}
    \bar{m}_c^2
    = &\partial_\rho \bar{u} + 2\tilde{\rho} \partial_{\tilde{\rho}}^2 \bar{u}
    = \partial_{\tilde{\rho}}u(\tilde{\rho},h_0)  \nonumber \\    &+2\tilde{\rho}\partial^2_{\tilde{\rho}}u(\tilde{\rho},h_0) - \frac{2\tilde{\rho}}{\hat{\lambda}_h} (\partial_{\tilde{\rho}} \partial_h u)^2\,,
\end{align}
where all derivatives are evaluated at $h=h_0(\tilde{\rho})$ and $\tilde{m}^2_c$ is given by eq.~\eqref{eq:SF35}.
This may be compared with the eigenvalue $\tilde{m}_-^2$ in eq.~\eqref{eq:SF36} for the case $\tilde{m}_{rc}^4 \ll (\hat{m}_r^2 - \tilde{m}_c^2)^2$,
\begin{equation}
    \label{eq:RCP52}
    \tilde{m}_-^2 = \tilde{m}_c^2 - \frac{\tilde{m}_{rc}^4}{\tilde{m}_r^2 - \tilde{m}_c^2}\,,
\end{equation}
where $\tilde{m}_c^2$ is converted to $\tilde{\rho}$-derivatives at fixed $h$,
\begin{equation}
    \label{eq:74A}
    \tilde{m}_c^2 = \left(\partial_{\tilde{\rho}} + 2\tilde{\rho} \partial_{\tilde{\rho}}^2\right) u(\tilde{\rho},h_0) + h_0 \tilde{m}_r^2 - 4 h_0 \partial_{\tilde{\rho}} \partial_h u(\tilde{\rho},h_0)\,.
\end{equation}
For $\tilde{m}_c^2 \ll \tilde{m}_r^2$, $\tilde{m}_r^2 = 2\hat{\lambda}_h h_0 / \tilde{\rho}$ the mass eigenvalue $\tilde{m}_-^2$ coincides with $\bar{m}_c^2$.
In the limit where $\tilde{m}_r^2$ is the dominant entry in the mass matrix \eqref{eq:SF35} we can replace $\tilde{m}_-^2$ by $\bar{m}_c^2$ and $\tilde{m}_+^2$ by $\tilde{m}_r^2$.
In this region, which is the relevant one for the IR-limit, we can solve self-consistently the scaling equation for $\bar{u}(\tilde{\rho})$. For the contribution for fluctuations of fermions or gauge bosons one inserts for $h$ the partial minimum $h_0(\tilde{\rho})$.

\indent For cosmological epochs long after the electroweak phase transition the invariant $h$ has settled to the partial minimum at $h_0(\tilde{\rho})$.
One can therefore work with the reduced cosmon potential.
The flow equation for the reduced cosmon potential $\bar{u}(\tilde{\rho})$ equals the flow equation for $u(\tilde{\rho},h)$ evaluated for $h=h_0(\tilde{\rho})$.
The same holds for the differential equation for the scaling solution which evaluates eq.~\eqref{eq:FESS12} for $h=h_0(\tilde{\rho})$.
Inserting $h_0(\tilde{\rho})$ in eq.~\eqref{eq:FESS7} the fermion mass term can be cast into an effective Yukawa coupling for $\chi$ \cite{DEQG},
\begin{equation}
    \label{eq:RCP53}
    \tilde{m}_f^2 = y_f^2 h_0 \tilde{\rho} = 2 h_f^2 \tilde{\rho}\,.
\end{equation}
This holds in a similar way for the massive gauge bosons, while for the Higgs boson one employs $\tilde{m}_r^2 = 2 \lambda_h h_0 \tilde{\rho}$.

\indent The scaling solution for the reduced cosmon potential $\bar{u}(\tilde{\rho})$ has been discussed in detail in ref \cite{DEQG}.
Extending the range of $\tilde{\rho}$ to values of the order one the contribution of the metric fluctuations becomes more complicated than the simple term $1/(64\pi^2)$ in $c_U$.
Also the scaling solution for $U$ has to be combined with a scaling solution for $F$.
Furthermore, for $h_0\tilde{\rho}\lesssim 1$ the partial minimum $h_0(\tilde{\rho})$ may no longer be taken to be approximately constant.
Nevertheless, for $\tilde{\rho} < 1$ only the fluctuations of the standard model particles with their known couplings are needed.
The scaling solution $\bar{u}(\tilde{\rho})$ gives rise to an interesting cosmology with dynamical dark energy \cite{DEQG,CWQGSS}.

\subsection*{Boundary conditions}
\label{sec:BC}
Possible predictions of the scaling solution for the cosmological constant, the Fermi scale or the quartic scalar coupling depend on restrictions for the boundary function $L(h)$.
For the reduced cosmon potential this reduces to the constant $L_0 = L(h_0)$.
General properties of the scaling solution for the reduced cosmon potential in the limit of large $\tilde{\rho}$ are very similar to dilaton quantum gravity \cite{DQG1, DQG2}.
Only the fluctuations of the massless photon have to be added for very large $\tilde{\rho}$.
A common scaling solution for $\bar{u}(\tilde{\rho})$ and $\bar{f}(\tilde{\rho}) = F(\tilde{\rho}, h_0(\tilde{\rho})) / k^2$ has only been found for $L_0 = 0$.
This reflects a general restriction that for $\chi\to\infty$ the potential cannot grow faster than $F$ \cite{CWGFC}

\indent The condition $L_0 = 0$ translates to a relation between the couplings $\bar{\lambda}_\chi$, $\bar{\lambda}_m$ and $\bar{\lambda}_h$ in eq.~\eqref{eq:FS29}, namely $\delta = 0$ in eq.~\eqref{eq:FS33}.
As a consequence, the reduced cosmon potential becomes flat for $\tilde{\rho} \to \infty$,
\begin{equation}
    \label{eq:BC54}
    \lim_{\tilde{\rho} \to \infty} \bar{u}(\tilde{\rho}) = u_\infty = \frac{5}{128\pi^2}\,,
\end{equation}
where the factor $5$ accounts for the massless graviton, photon and cosmon.
We emphasize that the flat direction reflected by eq.~\eqref{eq:BC54} corresponds precisely to the partial minimum $h=h_0(\tilde{\rho})$.
For more general $h$ there is no similar property requiring $L(h)=0$.
Only the increase of $L(h)$ with $h$ will be bounded for $h\to\infty$.

\indent A potential $U=u_\infty k^4$ combined with $F = \xi \chi^2$ results in a decrease of the potential in the Einstein frame with increasing $\chi$,
\begin{equation}
    \label{eq:BC55}
    U_E = \frac{u_\infty h^4 M^4}{\xi^2\chi^4}\,.
\end{equation}
"Runaway cosmologies" for which $\chi$ diverges in the infinite future lead then to a dynamical solution of the cosmological constant problem.

\indent For very large $\tilde{\rho}$ or $\tilde{h} \gg 1$ the potential is dominated by the boundary term.
With our boundary condition $\bar{m}_c^2 = \tilde{m}_-^2$ vanishes.
At the partial minimum of $u$ with respect to variations of $\tilde{h}$ or $h$ the potential does not depend on $\tilde{\rho}$ , corresponding to a flat direction in the space of the two field-invariants $\tilde{\rho}$ and $h$ or $\tilde{\rho}$ and $\tilde{h}$.
In a cosmological context the cosmon will acquire an additional mass due to the coupling of $\chi$ to the curvature scalar $R$.
This cosmon mass is proportional to the Hubble parameter and therefore tiny.

\indent A second set of boundary restrictions arises from the extension of the scaling solution to $\tilde{\rho} \to 0$.
For fixed fields $\chi$ and $H$ this corresponds to the ultraviolet (UV) limit $k\to\infty$.
We define the dimensionless quartic couplings at the UV-fixed point by
\begin{align}
\label{eq:BC56}
    \lambda_{\chi,0} &= 4 \frac{\partial^2 U}{(\partial\chi^2)^2}\,,\quad
    \lambda_{m,0} = 2 \frac{\partial^2 U}{\partial \chi^2 \partial (H^\dagger H)}\,,\nn\\
    \lambda_{h,0} &= \frac{\partial^2 U}{\partial (H^\dagger H)^2}\,,
\end{align}
evaluated at $\chi = 0$, $H^\dagger H = 0$.
In perturbation theory, these couplings would be marginal.
The graviton fluctuations induce, however, an anomalous dimension which renders all quartic scalar couplings irrelevant.
For $k\to \infty$ they take the values at the UV-fixed point.
If gauge and Yukawa couplings vanish in the ultraviolet limit for $k\to \infty$ or $\tilde{\rho} \to 0$ the couplings $\lambda_{\chi,0}$, $\lambda_{m,0}$ and $\lambda_{h,0}$ vanish.
For a fixed point with non-zero gauge and Yukawa couplings the fixed point value of $\lambda_h$ differs from zero.
It remains very small, however, of the size $y_t^4/(16\pi^2)$.
Computing the flow of $\lambda_h$ from $\tilde{\rho} = 0$ to $\tilde{\rho} \to \infty$ yields $\bar{\lambda}_h$: 
This gives rise to the quantum gravity prediction for the Higgs boson mass \cite{SHAW}.

\indent What remains open at this stage is the relation between $\lambda_{m,0}$ and $\bar{\lambda}_m$.
This key issue for the understanding of the Fermi scale will be addressed in the later parts of this paper.

\section{Field-dependent Yukawa and gauge couplings}
\label{sec:V}
The zero-temperature transition in the standard model is not precisely a phase transition of second order.
Due to the running of dimensionless couplings the Fermi scale $\varphi_0$ cannot be arbitrarily small.
Chiral symmetry breaking in QCD implies quark-antiquark condensates which break spontaneously the electroweak symmetry.
The physical Higgs doublet becomes a mixture of the ``fundamental'' Higgs field and mesons with the same quantum number.
This effect is small if the Fermi scale $\varphi_0$ is much larger than the scale of chiral symmetry breaking in QCD or the related confinement scale $\Lambda_{\QCD}$.
It becomes important however, for $\varphi_0$ close to $\Lambda_{\QCD}$, and finally leads to a lower bound for $\varphi_0$.

\indent A second important effect of running gauge and Yukawa couplings is the generation of new interactions. 
These vertices result from the field dependence of these couplings. 
This field dependence arises necessarily for the scaling solution and does not depend on a given truncation or other approximation.
For zero momentum, the fluctuations which induce running couplings are cut off by the infrared scale $k$. 
As a consequence, the couplings depend on $k$ as long as $k$ is in a range where mass terms can be neglected. 
For small couplings, the running with $k$ at fixed fields $H$ and $\chi$ is governed by the standard perturbative $\beta$-function.

For a scaling solution, the couplings can only depend on the
dimensionless quantities $\frac{k^2}{\chi^2}$ and $ \frac{k^2}{H^\dagger H}$ --
no further explicit $k$-dependence is allowed by the very definition of the scaling solution. 
If couplings depend on $k$ at fixed fields, they will depend on $\tilde{\rho}$ or $\tilde h$ and therefore be field-dependent.
For fixed $\tilde h$, the $\tilde{\rho}$-dependence of the gauge and
Yukawa couplings follows directly from their $k$-dependence. 
We therefore only need to understand the $h$-dependence at fixed $\tilde{\rho}$ or $\htilde$. 
In other words, one has to determine which part of the flow arises from the $\tilde{\rho}$-dependence at fixed $\tilde h$, and which part is due to the $\tilde h$-dependence at fixed $\tilde{\rho}$. 
This issue may involve assumptions about the ultraviolet behavior which we will discuss in the next section.

\indent If gauge or Yukawa couplings depend both on $\tilde{\rho}$ and $\tilde h$, this will involve effective couplings between $\chi$ and $H$. 
These couplings contribute to the flow of the cosmon-Higgs coupling $\lambda_m$. 
They will affect the precise location of the critical surface for the (almost) second order electroweak quantum phase transition.
Since running couplings are responsible for turning the phase transition into a crossover, one has to investigate whether this violation of particle scale symmetry
could induce a departure from the critical surface already for scales substantially above the confinement scale of QCD. 
In principle, the violation of particle scale symmetry by running gauge and Yukawa couplings could even generate the observed Fermi scale. 
Our findings indicate that this is not the case. 
As long as the running couplings remain perturbative, they only induce a modification of the location of the critical surface in coupling constant space.

\indent We discuss here only vertices which arise from terms as $Z_F(\tilde{\rho},\tilde h) F^{\mu\nu}F_{\mu\nu}$ for the gauge couplings, and similar terms for Yukawa couplings. 
In our truncation they can be incorporated in a rather direct way into the flow equation of the effective potential by allowing the gauge and Yukawa couplings to depend on $\tilde{\rho}$ and $\tilde h$. 
In principle, other terms with more complex structures of derivatives are generated by the functional flow as well. 
We see no good reason why such additional terms should change the qualitative behavior. 
We actually find that their effect on the flow of $u$ is likely to be subleading compared with the vertices retained here.
The one-loop structure of the exact flow equation for $u$ and the decomposition into separate contributions from different representations of the Lorentz and gauge groups largely fixes the structure. 
Beyond field-dependent mass terms, one may have corrections to the momentum
dependence -- a sort of field-dependent and momentum dependent anomalous dimension. 
These anomalous dimensions are small already in leading order perturbation theory.

\indent In summary, one expects that the effects of running couplings lead to deviations of the scaling solution for the effective potential from the form \eqref{eq:PTO1}.
These effects should become important as $\varphi_0$ comes close to $\Lambda_{\QCD}$.
We will see, however, that they remain rather moderate in the range where $H^\dagger H \gg \Lambda_{\QCD}^2$, as relevant for the observed Fermi scale.
In this range the main modification will simply replace in eq.~\eqref{eq:N2} the constant dimensionless couplings by field-dependent Yukawa and gauge couplings.
In the following we include the effects of running dimensionless couplings for the scaling form of the effective potential.

\subsection*{Strong gauge coupling}
Let us start with the gauge coupling $g_3$ of QCD.
At fixed fields $H$ and $\chi$ its dependence on the renormalization scale $k$ follows for $k^2 \ll \xi \chi^2$ the perturbative flow of the standard model.
In the one loop approximation one finds
\begin{equation}
    \label{eq:GCA1}
    k \partial_k g_3^2 = \bar{\beta}_3 g_3^4\,,
\end{equation}
with
\begin{equation}
    \label{eq:GCA2}
    \bar{\beta}_3 = -\frac{33}{24\pi^2} + \frac{1}{12\pi^2} \sum_q \frac{k^2}{k^2 + m_q^2}\,.
\end{equation}
In this approximation the only "non-perturbative" effect of functional renormalization is the decoupling of the quarks from the flow once $k$ is much smaller than the quark mass $m_q$.
As an effect, the effective gauge coupling will be field-dependent.
The threshold function for the decoupling of heavy particles used in eq.~\eqref{eq:GCA2} is only approximative. (The true threshold function leads to a stronger suppression for $m^2_q/k^2 \gg 1$.) We use it here since it accounts for all qualitative features and permits a simple analytic treatment. 

\indent Within functional renormalization the gauge couplings become field-dependent functions. One adds to the effective action \eqref{eq:ESP1} a gauge invariant kinetic term for the gauge bosons
\begin{equation}
    \label{eq:GCA}
    \Gamma_k^{(g)} = \frac{1}{4}\int_x\sqrt{g'}Z_F F^{\mu\nu} F_{\mu\nu},
\end{equation}
with $F_{\mu\nu}$ the gauge invariant field strength tensor. The function $Z_F(\chi, H)$ is a function of the scalar fields. In turn, the field-dependent gauge coupling is defined by
\begin{equation}
    \label{eq:GCA2b}
    Z_F(\chi,H) = g^{-2}(\chi,H),
\end{equation}
Similar to the scalar effective potential the scaling solution for $Z_F$ or $g^{-2}$ depends on $\tilde{h}$ and $h$ or $\tilde{\rho}$ and $h$, but not separately on $k$.

\indent For finding the scaling solution one may investigate the flow of $g_3(k;\tilde h, h)$ at fixed $\tilde h$ and $h$.
The scaling solution $g_3(\tilde h,h)$ obtains from $k\partial_k g_3(k,\tilde h,h)=0$.
Thus the scaling solution is only a function of $\htilde$ and $h$.
It obeys at fixed $h$ the differential equation
\begin{equation}
    \label{eq:GC1}
    \tilde h \partial_{\tilde h} g_3^2 = -\frac{1}{2} \bar{\beta}_3 g_3^4\,.
\end{equation}
At fixed $h$ we may turn to $g_3(\tilde \rho, h)$ using $\tilde h\partial_{\tilde h} = \tilde \rho \partial_{\tilde \rho}$.
For $g_3(\tilde \rho, h)$ eq.~\eqref{eq:GC1} yields then the form of the scaling equation of ref.~\cite{CWSSFGC}.

\indent For $\tilde h\gg h$ or $\tilde \rho \gg 1$ the fluctuations of the metric and possible other heavy particles (as for grand unified theories or other extensions of the standard model) are already decoupled.
In this range $\bar{\beta}_3$ can be well approximated by the known perturbative $\beta$-function for the running strong gauge coupling, whereby functional renormalization provides for the appropriate thresholds.
In one loop order one has
\begin{align}
    \label{eq:GC2}
    \bar{\beta}_3 &= -\frac{33}{24\pi^2} + \frac{1}{12\pi^2} \sum_q \frac{1}{1 + \tilde{m}_q^2} \nn\\
		  &= -\frac{1}{24\pi^2} \left(33-2\sum_q \frac{1}{1 + y_q^2 \tilde h}\right)\,.
\end{align}

\indent At fixed $h$ we can solve the flow equation \eqref{eq:GC1} starting at $\tilde h = \bar h$, with $\bar h$ and $\tilde{h}$ much larger than $h$ and much smaller than one, such that neither Planck-scale physics nor the quark mass thresholds are important.
In this range one has
\begin{equation}
    \label{eq:GC3}
    \tilde h \partial_{\tilde h} g_3^{-2} = -\frac{7}{16 \pi^2}\,.
\end{equation}
The solution,
\begin{equation}
    \label{eq:GC4}
    \frac{1}{g_3^2} = \frac{1}{\bar{g}_3^2} - \frac{7}{16\pi^2} \ln \frac{\tilde h}{\bar h} = \frac{7}{16\pi^2}\ln \frac{\htilde_c}{\tilde h}\,,
\end{equation}
is parametrized by $\bar{g}_3^2 = g_3^2(\tilde h = \bar h, h)$.
Here
\begin{equation}
    \label{eq:GC5}
    \htilde_c = \exp \left(\frac{16\pi^2}{7\bar{g}_3^2}\right)\bar{h}
\end{equation}
is the value of $\htilde$ at which $g_3(\tilde h, h)$ diverges in this approximation.
This approximation requires $\tilde{h}_c \ll 1$.
For a realistic confinement scale below the Fermi scale the quark mass threshold matter, see below.

\subsection*{Confinement scale}
The ``confinement scale'' $k_c$ at which the strong gauge coupling diverges is proportional to $H$,
\begin{equation}
    \label{eq:GCA3}
    k_c^2 = \frac{H^\dagger H}{\tilde h_c} = \frac{H^\dagger H}{\bar{h}} \exp \left(- \frac{16\pi^2}{7 \bar{g}_3^2}\right)\,.
\end{equation}
Using $\tilde{\rho}_c = \htilde_c / h$ one sees that for fixed $h$ the confinement scale is also proportional to $\chi$ \cite{CWQ},
\begin{equation}
    \label{eq:GC6}
    k_c^2 = \frac{\chi^2}{2\tilde{\rho}_c}\,.
\end{equation}
This proportionality of the confinement scale and $\chi$ is a general property of the scaling solution.
A dependence of $\bar{g}^2_3$ on $h$ introduces a further dependence of $\Lambda_\QCD$ on $H^\dagger H/\chi^2$.
This will not change the proportionality $k_c \sim \chi$ once $h$ settles to a fixed value $h_0$.

\indent If $\bar{g}_3^2$ is independent of $h$ the scaling solution for $g_3^2$ only depends on $\tilde h$, and does not depend on $h$ separately.
This is similar for the other dimensionless couplings. 
For Yukawa couplings $y_q(\tilde h)$ not depending on $h$ separately the mass term $y_q^2 \tilde h$ in eq.~\eqref{eq:GC2} depends only on $\tilde h$.
As a consequence, the scaling solution for $g_3^2(\tilde h)$ remains only a function of $\tilde h$ even for the region where the mass thresholds for the quarks matter.
Important modification of this situation will be connected to the $h$-dependence of $\bar{g}^2_3$.

\indent For a more quantitative treatment of the thresholds we approximate $y_q(\tilde h)$ by its value for $\tilde{m}_q = 1$, e.g.
\begin{equation}
    \label{eq:GCA4}
    \bar{y}_q = y_q\left(\tilde h= \bar{y}_q^{-2}\right)\,.
\end{equation}
This yields
\begin{equation}
    \label{eq:GCA5}
    \tilde h\partial_{\tilde h} g_3^{-2} = -\frac{33}{48\pi^2} + \frac{1}{24\pi^2} \sum_q \frac{1}{1 + \bar{y}_q^2 \tilde h}\,.
\end{equation}
(For a suitable choice of $\bar{y}^2_q$ eq.~\eqref{eq:GCA5} may also be used for a more accurate threshold function in eq.~\eqref{eq:GCA2}.) The solution reads
\begin{equation}
    \label{eq:GCA6}
    g_3^{-2}(\tilde h) = \frac{1}{\bar{g}_3^2} - \frac{7}{16\pi^2}\ln \bigg(\frac{\tilde h}{\bar h}\bigg) - \frac{1}{24\pi^2} \sum_q \ln \left(\frac{1 + \bar{y}_q^2 \tilde h}{1 + \bar{y}_q^2 \bar h}\right).
\end{equation}
For a given $\bar{g}_3^2 = g_3^2 (\tilde h = \bar h)$ we can again determine the value $\htilde_c$ for which $g_3^2$ diverges, and determine the confinement scale
\begin{equation}
    \label{eq:GCA7}
    \Lambda_{\QCD}^2 = \frac{H^\dagger H}{\htilde_c}\,.
\end{equation}

In the limit where $\bar{g}_3^2$ does not depend on $h$ or $\tilde{\rho}$ the value $\tilde{h}_c$ does not depend on $h$ or $\tilde{\rho}$. Then the confinement scale $\Lambda_{\QCD}$ is proportional to $H$.
\indent As long as the gauge and Yukawa couplings only depend on $\tilde h$ without explicit dependence on $h$ the running couplings will not induce deviations of the scaling potential from the form \eqref{eq:PTO1}
\begin{equation}
    \label{eq:GC8}
    u = \bar{u}_1(\tilde h) + \bar{u}_2(\tilde\rho) + \bar{\lambda}_m \htilde \tilde{\rho}\,.
\end{equation}
In this limit the coupling $\lambda_m$ will be independent of $\tilde h$ or $\tilde \rho$, given by the boundary value $\bar{\lambda}_m$.
The value $\bar{\lambda}_m = 0$ determines again the critical surface of a second order phase transition.
In turn, the Fermi scale vanishes for $\bar{\lambda}_m = 0$.
For $k\to 0$ the scaling potential is determined in this case uniquely by the boundary terms, similar to eq.~\eqref{eq:FS30}
\begin{equation}
    \label{eq:GCA8}
    U = \frac{1}{8} \bar{\lambda}_\chi \chi^4 + \frac{1}{2} \bar{\lambda}_m \chi^2 H^\dagger H + \frac{1}{2} \bar{\lambda}_h (1+\Delta) (H^\dagger H)^2\,.
\end{equation}

\subsection*{Field-dependence of couplings and confinement scale}
\indent A modification of this situation, leading to a non-trivial field-dependence of $\lambda_m$, arises from the dependence of the dimensionless couplings on $h$ besides their dependence on $\htilde$.
We may demonstrate this by the dependence of the strong gauge coupling and the associated confinement scale on $h$.
This is connected to the mass thresholds.

\indent Let us consider a value $\tilde{\rho}_0$ sufficiently large as compared to one, such that gravitational fluctuations can be neglected, and not too large, such that the mass thresholds do not play a role for the beta function.
The scaling solution is parametrized by a fixed value of $g_3$ at $\tilde{\rho}_0$ which is independent of $h$,
\begin{equation}
    \label{eq:LAM1}
    g_3^{-2}(\tilde{\rho}_0) = g_{3,0}^{-2}\,.
\end{equation}
This assumption plays an important role for determining the field-dependence of couplings and the induced flow of the cosmon-Higgs coupling $\lambda_m$. 
In a region of $k^2$ much above the field value $H^\dagger H$ it seems natural that the value of $H$ or $\tilde h$ plays no role. 
The gauge coupling then depends only on the ratio of $k$ over the effective Planck mass $\chi$, and therefore on $\tilde{\rho}$.
In other words, a change of the value of $H^\dagger H/\chi^2$ plays no role and one has in this region
\begin{equation}
\partial_{h}\,
g_3^{-2}(\tilde{\rho},h)
\approx 0\,.
\label{eq:140A}
\end{equation}
We can then fix parameterize the gauge coupling by its value at some fixed $\tilde{\rho}_0$ in this region, which is the content of assumption \eqref{eq:LAM1}. 
We will further motivate this relation by a discussion of the ultraviolet physics in sect.~\ref{sec:VI}.

\indent From \eqref{eq:LAM1} we can extrapolate the value of $g_3^2(\tilde{\rho}, h)$ to larger values of $\tilde{\rho}$. 
A dependence on $h$ will be induced once $\tilde{\rho}$ grows large enough such that the particle masses proportional $\sim \tilde h$ start to matter.
One expects $g_3^2(\tilde{\rho},\tilde h)$ to be approximately independent of $\tilde h$ for $\tilde h\gg1$, and to develop a $\tilde h$-dependence for $\tilde h\lesssim1$.

\indent We approximate the mass thresholds by taking $\bar{\beta}_3$ stepwise constant,
\begin{equation}
    \label{eq:LAM2}
    \bar{\beta}_3 = - \frac{c}{24\pi^2}\,,
\end{equation}
with $c=21$ as long as $\tilde{m}_t^2 < 1$, $c=23$ for the range where $\tilde{m}_b^2 < 1$, $\tilde{m}_t^2 > 1$, $c=25$ if $\tilde{m}_{ch}^2 < 1$, $\tilde{m}_b^2 > 1$, and $c=27$ if $\tilde{m}_{ch}^2 > 1$.
The value of $\tilde{\rho}$ where $\tilde{m}_t^2 = 1$ corresponds to
\begin{equation}
    \label{eq:LAM3}
    y_t^2 h \tilde{\rho}_t = 1\,,\quad
\tilde{\rho}_t = \frac{1}{y_t^2 h}\,,
\end{equation}
and similar for $\tilde{\rho}_b$ and $\tilde{\rho}_{ch}$.
For $\htilde \ll 1$ or $\tilde{\rho}\ll \tilde{\rho}_t$ the coupling $g^2_3$ depends only on $\tilde{\rho}$.
A dependence on $h$ is only introduced by the quark mass threshold for $\tilde{\rho} \gtrsim \tilde{\rho}_t$.

\indent The value of the strong gauge coupling for $\tilde{\rho} > \tilde{\rho}_{ch}$ is given by
\begin{align}
    \label{eq:LAM4}
    g_3^{-2}(\tilde{\rho}) = g_{3,0}^{-2} &- \frac{7}{16\pi^2} \ln \frac{\tilde{\rho}_t}{\tilde{\rho}_0} - \frac{23}{48\pi^2} \ln \frac{\tilde{\rho}_b}{\tilde{\rho}_t} \nn\\
			    &- \frac{25}{48\pi^2} \ln \frac{\tilde{\rho}_{ch}}{\tilde{\rho}_b} - \frac{9}{16\pi^2} \ln \frac{\tilde{\rho}}{\tilde{\rho}_{ch}}\,.
\end{align}
With the definition
\begin{align}
    \label{eq:LAM5}
    \hat{g}_3^{-2} = g_{3,0}^{-2} &+ \frac{7}{16\pi^2} \ln \frac{y_t^2}{y_{ch}^2} - \frac{23}{48\pi^2} \ln \frac{y_t^2}{y_b^2} \nn\\
		   &-\frac{25}{48\pi^2} \ln \frac{y_b^2}{y_{ch}^2} - \frac{1}{8\pi^2} \ln y_{ch}^2\,,
\end{align}
one has
\begin{align}
    \label{eq:LAM6}
    g_3^{-2}(\tilde{\rho},h) = \hat{g}_3^{-2} - \frac{1}{8\pi^2} \ln (h\tilde{\rho}_0) - \frac{9}{16\pi^2} \ln \frac{\tilde{\rho}}{\tilde{\rho}_0}\,.
\end{align}
Here we take for the Yukawa couplings the values they have at the corresponding mass threshold.
These values depend on $h$, but the corresponding $h$-dependence of $\hat{g}_3^{-2}$ is subleading.

\indent In the approximation of massless $u$, $d$, $s$ quarks the critical $\tilde{\rho}_c$ at which $g_3$ diverges is given by
\begin{align}
    \label{eq:LAM7}
    \ln \frac{\tilde{\rho}_c}{\tilde{\rho}_0} &= \frac{16\pi^2}{9\hat{g}_3^2} - \frac{2}{9} \ln(h\tilde{\rho}_0)\,,\nn\\
    \tilde{\rho}_c &= \exp \left(\frac{16\pi^2}{9\hat{g}_3^2}\right) h^{-\frac{2}{9}} \tilde{\rho}_0^{\frac{7}{9}}\,.
\end{align}
It decreases with increasing $h$.
The corresponding confinement scale obeys
\begin{equation}
    \label{eq:LAM8}
    \Lambda_{\QCD}^2 = \frac{\chi^2}{2 \tilde{\rho}_c} = \exp \left(-\frac{16\pi^2}{9\hat{g}_3^2}\right)(H^\dagger H)^{\frac{2}{9}} k_0^{\frac{14}{9}}\,,
\end{equation}
with normalization scale $k_0$ proportional to $\chi$ as defined by
\begin{equation}
    \label{eq:LAM9}
    \tilde{\rho}_0 = \frac{\chi^2}{2 k_0^2}\,.
\end{equation}
The confinement scale increases with increasing $H^\dagger H$ since the heavy quarks decouple at a higher scale.
At fixed $\chi$ and $H$ the scale $k_0$ is the one where $g_3$ is fixed
\begin{equation}
    \label{eq:LAM10}
    g_3^2(k_0) = g_{3,0}^2\,.
\end{equation}

\indent In summary, the field-dependent confinement scale is given approximatively by
\begin{equation}
    \label{eq:145A}
    \Lambda^2_\QCD = \exp\left( -\frac{16\pi^2}{9g^2_{3,0}} \right) \left( 2\tilde{\rho}_0 \right)^{-\frac{7}{9}} A_q \left( H^\dagger H \right)^{\frac{2}{9}} \chi^{\frac{14}{9}}\, ,
\end{equation}
with
\begin{align}
A_q  
&= \exp\Bigg\{
-\frac{1}{27}\Big(
  21\ln \frac{y_t^2}{y_{ch}^2}
 -23\ln \frac{y_t^2}{y_b^2}
 -25\ln \frac{y_b^2}{y_{ch}^2}
\notag\\
&\qquad\qquad\qquad\quad
 -6\ln y_{ch}^2
\Big)
\Bigg\}\,.
\label{eq:145B}
\end{align}
The dependence $\sim \left(H^\dagger H\right)^{2/9}$ is valid for the range of $h$ for which the charm-quark mass is above the confinement scale, while the strange-quark mass is of the order of or below the confinement scale. 
For this region, the power $2/9$ does not depend on the precise truncation or thresholds.
For values of $h$ outside this region one can modify the discussion, starting again from \eqref{eq:LAM1} and taking mass thresholds into account as $\tilde{\rho}$ increases beyond $\tilde{\rho}_0$.

\indent The important general lesson is that the gauge and Yukawa couplings depend both on $\tilde{\rho}$ and on $\tilde h$ in the region $\tilde h\lesssim1$, while they become essentially independent of $\tilde h$ for $\tilde h\ll1$.
The matching to the observed values may be done for $k$ at the top-quark mass, $\tilde m_t=1$, or $\tilde h\approx1$.
For $\tilde h\ll1$ the effective top-quark mass depends on $\tilde{\rho}$ through the $\tilde{\rho}$-dependence of the Yukawa coupling, $\tilde m_t^2 (\tilde \rho, \htilde)=y_t^2(\tilde{\rho})\,\tilde h$.
The $\tilde{\rho}$-dependence of the Yukawa coupling will lead to a non-trivial dependence of the cosmon-Higgs coupling $\lambda_m$ on $\tilde{\rho}$ and therefore modify the precise location of the critical surface for the electroweak quantum phase transition. 
For $\tilde h\ll1$, the $\tilde{\rho}$-dependence of the gauge and Yukawa couplings will be given by the perturbative $\beta$-functions and is therefore rather robust. 
We also observe that assumption \eqref{eq:LAM1} permits a smooth transition to the symmetric point $\htilde=0$, $h=0$.
The $\tilde{\rho}$-dependence of the couplings corresponds then to the standard model with massless quarks and leptons.

\subsection*{Non-perturbative strong interaction effects}
\indent The strong interactions of QCD break the electroweak gauge symmetry non-perturbatively. 
They induce a lower bound for the masses of W- and Z-bosons. 
This is seen most directly by the chiral condensate of quark-antiquark pairs for the light quarks. These condensates have the same transformation properties as the Higgs doublet. 
A non-zero condensate breaks the electroweak symmetry spontaneously. 
This induces constituent quark masses of the order of the confinement scale, which are substantially larger than the current masses for the up and down quarks. 
Couplings to gauge bosons are universal via covariant derivatives. The kinetic term for the $\bar{q}q$-composites therefore produces contributions to the W- and Z-boson masses.

\indent The nonvanishing masses of W- and Z-bosons indicate that the electroweak transition cannot be a second order phase transition. There exists no symmetric phase for which the W- and Z-bosons would be massless. We rather deal with a crossover from a SSB-regime, where the W- and Z-boson masses are set by the negative scalar mass term $\mu_h$, to a regime where they are given by $\Lambda_{\text{QCD}}$ independently of the parameters. This generic behaviour has been demonstrated for one-flavour QCD by the use of $k$-dependent field transformations or "flowing bosonisation" \cite{Gies_2002, Gies_2004}. This method can account for the interplay between quark-antiquark condensates and the Higgs doublet.

\indent For the standard model one expects a lower bound for the Fermi scale $\varphi_0$ and the associated particle masses. This can be established by employing flowing bosonisation for the coupled system of top-antitop composite and the Higgs doublet. 
We will find that the lower bound for $\varphi_0$ is substantially below the observed Fermi scale. 
This lower bound for $h_0$ is found approximately at $(\Lambda_\QCD^{(6)}/\chi)^2$, where $\Lambda_\QCD^{(6)}$ is the confinement scale for six massless quark flavours, which is somewhat smaller than the confinement scale of the standard model.
With $\Lambda_\QCD^{(6)}/\chi \approx 50 \, \text{MeV}/M_p$ the lower bound for $h_0$ is smaller than the observed value by a factor $\approx 10^{-7}$.
In consequence, for $h_0$ in the region corresponding to observation the non-perturbative QCD-effects from flowing bosonisation are a small correction. 
The simplified picture of a second order quantum phase transition remains a good approximation in this case. Only for $h_0\approx \Lambda^2_{\text{QCD}}/M_p^2$ the strong QCD effects have an important influence. Nevertheless, for $h_0$ in the observed range the QCD-induced top-antitop composite may lead to a small correction of the predicted ratio of Higgs boson mass over top quark mass. This may matter for the quantum gravity prediction of this ratio \cite{SHAW} and the value of "vacuum stability".

\indent The understanding of non-perturbative QCD effects is important because they affect the general nature of the electroweak transition. 
Since at the end the effect is found to be small in the region of $h$ relevant for the observed Fermi scale we deal with the quantitative features of flowing bosonisation in Appendix~\ref{app:A}. 
This also accounts for the non-perturbative masses of the light quarks and the $W$,$Z$-gauge bosons.
The generation of the gluon mass is more complex. 
One possibility is spontaneous color symmetry breaking in the Higgs picture of QCD \cite{CWSBC, CWHP}.

\indent For the region of $h$ corresponding to the observed Fermi scale the non-perturbative QCD-effects are expected to be small for all values of $\tilde\rho$ or $\tilde h$. 
They are further suppressed for $\tilde h\ll 1$, where quarks and gauge bosons are effectively massless.
In view of the qualitative change from a second order phase transition to a crossover we nevertheless want to deal with these effects quantitatively. 
This may also be used to explore the region of field space where $h$ is much smaller than the observed value. 
We employ here a rough approximation which covers the qualitative features.

\indent Non-perturbative strong interaction effects will generate a mass gap through chiral symmetry breaking. These non-perturbative masses are proportional to $\Lambda_{\QCD}$.
Instead of massless gluons and light quarks we take for simplicity a common mass,
\begin{equation}
    \label{eq:LAM11}
    m_{\QCD}^2 = c_s \Lambda_\QCD^2\,,\quad
    \tilde{m}_\QCD^2 = \frac{c_s\tilde{\rho}}{\tilde{\rho}_c}\,.
\end{equation}
In this rough approximation we may take for $m_{\QCD}$ the mass of the $\rho$-meson, $m_{\QCD}\approx 4 \Lambda_\QCD$.
Using this mass in the threshold functions account qualitatively for non-perturbative QCD effects.
This leads to our final approximation for the differential equation defining the scaling solution for the effective potential \eqref{eq:FESS10},
\begin{align}
    \label{eq:LAM12}
    c_U = -\frac{1}{128\pi^2} &\bigg\{ \frac{12}{1 + y_t^2 \htilde} + \frac{12}{1+y_b^2\htilde} + \frac{12}{1+y_{ch}^2\htilde} \nn\\
	+&\, \frac{4}{1+y_{\tau}^2\htilde} + \frac{12}{1+\tilde{m}_s^2} - \frac{6}{1+\tilde{m}^2_W} - \frac{3}{1+\tilde{m}_Z^2} \nn\\
    -&\, \frac{1}{1+\tilde{m}_r^2} - \frac{3}{1+\partial_h u / \tilde{\rho}} + 20\bigg\}\,.
\end{align}
Here we approximate muons, electrons, neutrinos and the cosmon as massless, and employ
\begin{align}
    \label{eq:LAM13}
    \tilde{m}_W^2 &= \frac{g_2^2 \tilde{h}}{2}\,,&
    \tilde{m}_Z^2 &= \left(\frac{g_2^2}{2} + \frac{3g_1^2}{10}\right)\tilde{h}\,, \nn\\
    \tilde{m}_r^2 &= \frac{\partial_h u + 2h\partial_h^2 u}{\tilde{\rho}}\,,&
    \tilde{m}_s^2 &= \tilde{m}_\QCD^2 + y_s^2 \tilde{h}\,.
\end{align}
The contributions of the gluons cancel with the contributions of the up and down quarks.
The role of the strange quark is particular.
For small enough $h$ the current quark mass contribution $y_s^2 \tilde{h}$ is small and $\tilde{m}_s^2$ can be approximated by $\tilde{m}_\QCD^2$.
On the other hand, for large enough $h$ one has $\tilde{m}_s^2 \approx y_s^2 \tilde{h}$.
The actual strange quark mass is found in the transition between the two regimes.

\subsection*{Field dependence of Yukawa coupling}
The field-dependence of the Yukawa coupling for the scaling solution can be estimated in close analogy to the strong gauge coupling.
The flow equation for the dependence of $y_t^2$ on $k$ at fixed $\chi$ and $H$ can be approximated by
\begin{align}
    \label{eq:YU1}
    k \partial_k y_t^2 = \beta_t =& \frac{k^2}{16\pi^2(k^2 + m_t^2)} \bigg[9y_t^4 \nn\\
    &- \left(16 g_3^2 + \frac{9}{2}g_2^2 + \frac{17}{10}g_1^2\right) y_t^2\bigg].
\end{align}
This is the perturbative $\beta$-function, modified by a threshold function for $\tilde{m}_t^2 \gtrsim 1$.

\indent A very rough approximation, which nevertheless covers all qualitative features, takes the ratio
\begin{equation}
    \label{eq:YU2}
    R_t = \left(g_3^2 + \frac{9}{32}g_2^2 + \frac{17}{160}g_1^2\right)/y_t^2
\end{equation}
as constant.
Indeed, in the limit $g_2^2 = g_1^2 = 0$ and neglecting the mass threshold the one loop evolution of the ratio $y_t^2 / g_3^2$ obeys
\begin{equation}
    \tilde{\rho} \partial_{\tilde{\rho}} \left(\frac{y_t^2}{g_3^2}\right) = - \frac{9y_t^2}{32\pi^2}\left(\frac{y_t^2}{g_3^2} - \frac{2}{9}\right)\,,
\end{equation}
with an infrared attractive fixed point at $R_t = 9/2$.
The observed value $R_t \approx 1.9$ at the Z-boson mass is somewhat below this fixed point. 
(For a more detailed justification of approximately constant $R_t$ see ref.~\cite{CWQS}.)

\indent For constant $R_t$ the scaling solution for $y_t^2$ obeys the equation
\begin{equation}
    \label{eq:YU3}
    \tilde{\rho} \partial_{\tilde{\rho}} y^2_{t\;|h} = -\frac{b_t y_t^4}{2(1 + y_t^2 h \tilde{\rho})}\,,\quad
    b_t = \frac{9-16 R_t}{16\pi^2}\,.
\end{equation}
For the observed values of the dimensionless couplings one finds negative $b_t$.
Replacing $1+y_t^2 h \tilde{\rho}$ by $1+\bar{y}_t^2 h \tilde{\rho}$ and taking $b_t$ constant the solution of eq.~\eqref{eq:YU3} is given by
\begin{align}
    \label{eq:YU4}
    y_t^{-2} (\tilde{\rho},h) =&\; y_t^{-2} (\tilde{\rho}_0,h)  \\
			      &+ \frac{b_t}{2} \left[\ln \left(\frac{\tilde{\rho}}{1+\bar{y}_t^2 h \tilde{\rho}}\right) - \ln \left(\frac{\tilde{\rho}_0}{1 + \bar{y}_t^2 h \tilde{\rho}_0}\right)\right]\,,\nn
\end{align}
where we take $\tilde{\rho}_0$ with $\bar{y}_t^2 h \tilde{\rho}_0 \ll 1$, $\tilde{\rho}_0 < \tilde{\rho}$. For a numerical solution we employ eq.~\eqref{eq:YU1}.

\indent Similar to eq.~\eqref{eq:LAM1} we assume that $y_t^2 (\tilde\rho_0, h)$ is independent of $h$ for $\tilde{\rho}_0$ in a range where $\tilde h_0 = h \tilde{\rho}_0 \ll 1$.
For all values $\tilde \rho \ll \left( \bar{y}_t^2 h  \right)^{-1}$ the Yukawa coupling depends on $\tilde\rho$, but not on $h$ or $\htilde$.
The $\tilde\rho$-dependence is given by the perturbative $\beta$-function \eqref{eq:YU1}, where we may add the contribution from the scalar anomalous dimension $\eta_H$.

\subsection*{Effective vertices from field-dependent gauge and Yukawa couplings}
The field-dependence of the gauge couplings induces effective vertices for the interaction of the Higgs boson and the cosmon with the gauge fields.
This can be seen directly from eq.~\eqref{eq:GCA} with a field-dependent $Z_F$.
An interaction involving two cosmons and two gluons is proportional to $\partial Z_F/\partial \rho$, and similar for higher couplings.
Similarly, one obtains a quartic interaction between two Higgs bosons and two gluons by $\partial Z_F/\partial(H^\dagger H)$.
These interactions are proportional to the beta-function for $g^2_3$.
Similarly, the dependence of the Yukawa coupling on $\chi$ induces on effective five-point vertex involving two cosmons, one Higgs doublet, and two top quarks.

\indent The effective $\chi^2\bar{t}tH$-vertex is of the size
\begin{equation}
    \label{eq:153A}
    \Gamma_{2\chi,2t,H}= \frac{\partial y_t}{\partial \rho} = \frac{2}{\chi^2} \tilde{\rho} \partial_{\tilde{\rho}}y_t = - \frac{1}{2y_t \chi^2}\beta_t \,.
\end{equation}
It is suppressed by two powers of the inverse Planck mass $\chi^{-2}$.
The corresponding interaction is even weaker than gravitational strength by the factor $\beta_t$.
Nevertheless, it contributes to the field-dependence of the cosmon-Higgs coupling $\lambda_m$.
Since the top quark couples to both the cosmon and the Higgs boson a top-quark loop contributes to a vertex $\sim \chi^2 H^\dagger H$ and therefore to $\lambda_m$.
In the absence of the coupling \eqref{eq:153A}, and similar induced couplings for other particles, only graviton loops would connect the cosmon and the Higgs doublet.
In the absence of gravity $\lambda_m$ would not flow or be field-dependent. The field dependence of the scaling solution for $\lambda_m$ which will be discussed in sect.~\ref{sec:VIII} can be directly connected to the induced vertices as in eq.~\eqref{eq:153A}.
The dimensionless vertex $k^2\Gamma_{2\chi,2t,H}$ is proportional $\tilde{\rho}^{-1}$.
It is therefore dominated by ultraviolet physics with small values of $\tilde{\rho}$.

\indent The quartic vertex involving two cosmons and two gluons, evaluated for typical momenta $q^2\approx k^2$, is of the size
\begin{equation}
    \label{eq:153B}
    \Gamma_{2\chi,2A} \sim k^2 \frac{\partial Z_F}{\partial \rho} = \frac{\partial Z_F}{\partial \tilde{\rho}} = \frac{1}{\tilde{\rho}}\left( \tilde{\rho} \partial_{\tilde{\rho}} g_3^{-2} \right) = \frac{1}{2\tilde{\rho}} \left(\bar{\beta}_3 + \cdots \right)\,,
\end{equation}
where the dots denote a contribution from the $h$-dependence of $g_3^{-2}$.
For large $\tilde{\rho}$ this interaction is again very small.
For the present epoch it is suppressed by the second power of the inverse Planck mass $M^{-2}$ and its strength is weaker than gravity.

\indent For large $\tilde\rho$ the additional effective interactions are very small.
On the other side, the observed value of the cosmon-Higgs coupling $\lambda_m \approx 10^{-34}$ is also tiny.
The $\tilde{\rho}$-dependence of the gauge and Yukawa couplings is the only important contribution to the flow of $\lambda_m$ besides gravity.
We therefore need to take it into account.
Furthermore, we have to follow the scaling solution to $\tilde{\rho} \to 0$ where the suppression $\sim \tilde{\rho}^{-1}$ no longer operates.

\section{Ultraviolet limit for gauge and Yukawa couplings}
\label{sec:VI}
\indent So far we have investigated the field-region $\tilde{\rho} \gg 1$, corresponding to $k$ sufficiently below the Planck mass. 
One finds a family of local scaling solutions parametrized by boundary functions as $L(h)$.
For a UV-complete theory these local scaling solutions have to be continued to $\tilde{\rho}\xrightarrow{}0$.
This will select among the local scaling solutions a discrete set for which this is possible.
For $\tilde{\rho}<1$ one deals with scales $k$ above the Planck mass.
The gravitational fluctuations become important.
They will decide on the selection of the global scaling solution.
One expects that predictions for the couplings of the standard model, and in particular for the Fermi scale, may depend on the precise setting for the gravitational fluctuations.

\indent The fluctuations of the gravitational degrees of freedom affect the dependence of dimensionless couplings as $g^2_3(\tilde{\rho})$ for small values of $\tilde{\rho}$.
They matter for the "initial value" of the "infrared flow" $g^2_3(\tilde{\rho}_0)$.
This is the reason why this section discusses the ultraviolet limit for
the gauge and Yukawa couplings before we address the local scaling
solution in the presence of field-dependent couplings in the next section.
For the moment we may consider the results of this section as an
argument motivating eq.~\eqref{eq:LAM1} and a similar assumption for the
Yukawa coupling. 
Once we turn to the general UV-limit for $\tilde\rho\to0$ in later parts of this work the UV-flow of the gauge and Yukawa couplings will be an important part of the picture.
We will then discuss the effect of the gravitational fluctuations on the effective scalar potential. 
For $\tilde\rho \lesssim 1$ the gravitational fluctuations also add further terms to the scaling equation \eqref{eq:FESS12} for the effective potential $u(\tilde{\rho},h)$.
Those will be addressed in sect.~\ref{sec:IX}.

\subsection*{Ultraviolet-infrared connection}
For very small $h$ the Fermi scale $\varphi_0$ and the Planck scale $\chi$ correspond to field values separated by many orders of magnitude.
In a rough sense we may associate the Higgs field $H$ with the ``infrared physics'', and the singlet field $\chi$ with the ``ultraviolet physics''.
For the infrared physics the scalar field $\chi$ plays no direct role for the flow equation, and we correspondingly have formulated the scaling solution by functions of $\tilde{h}$.
On the other hand, for Planck scale physics the Fermi scale $\varphi_0$ is unimportant.
In the field region relevant for Planck scale physics a formulation of the scaling solution in terms of $\tilde{\rho}$ seems more appropriate.
The flow of couplings has to match the ultraviolet (UV) and infrared (IR) regions.
This will provide information on the ``boundary values'' of the local scaling solution as $\bar{g}_3(h)$ or $L(h)$ for the potential.
The dependence of the boundary values on $h$ reflects a dependence on the scale separation between the UV- and IR-regions.

\indent The ratio $h$, or on a logarithmic scale - $\ln(h)$, is a measure for the separation of the IR-physics from the UV-physics.
Variation of $h$ measures the dependence of quantities as the gauge couplings or the effective potential on the UV-IR-separation.
The issue is directly related to the cosmon-Higgs coupling $\lambda_m$.
A very small $\lambda_m$ results in a large separation for $h$ taken at the partial minimum of $u$, and vice versa.

\subsection*{Field dependence of gauge couplings in the ultraviolet}
The fluctuations of the gravitational degrees of freedom (metric or vierbein) induce an anomalous dimension $B_g$ in the flow of the gauge couplings \cite{DHR, FDP, HARE, CHE, CLPR, EVER, CWSSFGC}.
We parametrize the flow in the UV-region at fixed $\chi$ and $H$ by
\begin{equation}
    \label{eq:UV1}
    k \partial_k g^2 = - B_g g^2 - B_F g^4\,,
\end{equation}
where $B_F$ arises from the fluctuations of gauge bosons, fermions and scalars.
The gravitational fluctuations decouple once $k$ becomes much smaller than the flowing Planck mass,
\begin{equation}
    \label{eq:UV2}
    M_p^2(k) = 2w_0 k^2 + \xi \chi^2 = 2w(\tilde{\rho})k^2\,.
\end{equation}
One finds the approximative behavior \cite{CWSSFGC}
\begin{equation}
    \label{eq:UV3}
    B_g \approx \frac{5k^2}{144\pi^2 M_p^2(k)} \left(\frac{4}{1-v(k)} - \frac{3}{(1-v(k))^2}\right)\,,
\end{equation}
with
\begin{equation}
    \label{eq:UV4}
    v(k) = \frac{u(k)}{w(k)}\,.
\end{equation}
For large $\tilde{\rho}$ the dimensionless squared Planck mass increases $w \approx \xi \tilde{\rho}$, such that $v$ vanishes $\sim \tilde{\rho}^{-1}$.
Then $B_g$ decreases
\begin{equation}
    \label{eq:UV5}
    B_g(\tilde{\rho} \gg1)\approx \frac{5}{288\pi^2 \xi \tilde{\rho}}\,,
\end{equation}
and $B_g g^2$ can be neglected as compared to $B_F$.

\indent Towards the UV for $k\to \infty$ or $\tilde{\rho} \to 0$ one has
\begin{equation}
    \label{eq:UV6}
    B_g \approx \frac{5}{288\pi^2 w_0} \left(\frac{4}{1-v_0} - \frac{3}{(1-v_0)^2}\right)\,,\quad
    v_0 = \frac{u_0}{w_0}\,.
\end{equation}
The part $B_F g^4$ involves the contribution of the particles of the standard model $-\bar{\beta}$, plus additional contributions (for example from additional particles with mass $\sim\chi$) which decouple for large $\tilde\rho$,
\begin{equation}
    \label{eq:UV7}
    B_F = -\bar{\beta} - \frac{D_F}{1 + c_F \tilde{\rho}}\,.
\end{equation}
For the example of the strong gauge coupling $\bar{\beta}$ is given by $\bar{\beta}_3$ in eq.~\eqref{eq:GCA2}.

\subsection*{UV-fixed point for gauge couplings}
\indent For $\tilde{\rho} \to 0$ both $B_g$ and $B_F$ reach constant values.
Those depend on details of the microscopic model.
We will therefore treat them as unknown constants.
For $B_g > 0$ the coupling $g$ is asymptotically free, with a fixed point at $g_* = 0$ that is approached in the UV for $k\to \infty$.
Another fixed point is present if $B_g$ and $B_F$ have opposite sign, with value
\begin{equation}
    \label{eq:UV8}
    g_*^2 = - \frac{B_g}{B_F}\,.
\end{equation}
For $B_g < 0$, $B_F > 0$ this fixed point is approached in the UV for $k\to\infty$.
In contrast, for $B_g > 0$, $B_F < 0$ the fixed point \eqref{eq:UV8} is approached as $k$ is lowered.
In this case the coupling $g$ has two fixed points, at $g^2 = 0$ and $g^2 = -B_g / B_F$.
For both $B_g < 0$ and $B_F < 0$ the coupling $g$ diverges for $k \to \infty$ and the model is not ultraviolet complete

\indent For the scaling solution we can convert eq.~\eqref{eq:UV1} to the scaling equation
\begin{equation}
    \label{eq:UV9}
    \tilde{\rho} \partial_{\tilde{\rho}} g^{-2}_{|h} = -\frac{B_g}{2} g^{-2} - \frac{B_F}{2}\,.
\end{equation}
In the limit of constant $B_g$ and $B_F$ for $\tilde{\rho} \to 0$ the solution reads for $B_g > 0$
\begin{equation}
    \label{eq:UV10}
    g^{-2} (\tilde{\rho}) =C_g \left(\frac{\tilde{\rho}}{\hat{\rho}}\right)^{-\frac{1}{2}B_g} - \frac{B_F}{B_g}\,.
\end{equation}
For positive $B_F$ this solution implies the existence of a critical $\tilde{\rho}$ where $g^{-2}$ reaches zero.
In contrast, for negative $B_F$ the fixed point \eqref{eq:UV8} is approached for $\tilde{\rho} \to \infty$.
For $B_g < 0$, $B_F > 0$ the solution \eqref{eq:UV10} holds for $g^{-2} > g_*^{-2}$.
In this case the fixed point is approached for $\tilde{\rho} \to 0$, while $g^{-2}$ diverges for $\tilde{\rho} \to \infty$.
In the range $g^{-2} < g_*^{-2}$ one has instead
\begin{equation}
    \label{eq:UV11}
    g^{-2}(\tilde{\rho}) = - \left(\frac{\tilde{\rho}}{\hat{\rho}}\right)^{-\frac{1}{2} B_g} - \frac{B_F}{B_g}\,,
\end{equation}
such that $g^{-2}$ vanishes for $\tilde{\rho}$ approaching a critical $\tilde{\rho}_c$.

\subsection*{Field-dependent gauge coupling below the Planck mass}
\indent As $\tilde{\rho}$ grows larger, the gravitational fluctuations decouple according to eq.~\eqref{eq:UV5}.
The term $\sim B_F$ dominates eq.~\eqref{eq:UV9}, leading to a logarithmic change.
At this point $g^2$ is a function of $\tilde{\rho}$.
\begin{equation}
    \label{eq:UV12}
    g^{-2} = \hat{g}^{-2} - \frac{B_F}{2} \ln \frac{\tilde{\rho}}{\hat{\rho}}\,.
\end{equation}
It is independent of $h$ if $\hat{g}^{-2}$ is independent of $h$.
The coefficient $B_F$ may undergo some change as additional heavy particles decouple.
If these masses do not involve the Higgs doublet, this change occurs as a function of $\tilde{\rho}$, independently of $h$.
This question decides on a possible dependence on $h$ for the gauge coupling $g^2_{3,0}$ in eq.~\eqref{eq:LAM1}.

\indent Let us focus on a scenario where the UV-fixed point value for $\tilde{\rho} \to 0$ is independent of $h$.
This rather natural assumption implies
\begin{equation}
    \label{eq:UVO13}
    \partial_h g^{-2}(\tilde{\rho} \to 0) = 0\,.
\end{equation}
We further assume that a possible $h$-dependence of the decoupling of the fluctuations of the gravitational degrees of freedom and heavy particles can be neglected.
A dependence on $h$ only arises then from the masses of particles generated by electroweak symmetry breaking for $H \neq 0$.
For simplicity, we consider here the contribution of the top quark to the running of the strong gauge coupling
\begin{align}
    \label{eq:UV13}
    \tilde{\rho} \partial_{\tilde{\rho}} \partial_h g_3^{-2} &= -\frac{1}{2} \partial_h B_F = \frac{1}{2} \partial_h \bar{\beta}_3 \nn\\
							     &= \frac{1}{24\pi^2}\partial_h \left(1 + y_t^2 h \tilde{\rho}\right)^{-1}\,.
\end{align}
In the approximation of $h$-independent $y_t^2$ this becomes
\begin{equation}
    \label{eq:UV14}
    \tilde{\rho} \partial_{\tilde{\rho}} \partial_h g_3^{-2} = -\frac{y_t^2 \tilde{\rho}}{24\pi^2 (1 + y_t^2 h \tilde{\rho})^2}\,.
\end{equation}
As long as $y_t^2 h \tilde{\rho} \ll 1$ this results in 
\begin{equation}
    \label{eq:UV15}
    \partial_h g_3^{-2} = - \frac{\bar{y}_t^2 \tilde{\rho}}{24\pi^2}\,,
\end{equation}
with $\bar{y}_t^2$ an appropriate $\tilde{\rho}$-average of $y_t^2$.
One concludes
\begin{equation}
    \label{eq:UV16}
    g_3^{-2} (\tilde{\rho},h) = g_3^{-2}(\tilde{\rho},0) - \frac{\bar{y}_t^2 h \tilde{\rho}}{24\pi^2}\,.
\end{equation}
For the tiny values of $h$ corresponding to the observed Fermi scale  the dependence of $g_3^{-2}$ on $h$ is negligible at a scale $\tilde\rho_0$ which corresponds to $k^2$ much larger than $H^\dagger H$. 
This justifies eq.~\eqref{eq:LAM1}.

\indent This dependence on $h$ involves the combination $\tilde{h}$.
The shift in the gauge coupling due to its dependence on $h$ is tiny as long as $\tilde{m}_t^2 = y_t^2 h \tilde{\rho}$ is a tiny quantity.
More generally, the integration of eq.~\eqref{eq:UV14} yields for $y_t^2(\tilde\rho,h)$ replaced by $\bar{y}_t^2$,
\begin{equation}
    \label{eq:UV16A}
    \partial_h g_3^{-2} = - \frac{\bar{y}_t^2 \tilde{\rho}}{24\pi^2 ( 1+\bar{y}_t^2 h \tilde\rho)}\,,
\end{equation}
and therefore
\begin{equation}
    \label{eq:UV16B}
    g_3^{-2}(\tilde{\rho},h) = g_3^{-2}(\tilde{\rho},0) - \frac{1}{24\pi^2} \ln \left(1+\bar{y}_t^2 h \tilde{\rho}\right)\,.
\end{equation}
We observe that $\partial_h g_3^2$ remains finite for all $\tilde{\rho}$ for which $g_3^2$ remains finite.
For an estimate of the relative size of the $\tilde h$-dependence as compared to the $\tilde\rho$-dependence one has to compare $h\,\partial_{h}g_3^{-2}$ with $\tilde\rho\,\partial_{\tilde\rho}g_3^{-2}$.
One finds a suppression factor $\tilde h$ which is very small at $\tilde\rho_0$ by assumption.

\subsection*{Boundary value for the gauge coupling of the effective low energy theory}

\indent The effective low energy theory is formulated in terms of $g^2_3$ as a function of $\htilde$ and $h$.
For the solution \eqref{eq:GC4} we need the dependence of the boundary $\bar{g}^2_3$ on $h$.
The switch from $g_3^2(\tilde\rho,h)$ to $g_3^2(\tilde h,h)$ is a pure translation of variables. 
No physical ``matching condition'' is involved. 
For a function depending only on $\tilde\rho=\htilde/h$ the logarithmic dependence on $h$ is the negative one of the logarithmic dependence on $\tilde h$. 
We therefore expect that $g_3^2(\tilde h,h)$ depends substantially on $h$ in the region $\tilde h\ll1$.

\indent The switch from $g_3^2(\tilde{\rho}, h)$ to $g_3^2(\tilde{h},h)$ involves typically a logarithm of $h$ since $\tilde{\rho} = \tilde{h}/h$.
In particular, if one solves the evolution equation for $g_3^2(\tilde{h},h)$ with some initial value $\bar{g}_3^2(h) = g_3^2(\tilde{h} = \bar{h},h)$, this initial value involves $\ln h$.
The translation from $g_3^2(\tilde{\rho},h)$ to $g_3^2(\tilde{h},h)$ is done according to
\begin{equation}
    \label{eq:UV17}
    g_3^2(\tilde{h} = h\tilde{\rho}, h) = g_3^2(\tilde\rho, h)\,.
\end{equation}
We may switch the variable at some given $\tilde{\rho}_0 \gg 1$, with $h\tilde{\rho}_0 \ll 1$.
At $\tilde{\rho}_0$ the gauge coupling has the value
\begin{equation}
    \label{eq:UV18}
    g_{3,0}^2 = g_3^2 (\tilde{\rho}_0,h) = g_3^2 (\tilde{h} = h\tilde{\rho}_0,h)\,.
\end{equation}
We want to translate to a fixed $\bar{h}$, corresponding to $\bar{\rho} = \bar{h}/h$,
\begin{equation}
    \label{eq:UV19}
    \bar{g}_3^2 = g_3^2(\tilde{h}=\bar{h},h)\,.
\end{equation}
The value of $\bar{g}_3^2$ depends on $h$ according to
\begin{equation}
    \label{eq:UV20}
    \bar{g}_3^{-2} = g_{3,0}^{-2} + \frac{1}{2} \bar{\beta}_3 \ln \left(\frac{\bar{h}}{h \tilde{\rho}_0}\right)\,.
\end{equation}
The corresponding $h$-dependence of $\bar{g}_3^2$,
\begin{equation}
    \label{eq:UV21}
    h \partial_h \bar{g}_3^{-2} = -\frac{1}{2} \bar{\beta}_3\,,
\end{equation}
is much bigger than the one from eq.~\eqref{eq:UV15}, which is suppressed by $\tilde{h}$.
In other words, the relation of the coupling at $k^2 = H^\dagger H / \bar{h}$ to the short distance coupling at $k^2 = c \chi^2$ depends on the scale separation between $H^\dagger H$ and $\chi^2$, as encoded in $h$.
It is the dependence of $\bar{g}_3^2$ on $h$ which induces the dependence of the confinement scale on $\chi$.

\indent Similar arguments apply to the other dimensionless couplings.
The dependence of their values at $\tilde{h} = \bar{h}$ on $h$ is determined by the beta-functions at $\tilde{h} = \bar{h}$.
It does not involve any additional suppression by small dimensionless particle masses.
An explicit dependence of the dimensionless couplings on $h$ at fixed $\tilde{h}$ will lead to a non-trivial dependence of $\lambda_m(\tilde{\rho},h)$ on $\tilde{\rho}$.

\indent The scale of the infrared physics is set by $\varphi_0(k)$. 
The particle scale symmetry associated to a second order phase transition states that for a fixed Planck mass or fixed $\chi$ the effective low-energy theory does not involve an intrinsic scale. 
In this case the dimensionless couplings can only depend on $\varphi_0/k$ or $\tilde h$.
A dependence of the gauge or Yukawa couplings on $h$ at fixed $\tilde h$ violates particle scale symmetry. 
We have seen that this violation is due to the logarithmic running of the dimensionless couplings. 
It ultimately leads to the crossover instead of a second order phase transition.
The infrared physics keeps some memory of the ultraviolet physics which induces a scale proportional to $\chi$. 
Observation tells us that this violation of particle scale symmetry does not induce a large mass scale.
The mass scale generated by the violation of particle scale symmetry is not minimal, however. 
For a minimal mass scale $\varphi_0$ would be around $\Lambda_{\mathrm{QCD}}$.

\subsection*{Validity of assumptions}
\indent The dependence of the gauge and Yukawa couplings on $\tilde{\rho}$ for fixed $\htilde$ is crucial for the induced effective vertices of the cosmon and the scaling solution for $\lambda_m$.
One may therefore question the robustness of our assumptions.
The effective squared Planck mass involves a term $\sim H^\dagger H$ in addition to the one $\sim \chi^2$.
The threshold behaviour for the decoupling of the gravitational fluctuations for the flow of the field-dependence of $u$ will therefore depend on $h$, leading to an $h$-dependence of $g_3^2(\tilde{\rho})$.
For small $h$ it seems reasonable to linearize in $h$, such that the $h$-dependence of $g_3^{-2}$ is linear in $h$.
A similar argument applies for the $h$-dependence from the decoupling of fluctuations of additional particles with mass $\sim\chi$.
Given the tiny scale of the IR-physics related to the Fermi scale for $\htilde \ll 1$ it seems reasonable that the shift in couplings due to $h \neq 0$ can be linearised in $h$.
It is then very small if one takes boundary values at $\htilde \ll 1$.

\indent While reasonable, our assumptions are not proven.
From the point of view of the local solution for the differential equation defining the scaling solution for $g_3^2$, the initial value $g_3^2(\tilde{h}=\bar{h})$ is an arbitrary function $\bar{g}_3^2(h)$ of $h$,
\begin{equation}
    \label{eq:UV22}
    g_3^2(\tilde{h} = \bar{h},h) = \bar{g}_3^2(h)\,.
\end{equation}
Every function $\bar{g}_3^2(h)$ defines a member of the family of local scaling solutions.
Restrictions arise only from the requirement of a continuation of the local solution to $\tilde{h} \to \infty$ and $\tilde{h} \to 0$.
Only a detailed investigation  of the UV-behavior for $\tilde{\rho} \to 0$ can definitely establish our assumptions.

\indent For example, one could envisage the alternative possibility that $\bar{g}_3^2(h)$ is approximately independent of $h$,
and similar for the other dimensionless couplings.
For such a scenario the ratio of the strong interaction confinement scale over the Fermi scale, $\Lambda_\QCD^2 / \varphi_0^2$,
is predicted as a function of $\bar{h}$ and $\bar{g}_3^2$, without further dependence on $\chi$.
The cosmon-Higgs coupling $\lambda_m(\tilde{\rho}, h)$ would be independent of $\tilde{\rho}$ for all $h$ in this case.
It is not disproven so far that this alternative local scaling solution could have a continuation to $\tilde{\rho} \to 0$.
Nevertheless, our considerations of the UV-fixed point for the dimensionless couplings make this scenario rather unlikely. We will therefore work with the approximation that $g^2_{3,0}$ in eq.~\eqref{eq:LAM1} or \eqref{eq:UV20} is independent of $h$.

\section{Scaling potential for field-dependent dimensionless couplings}
\label{sec:VII}

\indent Our truncation for the flow of the effective potential, together with the
perturbative running of the gauge and Yukawa couplings with given initial
values at $\tilde\rho_0$ independent of $h$, defines a closed system. 
We can therefore establish the general local solution for this system for $\tilde\rho>\tilde\rho_0$. 
The general local solution of the scaling equations \eqref{eq:FESS12}, \eqref{eq:FESS5} yields for the region of small $h$ a
rather detailed and robust picture for the dependence of $u$ on $\tilde\rho$ and $\tilde h$.
This local scaling solution involves two free parameters $\bar\lambda_m$ and $\bar\lambda_h$. 
They correspond to the renormalizable couplings for the Higgs potential in the standard model, namely the mass term and the quartic coupling. 
The validity of the local scaling solution extends to values of $\tilde\rho$ near one where
gravity and perhaps additional fields as for grand unified theories become important.
We will below include these additional fluctuations in order to perform the ``ultraviolet completion'' by extending the local scaling solution to $\tilde\rho\to0$. 
This extension to a global scaling solution will fix $\bar\lambda_m$ and $\bar\lambda_h$ and allow for predictivity for the Fermi scale.

\indent Control over the $\tilde\rho$-dependence of $u$ and its
$\tilde h$-derivatives over a large range between
$\tilde\rho\approx1$ and $\tilde\rho\to\infty$ is necessary to connect the
physics below the Planck mass with the physics above the Planck mass.
In the regions where the number of particles with masses $\tilde m_i^2\ll1$
does not change with $\tilde\rho$, the simple truncation
\eqref{eq:FESS5} produces correctly the running of the couplings, including the anomalous mass dimension. 
It is one-loop exact if one treats the scalar fluctuations more carefully -- this will be done in the next
section -- and includes the scalar-field anomalous dimension $\eta_H$ discussed in Sect.~\eqref{sec:II}. 
Furthermore, it deals coherently with all mass thresholds.

\indent The field-dependent gauge and Yukawa couplings appear in the scaling equation \eqref{eq:FESS12} for the effective potential $u(\tilde{\rho},h)$.
In this and the next section we focus on local scaling solutions for $\tilde{\rho} \gg1$.
Additional terms in eq.~\eqref{eq:FESS12} for the contribution of the metric fluctuations for $\tilde{\rho} \lesssim 1 $ will be included in the sect.~\ref{sec:IX}.
We first discuss the approximation where $c_U$ does not depend explicitly on $u$ or its field-derivatives.
The field-dependence of the dimensionless couplings does not change the structure of the solution in this case.
Only the threshold functions $t_U$ are modified.
We are mainly interested in the dependence of $u$ on $h$, or the quantity $\tilde{\nu}(\tilde{\rho},h)=\partial_h u(\tilde{\rho},h)$.
The second part of this section establishes a scaling equation for $\tilde{\nu}(\tilde{\rho},h)$ which continues to hold if $c_U$ depends on $u$.
We derive both the formal and an approximate solution.

\subsection*{Threshold functions for field-dependent couplings}
For $c_U$ independent of $u$ the general scaling solution for $u$ takes the form 
\begin{equation}
    \label{eq:G1}
    u = L(h) \tilde{\rho}^2 + \frac{1}{128\pi^2} \sum_i \hat{n}_i t_u^{(i)}(\tilde{m}_i^2)\,.
\end{equation}
Here we work in a basis of mass eigenvalues $m_i$.
This is the same expression as for eq.~\eqref{eq:FS43B} - only the threshold functions $t^{(i)}_u$ are modified.
The general threshold functions $t_u^{(i)}(\tilde{m}_i^2)$ obey the defining equation
\begin{equation}
    \label{eq:G2}
    \tilde{\rho} \partial_{\tilde\rho} t_u^{(i)} = 2\left(t_u^{(i)} - \frac{1}{1+\tilde{m}_i^2}\right)\,,
\end{equation}
where $\partial_{\tilde\rho}$ is taken at fixed $h$.
We recall that the factors $\hat{n}_i$ count the number of degrees of freedom for a given mass eigenvalue $m_i$, $\tilde{m}_i^2 = m_i^2 / k^2$, with a minus sign for fermions. 
With $t_u^{(i)}(0) = 1$ eq.~\eqref{eq:G1} also includes the constant contribution of massless particles.
For the particular case where $\tilde{m}_i^2 \sim \tilde\rho$ the threshold functions $t_u^{(i)}$ equal the ones given by eq.~\eqref{eq:TQF15} since $\tilde\rho \partial_{\tilde\rho} = \tilde{m}^2 \partial_{\tilde{m}^2}$.
In the presence of running gauge couplings the $\tilde\rho$-dependence of $\tilde{m}_i^2$ may deviate from a strictly linear behavior, such that the threshold functions are modified \cite{DEQG}.

\indent We display details on the threshold functions $t_u^{(i)}$ for field-dependent couplings in appendices \ref{app:B} and \ref{app:C}.
The main outcome is that the threshold functions continue to vanish $\sim \tilde{m}_i^{-2}$ for $\tilde{m}^2_i \gg 1$. 
In a reasonable approximation they are again given by eq.~\eqref{eq:TQF26}, with $\tilde{m}^2_i$ now involving the field-dependent dimensionless couplings.
This approximation yields a valid overall picture of the scaling equation for $u$.
When one needs more precision on the derivatives of $u$ it is advantageous to define threshold functions associated to the derivatives of $t_u^{(i)}$ with respect to the mass $\tilde{m}^2_i$.

\subsection*{Dependence of effective potential on the Higgs field}
For a discussion of the Fermi scale we are mainly interested in the dependence of $U$ on the Higgs scalar, or in the dependence of $u$ on $h$.
Instead of a detailed investigation of the threshold functions $t_u^{(i)}$ we can focus directly on the quantity
\begin{equation}
    \label{eq:NYO1}
    \tilde{\nu} (\tilde\rho, h) = \partial_h u(\tilde\rho,h)=\tilde{\rho}\mu_h(\tilde\rho,h)\,,
\end{equation}
with
\begin{equation}
    \label{eq:NYO1b}
    \tilde{\nu}(\tilde\rho) = \tilde{\nu}(\tilde{\rho},0)\,.
\end{equation}
Up to a factor $\tilde\rho$ the quantity $\tilde\nu$ corresponds to $\mu_h = \partial_{\htilde} u (\tilde\rho, \htilde)$.
It is related to the cosmon-Higgs coupling function $\lambda_m$ by
\begin{equation}
    \label{eq:NYO2}
    \lambda_m(\tilde\rho,h) = \frac{1}{\tilde{\rho}^2} \left(\tilde{\rho} \partial_{\tilde\rho} - h\partial_h -1\right) \tilde{\nu}(\tilde\rho,h)\,.
\end{equation}
In particular, one has for $h=0$ the cosmon-Higgs coupling
\begin{equation}
    \label{eq:NYO3}
    \lambda_m(\tilde\rho) 
    = \lambda_m(\tilde\rho,0) 
= \left(\frac{1}{\tilde\rho} \partial_{\tilde\rho} - \frac{1}{\tilde{\rho}^2}\right) \tilde{\nu}(\tilde\rho) = \partial_{\tilde\rho} \left(\frac{\tilde{\nu}(\tilde\rho)}{\tilde\rho}\right)\,.
\end{equation}

\indent As long as $\tilde{\nu}(\tilde{\rho},h)$ remains positive for all $h$ a partial minimum of $u(\tilde{\rho},h)$ with respect to $h$ occurs for $h=0$.
A possible non-trivial partial minimum (or maximum) at $h_0(\tilde{\rho})$ is realized for
\begin{equation}
    \label{eq:112A}
    \tilde{\nu}(\tilde{\rho},h_0) = 0\,.
\end{equation}
In principle, the location of an extremum at $h_0(\tilde{\rho})$ depends on $\tilde{\rho}$.
The Fermi scale obtains for a partial minimum from $h_0(\tilde{\rho}\to\infty) \equiv h_0$.
Thus the determination of the Fermi scale investigates a possible zero of $\tilde{\nu}(\tilde{\rho}, h)$ for $\tilde{\rho} \to \infty$.

\indent In particular, for negative values of $\tilde{\nu} = \tilde{\nu}(\tilde\rho,0)$ the value $h=0$ is not a partial minimum of $u$ with respect to $h$ at fixed $\tilde \rho$. 
This partial minimum occurs rather for $h_0(\tilde \rho) > 0$ according to
\begin{equation}
    \label{eq:94A}
	\partial_h u_{|\tilde \rho} = \tilde \nu + \lambda_h h_0 \tilde{\rho}^2 = 0\,,
\end{equation}
where we have assumed the validity of an expansion in the small quantity $h_0(\tilde \rho)$.
The Fermi scale is obtained for $k\to 0$ or $\tilde \rho \to \infty$,
\begin{align}
    \label{eq:94B}
	\varphi_0^2 &= \frac{1}{2} h_0(\tilde \rho \to \infty) \chi^2\,,\nn\\
	h_0 &= \frac{2\varphi_0^2}{\chi^2} = -\frac{\tilde \nu(\tilde\rho)}{\lambda_h (\tilde\rho) \tilde{\rho}^2}\,\, (\tilde\rho \rightarrow \infty) \,.
\end{align}

\indent In general, the threshold functions $t_u^{(i)}$ defined by eq.~\eqref{eq:G2} depend on the flow of the dimensionless couplings.
In addition to their dependence on $\tilde{m}_i^2$ they typically depend on the beta-functions of these dimensionless couplings.
It is instructive to discuss $\tilde{\nu}$ in the limit where $t_u^{(i)}$ are only functions of $\tilde{m}_i^2$, such that
\begin{equation}
    \label{eq:NY1}
    \nu(\tilde{\rho},h) = \tilde{\nu}(\tilde{\rho},h) - \partial_h L(h) \tilde{\rho}^2 = \frac{1}{128\pi^2} \sum_i \hat{n}_i \partial_h \tilde{m}_i^2 \frac{\partial t_u^{(i)}}{\partial \tilde{m}_i^2}\,.
\end{equation}
For eq.~\eqref{eq:TQF26} this yields
\begin{equation}
    \label{eq:NY2}
    \nu = - \frac{1}{64\pi^2} \sum_i \hat{n}_i \partial_h \tilde{m}_i^2 s^{(i)}_u (\tilde{m}_i^2)\,,
\end{equation}
with
\begin{equation}
    \label{eq:NY3}
    s^{(i)}_u (\tilde{m}_i^2) = -\frac{1}{2} \frac{\partial t_u^{(i)}}{\partial \tilde{m}_i^2} = 1 + \frac{\tilde{m}_i^2}{1 + \tilde{m}_i^2} - 2\tilde{m}_i^2 \ln \frac{1 + \tilde{m}_i^2}{\tilde{m}_i^2}\,.
\end{equation}
With
\begin{equation}
    \label{eq:NY4}
    \lim_{\tilde{m}_i^2 \to \infty} s^{(i)}_u (\tilde{m}_i^2) = \frac{1}{3\tilde{m}_i^4}\,,
\end{equation}
and diverging $\tilde{m}_i^2(\tilde{\rho} \to \infty)$ the \rhs of eq.~\eqref{eq:NY2} vanishes for $\tilde{\rho} \to \infty$, implying
\begin{equation}
    \label{eq:NY5}
    \lim_{\tilde{\rho} \to \infty} \nu = 0\,,\quad
    \lim_{\tilde{\rho}\to\infty} \tilde{\nu}(\tilde{\rho},h) = \partial_h L \tilde{\rho}^2\,.
\end{equation}
This result is independent of the detailed shape of threshold functions, since only the decrease of $s^{(i)}_u$ for increasing $\tilde{m}_i^2$ matters.
This property remains valid if the field-dependence of dimensionless couplings is taken into account.
We recall, however, that the general structure \eqref{eq:G1} holds only in the approximation for which the explicit dependence of $c_U$ on $u$ is neglected.

\indent The rather complex field-dependence of $\nu$ is discussed in appendix \ref{app:D}.
One uses in eq.~\eqref{eq:NY2} the relation $\partial_h \tilde m^2_i = \tilde\rho \partial_{\htilde} \tilde m^2_i$ where both derivatives are taken at fixed $\tilde\rho$.
In turn, $\partial_{\htilde} \tilde m ^2_i$ involves dimensionless couplings, as $y^2_t$ for the top-quark fluctuations.
One finds that for $\bar\lambda_m \geq 0 $ the function $\tilde\nu (\tilde\rho,h)$ remains positive for the observed values of the gauge and Yukawa couplings. In this case the partial minimum of the scaling potential occurs for $h_0 (\tilde\rho) =0$.
A value $h_0>0$ requires $\bar\lambda_m<0$.

\subsection*{Scaling equation for $\tilde{\nu}(\tilde{\rho},h)$}
For field-dependent couplings it is possible to compute directly the $\tilde{\rho}$-dependence of $\tilde{\nu}$ without an explicit solution for the threshold functions $t_u^{(i)}$.
For fixed $\chi$ the flow with $\tilde{\rho}$ is transmuted to the flow with $k$, such that the $\tilde{\rho}$-dependence of $\tilde{\nu}$ corresponds to the $k$-dependence of the derivative $\partial U / \partial (H^\dagger H)$.
Furthermore, the cosmon-Higgs coupling $\lambda_m$ can be related directly to the flow of $\tilde{\nu}$ with $\tilde{\rho}$.

\indent The $\tilde{\rho}$-dependence of $\tilde{\nu}(\tilde\rho,h)$ is given by
\begin{equation}
    \label{eq:NYO4}
	\tilde\rho \partial_{\tilde\rho} \tilde\nu(\tilde\rho,h)
	= \partial_h (\tilde\rho \partial_{\tilde\rho} u)
	= 2\partial_h (u - c_U)
	= 2\tilde\nu - 2\partial_h c_U.
\end{equation}
If the dependence of $c_U$ on $u$ is neglected we may write this in the form
\begin{equation}
    \tilde{\rho} \partial_{\tilde{\rho}} \tilde{\nu} (\tilde{\rho}, h) = 2\partial_h L + \frac{1}{64\pi^2} \sum_i \hat{n}_i \partial_h \left(t_u^{(i)} - \frac{1}{1 + \tilde{m}_i^2}\right)\,,
\end{equation}
in order to see that massless particles do not contribute to the sum.
This approximation is not needed, however.
Eq.~\eqref{eq:NYO4} follows directly from our ansatz for the scaling equation for $u$ without further approximations.

\indent It is convenient to work with the expression
\begin{equation}
    \label{eq:193A}
        \tilde{\rho} \partial_{\tilde{\rho}} \tilde{\nu} (\tilde{\rho},h) = 2\tilde{\nu} (\tilde{\rho},h) - B(\tilde{\rho},h)\,,
\end{equation}
with
\begin{align}
    \label{eq:115A}
    B = 2\partial_h c_U &= \frac{1}{64\pi^2} \sum_i \hat{n}_i \partial_h \left(1 + \tilde{m}_i^2\right)^{-1} \nn\\
    &= -\frac{1}{64\pi^2} \sum_i \hat{n}_i \partial_h \tilde{m}_i^2 \left(1 + \tilde{m}_i^2\right)^{-2}\,.
\end{align}
Omitting for a moment the scalar fluctuations (which will be treated more accurately below) we employ
\begin{equation}
	\label{eq:95B}
	\tilde{m}_i^2 = \alpha_i (\tilde h, h) \tilde h = \alpha_i h \tilde\rho \, ,
\end{equation}
with
\begin{align}
    \label{eq:95C}
	\partial_h \tilde{m}_{i\;|\tilde\rho}^2 &= \partial_h \tilde{m}_{i\;|\tilde h}^2 + \tilde\rho \partial_{\tilde h} \tilde{m}_{i\;|h}^2 \nn\\
											&= \alpha_i \left(1 + \frac{\partial \ln \alpha_i}{\partial \ln \tilde h} + \frac{\partial \ln \alpha_i}{\partial \ln h }\right) \tilde \rho\,.
\end{align}
Equivalently, we may use $\tilde{m}_i^2 = \alpha_i(\tilde{\rho}, h) h\tilde{\rho}$ with
\begin{equation}
    \label{eq:118A}
    \partial_h \tilde{m}_i^2(\tilde{\rho},h) = \alpha_i (\tilde{\rho},h) \left(1 + \frac{\partial \ln \alpha_i (\tilde{\rho}, h)}{\partial \ln h}\right)\tilde{\rho}\,.
\end{equation}

\subsection*{General scaling solution for $\tilde\nu(\tilde\rho,h)$}
For the general solution of eq.~\eqref{eq:193A} we define
\begin{equation}
    \label{eq:NYA}
    \beta_\nu = - \frac{B}{\tilde{\rho}}\,.
\end{equation}
This implies
\begin{equation}
    \label{eq:NYB}
    \partial_{\tilde{\rho}} \left(\frac{\tilde\nu}{\tilde{\rho}^2}\right) = \frac{\beta_\nu}{\tilde{\rho}^2}\,,
\end{equation}
with general local solution
\begin{equation}
    \label{eq:NYC}
    \tilde{\nu}(\tilde{\rho},h) = \partial_h L \tilde{\rho}^2 - \tilde{\rho}^2 \int_{\tilde{\rho}}^{\infty} \mathrm{d}\rho' \;\frac{\beta_\nu(\rho',h)}{\rho'^2}\,.
\end{equation}
In its general form eq.~\eqref{eq:NYC} is an implicit solution since $\beta_\nu$ may depend on $\tilde{\nu}$.
It provides in this form a good view on how $\tilde{\nu}(\tilde{\rho},h)$ at finite $\tilde{\rho}$ is connected to the IR-values for $\tilde{\rho} \to \infty$.
The boundary condition is defined by
\begin{equation}
    \label{eq:200A}
    \partial_h L = \lim_{\tilde{\rho} \to \infty} \frac{\tilde{\nu}(\tilde{\rho},h)}{\tilde{\rho}^2}\,.
\end{equation}

\indent The dependence of $\beta_\nu$ on $\tilde{\nu}$ arises from the scalar fluctuations.
Their role will be discussed separately below.
We first focus on an approximation where the mass of the Goldstone modes is set to zero and the mass of the radial mode is given by $2\lambda_h \tilde{h}$, with $\lambda_h$ independent of $\tilde{\nu}$.
These relations hold for a partial minimum at $h_0(\tilde{\rho})$, while for general $h$ and for derivatives of $\tilde{\nu}$ modifications are expected for a more accurate treatment of the scalar fluctuations.
For this approximation for the scalar fluctuations $\beta_\nu$ is simply a function of $\tilde{\rho}$ and $h$, as determined by the field-dependent dimensionless couplings. With $\alpha_i$ being simple functions of these couplings, perturbation theory becomes a good guide for eq.~\eqref{eq:118A}.
In leading order one can omit the term $\sim \partial\ln\alpha_i(\tilde{\rho}\,h)/\partial\ln h$, resulting in
\begin{equation}
    \label{eq:118C}
    \beta_\nu = \frac{1}{64\pi^2} \sum_i \hat{n}_i \alpha_i \left(1 + \alpha_i h \tilde{\rho}\right)^{-2}\,.
\end{equation}

\indent In the range where $\alpha_i h \tilde{\rho} \ll 1$ the quantity $\beta_\nu$ is dominated by the top quark fluctuations.
It is negative and almost constant.
For constant $\beta_\nu$ the general solution \eqref{eq:NYC} reads
\begin{equation}
    \label{eq:118E}
    \tilde{\nu} = - \beta_\nu \tilde{\rho} + \partial_h L \tilde{\rho}^2 \,.
\end{equation}
On the other side, for $\alpha_i h \tilde{\rho} \gg 1$, one can approximate
\begin{equation}
    \label{eq:118F}
    \beta_\nu = \frac{1}{64\pi^2} \sum_i \frac{\hat{n}_i}{\alpha_i h^2 \tilde{\rho}^2}\,.
\end{equation}
Neglecting the $\tilde{\rho}$-dependence of $\alpha_i$ this yields
\begin{equation}
    \label{eq:118G}
    \tilde{\nu}(\tilde{\rho},h) = \partial_h L \tilde{\rho}^2 - \frac{1}{192\pi^2}\sum_i \frac{\hat{n}_i}{\alpha_i h^2 \tilde{\rho}}\,.
\end{equation}
This confirms that the ``integration constant'' $L(h)$ corresponds to the boundary value for $\tilde{\rho}\to \infty$.

\subsection*{Approximate scaling solution for $\tilde{\nu}(\tilde{\rho},h)$}
\indent For an approximate computation of $\tilde{\nu}(\tilde{\rho},h)$ we define $\tilde{\beta}_\nu$ by
\begin{equation}
    \label{eq:118H}
    \tilde{\nu}(\tilde{\rho},h) = \partial_h L \tilde{\rho}^2 - \tilde{\beta}_\nu (\tilde{\rho},h)\tilde{\rho}\,.
\end{equation}
Here $\tilde{\beta}_\nu$ is an appropriately weighed integral \eqref{eq:NYC} over $\beta_\nu$, obeying
\begin{equation}
    \label{eq:118I}
    \tilde{\rho} \partial_{\tilde\rho} \tilde{\beta}_\nu = \tilde{\beta}_\nu - \beta_\nu\,.
\end{equation}
The integral \eqref{eq:NYC} for $\tilde{\beta}_\nu$ can be written in terms of $\beta_\nu(\tilde{h}, h)$ as
\begin{align}
    \label{eq:118L}
    \tilde{\beta_\nu}(\tilde{h},h) &= \tilde{h} \int_{\tilde{h}}^{\infty} \frac{\mathrm{d} h'}{h^{\prime 2}}\beta_\nu(h',h) \nn\\
			       &\approx \frac{\tilde{h}}{64\pi^2} \sum_i \hat{n}_i \int_{\tilde{h}}^{\infty} \frac{\mathrm{d} h'\; \alpha_i}{h'^2 (1 + \alpha_i h')^2}\,.
\end{align}
Approximating $\alpha_i(h',h)$ by $\alpha_i(\tilde{h},h)$ one finds
\begin{align}
    \label{eq:118M}
    \tilde{\beta}_\nu &= \frac{1}{64\pi^2} \sum_i \hat{n}_i \alpha_i \!\left[1 + \alpha_i \tilde{h}\!\left(\!2\ln \!\left(\!\frac{\alpha_i \tilde{h}}{1 + \alpha_i \tilde{h}}\!\right)\! + \frac{1}{1 + \alpha_i\tilde{h}}\!\right)\!\right]\! \nn \\
		      &= \frac{1}{64\pi^2}\sum_i \hat{n}_i \alpha_i s_u\left(\tilde{m}_i^2\right) \,,
\end{align}
and
\begin{align}
    \label{eq:118N}
    \partial_h \tilde{\beta}_{\nu\;|\tilde{h}} = \frac{1}{64\pi^2} \sum_i \hat{n}_i \partial_h \alpha_{i\;|\tilde{h}} \Bigg[1 + 4\alpha_i \tilde{h} \ln \bigg(\frac{\alpha_i \tilde{h}}{1 + \alpha_i \tilde{h}}\bigg) \nn\\
    + \frac{4\alpha_i \tilde{h}}{1 + \alpha_i \tilde{h}} - \frac{\alpha_i^2 \tilde{h}^2}{(1 + \alpha_i \tilde{h})^2}\Bigg]\,.
\end{align}
For $\alpha_i \tilde{h} \gg 1$ the quantity $\tilde{\beta}_\nu$ decreases $\sim \tilde{h}^{-2}$,
\begin{align}
    \label{eq:118O}
    \tilde{\beta}_\nu &= \frac{1}{192\pi^2 \tilde{h}^2} \sum_i \frac{\hat{n}_i}{\alpha_i} \,,\nn\\
    \partial_h \tilde{\beta}_{\nu\;|\tilde{h}} &= -\frac{1}{192\pi^2 \tilde{h}^2} \sum_i \frac{\hat{n}_i \partial_h \alpha_{i\;|\tilde{h}}}{\alpha_i^2}\,.
\end{align}

\subsection*{Dependence of scaling potential on the Higgs field}
\indent From $\tilde{\beta}_\nu$ and its derivatives we can compute the field derivatives of the effective potential at fixed $\chi$ and $k$ as
\begin{equation}
    \label{eq:PD1}
    \frac{\partial U}{\partial (H^\dagger H)} = \frac{1}{2} \chi^2 \partial_h L - \tilde{\beta}_\nu k^2\,,
\end{equation}
and
\begin{equation}
    \label{eq:PD2}
    \lambda_h = \frac{\partial^2 U}{\partial (H^\dagger H)^2} = \partial_h^2 L - \frac{1}{\tilde{\rho}} \partial_h \tilde{\beta}_{\nu\;|\tilde{\rho}}\,.
\end{equation}
The term $-\tilde{\beta}_\nu k^2$ determines the quadratic term for the Higgs field for the region of large $k$, $k^2 \gg m_t^2$.
In this region of $k$ one finds at the origin $H=0$
\begin{equation}
    \label{eq:PD2A}
    \frac{\partial U}{\partial(H^\dagger H)} \approx \frac{\bar\lambda_m}{2}\chi^2- \frac{k^2}{64\pi^2}\sum_i \hat{n}_i \alpha_i\,.
\end{equation}
This extends the relation \eqref{eq:N11} to the case of running dimensionless couplings.
This result for the $k$-dependence of the mass term for the Higgs scalar is very simple.
The term $\sim k^2$ is the remnant of the "quadratic divergences" in perturbation theory \cite{CWQR}.
We will improve this result below by a more accurate treatment of the scalar field dimension $\eta_H$.
What is not settled at this stage is the boundary value $\bar\lambda_m$ for the local scaling solution which will finally determine the Fermi scale for $k\rightarrow0$.
For large enough $k^2$ the term $\sim \bar\lambda_m \chi^2$ can be neglected.
For realistic values of the dimensionless couplings the contribution from the top-quark fluctuations dominates.
This results in a positive quadratic term.
For large $k$ the partial minimum of $U$ is located at $H=0$.

\indent We can also infer the $k$-dependence of the quartic Higgs coupling.
The logarithmic field-dependence of $\lambda_h$ is recovered by
\begin{align}
    \label{eq:PD3}
    \frac{1}{\tilde\rho}&\partial_h \tilde{\beta}_\nu (\tilde{\rho},h) = \lambda_h(\tilde{\rho},h) - \bar{\lambda}_h = \\
    &\frac{1}{64\pi^2} \sum_i  \hat{n}_i \bigg\{\alpha_i^2 \left[2\ln \left(\frac{\tilde{m}_i^2}{1 + \tilde{m}_i^2}\right) + \frac{3 + 2\tilde{m}_i^2}{(1+\tilde{m}_i^2)^2}\right] \nn\\
			     &+\frac{1}{\tilde{\rho}} \partial_h \alpha_{i\;|\tilde{\rho}} \bigg[1 + \tilde{m}_i^2\left(4\ln \left(\frac{\tilde{m}_i^2}{1 + \tilde{m}_i^2}\right) + \frac{4+3\tilde{m}_i^2}{(1 + \tilde{m}_i^2)^2}\right)\!\bigg]\bigg\},\nn
\end{align}
with leading term $\sim\alpha_i^2$.
The expression remains finite for $h\neq 0$.
Taking the logarithmic $k$-derivative at fixed fields one recovers for $\tilde m ^2_i \ll 1$ the perturbative renormalization group equation for $\lambda_h$ if one adds the contribution from the anomalous dimension $\eta_H$.
The $\beta$-function for $\lambda_h(k)$ involves the renormalized couplings at the scale $k$.
Furthermore, eq.~\eqref{eq:PD3} accounts for the decoupling of fluctuations once $\tilde m^2_i  \gg 1$.
The coupling $\lambda_h (\tilde \rho, \htilde)$ involves the integration constant $\bar{\lambda}_h$.
Its field-dependence is fixed by eq.~\eqref{eq:PD3}.

\indent One may want to evaluate more precisely the contribution of the top quark fluctuations to the scaling form of the effective potential.
With $\alpha_t = y_t^2$ we need the dependence of $y_t^2$ on $\tilde\rho$ and $h$.
For the dependence of the Yukawa coupling on $h$ eq.~\eqref{eq:YU4} implies
\begin{equation}
    \label{eq:YU5}
    \frac{1}{\tilde{\rho}} \partial_h y_{t\;|\tilde{\rho}}^2 \approx \frac{b_t \bar{y}_t^2 y_t^4}{2(1+\bar{y}_t^2 h \tilde{\rho})}\,.
\end{equation}
Insertion of this expression in eqs.~\eqref{eq:118N}, \eqref{eq:PD2} leads to a subleading contribution to $\lambda_h$ proportional to $y_t^6$.
The functional flow equation is non-perturbative and not limited to small couplings. 
In the limit of small couplings a given truncation produces beyond the perturbative one-loop results already many higher loop contributions. 
The correction \eqref{eq:YU5} may be interpreted as one of these higher loop contributions.

\indent In summary, the field-dependence of the top quark Yukawa coupling results to a good approximation in the insertion of the field-dependent coupling into the formula we have found previously for a field independent coupling.
This generalizes to the other dimensionless couplings.
We can use eq.~\eqref{eq:108A} by inserting the field-dependent Yukawa and gauge couplings.
This provides for a rather simple expression for the dependence of the scaling potential on the Higgs field.
The dominant correction to this simple picture expands in eq.~\eqref{eq:118L} $\alpha_i(h',h)$ around $\alpha_i(\tilde{h},h)$.
This involves the $\beta$-functions for $\alpha_i$ which are much smaller than $\alpha_i$.
A similar approximation can be used for the scaling form of the effective potential \eqref{eq:N2}, with field-dependent dimensionless couplings used for the threshold functions $t_u(\tilde{m}^2)$.

\indent We have obtained a rather robust picture for the overall features of the scaling potential for the region of small $h$ and $\tilde\rho \ll1$.
This will only be mildly modified by a more accurate treatment of the scalar fluctuations.
What remains open at this stage are the "integration constants" $\bar\lambda_m$ and $\bar\lambda_h$.

\subsection*{Global $h$-dependence of boundary term}
\indent In this paper we focus on very small $h$ for which a polynomial expansion of the boundary term around some small $h_0$ is justified.
The overall $h$-dependence of the boundary term $L(h)$ is more complex, being far from a simple polynomial.
This may be demonstrated by the effect of the top-quark fluctuations.

\indent Taking into account only the top quark fluctuations with constant Yukawa coupling $y_t$ we can derive $\tilde{\nu}$ directly from eq.~\eqref{eq:TQF25}
\begin{equation}
    \label{eq:GH1}
    \tilde{\nu}(\tilde{\rho},h) = \partial_h L(h) \tilde{\rho}^2 + \frac{3y_t^2}{16\pi^2} \tilde{\rho}s_u(y_t^2 h\tilde{\rho})\,,
\end{equation}
which agrees with eq.~\eqref{eq:118M}.

\indent If one has information on $\tilde{\nu}(\tilde{\rho}_0,h)$ for small $\tilde{\rho}_0$ one can extract information on $L(h)$ by 
\begin{equation}
    \label{eq:GH2}
    \partial_h L(h) = \frac{\tilde{\nu}(\tilde{\rho}_0,h)}{\tilde{\rho}_0^2} - \frac{3y_t^2}{16\pi^2 \tilde{\rho}_0} s_u(y_t^2 h\tilde{\rho}_0)\,.
\end{equation}
For example, one has, for $\tilde{m}_{t,0}^2 = y_t^2 h \tilde{\rho}_0$,
\begin{align}
    \label{eq:GH3}
    \bar{\lambda}_h(h) &= \partial_h^2 L(h) = \frac{\partial_h \tilde{\nu}(\tilde{\rho}_0,h)}{\tilde{\rho}_0^2} - \frac{3 y_t^4}{16\pi^2} \frac{\partial s_u(\tilde{m}_t^2)}{\partial \tilde{m}_t^2} \nn\\
		       &= \frac{\partial_h \tilde{\nu}(\tilde{\rho}_0,h)}{\tilde{\rho}_0^2} - \frac{3y_t^4}{8\pi^2} \bigg[ \ln \left(\frac{\tilde{m}_{t,0}^2}{1+\tilde{m}_{t,0}^2}\right) + \frac{1}{1+\tilde{m}_{t,0}^2} \nn\\
		       &\hspace{93pt}+ \frac{1}{2(1+\tilde{m}_{t,0}^2)^2}\bigg]\,.
\end{align}
If there is a value $\tilde{\rho}_0$ for which $\partial_h \tilde{\nu}(\tilde{\rho}_0,h) = 0$ one infers
\begin{align}
    \label{eq:GH4}
    \bar{\lambda}_h(h) = \frac{3y_t^4}{8\pi^2} \bigg[ &\ln \left(\frac{1+y_t^2 h\tilde{\rho}_0}{y_t^2 h \tilde{\rho}_0}\right) - \frac{1}{1+y_t^2 h \tilde{\rho}_0} \nn\\
    & - \frac{1}{2(1+y_t^2 h \tilde{\rho}_0)^2}\bigg] \,.
\end{align}
For $h \ll (y_t^2\tilde{\rho}_0)^{-1}$ one infers the logarithmic dependence on $h$
\begin{equation}
    \label{eq:GH5}
    \bar{\lambda}_h(h) \approx \frac{3 y_t^4}{8\pi^2} \ln \left(\frac{1}{y_t^2 h \tilde{\rho}_0}\right)\,.
\end{equation}
On the other hand, for $h \gg (y_t^2 \tilde{\rho}_0)^{-1}$ one uses $\partial s_u / \partial \tilde{m}^2 = -2/(3\tilde{m}^6)$, implying a fast decay
\begin{equation}
    \label{eq:GH6}
    \bar{\lambda}_h(h) = \frac{1}{8\pi^2 y_t^2 h^3 \tilde{\rho}_0^3}\,.
\end{equation}
One concludes that the global dependence of the boundary term on $h$ is far from being a polynomial.

\indent For a polynomial expansion for small $h$ eq.~\eqref{eq:GH2} relates the
short distance value of the couplings at $\tilde\rho_0$, specified by
$\tilde \nu/\tilde\rho_0^2$, to the infrared values encoded in $\partial_hL(h)$. 
In order to avoid infrared complications we may consider an expansion around some fixed small value $\bar h_0$, which may be identified with the location of the potential minimum at the end.
Defining
\begin{equation}
\lambda_h(\tilde\rho)
=
\partial_{\tilde h}^2u(\tilde\rho,\bar h_0)
=
\frac{
\partial_{h}\tilde \nu(\tilde\rho,\bar h_0)
}{
\tilde\rho^2}\,,
\quad
\bar\lambda_h
=
\lambda_h(\tilde\rho\to\infty)\,,
\label{eq:232A}
\end{equation}
one obtains
\begin{equation}
\bar\lambda_h
=
\lambda_h(\tilde\rho_0)
-
\frac{3y_t^4}{16\pi^2}
\left.
\frac{\partial s_u(\tilde m_t^2)}
{\partial\tilde m_t^2}
\right|_{\tilde m_t^2(\tilde\rho_0)}\,.
\label{eq:232B}
\end{equation}
Here $\tilde m_t^2(\tilde\rho_0) = y_t^2\,\bar h_0\tilde\rho_0$ is evaluated at $h=\bar h_0$. 
For a realistic $\bar h_0$ this quantity corresponds to $m_t^2/k^2$, with $m_t$ the physical top quark mass. 
With eq.~\eqref{eq:GH5} the logarithmic dependence describes the running of $\lambda_h$ due to the top quark fluctuations. 
One may include the contribution of other fluctuations and consider $\tilde\rho\,\partial_{\tilde\rho}\lambda_h(\tilde\rho)$ for an improved estimate.

\indent Let us consider the next term in the expansion,
\begin{equation}
\gamma_h(\tilde\rho)
=
\frac{\partial_{\tilde h}^2\tilde \nu(\tilde\rho,\bar h_0)}{\tilde\rho^2}
=
\tilde\rho\,\partial_{\tilde h}^3 u(\tilde\rho,\bar h_0)\,.
\label{eq:232C}
\end{equation}
The infrared value,
\begin{equation}
\gamma_h(\tilde\rho\to\infty)
=
\partial_h^3L
=
\bar\gamma_h\,,
\label{eq:232D}
\end{equation}
parameterizes the polynomial expansion of the effective potential at $H=H_0$,
\begin{equation}
\frac{\partial^3U}
{\partial(H^\dagger H)^3}
=
\frac{2\bar\gamma_h}{\chi^2}\,.
\label{eq:232E}
\end{equation}
Equation~\eqref{eq:GH2} yields for constant $y_t$
\begin{align}
\bar\gamma_h
&=
\gamma_h(\tilde\rho_0)
-
\frac{3y_t^6\tilde\rho_0}{16\pi^2}
\left.
\frac{
\partial^2 s_u(\tilde m_t^2)
}{
\left(\partial\tilde m_t^2\right)^2
}
\right|_{\tilde m_t^2(\tilde\rho_0)}
\nonumber\\
&=
\gamma_h(\tilde\rho_0)
-
\frac{3y_t^4}{8\pi^2\bar h_0}
\left(
1+\tilde m_t^2(\tilde\rho_0)
\right)^{-3}\,.
\label{eq:232F}
\end{align}
For $\tilde m_t^2(\tilde\rho_0)\ll1$ the second term dominates $\gamma_h(\tilde\rho_0)$ by many orders of magnitude if $\bar h_0$ is very small. 
The coupling $\bar\gamma_h$ is infrared dominated and we may omit a small correction $\gamma_h(\tilde\rho_0)$. 
In consequence the boundary term $\partial_h^3 L$ is no free parameter. 
It is predicted by the local scaling solution. 
The corresponding term in the potential is given by
\begin{equation}
\frac{\partial^3U}
{\left(\partial(H^\dagger H)\right)^3}
=
-\frac{3y_t^4}
{8\pi^2H_0^\dagger H_0}\,.
\label{eq:232G}
\end{equation}

\indent Similar features are found for higher orders in a polynomial expansion of $L(h)$ -- they are all infrared dominated and predicted by the local scaling solution.
This generalizes if the contribution from other fluctuations is included. 
One may improve the estimate by following the scaling equation for $\tilde\rho\,\partial_{\tilde\rho}\bar\gamma_h$. 
We conclude that $\bar\lambda_m$ and $\bar\lambda_h$ are the only free parameters for the local scaling solution in the region of small $h$.

\subsection*{Overall picture of the local scaling solution}

\indent In the region of small $h$ the $\tilde\rho$-dependence of $u$ according
to the local scaling solution is rather similar to the dependence on $k^{-2}$
of $u$, as computed by the flow equation in a model without the scalar cosmon field. 
The two relevant parameters in the $k$-flow correspond to the integration constants $\bar\lambda_m$ and $\bar\lambda_h$. 
These are the free renormalizable couplings. 
All features of $u$ can be computed in terms of these two free parameters.
The local scaling solution encompasses in this respect the predictivity of a renormalizable theory.

\indent The effective potential according to the local scaling solution is rather smooth. 
For small $h$ it can well be approximated by a polynomial potential, with running ($\tilde\rho$-dependent) couplings which can involve logarithms. 
For the small couplings of the standard model the dominant running is given by the one-loop $\beta$-functions.
In one loop order those are the same for the $\tilde\rho$-dependence of the local scaling solution and the $k^{-2}$-dependence of the model without the cosmon.
As a result, the overall picture of the potential for small $h$ and $\tilde\rho\ll1$ or $k^2\ll M_p^2$ is rather similar for the two scenarios. 
Since both are based on functional flow equations they account both for the important threshold functions describing the decoupling of heavy particles. 
It is rather impressive to see how these threshold functions produce correctly the one-loop running.

\indent The difference between the $\tilde\rho$-flow of the local scaling solution and the $k^{-2}$-flow of the standard model without the cosmon originates in the fact that $\tilde\rho$-dependent couplings induce new vertices which in turn affect the flow. 
These effects are of higher loop order and therefore modify the overall shape of the potential only modestly. 
Nevertheless, for tiny quantities such as the cosmon-Higgs
coupling $\lambda_m$ these effects have to be taken into account. 
We will see in the next section how they affect the precise location of the critical surface for the second order phase transition.

\indent The flow equation involves a choice of an infrared cutoff function.
Different cutoff functions lead to different details of the threshold
functions which retain, however, the qualitative features. 
One could consider this as a source of uncertainty.
In addition, our truncation captures well all the physics relevant for our problem but it is, of course, not perfect. 
It is particularly well adapted to the simplified flow equation for which the threshold functions are fixed.
For $k\to0$ the flowing action becomes the quantum effective action. 
In this limit there is no longer an uncertainty from the choice of cutoff functions for an idealized situation without truncation.
For fundamental scale invariance (or LIMS very small) the local scaling solution describes already the limit $k \to 0$. 
One may therefore suspect that the effective potential computed from the local scaling solution is a particularly robust quantity. 
It involves quantum vertices in the limit $k \to 0$, which obtain from $\tilde{\rho} \to \infty$. 
These vertices are directly related to observables.

\indent To sum up the status at this stage, the local scaling solution for the dependence of the effective potential on the Higgs field is well under control, in the region of small $h$ and $\tilde\rho\gg1$. 
It retains the predictivity of a renormalizable quantum field theory, with renormalizable couplings associated to the two free parameters $\bar\lambda_m$ and $\bar\lambda_h$.
In this sector, the only additional predictivity from the global scaling solution concerns a prediction for $\bar\lambda_m$ and $\bar\lambda_h$. 
In addition, the local scaling solution fixes completely the dependence of the potential on the cosmon field, including the determination of $\bar\lambda_\chi$. 
This can be used directly for cosmology.

\section{Cosmon-Higgs coupling and electroweak phase transition}
\label{sec:VIII}

\indent This section discusses the cosmon-Higgs coupling $\lambda_m$ in a field range where the gravitational fluctuations can be omitted.
We derive the scaling equation for a coupling function $\lambda_m(\tilde\rho, \htilde)$ with boundary value $\lambda_m(\tilde\rho \rightarrow \infty, \htilde = 0) = \bar\lambda_m$.
The critical surface of the quantum electroweak phase transition is associated with a critical trajectory for $\lambda_m(\tilde\rho,0)$.
Due to the field-dependence of the gauge and Yukawa couplings this critical cosmon-Higgs coupling $\lambda^{(\text{cr})}_m (\tilde{\rho})$ vanishes only for $\tilde{\rho} \to \infty$, while it differs from zero for finite $\tilde{\rho}$.
An anomalous dimension for $\lambda_m$ and the associated critical exponent obtain by including the dependence of the scalar mass term on the derivatives of the potential.

\subsection*{Cosmon-Higgs coupling}
\indent The cosmon-Higgs coupling $\lambda_m$ is a central quantity for our investigation.
Its value $\bar{\lambda}_m$ for $\tilde{\rho} \to \infty$ determines the ratio of Fermi scale over Planck scale.
This coupling is also closely related to deviations from a second order quantum electroweak phase transition, as we have seen for the simple approximation for the top quark fluctuations with constant Yukawa couplings.
For general $\tilde{\rho}$ and $\tilde{h}$ the cosmon-Higgs coupling function $\lambda_m(\tilde{\rho},\tilde{h})$ is defined by
\begin{equation}
    \label{eq:97A}
    \lambda_m(\tilde{\rho},\tilde{h}) = \partial_{\tilde\rho}\partial_{\tilde h} u(\tilde{\rho},\htilde)\,.
\end{equation}
Here the $\tilde{\rho}$-derivative is taken at fixed $\tilde{h}$, and the $\tilde{h}$-derivative at fixed $\tilde{\rho}$.
Thus $\lambda_m(\tilde{\rho},\tilde{h})$ is given by the $\tilde{\rho}$-dependence of the dimensionless Higgs mass function $\mu_h$ in eq.~\eqref{eq:PP1}
\begin{equation}
    \label{eq:97B}
    \lambda_m(\tilde{\rho},\tilde{h}) = \partial_{\tilde{\rho}} \mu_h (\tilde{\rho},\tilde{h})\,,\quad
    \mu_h (\tilde{\rho},\tilde{h}) = \partial_{\tilde{h}} u (\tilde{\rho},\tilde{h})\,.
\end{equation}
We may also employ the function $\lambda_m(\tilde{\rho},h)$ which is related to $u(\tilde{\rho},h)$ by
\begin{equation}
    \label{eq:BC58}
    \lambda_m(\tilde{\rho},h) = \frac{1}{\tilde{\rho}^2} \partial_h \left(\tilde\rho \partial_{\tilde\rho} - h \partial_h\right) u(\tilde \rho, h)\,.
\end{equation}
In eq.~\eqref{eq:97A} the derivative $\partial_{\tilde{\rho}}$ is taken at fixed $\htilde$, and in eq.~\eqref{eq:BC58} at fixed $h$, with
\begin{equation}
    \label{eq:BC58A}
    \partial_{\tilde \rho | \tilde h} = \partial_{\tilde \rho | h} + \frac{\partial h}{\partial \tilde \rho | \tilde h} \partial_{h | \tilde \rho} = \partial_{\tilde \rho | h} - \frac{h}{\tilde \rho} \partial_{h | \tilde\rho}\,.
\end{equation}

\indent The off diagonal elements in the scalar mass matrix \eqref{eq:SF35} are proportional to $\lambda_m$,
\begin{equation}
    \label{eq:97C}
    \tilde{m}_{rc}^2 = 2\sqrt{\tilde{h} \tilde{\rho}}\, \lambda_m (\tilde{\rho}, \tilde{h})\,.
\end{equation}
This dependence of scalar masses or their derivatives on $\lambda_m$ will lead to an anomalous dimension in the $\tilde{\rho}$-flow of $\lambda_m$, which is related to the anomalous mass dimension.

\indent For the Fermi scale $\varphi_0$ we are interested in the field dependence of $\lambda_m$ with the limits 
\begin{align}
    \label{eq:BC57}
    \lambda_m(\tilde{\rho}\to 0,h=0) &= \lambda_{m,0} \,,\\
    \lambda_m(\tilde{\rho}\to \infty, h=0) &= \bar{\lambda}_{m} \,.\nn
\end{align}
Our main focus will therefore be on the scaling solution for the cosmon-Higgs coupling 
\begin{equation}
    \label{eq:243A}
    \lambda_m(\tilde\rho) = \lambda_m(\tilde\rho, h=0)\,.
\end{equation}
The scaling equation for $\lambda_m(\tilde\rho)$ relates a given $\bar\lambda_m$ to a value $\lambda_m(\tilde\rho_0)$ with $\tilde\rho_0$ in the vicinity of the onset of the gravitational fluctuations.
The relation between $\lambda_m(\tilde\rho_0)$ and $\lambda_m(0)$ depends on the ultraviolet physics and has to include the quantum gravity effects.
This relation, which is at the origin of the predictivity for the Fermi scale, will be discussed in sect.~\ref{sec:IX}.

\subsection*{Critical trajectory for cosmon-Higgs coupling}
\indent In the presence of a second order quantum electroweak phase transition there exists a critical trajectory for the cosmon-Higgs coupling $\lambda_m(\tilde{\rho},h=0)$ which characterizes the critical surface.
It is characterized by the boundary value $\lambda_m(\tilde\rho\rightarrow\infty,0)=0$.
With $\lambda_m(\tilde\rho)=\lambda_m(\tilde\rho, h=0)$ the basic definition of the critical trajectory $\lambda_m^{(\text{cr})}(\tilde\rho)$ is given by the choice of the integration constant $\bar\lambda_m=0$,
\begin{equation}
\lambda_m^{(\text{cr})}(\tilde\rho\to\infty)=0\,.
\label{eq:239A}
\end{equation}
We may also use some extended definition of $\lambda_m^{(\text{cr})}(\tilde\rho,h)$ with $\lambda_m^{(\text{cr})}(\tilde\rho) = \lambda_m^{(\text{cr})}(\tilde\rho, h=0)$.

\indent We can extract $\lambda_m^{(\text{cr})}$ from the behavior of $\tilde{\nu}(\tilde{\rho}) = \tilde{\nu}(\tilde{\rho},h=0)$.
With $\mu_h = \tilde{\nu}/\tilde{\rho}$ the quantity $\tilde{\beta}_\nu$ is related to $\lambda_m(\tilde{\rho},h)$ by eq.~\eqref{eq:97B}
\begin{equation}
    \label{eq:247A}
    \lambda_m(\tilde{\rho},h)
    = \partial_{\tilde{\rho}\;|\tilde{h}}\left(\frac{\tilde{\nu}}{\tilde{\rho}}\right)
    = \left(\partial_{\tilde{\rho}\;|h} - \frac{h}{\tilde{\rho}}\partial_{h\;|\tilde{\rho}}\right)
      \left(\frac{\tilde{\nu}}{\tilde{\rho}}\right)\,.
\end{equation}
In the limit where the dependence of $c_U$ on $u$ is neglected we may define
\begin{equation}
    \label{eq:118J}
    \lambda_m(\tilde{\rho},h)
    = \partial_h L - h \partial_h^2 L + \lambda_m^{\cri}(\tilde\rho,h)\,.
\end{equation}
Using eq.~\eqref{eq:118H} this yields
\begin{equation}
    \label{eq:118K}
    \lambda_m^{\cri}(\tilde{\rho},h) = -\partial_{\tilde\rho} \tilde{\beta}_{\nu\;|\tilde{h}} = \frac{h}{\tilde\rho}\partial_h \tilde{\beta}_{\nu\;|\tilde{h}} = \frac{h^2}{\tilde{h}} \partial_h \tilde{\beta}_{\nu\;|\tilde{h}}\,.
\end{equation}
In particular, if $\tilde{\beta}_\nu$ is only a function of $\tilde{h}$ one finds that $\lambda_m$ is purely given by the boundary term $L(h)$ and therefore independent of $\tilde{\rho}$, $\lambda^{\cri}_m = 0$.

\indent The critical cosmon-Higgs coupling is defined by $\lambda_m^{\cri}(\tilde{\rho}) = \lambda_m^{\cri}(\tilde{\rho},h=0)$.
We will see that $\lambda_m^{\cri}(\tilde{\rho},h=0)$ determines the location of the critical surface for an (approximate) second order quantum phase transition.
For general field-dependent gauge and Yukawa couplings $\lambda_m^{\cri}(\tilde{\rho},h=0)$ does not vanish for finite $\tilde{\rho}$.

\indent For a first estimate of $\lambda_m^{\cri}$ we evaluate the r.h.s.~of eq.~\eqref{eq:118K}.
Focusing on $\lambda_m^{\cri}(\tilde{\rho})$ we take eq.~\eqref{eq:118N} for $\tilde{h} = 0$ and insert in eq.~\eqref{eq:118K} for $\tilde{h} = 0$
\begin{align}
    \lambda_m^{\cri} &= \frac{1}{64\pi^2 \tilde{\rho}} \sum_i \hat{n}_i h\partial_h \alpha_{i\;|\tilde{h}} \nn\\
		     &= -\frac{1}{64\pi^2\tilde{\rho}} \sum_i \hat{n}_i \tilde{\rho}\partial_{\tilde{\rho}} \alpha_{i\;|\tilde{h}=0} \nn\\
		     &= \frac{1}{128\pi^2\tilde{\rho}} \sum_i \hat{n}_i \beta_i \,,
\label{eq:241A}
\end{align}
where $\beta_i = -2\tilde{\rho} \partial_{\tilde{\rho}} \alpha_i = k \partial_k \alpha_{i\;|\rho}$ are the $\beta$-functions for the dimensionless combinations of couplings $\alpha_i$ evaluated at $\tilde{h}=0$.
In this approximation $\lambda_m^{\cri}$ approaches a finite value for $\tilde{\rho} \to 0$ if $\beta_i/\tilde{\rho}$ stays finite.
We will argue below that this condition has indeed to be met by the scaling solution.
The appearance of the $\beta$-functions in eq.~\eqref{eq:241A} shows directly that a non-vanishing $\lambda_m^{(\text{cr})}(\tilde\rho)$ is related to the field-dependence of the gauge and Yukawa couplings.
The factor $\tilde\rho^{-1} = 2k^2/\chi^2$ shows that $\lambda_m^{(\mathrm{cr})}(\tilde\rho)$ is suppressed by the inverse second power of the Planck mass.

\indent The field-dependence of the dimensionless couplings in $\tilde{\beta}_\nu$ induces a field-dependence of the quadratic term $\sim H^\dagger H k^2$ in the effective potential.
This induces a non-zero cosmon-Higgs coupling $\lambda_m$, which is, however, suppressed by $\chi^{-2}$.
The simplest view states that the quadratic term $\sim H^\dagger H k^2$ depends logarithmically on $\chi$.
As a consequence, the critical surface for the second order phase transition, as measured in terms of $\lambda_m(\tilde{\rho},h=0)$, depends on $\tilde\rho$, corresponding to a non-zero $\lambda_m^{\cri}$ in eq.~\eqref{eq:118K}.

\indent For the contribution of the top quark fluctuations to the field-dependence of $\lambda_m^{\cri}$ we insert eq.~\eqref{eq:118N} into eq.~\eqref{eq:118K},
\begin{equation}
    \label{eq:YU5A}
    \lambda_m^{\cri}(\tilde{\rho},h) = -\frac{1}{64\pi^2 \tilde{\rho}} \sum_i \hat{n}_i(\tilde{\rho} \partial_{\tilde\rho} - h \partial_h)\alpha_i T_i(\alpha_i \tilde{h})\,,
\end{equation}
with
\begin{equation}
    \label{eq:YU6}
    T_i (\alpha_i \tilde{h}) = 1 + 4\alpha_i \tilde{h} \left(\!\ln \frac{\alpha_i\tilde h}{1+\alpha_i\tilde{h}} + \frac{1}{1 + \alpha_i \tilde{h}}\!\right) - \frac{\alpha_i^2 \tilde{h}^2}{(1 + \alpha_i\tilde{h})^2}.
\end{equation}
For the top quark fluctuations we use eq.~\eqref{eq:YU4}
\begin{equation}
    \label{eq:YU7}
    (\tilde{\rho} \partial_{\tilde{\rho}} - h \partial_h) y_t^2 \approx - \frac{b_t}{2}y_t^4\,.
\end{equation}
As long as $\tilde{m}_t^2 \ll 1$ this implies the approximative solution
\begin{equation}
    \label{eq:YU8}
    \lambda_m^{\cri \,t} = - \frac{3\beta_t}{32\pi^2\tilde{\rho}}\,,
\end{equation}
with $\beta_t$given by eq.~\eqref{eq:YU1}.
The approach to zero for $\tilde{m}_t\gg 1$ is accounted for by the threshold function $T_i(\tilde{m}_t^2)$.
For the observed values of the gauge and Yukawa couplings one finds negative $\beta_t$ for a substantial range of $\tilde{h} \ll 1$.
This implies a positive contribution to $\lambda_m^{\cri}$.
Even for the range of very small $\tilde{h}$ where $\beta_t$ turns positive a negative $\lambda_m$ will not imply a negative mass term for the Higgs scalar.
This mass term is given by $\tilde{\beta}_\nu$ which remains positive.

\indent The precise location of the quantum electroweak phase transition continues to be given by the value of the ``integration constant'' $\bar{\lambda}_m = 0$.
The effect of the field-dependent couplings only affects the way how this value is reached as $\tilde{\rho}$ increases.
Typically, the value of $\lambda_m(\tilde\rho) = \lambda_m(\tilde{\rho},h=0)$ is larger in size than $\bar{\lambda}_m$ for $\tilde{\rho}$ sufficiently small.
According to eq.~\eqref{eq:YU8} or \eqref{eq:118J} its size decreases with increasing $\tilde{\rho}$ proportional to $\tilde{\rho}^{-1}$, until $\lambda_m$ settles finally at $\bar{\lambda}_m$.
In the ultraviolet limit $\tilde{\rho} \to 0$ the dimensionless couplings reach fixed points.
The vanishing of the $\beta$-functions $\beta_i$ is, however, not sufficient for obtaining a finite $\lambda_m(\tilde\rho \rightarrow0)$.
One needs finite values for $\beta_i/ \tilde{\rho}$.

\indent Up to this point we have found no constraint which fixes or limits $\bar{\lambda}_m$.
This integration constant for the local solution remains so far a free parameter specifying a particular member of a family of scaling solutions.
This situation has been obtained, however, in the approximation that the dependence of $\beta_\nu$ on $\tilde{\nu}$ is neglected in eq.~\eqref{eq:NYC}.
Only in this case the difference between $\lambda_m$ and $\lambda_m^{\cri}$ in eq.~\eqref{eq:118J} is a $\tilde{\rho}$-independent term $\partial_h L - h \partial_h^2 L$.
We will see that the scalar fluctuations induce an anomalous dimension for this difference.
Extending our truncation in order to include the fluctuations of the gravitational degrees of freedom for $\tilde{\rho} \lesssim 1$ will argue that a positive anomalous dimension persists for the whole range of $\tilde{\rho}$, rendering $\lambda_m(\tilde{\rho},h=0)$ an irrelevant parameter of the renormalization flow.

\subsection*{Scaling equation for cosmon-Higgs coupling}
Starting from the scaling equation for $u$ at fixed $h$ one can derive the scaling equation for the cosmon-Higgs coupling function $\lambda_m(\tilde\rho,h)$ by combining suitable derivatives.
One acts with $\tilde{\rho}\partial_{\tilde{\rho}\,|h}$ on eq.~\eqref{eq:BC58}.
At fixed $h$ the $\tilde\rho$-dependence of $\lambda_m$ obeys
\begin{equation}
    \label{eq:BC58C}
    \begin{aligned}
	\tilde\rho \partial_{\tilde\rho} \lambda_m
	&= -2\lambda_m + \frac{1}{\tilde\rho^2} \partial_h \left(\tilde\rho \partial_{\tilde\rho} - h\partial_h\right)\tilde\rho \partial_{\tilde\rho} u \\
	&= -2\lambda_m + \frac{2}{\tilde\rho^2} \partial_h \left(\tilde\rho \partial_{\tilde\rho} - h\partial_h\right)(u-c_U) \\
	&= - \frac{2}{\tilde\rho^2} \partial_h \left(\tilde\rho \partial_{\tilde\rho} - h\partial_h\right)c_U \\
	&= - 2 \left[\left(\frac{1}{\tilde\rho}\partial_{\tilde\rho} - \frac{1}{\tilde{\rho}^2}\right)\partial_h - \frac{h}{\tilde\rho^2}\partial_h^2\right] c_U\,.
    \end{aligned}
\end{equation}
If we express $c_U$ as a function of $\tilde\rho$ and $\tilde h$ one arrives at
\begin{equation}
    \label{eq:BC58D}
    \tilde\rho \partial_{\tilde\rho} \lambda_m(\tilde\rho,h) = -2\partial_{\tilde\rho} \partial_{\tilde h} c_U(\tilde\rho,\tilde h)\,.
\end{equation}
The r.h.s.\ of eq.~\eqref{eq:BC58D} vanishes if $c_U$ is only a function of $\tilde h$.

\indent Summing over the contribution of particles with various masses yields
\begin{align}
    \label{eq:BC58E}
	\tilde\rho \partial_{\tilde\rho} &\lambda_m (\tilde\rho,h) 
					 = \frac{1}{64\pi^2} \partial_{\tilde\rho} \Bigg\{ 6\, \partial_{\tilde h} \tilde{m}_W^2 (1+ \tilde{m}_W^2)^{-2} \nn\\
					 &\qquad + 3\partial_{\tilde h} \tilde{m}_Z^2 (1+\tilde{m}_Z^2)^{-2} + \sum_s \partial_{\tilde h} \tilde{m}_s^2 (1+\tilde{m}_s^2)^{-2} \nn\\
					 &\qquad\qquad- 2\sum_f n_f \partial_{\tilde h} \tilde{m}_f^2 (1+ \tilde{m}_f^2)^{-2} \Bigg\} \nn\\
					 &= \frac{1}{64\pi^2} \partial_{\tilde\rho} \sum_i \hat{n}_i \partial_{\tilde h} \tilde{m}_i^2 (1+\tilde{m}_i^2)^{-2} \nn\\
					 &= \frac{1}{64\pi^2}\sum_i \hat{n}_i\Bigg\{\partial_{\tilde\rho} \partial_{\tilde h} \tilde{m}_i^2 (1+\tilde{m}_i^2)^{-2} \nn\\
					 &\qquad\qquad\qquad\quad - 2 \partial_{\tilde h} \tilde{m}_i^2 \partial_{\tilde\rho} \tilde{m}_i^2 (1+\tilde{m}_i^2)^{-3} \Bigg\}\,.
\end{align}
Here $\partial_{\tilde\rho}$ is taken at fixed $h$ on the l.h.s.\ and at fixed $\htilde$ on the r.h.s..
The numbers $\hat{n}_i$ are the degrees of freedom for the different mass eigenstates, including a minus sign for the fermions, e.g.\ $\hat{n}_Z=3$ or $\hat{n}_t = -12$.

\indent For the fermions and gauge bosons we employ $\tilde{m}_i^2 = \alpha_i h \tilde{\rho} = \alpha_i \tilde{h}$, resulting in
\begin{align}
    \label{eq:FL1}
    \partial_{\tilde{\rho}} \tilde{m}_{i\;|\tilde{h}}^2 &= h \left(\tilde{\rho} \partial_{\tilde{\rho}} - h \partial_h\right) \alpha_i (\tilde{\rho}, h)\,, \nn\\ 
    \partial_{\tilde{h}} \tilde{m}_{i\;|\tilde{\rho}}^2 &= \alpha_i(\tilde{\rho}, h) + h \partial_h \alpha_i(\tilde{\rho}, h)\,, \nn\\
    \partial_{\tilde{\rho}} \partial_{\tilde{h}} \tilde{m}_{i}^2 &= \frac{1}{\tilde{\rho}} \left(\tilde{\rho} \partial_{\tilde{\rho}} - h \partial_h\right) \left(1 + h \partial_h\right) \alpha_i (\tilde{\rho},h)\,.
\end{align}
For $\tilde{m}_i^2 \ll 1$ the $h$-dependence of $\alpha_i$ can be neglected, resulting in
\begin{equation}
    \label{eq:FL2}
    \tilde{\rho}\partial_{\tilde{\rho}} \lambda_m = -\frac{1}{128\pi^2 \tilde{\rho}}\sum_i{}^\prime \hat{n}_i \beta_i + \tilde{\rho}\partial_{\tilde{\rho}}\lambda_m^{(g)} + \tilde{\rho}\partial_{\tilde{\rho}} \lambda_m^{(r)}\,.
\end{equation}
Here $\sum_i '$ sums only over the contributions of fermions and gauge bosons, and the two last terms involve the contribution of the fluctuations of the Goldstone and radial modes of the Higgs scalar.

\indent A special role is played by the scalar fluctuations.
For the Goldstone bosons one has
\begin{equation}
    \label{eq:SF1}
    \tilde{m}_g^2 = \partial_{\tilde{h}} u_{|\tilde{\rho}}\,.
\end{equation}
With
\begin{equation}
    \label{eq:SF2}
    \lambda_m(\tilde{\rho},h) = \partial_{\tilde{\rho}\;|\tilde{h}} \partial_{\tilde{h}\;|\tilde{\rho}} u\,,\quad
    \lambda_h(\tilde{\rho},h) = \partial_{\tilde{h}\;|\tilde{\rho}}^2 u
\end{equation}
one has
\begin{align}
    \label{eq:SF3}
    \partial_{\tilde{\rho}} \tilde{m}^2_{g\;|\tilde{h}} &= \lambda_m \,,\quad
    \partial_{\tilde{h}} \tilde{m}^2_{g\;|\tilde{\rho}} = \lambda_h\,,\nn\\
    \partial_{\tilde{\rho}} \partial_{\tilde h} \tilde{m}_g^2(\tilde{\rho},\tilde{h}) &= \frac{1}{\tilde{\rho}} (\tilde{\rho} \partial_{\tilde{\rho}} - h \partial_h) \lambda_h (\tilde{\rho},h)\,.
\end{align}
The Goldstone boson contribution to $\tilde{\rho} \partial_{\tilde{\rho}} \lambda_m (\tilde{\rho}, h)$ is therefore given by
\begin{align}
    \label{eq:SF4}
    \tilde{\rho} \partial_{\tilde{\rho}} \lambda_m^{(g)} =&\; \frac{3}{64\pi^2\tilde{\rho}} \left(\tilde{\rho} \partial_{\tilde{\rho}} - h \partial_h\right) \lambda_h \left(1 + \tilde{m}_g^2\right)^{-2} \nn\\
							 &- \frac{3}{32\pi^2} \lambda_m \lambda_h \left(1 + \tilde{m}_g^2\right)^{-3}\,.
\end{align}
Similarly, for the radial Higgs boson one has
\begin{equation}
    \label{eq:SF5}
    \tilde{m}_r^2 = \tilde{m}_g^2 + 2\lambda_h \tilde{h}\,,
\end{equation}
and
\begin{align}
    \label{eq:SF6}
    \tilde{\rho} \partial_{\tilde{\rho}} \lambda_m^{(r)} =&\; \frac{1}{64\pi^2 \tilde{\rho}} \left(\tilde{\rho} \partial_{\tilde{\rho}} - h \partial_h\right)\left(3 + 2h\partial_h\right) \lambda_h \left(1+\tilde{m}_r^2\right)^{-2} \nn\\
    &-\frac{1}{32\pi^2}\Big(\lambda_m + 2h\big(\tilde{\rho} \partial_{\tilde{\rho}} - h\partial_h\big) \lambda_h\Big)\nn\\
    & \quad\quad\quad\quad\times \big(3+2h\partial_h\big) \lambda_h \left(1+\tilde{m}_r^2\right)^{-3}\,.
\end{align}

\indent The scalar contributions to the flow of $\lambda_m$ involve $\lambda_m$.
Therefore the r.h.s.~of eq.~\eqref{eq:FL2} is not simply a function of $\tilde{\rho}$ and $h$.
If $\lambda_m^{\cri}(\tilde\rho,h)$ is a solution of the scaling equation, this does not hold for a constant shift $\lambda_m(\tilde\rho,h) = \lambda_m^{\cri}(\tilde{\rho},h)+\delta\lambda_m$.
This reflects the fact that $c_U$ is not independent of $u$ and its derivatives.
The family of local scaling solutions has no longer the simple form \eqref{eq:G1} or \eqref{eq:118H} with a boundary term $\sim L(h)$ valid for all $\tilde{\rho}$.
For the global features of the scaling solution this is a rather small effect.
For a precise understanding of the $\tilde{\rho}$-dependence of a possibly tiny quantity as $\lambda_m$ it needs to be taken into account, however.

\indent Let us focus on the scaling equation of the cosmon-Higgs coupling $\lambda_m(\tilde{\rho}) = \lambda_m(\tilde\rho,h = 0)$.
We write it in the form
\begin{equation}
    \label{eq:262A}
    \tilde{\rho}\partial_{\tilde{\rho}} \lambda_m  = -\frac{1}{2} \beta_m = -\frac{1}{2} \left( \tilde\beta_m + A_m\lambda_m  \right)\, , 
\end{equation}
with ($\tilde{\rho}\partial_{\tilde{\rho}} \lambda_h = -\beta_h/2$)
\begin{align}
    \label{eq:262B}
    \tilde\beta_m = &\frac{1}{64\pi^2\tilde\rho} \Bigg\{ \sum_i{}^\prime\hat n_i \frac{\beta_i}{(1+\tilde m_i^2)^2} \nn \\
    &+ 3 \beta_h \left( \frac{1}{(1+\tilde m_g^2)^2} + \frac{1}{(1+\tilde m^2_r)^2}  \right) \Bigg\}
\end{align}
and
\begin{equation}
    \label{eq:262C}
    A_m = \frac{3 \lambda_h}{16\pi^2} \left(  \frac{1}{(1+\tilde m^2_g)^3} + \frac{1}{(1+\tilde m_r^2)^3}    \right) + \eta_H \, .
\end{equation}
Here we added to $A_m$ the contribution of the anomalous dimension $\eta_H$ for the Higgs field.
For $h=0$ one has $\tilde m ^2_i = 0$, $\tilde{m}^2_g = \tilde{m}_r^2 = \mu_h$.
The part $\tilde{\beta}_m$ is proportional to $\tilde\rho^{-1}$.
For large $\tilde\rho$ it is a tiny quantity.
For $\tilde\rho^{-1} = h$, which corresponds to $k$ given by the observed Fermi scale, the suppression factor is of the order $10^{-33}$.
If $\lambda_m$ is very small, the $\tilde\rho$-dependence of $\lambda_m$ is dominated by the ultraviolet for small $\tilde\rho$.
Even for $\tilde\rho \approx 1 $, which corresponds to $k$ around the Planck mass, $\tilde\beta_m$ remains a small quantity due to the small factors $\beta_i/(64\pi^2)$.
The fate of $\lambda_m$ for $\tilde{\rho} \ll 1$ needs the effects of quantum gravity which will be incorporated in the next section.
For the moment we retain the picture of a very small value of $\lambda_m$, which tells us already that one expects $\bar\lambda_m$ to be very small, and therefore the Fermi scale to be much smaller than the Planck mass.

\subsection*{Location of electroweak quantum phase transition}
From the point of view of physics at the Planck scale the standard model is characterized by a second order quantum phase transition.
This is not exact, since non-perturbative effects of the strong interactions proportional to $\Lambda_\QCD$ modify the precise character of the transition.
Since $\Lambda_\QCD$ is much smaller than the Fermi scale we may first discard these strong interaction effects.
We can then proceed with the approximation of a second order phase transition at zero temperature.

\indent The quantum phase transition occurs as the parameters of the model are varied.
For the local scaling solution this translates to a variation of the boundary values, e.g. $\bar\lambda_m$ or the associated $\lambda_m(\tilde\rho_0)$.
At every $k$ a critical surface in the space of flowing couplings separates the phase with spontaneous symmetry breaking (SSB) from the symmetric phase.
(Strictly speaking gauge symmetries are not broken spontaneously, but the perturbative picture of spontaneous symmetry breaking is actually a good guide.)
In the symmetric phase one has four scalars with equal mass, corresponding to the unbroken $SU(2)\times U(1)$ symmetry acting on the Higgs doublet. 
The $W$- and $Z$-bosons, as well as the fermions are massless.
In the SSB phase the $W$- and $Z$-bosons and the charged fermions are massive, with masses proportional to $\varphi_0$.
There is only one massive Higgs boson.
A second order phase transition is continuous.
Exactly on the transition the particles of the standard model are massless.
The discussion of the phase transition concerns the quantum effective action for $k \to 0$.
At non-zero $k$ the critical surface separates the region in the space of couplings which ends for $k \to 0$ in the symmetric phase from the one which ends in the SSB-phase.

\indent The presence of the second order phase transition takes over to the local scaling solution.
One has now a critical surface for every $\tilde\rho$.
Instead of $k\to0$ one investigates the limit $\tilde{\rho} \to \infty$.
The transition separates again massless fermions and gauge bosons and massive Higgs scalars on the symmetric side from massive fermions and gauge bosons on the SSB side.
For the local scaling solution the location of the transition is characterized by
\begin{equation}
    \label{eq:PT1}
    \bar{\lambda}_m = \lambda_m (\tilde{\rho} \to \infty, h=0) = 0\,.
\end{equation}
For arbitrary $\tilde{\rho}$ a point on the critical surface corresponds to a trajectory in the space of $\tilde{\rho}$-dependent couplings which ends for $\tilde{\rho}\to \infty$ at $\bar{\lambda}_m = 0$.
If a second order phase transition is compatible with a global scaling solution, this guarantees that there is at least one trajectory in the space of field dependent couplings that covers the whole range from $\tilde{\rho} \to 0$ to $\tilde{\rho} \to \infty$, and ends at $\bar{\lambda}_m = 0$.
For an ultraviolet complete theory this trajectory denotes a crossover from a UV-fixed point for $\tilde{\rho} \to 0$ to an IR-fixed point for $\tilde{\rho} \to \infty$.
In our setting this is the critical trajectory given by $\lambda_m^{\cri}(\tilde{\rho})$.

\indent The critical surface can be multi-dimensional.
Not all couplings need to have unique values on the critical surface.
For the local scaling solution this is reflected by the values of the Yukawa and gauge couplings.
Such free parameters correspond to relevant (including marginally relevant) directions for the flow away from the UV-fixed point as $\tilde{\rho}$-increases.
One defines the dimension of the critical surface by the number of such free parameters.
Generically, most of the infinitely many couplings in functional renormalization take fixed values on the critical surface since they correspond to irrelevant directions at the UV-fixed point.
The dimension of the critical surface is then finite.

\indent For the local scaling solution the critical surface exists up to the small non-perturbative QCD effects which are negligible for $\Lambda^2_\QCD / k^2 \ll 1$ or $\tilde\rho \ll \tilde\rho_c$.
It is not guaranteed, however, that this critical surface can be continued to $\tilde{\rho} \rightarrow 0$, as required for a global scaling solution.
We will argue that the presence of quantum gravity selects a unique trajectory for $\lambda_m(\tilde\rho)$.
This will fix the value of $\bar\lambda_m$, which may come out different from zero.

\subsection*{Critical cosmon-Higgs coupling}
We will compute the critical cosmon-Higgs coupling $\lambda_m^{\cri}(\tilde\rho)$ by a particular iterative solution of the scaling equation \eqref{eq:262A}. 
In a first approximation it corresponds to the critical surface, but it may deviate from it by small effects leading to tiny non-zero $\bar\lambda_m$. 
For a general solution of the scaling equation we make the ansatz
\begin{equation}
    \label{eq:263A}
    \lambda_m = \frac{1}{2}(1+g)\tilde\beta_m\,.
\end{equation}
For $\tilde\rho \rightarrow \infty$ one has $\tilde\beta_m \rightarrow 0 $.
Thus $\lambda_m(\tilde\rho\rightarrow\infty) = \bar\lambda_m = 0$ if $g$ remains finite.
The first approximation for $\lambda_m^{\cri}$ takes $g=1$.
This resembles the estimate \eqref{eq:241A}, now with a better estimate of the contribution from the scalar fluctuations.

\indent We write
\begin{align}
    \label{eq:263B}
    \tilde\beta_m &= \frac{b_m}{64\pi^2 \tilde\rho} \, , \nonumber \\ 
    b_m &= \sum_i{}^\prime \hat n_i \beta_i + \frac{6\beta_h}{(1+\mu_h)^2}\,. 
\end{align}
The dependence of $\tilde\beta_m$ on $\tilde\rho$ obeys 
\begin{align}
    \label{eq:263C}
    \tilde{\rho} \partial_{\tilde{\rho}} \tilde\beta_m &= - \tilde\beta_m - \frac{d_m}{128\pi^2\tilde\rho}\,, \nonumber \\
    d_m &= -2 \tilde\rho \partial_{\tilde{\rho}} b_m \, .
\end{align}
Since $b_m$ typically contains products $\alpha_j \alpha_k$ the quantity $d_m$ involves corresponding products $\beta_j \alpha_k + \beta_k \alpha_j$.
For small $\alpha_i$ it is suppressed as compared to $b_m$.

\indent Inserting the ansatz \eqref{eq:263A} into the scaling equation \eqref{eq:262A} yields
\begin{equation}
    \label{eq:263D}
    \frac{1}{2}\left(1+g\right)\tilde{\rho} \partial_{\tilde{\rho}} \tilde{\beta}_m + \frac{1}{2}\tilde\beta_m\tilde{\rho} \partial_{\tilde{\rho}}g = -\frac{1}{2} \left( \tilde\beta_m  + A_m\frac{1+g}{2}\tilde{\beta}_m \right)
\end{equation}
or
\begin{equation}
    2\tilde\beta_m\tilde{\rho} \partial_{\tilde{\rho}}g = \left( 2g- A_m \left( 1+g \right) \right) \tilde\beta_m + \frac{(1+g)d_m}{64\pi^2\tilde\rho}\,.
\end{equation}
The scaling equation of $g$ no longer involves factors $\tilde{\rho}^{-1}$,
\begin{align}
    \label{eq:263F}
    \tilde{\rho} \partial_{\tilde{\rho}}g &= g + (1+g)f_m \,, \nonumber \\
    f_m &=\frac{1}{2}\left( \frac{d_m}{b_m} - A_m\right)\,.
\end{align}
In the limit of constant $f_m$ the solution reads
\begin{equation}
    \label{eq:263G}
    g_1 (\tilde{\rho}) = \left( g(\tilde{\rho}_0) + \frac{f_m}{1+f_m} \right) \left( \frac{\tilde\rho}{\tilde{\rho}_0} \right)^{1+f_m} - \frac{f_m}{1+f_m} \, .
\end{equation}
For decreasing $\tilde\rho$ the function $g_1(\tilde\rho)$ approaches the particular solution 
\begin{equation}
    \label{eq:263H}
    g_1^{\cri}(\tilde\rho) = - \frac{f_m(\tilde\rho)}{1+f_m(\tilde\rho)}\, .
\end{equation}
Our first iteration step for $\lambda_m^{\cri}(\tilde\rho)$ is given by
\begin{equation}
    \label{eq:263HA}
    \lambda_m^{\cri}(\tilde\rho) = \dfrac{b_m (\tilde\rho)}{128\pi^2 \tilde{\rho} \left( 1 + f_m(\tilde{\rho}) \right)}\, .
\end{equation}

\indent This approximation breaks down if $f_m(\tilde{\rho})$ approaches the value $-1$.
As long as $f_m(\tilde{\rho}\rightarrow \infty)$ remains finite one has $\bar\lambda_m = \lambda_m^{(\cri)} (\tilde{\rho}\rightarrow \infty) =0$.
We can continue the iteration by making the ansatz
\begin{equation}
    \label{eq:263HB}
    \lambda_m (\tilde{\rho}) = \frac{1}{2} \tilde{\beta}_m \left( 1-\frac{f_m(\tilde\rho)}{1+f_m(\tilde\rho)} + h(\tilde{\rho}) \right) \, .
\end{equation}
The next iteration step yields
\begin{equation}
    \label{eq:263HC}
    \lambda_m^{\cri}(\tilde\rho) = \dfrac{\tilde{\beta}_m (\tilde\rho)}{1-f_m(\tilde\rho)} \left( 1-\dfrac{\tilde{\rho}\partial_{\tilde{\rho}}f_m(\tilde\rho)}{(1-f_m(\tilde\rho))^2}\right)\, . 
\end{equation}
Indeed, the scaling equation for $h(\tilde\rho)$ involves the derivative $\tilde{\rho} \partial_{\tilde{\rho}} f_m$ which is further suppressed by small $\beta$-functions for small gauge and Yukawa couplings.
This iteration scheme breaks down as one approaches $\Lambda_\QCD$ with large values for the $\beta$-functions.

\indent Let us estimate $b_m$ and $f_m$ in the limit where we keep the gauge couplings, the top-quark Yukawa coupling and the quartic scalar coupling and take $\mu_h=0$.
One obtains
\begin{align}
    \label{eq:263I}
    b_m &= -12\beta_t + 6\beta_h +\frac{9}{2}\beta_2 + \frac{9}{10}\beta_1 \nonumber \\
    &= \frac{1}{16\pi^2}\Bigg\{-12 \left[ 9y^4_t - \left( 16g_3^2 + \frac{9}{2}g^2_2 + \frac{17}{10} g_1^2 \right)y^2_t \right] \nonumber \\
    &+ 6\bigg[ 12\lambda^2_h -12y^4_t + \frac{9}{4}g^4_2 + \frac{9}{10} g_2^2 g_1^2 + \frac{27}{100} g^4_1 +12 y^2_t \lambda_h \nonumber \\
    &-9g^2_2 \lambda_h -\frac{9}{5} g^2_1 \lambda_h \bigg]  + \frac{9}{2} \left[ -\frac{19}{3}g^4_2 \right] + \frac{9}{10}\left[ \frac{41}{5} g^4_1 \right]  \Bigg\} \nonumber \\
    &= \frac{1}{16\pi^2}\bigg\{ -180 y^4_t + 192y^2_t g_3^2 + 54 y^2_t g^2_2 + \frac{102}{5} y^2_t g_1^2 \nonumber \\
    &\quad\; +72 \lambda^2_h + 72 y^2_t \lambda_h -54 g_2^2 \lambda_h - \frac{54}{5}g^2_1 \lambda_h \nonumber \\
    &\quad\; -15 g_2^4 + 9 g_1^4 + \frac{27}{5} g_2^2 g_1^2  \bigg\} \, .
\end{align}
For realistic values of the couplings $b_m$ is positive, such that the lowest order value for $\lambda_m$ is positive,
\begin{equation}
    \label{eq:263J}
    \lambda_m^{(1)} = \frac{b_m}{128\pi^2 \tilde{\rho}}\, .
\end{equation}

\indent The function $d_m$ involves again $\beta$-functions for the gauge, Yukawa and quartic scalar couplings
\begin{align}
    \label{eq:263K}
    d_m = &\frac{1}{16\pi^2} \Bigg\{ 192y^2_t \beta_3 \nonumber \\
    &+ \bigg( 192g^2_3 - 360 y^2_t +54 g^2_2 + \frac{102}{5}g^2_1 + 72 \lambda_h\bigg) \beta_t
    \nonumber \\
    &+ \bigg( 72y^2_t - 54 g^2_2 - \frac{54}{5}g^2_1 + 144 \lambda_h \bigg) \beta_h \nonumber \\
    &+ \bigg( 54y^2_t -30 g^2_2 + \frac{27}{5} g^2_1 -54\lambda_h \bigg)\beta_2 \nonumber \\
    &+ \bigg(  \frac{102}{5}y^2_t + \frac{27}{5}g^2_2 + 18g_1^2 -\frac{54}{5}\lambda_h \bigg)\beta_1 \Bigg\}\,.
\end{align}
This quantity is negative.
The negative region of the ratio $d_m/b_m$ reflects the increase of $b_m$ with increasing $\tilde\rho$ according to eq.~\eqref{eq:263C}.

\begin{figure}[ht]
    \centering
    \includegraphics[width=0.9\linewidth]{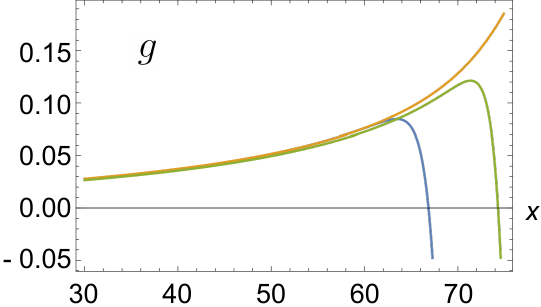}
    \caption{Critical trajectory. We plot $g$ as a function of $x = \ln \tilde\rho$, for two initial values $\sigma_m (x_F)=-0.1$ (blue) and $\sigma_m (x_F)= 0.0009$ (orange). The green curve shows the critical curve.}
    \label{fig:FA}
\end{figure}

\indent We have solved the system of flow equations numerically, see sect.~\ref{sec:XI}.
From the flow of $\lambda_m$ one can extract $g$ according to eq.~\eqref{eq:263A}.
We plot $g$ as a function of $x = \ln \tilde\rho$ for the observed values of the couplings of the standard model in Fig.~\ref{fig:FA}.
Close to the observed Fermi scale corresponding to $x_F = 75$ the value of $g$ depends on the "initial value" of $\sigma_m = \tilde{\rho} \lambda_m$ at $x_F$.
We show curves for two initial values, $\sigma_m (x_F) = -0.1$ (lower curve, blue) and $\sigma_m (x_F) = 0.0009$ (upper curve, orange). 
The universal critical curve is approached rapidly for smaller values of $x$.
We also show the critical $g$ according to the approximation \eqref{eq:263H} (middle curve, green).
The good agreement shows that the critical trajectory for the second order phase transition is well under control, both analytically and numerically.
A divergence of $g$ for $x>x_F$ indicates a nonzero $\bar \lambda_m$.

\subsection*{Sensitivity coefficient and critical exponent}
The critical behavior of a second order phase transition is characterized by critical exponents.
In the limit where the quantum electroweak phase transition is a second order phase transition one expects small deviations from the critical surface to be governed by such critical exponents.
We consider here the critical surface as a special solution among the more general class of local scaling solutions.
A critical exponent is related to the question how scaling solutions that start in the ultraviolet at some small $\tilde{\rho}_{0}$ in the neighborhood of the critical surface behave in the infrared.
This is analogous to phase transitions in classical statistical thermal equilibrium systems.
The quantity which corresponds to the difference from the critical temperature $T-T_c$ may be taken as
\begin{equation}
    \label{eq:PT2}
    \hat{\delta}_m = \lambda_m(\tilde{\rho}_0) - \lambda_m^{\cri}(\tilde{\rho}_0)\,.
\end{equation}
For $\hat{\delta}_m > 0$ (``$T > T_c$'') one ends in the IR in the symmetric phase, and for $\hat{\delta}_m < 0$ (``$T < T_c$'') in the SSB phase.
One is interested how the ratio of Fermi scale over Planck mass or $h_0$ depends on $\hat{\delta}_m$.

\indent For this purpose we generalize the parameter measuring the deviation from the critical surface to a $\tilde{\rho}$-dependent function,
\begin{equation}
    \label{eq:PT3}
    \delta_m(\tilde{\rho}) = \lambda_m(\tilde{\rho}) - \lambda_m^{\cri}(\tilde{\rho})\,,
\end{equation}
with
\begin{equation}
    \label{eq:TP3A}
    \delta_m(\tilde{\rho}_0) = \hat{\delta}_m\,, \quad
    \delta_m(\tilde{\rho} \to \infty) = \bar{\lambda}_m\,.
\end{equation}
For small enough $\delta_m$ we can linearize the scaling equation
\begin{align}
    \label{eq:PT4}
    \tilde{\rho} \partial_{\tilde{\rho}} \lambda_m &= -\frac{1}{2}\beta_m\,, \nn\\
    \tilde{\rho} \partial_{\tilde{\rho}} \delta_m &= -\frac{1}{2}\frac{\partial \beta_m}{\partial \lambda_m} \delta_m\,.
\end{align}
We consider first the approximation
\begin{equation}
    \label{eq:PT5}
    \beta_m = \tilde{\beta}_m(\tilde{\rho}) - A_m(\tilde{\rho}) \lambda_m\,,
\end{equation}
with $\tilde{\beta}_m(\tilde{\rho})$ and $A_m(\tilde{\rho})$ independent of $\lambda_m$.
One concludes in this case
\begin{equation}
    \label{eq:PT6}
    \tilde{\rho} \partial_{\tilde{\rho}} \delta_m = -\frac{1}{2} A_m (\tilde{\rho}) \delta_m \,,
\end{equation}
and therefore
\begin{equation}
    \label{eq:PT7}
    \delta_m(\tilde{\rho}) = C_m(\tilde{\rho}) \hat{\delta}\,,
\end{equation}
with
\begin{equation}
    \label{eq:PT8}
    C_m(\tilde{\rho}) = \exp \left\{-\frac{1}{2} \int_{\tilde{\rho}_0}^{\tilde{\rho}} \mathrm{d}\rho'\; \frac{A_m(\rho')}{\rho'}\right\}\,.
\end{equation}

\indent The quantity $C_m(\tilde{\rho})$ is the sensitivity coefficient.
A small value of the sensitivity coefficient means that the infrared physics at large $\tilde{\rho}$ is comparatively insensitive to the ultraviolet physics at small $\tilde{\rho}_0$.
The sensitivity coefficient $C_m$ is related to the mean anomalous dimension of the cosmon-Higgs coupling $\bar{A}_m$,
\begin{equation}
    \label{eq:PT9}
    \bar{A}_m = \int_{\tilde{\rho}_0}^{\tilde{\rho}} \frac{\mathrm{d}\rho'}{\rho'} \frac{A_m(\rho')}{\ln \left(\frac{\tilde{\rho}}{\tilde{\rho}_0}\right)}\,,
\end{equation}
by
\begin{equation}
    \label{eq:PT10}
    C_m(\tilde{\rho}) = \left(\frac{\tilde{\rho}}{\tilde{\rho}_0}\right)^{-\frac{\bar{A}_m}{2}}\,.
\end{equation}

\indent In our approximation we infer from eqs.~\eqref{eq:SF4}, \eqref{eq:SF6} a positive $A_m$,
\begin{equation}
    \label{eq:PT11}
    A_m(\tilde{\rho}) = \frac{3 \lambda_h(\tilde{\rho})}{8\pi^2}\left(1+\partial_{\tilde{h}} u\right)^{-3}\,.
\end{equation}
The sensitivity coefficient is therefore smaller than one and neighboring trajectories have a tendency to approach the critical surface as $\tilde{\rho}$ increases.
If we take $\tilde{\rho}_0$ somewhat above one and evaluate $\tilde{\rho}$ for $k \approx \varphi_0$, one finds $\tilde{\rho}/\tilde{\rho}_0 \approx 10^{32}$.
For a present value $\lambda_h \approx 1/4$ one has $A_m(\tilde{\rho}) \approx 10^{-2}$, and $\bar{A}_m$ somewhat smaller since $\lambda_h$ increases with $\tilde{\rho}$.
The sensitivity coefficient $C_m(\tilde{\rho}) \approx 10^{-0.1}$ stays close to one.
This will not change qualitatively if effects of the anomalous dimension for the scalar field due to the flow of the kinetic term is included. We will see that this situation changes substantially once the gravitational fluctuations are included and the UV-limit $\tilde{\rho}_0\to 0$ is taken.

\indent The critical exponent $\beta$ defined by
\begin{equation}
    \label{eq:PT12}
    h_0 = D |\hat{\delta}|^{2\beta}\,,
\end{equation}
reflects that the relevant sensitivity coefficient $C_m(\tilde{\rho})$ depends on $h_0$ through the ratio $\tilde{\rho}/\tilde{\rho}_0$.
If we take $\tilde{\rho} = h_0^{-1}$ one obtains
\begin{equation}
    \label{eq:PT13}
    h_0 = -\frac{\lambda_m}{\lambda_h} = \frac{C_m(\tilde{\rho})}{\lambda_h} |\hat{\delta}| = \frac{|\hat{\delta}|}{\lambda_h} \tilde{\rho}^{\bar{A}_m /2} h_0^{\bar{A}_m/2}\,,
\end{equation}
or
\begin{equation}
    \label{eq:PT14}
    \beta = \frac{1}{2} \left(1 - \frac{\bar{A}_m}{2}\right)^{-1}\,.
\end{equation}
The critical exponent $\beta$ is close to the canonical value $0.5$.

\subsection*{Mixing effects for the linearized flow}
\indent The relation \eqref{eq:PT5} is an approximation. 
A more complete treatment takes into account that the scaling equation for the couplings in the standard model forms a coupled system.
For example, the beta-functions for the electroweak gauge couplings depend on $\mu_h$, and $\mu_h$ depends on $\lambda_m$.
In this way $\tilde{\beta}_m$ depends on $\lambda_m$ and therefore on $\delta_m$.
In a linear approximation we may add a term $(\partial \tilde\beta_m/\partial \lambda_m)$ to the anomalous mass dimension $A_m$.

\indent We may demonstrate this by keeping only one of the couplings $\alpha_i$, with obvious generalization to several couplings.
We write the coupled system equations for $\alpha$ and $\mu \equiv \mu_h$ in the form

\begin{align}
    \label{eq:CLA1}
    \tilde{\rho} \partial_{\tilde{\rho}} \mu &= \mu - \frac{1}{2}\tilde{\beta}_\mu (\mu,\alpha)\,, \nn\\
    \tilde{\rho} \partial_{\tilde{\rho}} \alpha &= -\frac{1}{2} \beta_\alpha(\mu,\alpha)\,.
\end{align}
In the linear approximation for small deviations $(\delta\mu, \delta\alpha)$ from a given scaling solution one obtains
\begin{equation}
    \label{eq:CLA2}
    \tilde{\rho} \partial_{\tilde{\rho}}
    \begin{pmatrix}
\delta\mu \\
\delta\alpha
\end{pmatrix}
= -\frac{1}{2}
\begin{pmatrix}
S_{\mu\mu} & S_{\mu\alpha} \\
S_{\alpha\mu} & S_{\alpha\alpha}
\end{pmatrix}
\begin{pmatrix}
\delta\mu \\
\delta\alpha
\end{pmatrix}
\end{equation}
where the elements of the stability matrix $S$ are given by
\begin{align}
    \label{eq:CLA3}
    S_{\mu\mu} &= -2 + \frac{\partial \tilde{\beta}_\mu}{\partial \mu} \,, \quad S_{\mu\alpha} = \frac{\partial \tilde{\beta}_\mu}{\partial \alpha}\,, \nn\ \\
    S_{\alpha\mu} &= \frac{\partial \beta_\alpha}{\partial \mu}\,, \quad\quad\quad\,\,   S_{\alpha\alpha} = \frac{\partial \beta_\alpha}{\partial \alpha}\,.  
\end{align}
We have to diagonalise $S$.
We employ $\tilde\beta_\mu = -2\beta_\nu = -n\alpha/(32\pi^2)$, with $n$ given by $\hat n$ including mass thresholds according to eqs.~\eqref{eq:193A}, \eqref{eq:115A}, \eqref{eq:95C}, \eqref{eq:NYA}, \eqref{eq:118C}.
With $\beta_\alpha = c\alpha^2$ the off-diagonal element $S_{\alpha\mu}$ vanishes for $h=0$ for fermions and gauge bosons.
An exception is the scalar contribution with $\alpha = 2\lambda_h$.
In this case $\beta_\alpha$ involves a mass threshold function $(1+\mu)^{-3}$, such that $(\beta_\alpha=2\beta_h)$
\begin{equation}
    \label{eq:CLA4}
    \frac{\partial\beta_\alpha}{\partial \mu} = - \frac{9\lambda_h^2}{2\pi^2(1+\mu)^4}\,, \quad \frac{\partial\beta_\alpha}{\partial \alpha} =  \frac{3\lambda_h}{2\pi^2(1+\mu)^3} \, .
\end{equation}
One also has
\begin{equation}
    \label{eq:CLA5}
    \frac{\partial\tilde\beta_\mu}{\partial \alpha} = - \frac{3}{32\pi^2(1+\mu)^2}\, , \quad \frac{\partial\tilde\beta_\mu}{\partial \mu} = -2+\frac{3\lambda_h}{8\pi^2(1+\mu)^3}\,,
\end{equation}
and we recognise in $\partial\tilde\beta/\partial\mu$ the contribution to the anomalous mass dimension $A_m$. (There is a similar contribution from the anomalous dimension $\eta_H$ of the Higgs field.)
The eigenvalues of $S$ read in second order of $Z= 3\lambda_h/(16\pi^2(1+\mu)^3)$,
\begin{align}
\label{eq:CLA6}
\lambda_{+}
&=
\frac{3\lambda_h}{2\pi^2(1+\mu)^3}
-
18
\left(
\frac{3\lambda_h}{16\pi^2(1+\mu)^3}
\right)^2 \,,\nn \\
\lambda_{-}
&=
-2
+
\frac{3\lambda_h}{8\pi^2(1+\mu)^3}
+
18
\left(
\frac{3\lambda_h}{16\pi^2(1+\mu)^3}
\right)^2\,. 
\end{align}
Comparison with the diagonal elements of $S$ shows that the effect of the off-diagonal terms in $S$ is small, contributing in second order in $Z$.

\indent The eigenstate of $\lambda_{-}$ is $\delta\mu - \tilde{c}Z\delta\alpha$, with $\tilde{c} = \mathcal{O}(1)$, while the eigenstate of $\lambda_{+}$ has a small admixture of $\delta\mu$, given by $\delta\alpha + \tilde{c} Z\delta\mu$.
In the IR-limit for $\tilde{\rho}\rightarrow\infty$ the eigenstate of $\lambda_{-}$ dominates, with
\begin{equation}
    \label{eq:CAL7}
    \delta\mu = \delta\mu (\tilde{\rho}_0) \left(
    \frac{\tilde{\rho}}{\tilde{\rho}_0}
    \right)^{-\frac{\lambda_{-}}{2}}\,.
\end{equation}
In contrast, for decreasing $\tilde\rho$ the eigenstate of $\lambda_{+}$ will finally dominate unless one has at $\tilde{\rho}_0$ a pure eigenstate of $\lambda_{-}$.
For example, if one starts at $\tilde\rho_0$ with $\delta\alpha(\tilde{\rho}_0)=0$ one finds
\begin{equation}
    \label{eq:CLA8}
    \delta\mu(\tilde\rho)=\delta\mu(\tilde{\rho}_0) \left[ (1-\tilde{c}Z) \left( \frac{\tilde{\rho}}{\tilde{\rho}_0}
    \right)^{-\frac{\lambda_{-}}{2}}
    +\tilde{c}Z\left( \frac{\tilde{\rho}}{\tilde{\rho}_0}
    \right)^{-\frac{\lambda_{+}}{2}}
    \right] \,.
\end{equation}
In view of the small value of $\lambda_{+}$ the deviation $\delta\mu$ settles almost to a constant for small $\tilde{\rho}$.

\indent This can be translated directly to small deviations $\delta_m$ of the cosmon-Higgs coupling $\lambda_m$ from a given scaling solution,
\begin{align}
    \label{eq:CLA9}
    \delta_m(\tilde{\rho}) =  \,&\partial_{\tilde{\rho}} \delta\mu(\tilde{\rho}) \nn \\
    = &-\frac{\delta\mu(\tilde{\rho}_0)}{2\tilde{\rho}_0} \Bigg[ \lambda_{-}(1-\tilde{c}Z) 
    \left( \frac{\tilde{\rho}}{\tilde{\rho}_0}
    \right)^{-\frac{1}{2}\hat{A}_m} \nn \\
    &+ \lambda_{+} \tilde{c}Z
    \left( \frac{\tilde{\rho}}{\tilde{\rho}_0}
    \right)^{-\left(1+\frac{\lambda_{+}}{2}\right)}
    \Bigg]\,,
\end{align}
with
\begin{equation}
    \label{eq:CLA10}
    \hat{A}_m = A_m \left(
    1 + \frac{27\lambda_h}{8\pi^2 (1+\mu)^3}
    \right)\, .
\end{equation}
For the flow towards the IR one obtains only a
small correction to $A_m$, reflecting the influence
of $\partial \tilde{\beta}_m/\partial\lambda_m$.
In the opposite direction for the flow towards the UV with decreasing $\tilde\rho$ the deviation $\delta_m$ will increase $\sim \tilde{\rho}^{-1}$ once $\tilde\rho$ has reached a value of the order $\lambda_{+}\tilde{c}Z\tilde{\rho}_0$.
This effective asymmetry in the flow towards the IR and towards the UV matters for the interpretation of numerical solutions. 
For the flow towards the UV the difference $\delta_m$ between two initial values of the cosmon-Higgs coupling follows eq.~\eqref{eq:PT6}, \eqref{eq:PT7}, \eqref{eq:PT10} only for a certain range in $\tilde\rho$, starting to increase very substantially $\sim\tilde\rho^{-1}$ after a certain value of $\tilde\rho$ is reached.

\indent Starting at some small $\tilde{\rho}_0$ with $\mu_h(\tilde{\rho}_0) > 0 $ the mass term $\mu_h(\tilde\rho)$ may reach zero at some value of $\tilde{\rho}>\tilde{\rho}_0$.
Increasing $\tilde{\rho}$ further will lead to negative $\mu_h$ for which the definition of couplings at $\htilde = 0$ is no longer a good choice.
One rather should continue the flow by defining couplings at the partial minimum $\htilde_0 (\tilde{\rho})>0$.
In this case the fermions and gauge bosons get masses.
We discuss the flow in this "SSB-regime" in appendix \ref{app:G}.
The qualitative features of the stability analysis remain similar to the case for $\mu_h \geq 0$.

\section{Quantum gravity and self-organized criticality}
\label{sec:IX}
In this section we show that the existence of a second order quantum electroweak phase transition in quantum gravity leads to the phenomenon of self-organized criticality.
For this setting fundamental scale invariance \cite{CWFSI}, or more generally a largest intrinsic mass scale much smaller than the Fermi scale, is a key ingredient.
For a LIMS larger than the Fermi scale this statement does not hold. 
By existence we mean that the critical surface of the local scaling solution extends to a global scaling solution which is analytic for $\tilde\rho \to 0$. 
We use ``second-order phase transition'' with the understanding that non-perturbative QCD effects finally turn the transition into a crossover.

\indent If the mass term in the effective potential for the Higgs field is given by a scalar field, the critical quantity for deviations from the phase transition is related to the dimensionless cosmon-Higgs coupling $\lambda_m$.
Self-organized criticality requires in this case only a positive anomalous dimension $A_m$ for this coupling.
More precisely, the average value $\bar{A}_m$ has to remain positive and non-zero in the UV-limit $\tilde{\rho}_0 \to 0$.

\indent This situation contrasts from a setting where the mass term $\mu_H^2$ in the Higgs potential is field-independent.
In this case the canonical dimension of $\mu_H^2$ equals two and self-organized criticality requires an anomalous mass dimension $A_\mu > 2$ \cite{CWFP, FTP, CWMYGH}.
The latter is quite far from a perturbative value, unless the canonical dimension of fields is changed \cite{GIPI}.
For $A_\mu < 2$ a tuning of the UV-parameters is required in order to realize criticality, as well known from the presence of a relevant parameter.

\indent For $k^2 > M_p^2$ the metric fluctuations induce a positive anomalous dimension $A_m > 0$.
For fundamental scale invariance self-organized criticality occurs then whenever the UV-completion of a model admits a second order quantum electroweak phase transition.
This is a property of the UV-fixed point for $\tilde{\rho} \to 0$ and not changed by the deviations from an exact phase transition in the IR (very large $\tilde{\rho}$) by  effects of the strong interactions.
Suppose that one can find a critical trajectory for which the particles of the standard model remain almost massless for the whole range $\tilde{h} < 0.1$, and which connects to finite $\lambda_m$ for $\tilde{\rho} \to 0$.
This corresponds to the continuation of $\lambda^{\cri}_m (\tilde{\rho})$ to $\tilde{\rho} \rightarrow 0$.
Fundamental scale invariance predicts then that such a critical trajectory is realized.
The ``tuning'' to the critical surface is enforced by the positivity of the anomalous dimension $\bar{A}_m$ and the finiteness of $\lambda_m$ for $\tilde{\rho} \to 0$.

\indent The predictivity of $\varphi_0/M_p$ extends beyond the case of a second order electroweak quantum phase transition. We find that it is a general property of analytic scaling solutions.
The predicted value of $\varphi_0/M_p$ is related to the question which value of $\bar{\lambda}_m = \lambda_m (\tilde{\rho} \rightarrow \infty)$ is connected to a finite value of $\lambda_m (\tilde{\rho} \rightarrow 0)$.
We discuss this in detail within a linearized scaling equation for $\lambda_m (\tilde{\rho})$.
Realistic values for the Fermi scale require that the scaling solution admits an approximate second order quantum electroweak phase transition, as associated to a tiny value of $\bar{\lambda}_m$.

\subsection*{Gravity induced anomalous dimension for cosmon-Higgs coupling}
For quantum gravity in the UV-range $k\to \infty$ the fluctuations or the graviton (transverse traceless metric fluctuations of corresponding fluctuations of the vierbein) constitute the dominant contribution to the anomalous dimension of all quartic scalar couplings.
Within the gauge invariant flow equation \cite{CWGIF} or simplified flow equation \cite{SFE} this contribution takes a rather simple form \cite{CWGFC, CWQS},
\begin{equation}
    \label{eq:SC1}
    A_m^{(\text{gr})} = \frac{5}{12\pi^2 f}\left(1- \frac{2u}{f}\right)^{-2}\,,\quad
    f = \frac{F}{k^2} = \frac{M_p^2(k)}{k^2}\,.
\end{equation}
In the UV-limit $f$ approaches a constant, $f\to 2w_0$, and $A_m^{(\text{gr})}$ therefore approaches a positive constant.
(One always has $v=2u/f < 1$.)
On the other side, in the IR limit the gravitational fluctuations decouple, since $f = 2\xi\tilde{\rho}$ diverges.

\indent The anomalous dimension $A_m^{(\text{gr})}$ follows from eqs.~\eqref{eq:PT4}, \eqref{eq:PT5}, \eqref{eq:BC58D},
\begin{equation}
    \label{eq:SC2}
    A_m^{(\text{gr})} \lambda_m = 4 \partial_{\tilde{\rho}} \partial_{\tilde h} c_U^{(\text{gr})} (\tilde{\rho},\tilde{h})\,.
\end{equation}
Here we complete in the flow equation \eqref{eq:AA1}
for $u$ the gravitational contribution \cite{CWGFC}
\begin{equation}
    \label{eq:SC3}
    \begin{aligned}
    &c_U^{(\text{gr})} = \\
    &\frac{1}{128\pi^2} \!\left\{\!\left(1 + \frac{1}{3}\left(1 - \frac{\partial \ln f}{\partial \ln \tilde{\rho}}\right)\!\right)\!\left(\frac{5}{1-v} + \frac{1}{1 - v/4}\right) -4\right\}\!.
    \end{aligned}
\end{equation}
Its interpretation is rather easy for the simplified flow equation.
One recognises the five components of the transverse traceless metric fluctuations.
The dimensionless mass term for an expansion around flat space is negative for $U > 0$, leading to an inverse graviton propagator of the form $Fq^2 - 2U + Fk^2 r_k(q^2/k^2)$, where the last term involves the infrared cutoff.
Similar to the other contributions in eqs.~\eqref{eq:FESS7}, \eqref{eq:FESS10} there is a threshold function,
\begin{align}
    \left(1 + \tilde{m}_{\text{gr}}^2\right)^{-1} &= (1-v)^{-1}\,, 
    \nn\\
    v &= \frac{u}{w} = \frac{2u}{f}\,.
    \label{eq:SC4}
\end{align}
The contribution $\sim \frac{1}{3}\left(1 - \frac{\partial \ln f}{\partial \ln \tilde{\rho}}{}_{\;|h}\right)$ arises from the proportionality of the IR-cutoff to $F$.
There is a similar contribution from the physical scalar fluctuation in the metric, with mass term $-U/2$ instead of $-2U$.
Finally, the last term $-4$ in the bracket is a negative measure contribution, accounting for the four components of the gauge symmetry associated to diffeomorphisms. 
In the IR-limit for large $\tilde{\rho}$ one has $\partial \ln f / \partial \ln \tilde{\rho} = 1$ and $v=0$, such that one recovers the two propagating degrees of freedom of the graviton in eq.~\eqref{eq:FESS5}.

\indent Eq.~\eqref{eq:SC3} is an approximation.
For example, there are subleading terms mixing the scalar degree of freedom in the metric with the cosmon \cite{PRWY}.
Nevertheless, the leading contribution from the five-component traceless tensor field are rather robust.
Early computations using different versions of the flow equation can be found in refs.~\cite{DP, NP, PV, DELP, EHLY}.
They typically find a positive sign of $A_m^{(\text{gr})}$, even though the situation is sometimes less clear.

\indent Since a positive sign of $A_m$ is crucial for self-organized criticality we may discuss the robustness of the sign in some more detail. 
This concerns the robustness of $c_U^{(\text{gr})}$.
In the flow equation for $u$ the contributions of sectors with different representations of the Lorentz and gauge symmetries are independent. 
The conceptually simplest contribution arises from the traceless transverse component of the metric or the vierbein. 
These fluctuations are invariant under diffeomorphism transformations, such that their contribution is not affected by the choice of gauge fixing.

As long as the momentum dependence of the inverse graviton propagator is well approximated by the form $q^2+m_g^2(\tilde\rho)$, the computation of the graviton fluctuations is completely analogous to other degrees of freedom. 
Since gravitons are bosons, the contribution to $c_U^{(\text{gr})}$ is positive. 
The prefactor in eq.~\eqref{eq:SC3} depends on the implementation of the IR-cutoff and its precise value may be debatable. 
Its sign is always positive.

\indent The effective graviton mass $m_g^2=-2U/M_p^2$
follows directly from the second functional derivative of the flowing action \eqref{eq:AL1}. 
Its sign is unambiguous. 
We are mainly interested in short distance theories where $u(\tilde\rho \to 0)>0$ since those give rise to interesting inflationary cosmology.
For the scaling solution this is realized if the bosonic contribution overwhelms the fermionic contribution, as typically realized in grand unified theories. 
Increasing $u$ increases $c_U^{(\text{gr})}$ and $A_m^{(\text{gr})}$. 
The tachyonic mass term for the graviton is actually the reason why no local scaling solutions with $u\sim\tilde\rho^2$ for $\tilde\rho\to\infty$ are found. 
For $u\sim\tilde\rho^2$ the term $\frac{2u}{f}$
would diverge $\sim\tilde\rho^2$, violating the condition $2u/f<1$.
For $u\sim-\tilde\rho^2$ the instability occurs in the scalar sector.

\indent For positive $u$ or $v$ the graviton contribution dominates the other
contributions to $c_U^{(\text{gr})}$, and this effect is enhanced for $\partial c_U^{(\text{gr})}/\partial v$, which enters in $A_m^{(\text{gr})}$. 
As a consequence, the contributions from the scalar and vector components of the metric cannot change the sign of $A_m^{(\text{gr})}$. 
This conclusion could only change for large negative values of $v$, for which the scalar and vector contribution may be comparable to the graviton contribution. 
The treatment of the scalar and vector contributions is more complex since it depends on the precise setting and the choice of gauge. 
It becomes particularly apparent for the gauge-invariant or simplified flow equation which decouples the physical scalar mode in the metric from the gauge degrees of freedom. 
The gauge degrees of freedom contribute only a measure factor which does not depend on the values of the scalar fields.
Then the main uncertainty will reside in the precise contribution of the physical scalar mode in the metric and its mixing with other scalar fields. 
As long as this contribution remains subleading it cannot affect the sign of $A_m^{(\text{gr})}$.
We may require $A_m^{(\text{gr})}>0$ as a condition for the short distance theory which leads to self-organized criticality. 
In view of the above discussion this is not a very restrictive criterion.

\indent Let us first consider the approximation where
\begin{equation}
    \label{eq:SC4A}
    \delta_g = \frac{1}{4} \frac{\partial \ln f}{\partial \ln \tilde{\rho}}{}_{|h}
\end{equation}
does not depend on $h$.
Taking a $\tilde{h}$-derivative of $c_U^{(\text{gr})}$ at fixed $\tilde{\rho}$ one finds
\begin{equation}
    \label{eq:SC5}
    \partial_{\tilde{h}} c_U^{(\text{gr})} (\tilde{\rho},\tilde{h}) = \frac{(1-\delta_g) \partial_{\tilde{h}} v}{96\pi^2} \left(\frac{5}{(1-v)^2} + \frac{1}{4(1-v/4)^2}\right)\,.
\end{equation}
The graviton contribution dominates by a factor of 20 over the scalar metric fluctuations (unless $v$ is negative and large).
Neglecting from there on the scalar metric fluctuation one obtains
\begin{equation}
    \partial_{\tilde{h}} c_U^{(\text{gr})} (\tilde{\rho},\tilde{h}) = \frac{1-\delta_g}{48\pi^2(1-v)^2 f} \left(\partial_{\tilde{h}} u - u \partial_{\tilde{h}} \ln f\right)\,.
    \label{eq:SC6}
\end{equation}
Taking a further $\tilde{\rho}$-derivative at fixed $\tilde{h}$ yields
\begin{equation}
    \label{eq:SC7}
    \partial_{\tilde{\rho}} \partial_{\tilde{h}} c_U^{(\text{gr})} (\tilde{\rho},\tilde{h}) = \frac{\lambda_m(1-\delta_g)}{48\pi^2(1-v)^2 f} - \frac{\partial_{\tilde{\rho}} u \partial_{\tilde{h}} u(1-\delta_g)}{12\pi^2 (1-v)^3 f^2} + \Delta_{\text{gr}}\,,
\end{equation}
where $\Delta_{\text{gr}}$ involves derivatives of $\ln f$ and does not contain $\lambda_m$.
For small $\delta_g$ the first term $\sim \lambda_m$ leads to the anomalous dimension \eqref{eq:SC1}.
Writing $u(\tilde{\rho},\tilde{h}) = \lambda_m \tilde{\rho} \tilde{h} + \bar{u}(\tilde{\rho},\tilde{h})$ the dependence of the second term on $\lambda_m$ follows from
\begin{align}
    \label{eq:SC8}
    \partial_{\tilde{\rho}} u \; \partial_{\tilde{h}} u &= \lambda_m^2 \tilde{\rho} \tilde{h} + \lambda_m \left(\tilde{\rho} \partial_{\tilde{\rho}} + \tilde{h} \partial_{\tilde{h}}\right) \bar{u} + \partial_{\tilde{\rho}} \bar{u} \partial_{\tilde{h}} \bar{u} \nn\\
    &= - \lambda_m^2 \tilde{\rho} \tilde{h} + 2\lambda_m (u - c_U) + \partial_{\tilde{\rho}} \bar{u} \partial_{\tilde h} \bar{u}\,.
\end{align}
Since we evaluate $\lambda_m$ at $\tilde{h} = 0$ the term $\sim \lambda_m^2$ does not contribute.
Furthermore, the scaling solution for $\tilde{\rho} \to 0$ is typically given by $u = c_U$, such that the term linear in $\lambda_m$ in eq.~\eqref{eq:SC8} does not contribute either.
The derivatives of $\bar{u}$, together with a term $\sim \partial_{\tilde{\rho}} \partial_{\tilde{h}} \bar{u}$ in $\Delta_{\text{gr}}$, contribute to $\tilde{\beta}_m(\tilde{\rho})$ in eq.~\eqref{eq:PT5}, but not to the anomalous dimension $A_m$.

\indent We finally need to discuss the field-dependence of
\begin{equation}
    \label{eq:SC9}
    f = 2w_0 + 2\xi\tilde{\rho} + \xi_H \tilde{h}\,,
\end{equation}
or
\begin{equation}
    \label{eq:SC10}
    \delta_g = \frac{2\xi \tilde{\rho} + \xi_H \tilde{h}}{4(2w_0 + 2\xi\tilde{\rho} + \xi_H \tilde{h})}\,.
\end{equation}
In this approximation one has
\begin{align}
    \label{eq:SC11}
    \partial_{\tilde{h}} \ln f = \frac{\xi_H}{f}\,,\quad
    \partial_{\tilde{\rho}} \ln f = \frac{2\xi}{f}\,, \quad
    \partial_{\tilde{h}} \partial_{\tilde{\rho}} \ln f = -2\frac{\xi\xi_H}{f^2}\,,
\end{align}
and
\begin{equation}
    \label{eq:SC12}
    \partial_{\tilde{h}} \delta_g = \frac{\xi_H}{4f}(1-4\delta_g)
\end{equation}
In the limit $\tilde{\rho} \to 0$, $\tilde{h} \to 0$ the function $f$ approaches a constant, $f = 2w_0$, and $\delta_g = 0$.
Nevertheless, the field derivatives $\partial_{\tilde{h}} \ln f = \xi_H / (2w_0)$, $\partial_{\tilde{\rho}} \ln f = \xi / w_0$ contribute to $\tilde{\rho} \partial_{\tilde{\rho}} \lambda_m = -\beta_m/2$.
In the limit where $\xi / w_0$ and $\xi_H / w_0$ can be considered as independent of $\lambda_m$ the field derivatives of $f$ do not contribute to the anomalous dimension $A_m$, however.

\indent For the computation of $A_m$ we have assumed implicitly that the other couplings as $y_t^2$, $g_3^2$, $\xi/w_0$ \etc are independent of $\lambda_m$.
The proper procedure considers small deviations of all couplings from their values on the critical surface.
In the linear approximation the flow (or $\tilde{\rho}$-dependence) of these deviations is given by the stability matrix.
The anomalous dimension $A_m$ corresponds to an eigenvalue of the stability matrix.
The diagonalization of the stability matrix is equivalent to a computation to which degree other couplings depend on $\lambda_m$.
Given the small role that the Higgs field plays in the ultraviolet region, and the smallness of $\lambda_m$, it seems unlikely that other couplings depend effectively on $\lambda_m$ in a substantial way.
This justifies the neglect of this effect for the computation of $A_m$.
Additional fields of the short distance model can also contribute to the scalar field anomalous dimension $\eta_H$. 
The contribution from gauge bosons is negative, which can somewhat lower the value of $A_m$.
These contributions are suppressed, however, by the small values of the gauge couplings.

\subsection*{Self-organized criticality}
The presence of a substantial positive gravity induced anomalous dimension for quartic scalar couplings implies that these couplings correspond to irrelevant parameters of the renormalization flow.
Let us consider in the region $\tilde{\rho} \to 0$ the general form of the local scaling solution for $\lambda_m(\tilde{\rho}) = \lambda_m(\tilde{\rho},h=0)$.
We suppose that there exists some regular local scaling solution $\lambda^{(r)}_m (\tilde{\rho})$ which remains finite for $\tilde{\rho}\rightarrow0$.
Local scaling solutions in the vicinity of $\lambda^{(r)}$ obey
\begin{equation}
    \label{eq:YA1}
    \lambda_m(\tilde{\rho}) = \lambda_m^{(r)}(\tilde{\rho}) + d_m \tilde{\rho}^{-\frac{A_m}{2}}\,,
\end{equation}
with $d_m$ a free integration constant.
For $\tilde{\rho}\rightarrow 0$ the anomalous dimension $A_m$ is dominated by the gravitational fluctuations and takes a constant value.
With $A_m > 0$ a finite value for $\lambda_m(\tilde{\rho})$ in the limit $\tilde{\rho} \to 0$ requires $d_m = 0$.
This is the meaning of an irrelevant coupling.
Starting at $\tilde{\rho}_0$ with an initial value $\lambda_m(\tilde{\rho}_0)$ not too far from $\lambda_m^{(r)}(\tilde{\rho}_0)$, the difference $\lambda_m(\tilde{\rho})-\lambda_m^{(r)}(\tilde{\rho})$ decreases with increasing $\tilde{\rho}$ proportional to $(\tilde{\rho}/\tilde{\rho}_0)^{-A_m/2}$.
For $\tilde{\rho}_0 \to 0$ the coupling $\lambda_m(\tilde{\rho})$ coincides at any finite nonzero $\tilde{\rho}$ with $\lambda_m^{(r)}(\tilde{\rho})$.
There is no free integration constant for the field dependence of $\lambda_m(\tilde{\rho})$.
For all $\tilde{\rho}$ the cosmon-Higgs coupling is given by the regular solution $\lambda_m^{(r)}(\tilde{\rho})$.
If $\lambda_m^{(r)}(\tilde{\rho}) = \lambda_m^{\cri}(\tilde{\rho})$ corresponds to the critical surface of a second order quantum phase transition, all scaling solutions in its vicinity have to be on the critical surface.
This property is called self-organized criticality.
No parameter has to be tuned in order to obtain the critical behavior.
If a scaling solution (or a family of scaling solutions) which corresponds to a second order phase transition exists, the other local scaling solutions will be attracted towards the critical surface as $\tilde{\rho}$ increases.
If the range in $\ln \tilde{\rho}$ for which this attraction is present grows to infinity, as for $\tilde{\rho} \to 0$, any small deviation from the critical surface vanishes at finite nonzero $\tilde{\rho}$.

\indent For the scaling solution the self-organized criticality in the quartic scalar coupling $\lambda_m$ extends to self-organized criticality for the dimensionless mass term of the Higgs scalar $\mu_h = \partial_{\tilde{h}} u (\tilde{\rho}, \tilde{h}=0)$.
Indeed, for the scaling solution $\lambda_m(\tilde{\rho})$ and $\mu_h(\tilde{\rho})$ are related by $\lambda_m(\tilde{\rho}) = \partial_{\tilde{\rho}} \mu_h (\tilde{\rho})$.
This implies that $\mu_h(\tilde{\rho})$ is determined by $\lambda_m(\tilde{\rho})$ up to a constant.
In particular, if $\lambda_m(\tilde{\rho})$ is given by $\lambda_m^{\cri}(\tilde{\rho})$, one concludes that $\mu_h(\tilde{\rho})$ has to take its value on the critical surface, $\mu_h^{\cri}(\tilde{\rho})$.
In this respect there is a crucial difference between the field-dependence of couplings for the scaling solution and the $k$-dependence of couplings for arbitrary solutions.
For the flow with $k$ at fixed $\rho$ and $H$ the deviations from the critical surface involve $\mu_h$ and $\lambda_m$ as two independent couplings.
While $\lambda_m$ is irrelevant in the presence of the gravitational fluctuations, $\mu_h$ is a relevant coupling for $A_m < 2$.
Then the initial value of $\mu_h$ has to be tuned in order to obtain the critical behavior.
For the scaling solution this tuning is enforced by the finiteness of $\lambda_m$ for $\tilde{\rho} \to 0$.

\indent For more detail, consider the general $k$-dependence of $\mu_h$ at fixed $\tilde{\rho}$ and $h=0$, where we do not impose the scaling solution, 
\begin{equation}
    \label{eq:YA2}
    k \partial_k \mu_h = 2 \tilde{\rho} \partial_{\tilde{\rho}} \mu_h - 2\mu_h + 4\partial_{\tilde{h}} c_U\,.
\end{equation}
Taking for $c_U$ the leading gravitational contribution one infers from eq.~\eqref{eq:SC3}
\begin{equation}
    \label{eq:YA3}
    k \partial_k \mu_h = 2 \tilde{\rho} \partial_{\tilde{\rho}} \mu_h + (A_\mu - 2) \mu_h\,,
\end{equation}
with anomalous mass dimension
\begin{equation}
    \label{eq:YA4}
    \partial_{\tilde{h}} c_U^{(\text{gr})} = \frac{1}{4} A_\mu^{(\text{gr})}\mu_h + \dots\,,\quad
    A_\mu^{(\text{gr})} = A_m^{(\text{gr})}\,.
\end{equation}
This implies for deviations $\delta_\mu$ from the critical surface
\begin{equation}
    \label{eq:YA5}
    \mu_h(k) = \mu_h^{\cri}(k) + \delta_\mu(k)\,,\quad
    \delta_\mu(k) \sim k^{A_m-2}\,.
\end{equation}
For $k \to 0$ one finds a diverging $\delta_\mu$, indicating a relevant parameter.

\indent For the local scaling solution, the $k$-dependence translates to the $\tilde{\rho}$-dependence
\begin{equation}
    \label{eq:YA6}
    \mu_h(\tilde{\rho}) = \mu_h^{\cri}(\tilde{\rho}) + \delta_\mu (\tilde{\rho})\,,\quad
    \delta_\mu(\tilde{\rho}) = d_\mu \tilde{\rho}^{1-\frac{A_m}{2}}\,.
\end{equation}
Again, $\delta_\mu$ diverges in the infrared for $\tilde{\rho} \to \infty$ if $d_\mu \neq 0$.
For the scaling solution the deviations from the critical surface $\delta_\mu(\tilde{\rho})$ and $\delta_m(\tilde{\rho})$ are related, however.
With
\begin{equation}
    \label{eq:YA7}
    \lambda_m^{\cri}(\tilde{\rho}) = \partial_{\tilde{\rho}} \mu_h^{\cri} (\tilde{\rho})\,,\quad
    \delta\lambda_m(\tilde{\rho}) = \partial_{\tilde{\rho}} \delta_\mu(\tilde{\rho})\,,
\end{equation}
eqs.~\eqref{eq:YA1} and \eqref{eq:YA6} imply
\begin{equation}
    \label{eq:YA8}
    d_m = d_\mu \left(1-\frac{A_m}{2}\right)\,.
\end{equation}
With $d_m = 0$ one concludes $d_\mu = 0$.
Thus $\mu_h(\tilde{\rho})$ has to assume its value $\mu_h^{\cri}(\tilde{\rho})$ on the critical surface.

\indent This argument generalizes to an arbitrary regular solution $\lambda_m^{(r)}(\tilde\rho)$ which does not necessarily coincide with a second order phase transition.
The corresponding scaling solution for $\mu_h (\tilde\rho)$ is unique, being given by a regular solution $\mu_h^{(r)}(\tilde\rho)$.
There is no free parameter for the scaling solution despite the fact that for general solutions of the flow with $k$ the mass term $\mu_h$ is a relevant parameter.
Scaling solutions fix the relevant parameter to zero.
A free parameter does not reappear in the form of a continuous family of scaling solutions.

\indent A crucial ingredient for this important conclusion is the standard assumption that an irrelevant coupling remains finite in the UV-limit.
The finiteness of $\lambda_m$ relates rather naturally to the analytic properties of $c_U$ at $h=0$.
For $\tilde{\rho} \to 0$ the scaling equation for $\mu_h(\tilde{\rho})$ takes the typical form
\begin{equation}
    \label{eq:YA9}
    \tilde{\rho} \partial_{\tilde{\rho}} \mu_h = -\frac{1}{2}\bar{\beta}_{\mu}(\tilde{\rho}) + \left(1 - \frac{A_\mu}{2}\right)\mu_h\,.
\end{equation}
The critical values for $\mu_h$ and $\lambda_m$ takes in this case the limiting value for $\tilde{\rho}\rightarrow0$
\begin{equation}
    \label{eq:YA10}
    \mu_h(0) = \frac{\bar{\beta}_\mu(0)}{2-A_\mu}\,.
\end{equation}
This corresponds to the regular solution and contains no free integration constant.

\indent We emphasize that the coupled system of cosmon and Higgs scalar is the basis for this version of self-organized criticality.
Without the cosmon there is no coupling $\lambda_m$, and therefore no restriction on $d_\mu$ in eq.~\eqref{eq:YA6}.

\subsection*{Predictivity for the Fermi scale}
\indent The relation between $\mu_h$ and $\lambda_m$ for the scaling solution implies that there is no free mass parameter for the dependence of the effective potential on $H$.
Also the cosmon-Higgs coupling $\lambda_m$, which determines the ratio of Fermi scale over Planck mass, does not involve a free integration constant.
As a consequence the ``integration constant'' $\bar{\lambda}_m = \lambda_m(\tilde{\rho} \to \infty)$ is predicted for a given short distance model.
Similarly, the quartic Higgs coupling $\bar{\lambda}_h = \lambda_h(\tilde{\rho} \to \infty)$ is fixed.
In consequence, the ratio $h_0 = \varphi_0^2/\rho$ is computable rather than being a free parameter.
This results in the prediction of the Fermi scale in units of the Planck mass,
\begin{equation}
    \label{eq:YA11}
    \frac{\varphi_0}{M_p} = \sqrt{\frac{h_0}{2\xi}} = \left(-\frac{\bar{\lambda}_m}{2\xi\bar{\lambda}_h}\right)^{\frac{1}{2}}\,.
\end{equation}
For the normalization of the cosmon field with $\xi=1$ and given quartic scalar coupling $\bar \lambda_h$ the value $\bar\lambda_{{m}}$ determines directly the ratio $\varphi_0 / M_p$.

\indent In general, the regular function $\lambda_m^{(r)}(\tilde{\rho})$ depends on other functions, as $y_t^2(\tilde{\rho})$ or $g_3^2(\tilde{\rho})$.
This could induce supplementary free parameters in the form of boundary values.
Since for $\tilde{\rho} \to \infty$ these coupling functions are measured, the corresponding boundary values are fixed.
In extensions of the standard model the precise form of $\lambda_m^{\cri}(\tilde{\rho})$ may also depend on couplings in the beyond standard model sector.
This could weaken the predictivity of the scaling solution for the Fermi scale.
We comment on theses issues below.
In any case, predictivity means here "predictivity in principle" in the sense that a given model has no free parameter to adjust the outcome for the Fermi scale.
"Predictivity in practice" may depend on the knowledge of the ultraviolet physics.

\indent In the limit where a second order quantum electroweak phase transition is realized by a scaling solution the prediction becomes unique, namely $\bar{\lambda}_m = \lambda_m^{\cri}(\tilde{\rho} \to \infty) = 0$.
In such a scenario it would be sufficient that a local scaling solution with boundary values $\bar{\lambda}_m = 0$ can be extended to all values of $\tilde{\rho}$ and $h$.
This would then constitute a global scaling solution.
Self organized criticality for $A_m > 0$ selects this particular solution within the family of neighbouring local scaling solutions with $\bar{\lambda}_m \neq 0$.
Usually, the problem is to explain a tiny ratio $\varphi_0/M_p$.
For an exact second order phase transition one would end with a prediction, but for this prediction $\varphi_0/M_p=0$ turns out to be too small.

\indent Due to the running of the dimensionless couplings the quantum electroweak transition is not precisely a second order phase transition.
The masses of $W$- and $Z$-boson cannot be smaller than the QCD-confinement scale $\Lambda_\QCD$.
This also holds for the quark masses.
A characteristic mass scale for non-perturbative QCD may be taken around $1\;\mathrm{GeV}$, the mass of the vector mesons or baryons.
The question arises if the deviations from the second order phase transition can induce a Fermi scale which is a factor $200$ higher than this characteristic strong interaction scale.
Our discussion in appendix \ref{app:A} indicates that the non-perturbative QCD-effects are too small for generating a realistic Fermi scale in the case where the critical scaling solution $\lambda_m^{\cri} (\tilde{\rho})$ can be extended to a finite value at $\tilde{\rho} = 0$.

\indent On the other hand, the violation of particle scale symmetry by running couplings is also present for $k$ around the Planck mass. 
It needs to be explored if this effect is responsible for a small value $\bar\lambda_m\neq0$.

\subsection*{Analytic scaling solution}
\indent The basis for the predictivity for the Fermi scale is the requirement that $\lambda_m (\tilde{\rho})$ remains finite for $\tilde{\rho}\xrightarrow{}0$. This type of requirement is the standard setting for irrelevant parameters. For the flow with the renormalization scale $k$ any deviation from the UV-fixed point value of an irrelevant parameter diverges for $k\xrightarrow{}\infty$ in the linear approximation. In turn, the assumption that for arbitrarily large $k$ a finite deviation from the fixed point remains within the range of validity of the linear approximation implies the prediction that the deviation is zero. There remains no freedom in the choice of an irrelevant parameter in the vicinity of the fixed point. This generalises to other couplings corresponding to irrelevant parameters. For example, the derivative $\partial_{\tilde{\rho}}\lambda_m (\tilde{\rho}=0)$ is such a coupling. In the standard setting for irrelevant parameters it is finite for the scaling solution and can be predicted.

\indent The property underlying the concept of irrelevant parameters is the analyticity of the scaling solution for $\tilde{\rho}\xrightarrow{}0$ and $\tilde{h}\xrightarrow{}0$.  This assumption of analyticity has important consequences. Consider the scaling potential $u(\tilde{\rho}, \tilde{h})$. Analyticity implies that $\mu_h(\tilde{\rho}=0)=\partial_{\tilde{h}}u|_{\tilde{\rho}=\tilde{h}=0}$, $\lambda_m(\tilde{\rho}=0)=\partial_{\tilde{\rho}}\partial_{\tilde{h}}u|_{\tilde{\rho}=\tilde{h}=0}$ or   $\lambda_h(\tilde{\rho}=0)=\partial_{\tilde{h}}^2u|_{\tilde{\rho}=\tilde{h}=0}$ all are finite. The same holds for higher couplings as $\partial_{\tilde{\rho}}\partial_{\tilde{h}}^2u|_{\tilde{\rho}=\tilde{h}=0}$ or $\partial_{\tilde{\rho}}^2\partial_{\tilde{h}}u|_{\tilde{\rho}=\tilde{h}=0}$. This finiteness imposes restrictions for the behaviour of beta-functions in the vicinity of the UV-fixed point. Consider the scaling equation for $\mu_h(\tilde\rho)$,

\begin{equation} \label{eq:ASS1}
    \begin{aligned}
        \mu_h (\tilde{\rho}) &=\partial_{\tilde{h}}u(\tilde{\rho}, h=0),\\
        \tilde{\rho}\partial_{\tilde{\rho}}\mu_h &= -\frac{1}{2}\beta_\mu.
    \end{aligned}
\end{equation}
Since $\partial_{\tilde{\rho}}\mu_h=\lambda_m$ remains finite for $\tilde{\rho}\xrightarrow{}0$, one infers that $\beta_\mu/\tilde{\rho}$ remains finite for $\tilde{\rho}\xrightarrow{}0$. This condition is stronger than the fixed point condition $\beta_\mu(\tilde{\rho}\xrightarrow{}0)=0$. We can also take a $\tilde{\rho}$-derivative of eq.~\eqref{eq:ASS1}

\begin{equation}
    \label{eq:ASS2}
    \tilde{\rho}\partial_{\tilde{\rho}}^2\mu_h+\lambda_m =-\frac{1}{2}\partial_{\tilde{\rho}}\beta_\mu.
\end{equation}
Since the l.h.s of eq.\eqref{eq:ASS2} remains finite, this also holds for $\partial_{\tilde{\rho}}\beta_\mu$. Typically, $\beta_\mu$ depends on the other couplings. 
Inserting the scaling solution for these couplings $\beta_\mu(\tilde{\rho})$ becomes an analytic function of $\tilde{\rho}$,

\begin{equation}
    \label{eq:ASS3}
    \beta_\mu(\tilde{\rho}) = \beta_\mu^{(1)}\tilde{\rho} + \cdots=-2\lambda_m \tilde{\rho} + \cdots
\end{equation}

\indent These conditions for beta-functions generalize to arbitrary coupling functions $\sigma(\tilde{\rho})$, with $\tilde{\rho}\partial_{\tilde{\rho}}\sigma = -\beta_\sigma/2$. 
Once $\beta_\sigma$ is expressed in terms of scaling solutions for couplings it becomes a function of $\tilde{\rho}$. The coefficients of an expansion of $\beta_\sigma(\tilde{\rho})$ in powers of $\tilde{\rho}$ are given by finite fixed point values for couplings. Similar arguments hold for an expansion in terms of $\tilde{h}$ at fixed $h$. 
With
\begin{equation}
    \label{eq:ASS4}
    \tilde{h}\partial_{\tilde{h}}\mu_h=-\frac{1}{2}\beta_\mu =\tilde{h}\lambda_h,
\end{equation}
and 

\begin{equation}
    \label{eq:ASS5}
    \tilde{h}\partial_{\tilde{h}}^2\mu_h + \lambda_h = -\frac{1}{2}\partial_{\tilde{h}}\beta_\mu,
\end{equation}
one finds for $\beta_\mu(\tilde{h})$ relations as 

\begin{equation}
    \label{eq:ASS6}
    \partial_{\tilde{h}}\beta_\mu|_{\tilde{h}=0} =-2\lambda_h(\tilde{h}=0).
\end{equation}

\indent A somewhat particular role is played by the gauge couplings. An analytic scaling solution may be realized if $Z_F=g^{-2}$ is analytic for $\tilde{\rho}\xrightarrow{}0$, $\tilde{h}\xrightarrow{}0$, cf. eq.~\eqref{eq:GCA}. This poses no problem for an UV-fixed point at finite $g^2_*$. On the other hand, for asymptotic freedom $g^{-2}$ diverges for $\tilde{\rho}\rightarrow0$. 
In this case one needs to employ renormalized gauge fields and an appropriate definition of coupling functions which remain analytic for $\tilde{\rho}\xrightarrow{}0$, $\tilde{h}\xrightarrow{}0$.
More generally, analyticity has to be imposed for renormalized fields.
We use this type of fields implicitly, which induces for the Higgs field the anomalous field dimension $\eta_H$.

\indent An interesting case is the beta-function for the cosmon-Higgs coupling $\lambda_m(\tilde{\rho})$,

\begin{equation}
    \label{eq:ASS7}
    \tilde{\rho}\partial_{\tilde{\rho}}\lambda_m = -\frac{1}{2}\beta_m = - \frac{1}{128\pi^2\tilde{\rho}}\sum_i \hat{n}_i \beta_i -\frac{1}{2}A_m\lambda_m. 
\end{equation}
For $\tilde{\rho}\xrightarrow{}0$ the analytic behaviour$\beta_i \sim c_i\tilde{\rho}$ is needed in order to find a finite fixed point for $\lambda_m^* = \lambda_m(\tilde{\rho}=0)$,

\begin{equation}
    \label{eq:ASS8}
    \lambda_m^* = - \dfrac{\sum_i \hat{n}_ic_i}{64\pi^2A_m}.
\end{equation}
If the analytic behaviour of $\beta_i$ similar to eq.~\eqref{eq:ASS3} were not realized, a finite value $\lambda_m^*$ would require strong cancellations of individual contributions $\beta_i$. The finiteness of $\partial_{\tilde{\rho}}\beta_m$ further restricts the behaviour of the r.h.s~of eq.~\eqref{eq:ASS7}. It is realized for finite $\partial_{\tilde{\rho}}A_m$ and 

\begin{equation}
    \label{eq:ASS9}
    \beta_i = c_i \tilde{\rho} + \frac{d_i}{2}\tilde{\rho}^2 +\cdots,
\end{equation}
with

 \begin{equation}
     \label{eq:ASS10}
     \partial_{\tilde{\rho}} \beta_m = \frac{1}{128\pi^2}\sum_i \hat{n}_i d_i + A_m\partial_{\tilde{\rho}} \lambda_m +\partial_{\tilde{\rho}}A_m \lambda_m.
 \end{equation}
 
\indent The analyticity of the scaling solution provides for a selection among the family of local scaling solutions in the UV-region of small $\tilde{\rho}$. Assume that one has found an analytic scaling solution $\sigma^{(r)} (\tilde{\rho})$ for some coupling function $\sigma(\tilde{\rho})$. Solutions of the scaling equation in the vicinity of this scaling solution obey the differential equation

\begin{equation} \label{eq:ASS11}
    \begin{aligned}
        \sigma(\tilde{\rho}) &= \sigma^{(r)}(\tilde{\rho}) + \delta_\sigma(\tilde{\rho}),\\
        \tilde{\rho}\partial_{\tilde{\rho}} \delta_\sigma &= -\frac{1}{2} A_\sigma \delta_\sigma,\\
        A_\sigma &= \partial_\sigma \beta_\sigma.
    \end{aligned}
\end{equation}
Here the critical exponent $A_\sigma$ is evaluated for the analytic scaling solution $\sigma^{(r)}(\tilde{\rho})$. Eq.~\eqref{eq:ASS11} may be generalised to the standard system of linear equations in case of several coupled coupling functions $\sigma_i(\tilde{\rho})$. In this case $\sigma$ corresponds to one of the eigenmodes of the stability matrix $A_{ij}$. In the vicinity of $\tilde{\rho}=0$ one can approximate $A_\sigma$ by a constant, with local scaling solution

\begin{equation}
    \label{eq:ASS12}
    \sigma(\tilde{\rho}) = \sigma^{(r)}(\tilde{\rho}) + c_\sigma \tilde{\rho}^{-\frac{A_\sigma}{2}}.
\end{equation}
Analyticity of $\sigma(\tilde{\rho})$ requires $c_\sigma=0$ unless $A_\sigma$ takes the special values $-2,-4\dots$. This selects the particular solution $\sigma^{(r)}(\tilde{\rho})$ among the family of local scaling solutions characterized by $c_\sigma$. No free parameter remains.

\indent For $A_\sigma>0$ one deals with an irrelevant parameter and retrieves the standard result that irrelevant parameters can be predicted. Analyticity requires, however, the condition $c_\sigma=0$ also for the relevant parameters with $A_\sigma<0$. This is the reason for self-organized criticality and predictivity of the Fermi scale, which follows from taking $\sigma  =\mu_h$. The analyticity argument is very general. An analytic scaling solution implies a discrete set of local scaling solutions for $\tilde{\rho}\xrightarrow{}0$ for all coupling functions. 
Both irrelevant and relevant parameters are fixed or "predicted". 
A continuous family of analytic scaling solutions is only possible for the exceptional cases $A_\sigma=-2,-4\dots$. In general this is not realized.

\subsection*{Linear approximation}

\indent In the limit where the dependence of $\tilde{\beta}_m$ and $A_m$ on $\lambda_m$ can be neglected eq.~\eqref{eq:262A} becomes a linear differential equation for $\lambda_m$.
This is not exact, since the gauge and Yukawa couplings and their $\beta$-functions depend of $\mu_h$, and $\mu_h$ depends of $\lambda_m$.
The complete system of differential equations couples all couplings to each other.
Nevertheless, the approximation of decoupling the $\tilde{\rho}$-flow of $\lambda_m$ from the other couplings can provide a good overview of the general behavior.
With $x=\ln(\tilde{\rho})$ we write eq.~\eqref{eq:262A} as 
\begin{equation}
    \label{eq:EG1}
    \frac{\partial \lambda_m}{\partial x} = - \frac{1}{2} \left(
    \tilde{\beta}_m (x) + A(x) \lambda_m
    \right)\, .
\end{equation}
We can interpret eq.~\eqref{eq:EG1} as a general expansion of the scaling equation for $\lambda_m$ in powers of $\lambda_m$.
This includes the indirect contribution from the $\lambda_m$-dependence of the other couplings in linear order in $\lambda_m$.
From there $A$ may receive additional terms beyond $A_m$ and similar for $\tilde{\beta}_m$.
The discussion of the solution to eq.~\eqref{eq:EG1} extends to $\tilde{\rho}\rightarrow 0$ where gravitational and other degrees of freedom contribute to $\tilde{\beta}$ or $A$.
It remains valid to the extend that non-linear terms $\lambda^2_m$ can be neglected.
Since $\lambda_m$ remains very small for the whole range of $\tilde{\rho}$ this will be a good approximation.
When it comes to effects of the order of $10^{-30}$, as for the observed present value of $\lambda_m$, it is not excluded, however, that non-linear effects can play a role.

\indent Eq.~\eqref{eq:EG1} has the general solution
\begin{equation}
    \label{eq:EG2}
    \lambda_m (x)  = \lambda_c(x) + \tilde{C}(x)\lambda_m(x_0)\,,
\end{equation}
with
\begin{equation}
    \label{eq:EG3}
    \tilde{C}(x) = \exp\left\{ 
    -\frac{1}{2} \int_{x_0}^x dx^\prime\, A(x^\prime)
    \right\} \, .
\end{equation}
The second term involves the integration constant or initial value $\lambda_m(x_0)$.
The part $\lambda_c(x)$ is given by
\begin{equation}
    \label{eq:EG4}
    \lambda_c (x) = -\frac{\tilde{C}(x)}{2}
    \int_{x_0}^xdx^\prime \, \frac{\tilde{\beta}_m(x^\prime)}{\tilde{C}(x^\prime)}\,.
\end{equation}
For large $x$ the integrand in eq.~\eqref{eq:EG4} decays exponentially
\begin{equation}
    \label{eq:EG5}
    \frac{1}{2} \tilde{\beta}_m (x) = \frac{b_m(x)}{128\pi^2}e^{-x}\,.
\end{equation}
The value of $\bar{\lambda}_m = \lambda_m(x\rightarrow\infty)$ is determined by the combination of the contribution from $\lambda_{m,0} = \lambda_m (x_0)$ and the integral \eqref{eq:EG4}, which is dominated by $x \lesssim 1$.

\indent We need information about the integration constant $\lambda_m(x_0)$ for $x_0$ in the region with $\tilde{\rho} \ll 1$, together with information about $A(x)$ and $\tilde{\beta}_m (x)$ in the region $x \lesssim 1$.
This is necessary for the computation of $\lambda_m (\bar{\rho}_0)$ in a region where $\bar{\rho}_0 \gg 1$.
For $\tilde{\rho} > \bar{\rho}_0$ one can then use the flow in standard model.
We will discuss here the general features of the $\tilde{\rho}$-dependence of $\lambda_m$ for $\tilde{\rho}$ near zero by an expansion in powers of $\tilde{\rho}$.
The values of the expansion coefficients will depend on the short-distance model.

\indent For $\tilde{\rho}\rightarrow 0$ or $x\rightarrow-\infty$ we assume an analytic scaling solution for $\mu_h(\tilde{\rho})$,
\begin{equation}
    \label{eq:EG6}
    \mu_h(\tilde{\rho}) = \mu_0 + \lambda_0 \tilde\rho + \cdots \,.
\end{equation}
The scaling equation for $\mu_h$ reads 
\begin{equation}
    \label{eq:EG7}
    \tilde{\rho} \partial_{\tilde{\rho}} \mu_h = \left(
    1-\frac{A}{2}
    \right) \mu_h + \beta_\nu \, ,
\end{equation}
with $\beta_\nu$ given by eq.~\eqref{eq:118C}, \eqref{eq:NYA} for $\tilde{\rho} \gg 1$, and taken more generally here.
We include in $A$ the anomalous field dimension $\eta_H$ and other possible contributions linear in $\mu_h$.
We expand
\begin{equation}
    \label{eq:EG8}
    \beta_\nu = \beta_0 + \beta_1 \tilde\rho + \cdots \,, \quad A = A_0 + A_1 \tilde\rho + \cdots \,.
\end{equation}
The UV-fixed point for $\tilde{\rho}\rightarrow0$ follows from eqs.~\eqref{eq:EG6}, \eqref{eq:EG7}, \eqref{eq:EG8} as
\begin{align}
    \label{eq:EG12}
    \mu_0 &= -\beta_0 \left( 1- \frac{A_0}{2}\right)^{-1} \,, \nn \\
    \lambda_0 &= \frac{2\beta_1 -A_1\mu_0}{A_0} = \frac{2}{A_0} \left( \beta_1 + \frac{A_1\beta_0}{2-A_0} \right) \,.
\end{align}
We identify $\lambda_{m,0}$ with $\lambda_0$.
The scaling equation for $\lambda_m (\tilde{\rho}) = \partial_{\tilde{\rho}} \mu_h(\tilde{\rho})$ follows from eq.~\eqref{eq:EG7}
\begin{equation}
    \label{eq:EG9}
    \tilde{\rho} \partial_{\tilde{\rho}} \lambda_m (\tilde{\rho}) = \partial_{\tilde{\rho}} \beta_\nu (\tilde{\rho}) -\frac{1}{2} \partial_{\tilde{\rho}} A(\tilde{\rho})\mu_h(\tilde{\rho}) -\frac{1}{2}A(\tilde{\rho})\lambda_m(\tilde{\rho})\,,
\end{equation}
and we identify
\begin{equation}
    \label{eq:Eg10}
    \tilde{\beta}_m(\tilde{\rho}) = -2 \partial_{\tilde{\rho}}\beta_\nu (\tilde{\rho}) + \partial_{\tilde{\rho}} A_m(\tilde{\rho})\mu_h(\tilde{\rho})\,.
\end{equation}
For $\tilde{\rho}\rightarrow 0$ analyticity results in 
\begin{equation}
    \label{eq:EG11}
    \tilde{\beta}_0 = \tilde{\beta}_m (\tilde{\rho} = 0) = -2\beta_1 + A_1 \mu_0 \,.
\end{equation}

\indent Let us take $x_0$ in the region of $\tilde{\rho} \ll 1$ for which the expansion applies.
One finds 
\begin{align}
    \label{eq:EG13}
    \ln \tilde{C} &= -\frac{1}{2} \int_{\tilde{\rho}_0}^{\tilde{\rho}}\frac{d\rho^\prime}{\rho^\prime}\,(A_0 + A_1\rho^\prime) \nn \\
    &= -\frac{1}{2} \left(
    A_0\ln \frac{\tilde{\rho}}{\tilde{\rho}_0} + A_1 \left( \tilde{\rho}-\tilde{\rho}_0 \right)
    \right)\,, \nn \\
    \tilde{C} &= \left(
    \frac{\tilde{\rho}}{\tilde{\rho}_0}
    \right)^{-\frac{A_0}{2}}\exp \left(
    -\frac{A_1}{2} \left( \tilde{\rho}-\tilde{\rho}_0\right)
    \right)
    \,.
\end{align}
For the approximate computation of $\lambda_c$ in this region we expand 
\begin{equation}
    \label{eq:EG14}
    \tilde{\beta}(\tilde{\rho})= \tilde{\beta}_0 + \tilde{\beta}_1 \tilde\rho + \cdots
\end{equation}
and linearise the exponential in eq.~\eqref{eq:EG13}.
This yields
\begin{align}
    \label{eq:EG15}
    \lambda_c(\tilde{\rho}) = &-\frac{1}{2}\left(
    \frac{\tilde{\rho}}{\tilde{\rho}_0}
    \right)^{-\frac{A_0}{2}} \exp\left\{
    -\frac{A_1}{2}\left( \tilde{\rho}-\tilde{\rho}_0\right)
    \right\} \nn \\
    &\times\int_{\tilde{\rho}_0}^{\tilde{\rho}}\frac{d\rho^\prime}{\rho^\prime}\, \left( \tilde{\beta}_0 + \tilde{\beta}_1 \rho^\prime \right) \left( \frac{\rho^\prime}{\tilde{\rho}_0} \right)^{\frac{A_0}{2}} \exp \left\{
    \frac{A_1}{2} \left( \rho^\prime-\tilde{\rho}_0 \right)
    \right\} \nn \\
    = &\frac{\tilde{\beta}_0}{A_0} \left[ 
    \left( \frac{\tilde{\rho}_0}{\tilde{\rho}} \right)^{\frac{A_0}{2}} -1
    \right]\left(
    1-\frac{A_1}{2}\tilde{\rho}
    \right) \nn \\
    &+ \frac{2\tilde{\beta}_1 + A_1\tilde{\beta}_0}{4+2A_0}\left(
    \tilde{\rho}_0^{1+\frac{A_0}{2}}\tilde{\rho}^{-\frac{A_0}{2}} - \tilde{\rho}
    \right)\,.
\end{align}
The scaling solution for $\lambda_m$ in the linear approximation \eqref{eq:EG1} therefore reads for small $\tilde{\rho}$ and $\tilde{\rho}_0$
\begin{align}
    \label{eq:EG16}
    \lambda_m(\tilde{\rho}) = &\Bigg(
    \frac{\tilde{\beta}_0}{A_0} + \frac{2\tilde{\beta}_1 + A_1 \tilde{\beta}_0}{4+2A_0}\tilde{\rho}_0 \nn \\
    &+ \lambda_m(\tilde{\rho}_0)\left( 
    1+\frac{A_1}{2}\tilde{\rho}_0
    \right)
    \Bigg)
    \left(
    \frac{\tilde{\rho}}{\tilde{\rho}_0}
    \right)^{-\frac{A_0}{2}} \nn \\
    &-\frac{A_1}{2}\left(
    \frac{\tilde{\beta}_0}{A_0} + \lambda_m(\tilde{\rho}_0)
    \right)\tilde{\rho}\left(
    \frac{\tilde{\rho}}{\tilde{\rho}_0}
    \right)^{-\frac{A_0}{2}} \nn \\
    &-\frac{\tilde{\beta}_0}{A_0} + \left(
    \frac{A_1\tilde{\beta}_0}{2A_0} - \frac{2\tilde{\beta}_1 + A_1 \tilde{\beta}_0}{4+2A_0}
    \right)\tilde{\rho}\,.
\end{align}

\indent A finite value of $\lambda_m(\tilde\rho)$ for $\tilde\rho\rightarrow0$ requires for the integration constant $\lambda_m(\tilde{\rho}_0)$ the value
\begin{equation}
    \label{eq:EG17}
    \lambda_m(\tilde{\rho}_0) = -\left(
    \frac{\tilde{\beta}_0}{A_0} + \frac{2\tilde{\beta}_1 + A_1 \tilde{\beta}_0}{4+2A_0} \tilde{\rho}_0
    \right)\left(
    1 + \frac{A_1}{2}\tilde{\rho}_0
    \right)^{-1}\,.
\end{equation}
This demonstrates the general result that an analytic scaling solution does not admit free integration constants.
In lowest order in $\tilde{\rho}_0$ the relation $\lambda_m(\tilde{\rho}_0) = -\tilde{\beta}_0/A_0$ corresponds to the fixed point value.
This also ensures that the second term in eq.~\eqref{eq:EG16} vanishes for the lowest order of the simultaneous expansion in $\tilde\rho$ and $\tilde{\rho}_0$ chosen here.
What remains is a linear approach of $\lambda_m(\tilde\rho)$ to the fixed point value for $\tilde{\rho}\rightarrow0$,
\begin{equation}
    \label{eq:EG18}
    \lambda_m(\tilde\rho) = -\frac{\tilde{\beta}_0}{A_0} + \left(
    \frac{A_1\tilde{\beta}_0}{2A_0} - \frac{2\tilde{\beta}_1 + A_1 \tilde{\beta}_0}{4+2A_0}
    \right)\tilde{\rho} \,,
\end{equation}
as required for an analytic scaling solution.

\indent We can use the information on $\lambda_m(\tilde{\rho}_0)$ for the computation of $\lambda_m(\tilde\rho)$ for arbitrary $\tilde\rho$.
We choose initial conditions for $\tilde{\rho}_0 \rightarrow 0$ and obtain from eqs.~\eqref{eq:EG2}, \eqref{eq:EG3}, \eqref{eq:EG4}
\begin{align}
\label{eq:EG19}
\lambda_m(\tilde{\rho}) &= -\lim_{\tilde{\rho}_0 \rightarrow 0} \Bigg[\exp \Bigg\{ 
- \frac{1}{2} \int_{\tilde{\rho}_0}^{\tilde{\rho}} \frac{d\rho^\prime}{\rho^\prime} A(\rho^\prime) 
\Bigg\}  \frac{\tilde{\beta}_0}{A_0} \nn \\
&\quad + \frac{1}{2} \int_{\tilde{\rho}_0}^{\tilde{\rho}} \frac{d\rho^\prime}{\rho^\prime} \tilde{\beta}_m(\rho^\prime) 
\exp \Bigg\{ \frac{1}{2} \int_{\tilde{\rho}}^{\rho^\prime} \frac{d\rho^{\prime\prime}}{\rho^{\prime\prime}} A(\rho^{\prime\prime}) \Bigg\}\Bigg] \,.
\end{align}
The first term vanishes for $\tilde{\rho}_0 \rightarrow 0$.
For both $\tilde{\rho}_0$ and $\tilde{\rho} > \tilde{\rho}_0$ in the region of validity of the expansion in powers of $\tilde{\rho}$ this yields
\begin{align}
    \label{eq:355A}
    \lambda_m (\tilde{\rho}) &= -\frac{1}{2} \lim_{\tilde{\rho}_0 \rightarrow 0} \int_{\tilde{\rho}_0}^{\tilde{\rho}} \frac{d\rho^\prime}{\rho^\prime} \left( \tilde{\beta}_0 + \tilde{\beta}_1 \rho^\prime \right) \left( \dfrac{\rho^\prime}{\tilde{\rho}} \right)^{\frac{A_0}{2}} \nn \\
    &= - \left( \frac{\tilde{\beta}_0}{A_0} + \frac{\tilde{\beta}_1}{2+A_0}\tilde{\rho} \right) \,,
\end{align}
in accordance with eq.~\eqref{eq:EG18} since we have omitted the contribution $\sim A_1 \tilde{\rho}$ in $A$.

\indent We may also express the general solution by choosing $x_0 \rightarrow \infty$, with $\lambda_m(x_0) = \bar{\lambda}_m$,
\begin{equation}
\label{355B}
\lambda_m(\tilde{\rho}) = \exp \Bigg\{ \frac{1}{2} \int_{\tilde{\rho}}^{\infty} \frac{d\rho^\prime}{\rho^\prime} A(\rho^\prime) \Bigg\} \left( \bar{\lambda}_m + \Delta_m (\tilde{\rho}) \right) \,,
\end{equation}
with
\begin{equation}
\label{eq:coisa}
 \Delta_m (\tilde{\rho})= \frac{1}{2} \int_{\tilde{\rho}}^{\infty} \frac{d\rho^\prime}{\rho^\prime} \tilde{\beta}_m(\rho^\prime) \exp \Bigg\{ - \frac{1}{2} \int_{\rho^\prime}^{\infty} \frac{d\rho^{\prime\prime}}{\rho^{\prime\prime}} A(\rho^{\prime\prime}) \Bigg\} \,. 
\end{equation}
For a given value of $\bar{\lambda}_m$ this solution should be continued to $\tilde{\rho} \rightarrow 0$.
The predicted value for $\bar{\lambda}_m$ is the one for which $\lambda_m (\tilde{\rho})$ according to eq.~\eqref{eq:coisa} reaches for $\tilde{\rho}$ the value $-\tilde{\beta}_0/A_0$.
For neighbouring values of $\bar{\lambda}_m$ the value $\lambda_m (\tilde{\rho}_0)$ will not obey eq.~\eqref{eq:EG17}, and the solution \eqref{eq:EG15} will violate analyticity for $\tilde{\rho} \rightarrow 0$, with $\lambda_m (\tilde{\rho})$ diverging $\sim \tilde{\rho}^{-(A_0/2)}$.

\indent The factor
\begin{equation}
    \label{eq:356A}
    D(\tilde{\rho}) =\exp \left\{ \frac{1}{2} \int_{\tilde{\rho}}^\infty \frac{d\rho^\prime}{\rho^\prime} A(\rho^\prime) \right\}
\end{equation}
diverges for $\tilde{\rho} \rightarrow 0$ proportional $\tilde{\rho}^{-(A_0/2)}$.
The condition for an analytic scaling solution is therefore given by
\begin{align}
    \label{eq:356B}
    \bar{\lambda}_m &= -\Delta_m (0)  \nn\\ 
    &=- \frac{1}{2} \int_0^\infty \frac{d\rho^\prime}{\rho^\prime} \tilde{\beta}_m (\rho^\prime) \exp \left\{ -\frac{1}{2} \int_{\rho^\prime}^\infty \frac{d\rho^{\prime\prime}}{\rho^{\prime\prime}} A(\rho^{\prime\prime}) \right\}\,.
\end{align}
The relation $\Delta_m (0)=0$ is singled out as the condition for an analytic scaling solution to realise a second order quantum electroweak phase transition.
This is associated with a "particle physics scale symmetry".

\indent We may define some type of renormalized cosmon-Higgs coupling
\begin{equation}
    \label{eq:356C}
    \lambda_m^{(R)} (\tilde{\rho}) = D^{-1} (\tilde{\rho}) \lambda_m (\tilde{\rho})\,.
\end{equation}
The field-dependence of $\lambda_m^{(R)}$ is given by 
\begin{equation}
    \label{eq:356D}
   \tilde{\rho} \partial_{\tilde\rho} \lambda_m^{(R)} = -\frac{1}{2} \tilde{\beta}_m^{(R)} = -\frac{1}{2} D^{-1}(\tilde\rho) \tilde{\beta}_m (\tilde{\rho})\,.
\end{equation}
For a finite value $\lambda_m (\tilde{\rho}=0)$ the renormalised coupling $\lambda_m^{(R)}(0)$ vanishes.
Eq.~\eqref{eq:356B} reads
\begin{equation}
    \label{eq:356E}
    \bar{\lambda}_m = -\frac{1}{2}\int_{-\infty}^{\infty} dx \tilde{\beta}_m^{(R)} (x) = -\frac{1}{2}\int_{0}^{\infty} \frac{d\rho^\prime}{\rho^\prime}\tilde{\beta}_m^{(R)} (\rho^\prime)\,.
\end{equation}
The function $\tilde{\beta}_m^{(R)}$ behaves for $\rho^\prime \rightarrow 0$ as
\begin{equation}
    \label{eq:356F}
    \tilde{\beta}_m^{(R)}(\rho^\prime) = \left( \tilde{\beta}_0 + \tilde{\beta}_1 \rho^\prime \right)
    \left( \dfrac{\rho^\prime}{\bar{\rho}} \right)^{\frac{A_0}{2}}
    D^{-1}(\bar{\rho})\,,
\end{equation}
with $\bar{\rho} \ll 1$ such that the expansion in powers of $\rho^\prime$ is valid for $\tilde{\beta}_m$ and $A$ in the region $\rho^\prime < \bar{\rho}$.
The integral \eqref{eq:356E} has for $\rho^\prime \rightarrow 0$ the same behavior as eq.~\eqref{eq:355A} and is therefore finite.

\indent From eq.~\eqref{eq:356E} a tiny value of $\bar\lambda_m$ requires that the negative contribution to the integral from the region $\tilde\rho\gtrsim1$ needs to be canceled by a positive contribution from the region $\tilde\rho\lesssim1$. 
Thus $\tilde\beta_m$ has to change from positive values for large $\tilde\rho$ to negative values for small $\tilde\rho$.
At this stage we do not know if a cancellation needs the particular selection of a short distance model, or if there exists some more general mechanism based on the fixed point behavior of the critical trajectory linked to particle scale symmetry.

\subsection*{Iterative global solution}

\indent Some lessons may be learned by an attempt to construct iteratively a global scaling solution corresponding to a second order electroweak quantum phase transition. 
Instead of $\lambda_m$ we choose the variable
\begin{align}
\gamma
&=
(c+\tilde\rho)
\left(
\lambda_m-\frac12\tilde\beta_m
\right)\,,
\nonumber\\
\lambda_m
&=
\frac{\gamma}{c+\tilde\rho}
+\frac12\tilde\beta_m\,.
\label{eq:I1}
\end{align}
It interpolates between
$\sigma_m-\frac12\tilde\beta_m$, $\sigma_m=\tilde\rho\lambda_m$, for large $\tilde\rho$, and $c(\lambda_m-\frac12\tilde\beta_m)$ for $\tilde\rho\ll c$.
We have subtracted from $\lambda_m$ the lowest order expression of the critical trajectory for large $\tilde\rho$. 
The constant $c$ is of the order one and indicates the onset of the short distance regime at $\tilde\rho\approx c$. 
We can choose $c$ freely and look for an optimal value later. 
The scaling equation \eqref{eq:BC58C} for $\lambda_m$ translates to
\begin{align}
\partial_x\gamma
=
&\left(
\frac{\tilde\rho}{c+\tilde\rho}
-\frac{A_m}{2}
\right)\gamma
\nn \\
&-\frac12(c+\tilde\rho)
\left[
\left(
1+\frac{A_m}{2}
\right)\tilde\beta_m
+\partial_x\tilde\beta_m
\right]\,,
\label{eq:I2}
\end{align}
with $\tilde\rho=e^x$. 
In the region $\tilde\rho\gg c$ we employ $\tilde\beta_m = \tilde b_m/\tilde\rho$ and neglect $c$, resulting in
\begin{equation}
\partial_x\gamma
=
\left(
1-\frac{A_m}{2}
\right)\gamma
-\frac12
\left(
\partial_x\tilde\beta_m
+\frac{A_m}{2}\tilde\beta_m
\right)\,.
\label{eq:I3}
\end{equation}
The lowest order solution sets the r.h.s. to zero
\begin{equation}
\gamma_{>}^{(1)}
=
\frac12
\left(
\partial_x\tilde\beta_m
+\frac{A_m}{2}\tilde\beta_m
\right)
\left(
1-\frac{A_m}{2}
\right)^{-1}\,.
\label{eq:I4}
\end{equation}

\indent A first iteration step for $\gamma$ sets the right hand side of eq.~\eqref{eq:I2} to zero. 
This corresponds to a fixed point in the limit where the $x$-dependence of the source term is small compared to the coefficient of the term linear in $\gamma$. 
In this approximation one finds
\begin{align}
\gamma^{(1)}
&=
\frac12(c+\tilde\rho)
\left[
\left(
1+\frac{A_m}{2}
\right)\tilde\beta_m
+\partial_x\tilde\beta_m
\right]D^{-1},
\nonumber\\
D
&=
\frac{\tilde\rho}{c+\tilde\rho}
-\frac{A_m}{2}\,.
\label{eq:I5}
\end{align}
For large $\tilde\rho\gg c$ one recovers eq.~\eqref{eq:I4} and $\lambda_m = \frac{\gamma}{\tilde{\rho}}+ \frac{1}{2}\tilde \beta_m$ results in
\begin{align}
\lambda_m^{(1)}
&=
\frac12\tilde\beta_m
\left(
1+
\frac{2\gamma_{>}^{(1)}}
{\tilde\beta_m\tilde\rho}
\right)
=
\frac12\tilde\beta_m(1+g)
\nonumber\\
&=
\frac12\tilde\beta_m
\left(
1+
\left(
\partial_x\ln\tilde b_m+\frac{A_m}{2}
\right)
\left(
1-\frac{A_m}{2}
\right)^{-1}
\right)\,.
\label{eq:I6}
\end{align}
We recognize the critical trajectory \eqref{eq:263H} in leading order in $\partial_x\ln\tilde b_m$.

\indent In the opposite limit $\tilde\rho\ll c$ one has, with
$\tilde\beta_m=\tilde\beta_0+\tilde\beta_1\tilde\rho$, $A_m=A_0+A_1\tilde\rho$,
\begin{align}
&\gamma_{<}^{(1)}
=
-\frac{c\tilde\beta_0}{A_0}
\left(
1+\frac12A_0
\right)
+E\tilde\rho\,,
\nonumber\\
&E
=
-\frac1{A_0}
\left[
2\tilde\beta_0
+ c \tilde\beta_1
+(2-cA_1)\frac{\tilde\beta_0}{A_0}
+\frac12A_0
\left(
\tilde\beta_0+2c\tilde\beta_1
\right)
\right].
\label{eq:I7}
\end{align}
This yields
\begin{align}
\lambda_m^{(1)}
&=
\frac12\tilde\beta_0
+
\frac{\gamma_{<}^{(1)}}{c}
\left(
1-\frac{\tilde\rho}{c}
\right)
=
-\frac{\tilde\beta_0}{A_0}
+F\tilde\rho\,,
\nonumber\\
F
&=
-\frac1{cA_0}
\Bigg[
\tilde\beta_0
\left(
1-\frac{A_0}{2}
\right)
+c\tilde\beta_1
+\frac{(2-cA_1)\tilde\beta_0}{A_0}
\nn \\
&\qquad\qquad+\frac12A_0
\left(
\tilde\beta_0+2c\tilde\beta_1
\right)
\Bigg]\,.
\label{eq:I8}
\end{align}
For $\tilde\rho\to0$ the function $\lambda_m^{(1)}$ is analytic and reaches the UV-fixed point of the exact solution. 
The coefficient $F$ depends on $c$ and differs, in general, from the exact solution.

\indent The coefficient of the term linear in $\gamma$ in eq.~\eqref{eq:I2} changes sign at $\tilde\rho=cA_m/(2-A_m)$.
It is positive for large $\tilde\rho$, and negative for $\tilde\rho\to0$. 
As a consequence, the factor $D^{-1}$ in eq.~\eqref{eq:I5} diverges at this value of $\tilde\rho$.

\indent For a very small linear coefficient of $\gamma$, the approximation of
the solution which sets the r.h.s. of eq.~\eqref{eq:I2} to zero is not valid. 
We therefore may modify our ansatz $\gamma^{(1)}$ by smoothing $D^{-1}$,
\begin{equation}
\bar D^{-1}
=
\left(
\frac{\tilde\rho}{c+\tilde\rho}
-\frac{A_m}{2}
\right)
\left[
\left(
\frac{\tilde\rho}{c+\tilde\rho}
-\frac{A_m}{2}
\right)^2
+
\frac{\varepsilon c\tilde\rho}
{(c+\tilde\rho)^2}
\right]^{-1}.
\label{eq:I9}
\end{equation}
The positive constant $\varepsilon$ can be adapted for later purposes. 
This form of smoothing does not affect the value of $\lambda_m^{(1)}$ at $\tilde\rho=0$ or in the limit $\tilde\rho\gg1$.

\indent As a next step we write
\begin{equation}
\gamma(\tilde\rho)
=
\gamma^{(1)}(\tilde\rho)\,\eta(\tilde\rho)\,.
\label{eq:I10}
\end{equation}
An analytic scaling solution for $\lambda_m$ with $\bar\lambda_m=0$ would be realized if for $\eta(0)=1$ the function $\eta(\tilde\rho)$ remains finite for $\tilde\rho\gg1$.
The scaling equation for $\eta$ reads
\begin{equation}
\partial_x\eta
=
\frac{\partial_x\gamma}{\gamma^{(1)}}
-\eta\,\partial_x\ln\gamma^{(1)}
=
G\eta-D\,,
\label{eq:I11}
\end{equation}
with
\begin{align}
G
&=
\frac{\tilde\rho}{c+\tilde\rho}
-\frac{A_m}{2}
-\partial_x\ln\gamma^{(1)}
\nonumber\\
&=
-\Bigg[
\frac{A_m}{2}
+\partial_x\ln\tilde\beta_m
+
\frac{
\partial_x^2\ln\tilde\beta_m
+\frac12\partial_xA_m
}{
1+\frac{A_m}{2}+\partial_x\ln\tilde\beta_m
}
\nn \\ 
&\qquad-\partial_x\ln D
\Bigg]\,.
\label{eq:I12}
\end{align}
We may consider the approximate solution
\begin{equation}
\eta=\frac{D}{G}\,.
\label{eq:I13}
\end{equation}
This replaces in eq.~\eqref{eq:I5} the factor $D^{-1}$ by $G^{-1}$,
\begin{align}
\gamma^{(2)}
&=
\frac12(c+\tilde\rho)
\left[
\left(
1+\frac{A_m}{2}
\right)\tilde\beta_m
+\partial_x\tilde\beta_m
\right]G^{-1}\,,
\nn\\
\lambda_m^{(2)}
&=
\frac12\tilde\beta_m
+
\frac12
\left[
\left(
1+\frac{A_m}{2}
\right)\tilde\beta_m
+\partial_x\tilde\beta_m
\right]G^{-1}\,.
\label{eq:I14}
\end{align}
Since the factor $D^{-1}$ cancels out we may first have a look what happens if we take $\varepsilon\to0$, resulting in
\begin{equation}
\partial_x\ln D
=
-\left(
\frac12\partial_xA_m
-\frac{c\tilde\rho}{(c+\tilde\rho)^2}
\right)
\left(
\frac{\tilde\rho}{c+\tilde\rho}
-\frac{A_m}{2}
\right)^{-1}\,.
\label{eq:I15}
\end{equation}
The value $G^{-1}(\tilde\rho=0)=-2/A_0$
is the same as for $D^{-1}$, such that $\lambda_m(\tilde\rho=0) = - \tilde\beta_0 /A_0$ remains unchanged. 
For $\tilde\rho\gg c$ one finds a further iteration for the critical trajectory according to
\begin{align}
G^{-1}
=
\Bigg(
&1-\frac{A_m}{2}
-\partial_x\ln\tilde\beta_m
-\frac{\partial_x^2\ln\tilde\beta_m
+\frac12\partial_xA_m}{\partial_x\ln\tilde\beta_m+\frac12A_m}
\nn \\
&+
\frac{\partial_xA_m}{2-A_m}
\Bigg)^{-1}\,.
\label{eq:I16}
\end{align}
The function $G(\tilde\rho)$ still has a zero, since for large $\tilde\rho$ it assumes a positive value close to one, and for $\tilde\rho=0$ it is negative. 
The position of the singularity of $\gamma^{(2)}$ has shifted somewhat as compared to $\gamma^{(1)}$.
The value of $\tilde\rho$ where $G$ diverges due to eq.~\eqref{eq:I15} corresponds to $\gamma^{(2)}=0$ and therefore $\lambda_m^{(2)} = \tilde\beta_m/2$.

\indent One could continue this iteration. 
It may produce more accurate estimates of the behavior for $\tilde\rho\gg c$ and $\tilde\rho\to0$. 
There will remain, nevertheless, a ``transition region'' in $\tilde\rho$ where the sign of the term linear in the coupling switches. 
This region will not be under control for an expansion which assumes that the inverse of this term is small in some sense.
For the full solution of the scaling equation for $\gamma$ with boundary condition either set at $\tilde\rho=0$ or at $\tilde\rho\to\infty$, this transition region is not problematic. 
For a numerical solution the vanishing of the term linear in the coupling is not an issue. 
The iterative solution investigated here involves, however, boundary values simultaneously at $\tilde\rho\to0$ and $\tilde\rho\to\infty$. 
The branches of the iterative solution for small $\tilde\rho$ and large $\tilde\rho$ may not match in the transition region.

\indent The variable $\gamma$ reveals a general structure. 
For large $\tilde\rho$ the critical trajectory is ultraviolet attractive due to a positive linear term $\sim\gamma$ on the r.h.s. of eq.~\eqref{eq:I2}. 
For small $\tilde\rho$ the analytic solution is infrared attractive due to the negative term $-A_m\gamma/2$. 
The transition region corresponds to this switch in attraction properties.
If there are no special properties associated with the second order electroweak phase transition the two branches will not meet precisely unless the parameters of the short distance model have appropriate values. 
This may be called a ``tuning'' of the short distance physics.
For an analytic scaling solution such tuning is possible only if the lack of knowledge of the precise short distance model is put into the form of free parameters. 
The amount of tuning necessary depends on how far the extension of the critical trajectory into the region of small $\tilde\rho$ results in an almost constant $\gamma$ with value not too far from the limit for $\tilde\rho=0$. 
There may be particular interesting cases as $\tilde\beta_0=0$ resulting in $\lambda_m(0)=0$.

\subsection*{Predictions for gauge and Yukawa couplings}

\indent Analytic scaling solutions are rather restricted. In general, there is at most a discrete set of such solutions, without free parameters. 
For example, if gauge couplings have an UV-fixed point \eqref{eq:UV8} at a non-zero value $g^2_\ast$, this fixed point has to be realized for $\tilde{\rho}\xrightarrow{}0$ exactly.
The scaling solution for $\tilde \rho \rightarrow 0$ cannot involve the term $\sim C_g$ unless $B_g$ is a negative even integer.
Otherwise the function $Z_F (\tilde{\rho})$ in eq.~\eqref{eq:GCA} would not be analytic for $\tilde \rho \rightarrow 0$ and some couplings involving $\tilde{\rho}$-derivatives of $Z_F$ would diverge.
The coefficient $B_g$ is a loop expression involving factors of $\pi$.
In the absence of $\tilde \rho$-dependent masses it is not expected to be an integer.
With analyticity of the scaling solution requiring $C_g=0$, this leads to a simple strong conclusion. 
There is no free parameter which permits the gauge coupling near the Planck mass to assume an arbitrary value.
Its value is fixed by the properties of the UV-fixed point for $\tilde\rho\to0$.
In consequence, the value of the gauge coupling at the Fermi scale can be predicted for a given model. 
This ``predictivity in principle'' is rather similar to the issue of the Fermi scale. 
Since the predicted value depends on the short distance model the comparison with observation either discards models which do not have the required fixed point structure, or the LIMS cannot be below the Fermi scale.

\indent The gauge couplings still flow away from their UV-fixed point as $\tilde\rho$ increases.
For the flow with $k$ the deviation of the gauge coupling from the fixed point value is a relevant or irrelevant parameter, depending on the sign of $B_g$.
It involves the free parameter $C_g$.
Predictivity is realized only if the gauge coupling is an irrelevant parameter.
In contrast, for the scaling solution the behavior in the vicinity of the fixed point for $\tilde \rho \rightarrow 0$  is a mass-threshold effect both for a relevant and irrelevant parameter, independently of the sign of $B_g$. 
Predictivity is realized in both cases if the LIMS is below the Fermi scale.

\indent The decoupling of the metric fluctuations or of massive particles allows for an evolution away from the fixed point once $\tilde{\rho}$ grows sufficiently large. 
This $\tilde \rho$-dependence of the gauge couplings is induced by the $\tilde \rho$-dependence of the Planck mass and particle masses.
For $\tilde \rho \rightarrow 0$ we can linearize the threshold functions and extend eq.~\eqref{eq:UV9} to
\begin{equation}
    \label{eq:364A}
    \tilde{\rho} \partial_{\tilde{\rho}} \,g^{-2} = -\frac{B_g}{2} g^{-2} \left( 1-r_g \tilde{\rho} \right) - \frac{B_F}{2} (1-r_F \tilde{\rho})\,.
\end{equation}
Expanding
\begin{equation}
    \label{eq:364B}
    g^{-2} = -\frac{B_F}{B_g} + \delta g^{-2} \,,
\end{equation}
yields
\begin{equation}
    \label{eq:364C}
    \tilde{\rho} \partial_{\tilde{\rho}} \,\delta g^{-2} = - \frac{B_g}{2} \delta g^{-2} \left( 1-r_g \tilde{\rho}\right) +\frac{B_F}{2} \left( r_F - r_g \right) \tilde{\rho}\,.
\end{equation}
An analytic scaling solution is realized for
\begin{equation}
    \label{eq:364D}
    \delta g^{-2} = \gamma_F \tilde{\rho} + \cdots \,,
\end{equation}
with $\gamma_F$ the fixed point value of a coupling involving two scalar fields and two, three our four gauge bosons.
The expansion in $\tilde{\rho}$ yields
\begin{equation}
    \label{eq:364E}
    \gamma_F = \frac{B_F}{2} \left( r_F-r_g \right) \left( 1 + \frac{B_g}{2} \right)^{-1}\,.
\end{equation}

\indent The coupling $\gamma_F$ is typically small, with $B_F \sim 1/(16\pi^2)$, $r_F \sim g^2$.
At the value $\tilde{\rho}_g$ where the metric fluctuations decouple effectively the relative change of the gauge coupling as compared to its fixed point value is given by the factor
\begin{equation}
    \label{eq:364F}
    \delta g^{-2} g^2_\ast = \frac{(r_g - r_F)B_g \tilde{\rho}_g}{2 + B_g} = \frac{B_g}{2+B_g} \left( t_g - \frac{r_F}{r_g} \right)\,,
\end{equation}
with $t_g = r_g \tilde{\rho}_g$ close to one.
For small $\delta g^{-2}g^2_\ast$ the fixed point value $g^2_\ast/(4\pi)$ has to be close to the value extrapolated from the observed value at the Fermi scale which yields $\alpha_g \approx 1/40$.
Such a scenario requires for small $\tilde{\rho}$ the presence of particles beyond the standard model.

\indent Similar arguments apply to the Yukawa couplings and the quartic Higgs coupling $\lambda_h$.
An analytic scaling solution predicts the value of $\lambda_h$ near the Planck mass, or more precisely at $\tilde{\rho}_g$ where the gravitational fluctuations decouple.
It has to be close to the fixed point value $\lambda^\ast_h$.
The value of $\lambda^\ast_h$ is rather small, leading to the successful prediction of the ratio between Higgs mass and top quark mass \cite{SHAW}.
An analytic scaling solution further implies a prediction for the value of the $\beta$-function $\beta_h$ for $\lambda_h$ at $\tilde{\rho}_g$.
It is given by the value of the six point function for the Higgs scalar,
\begin{align}
    \label{eq:264G}
    \gamma_h (\tilde{\rho}_c) &= \frac{\partial \lambda_h (\tilde{\rho})}{\partial \tilde{\rho}}|_{\tilde{\rho}} = \tilde{\rho}_c \,, \nn \\
    \beta_\lambda (\tilde{\rho}_c) &= -2 \tilde{\rho} \partial_{\tilde{\rho}} \, \lambda_h (\tilde{\rho}_c) = -2 \gamma_h (\tilde{\rho}_c) \tilde{\rho}_c\,.
\end{align}
For small $\gamma_h$ close to its value for $\tilde \rho =0$ the value $\beta_\lambda (\tilde{\rho}_c)$ remains small, close to its zero for $\tilde{\rho} =0$.
Suppose that the six-point function $\gamma_h$ remains very small at a scale $\tilde\rho_{\mathrm{SM}}$ for which the standard model can already be used as an effective theory. 
If $\tilde\rho_{\mathrm{SM}}$ is not very large, this predicts a small value for $\beta_\lambda(\tilde\rho_{\mathrm{SM}})$.
On the other hand, one can compute $\beta_\lambda(\tilde\rho_{\mathrm{SM}})$ perturbatively as a function of the Yukawa and gauge couplings. 
Thus a small $\beta_\lambda(\tilde\rho_{\mathrm{SM}})$ implies a relation between Yukawa and gauge couplings. 
In turn, this results in a prediction for the value of the top quark mass which corresponds to a small $\beta_\lambda(\tilde\rho_{\mathrm{SM}})$. 
This relation works well and is compatible with the observed mass of the top quark.

\indent If the largest intrinsic mass scale (LIMS) is far below the Fermi scale, all couplings of the standard model are given by the scaling solution, except perhaps the masses of neutrinos. For an analytic scaling solution they all become predictable in principle in the UV. In case of a discrete set of analytic scaling solutions with more than one solution one shall find a discrete set of predictions. 
If for a given model the computational possibilities permit the estimate of the predictions with sufficient accuracy, the hypothesis of the LIMS below the Fermi scale, or of fundamental scale invariance, can be falsified by comparison with the observed values of couplings. With this hypothesis one can then judge if an UV-model with a given particle content is viable or not. Of course, predictability in  principle is not yet a prediction in practice.

\indent The predictions of fundamental scale invariance with an analytic scaling solution are based on the assumption that a given model can be continued to infinitely short distances or $k\xrightarrow{}\infty$. It is also conceivable that a model with a given particle content remains only valid up to some scale $\bar{k}_{UV}$ much above the Planck scale, with scaling solution only valid for $\tilde{\rho}>\bar{\rho}_{UV}$, $\bar{\rho}_{UV}\ll1$. In this case the predictions are weakened. One typically obtains "infrared intervals" around the predictions which would hold for $\bar{\rho}_{UV}=0$. In the limit $\bar{\rho}_{UV}\xrightarrow{}0$ these intervals shrink to points.

\indent For the present investigation we will assume that the particle content of the UV-model yields fixed point values of gauge and Yukawa couplings which lead to infrared values compatible with the observed values.

\section{Numerical solution}
\label{sec:X}

\indent Our analytic results can be checked by a numerical solution of the scaling equation.
For the scaling equation in our truncation this does not involve any further approximation.
We are mainly interested here in the general issue of predictivity for the Fermi scale.
For this purpose we do not aim for a realistic model of the ultraviolet physics.
We simply adjust the "particle content" by choosing for $k$ beyond the Planck mass $(\tilde{\rho} \ll 1)$ a form of the $\beta$-functions for gauge and Yukawa couplings such that these fixed point values are compatible with observation.
We focus on the $\tilde{\rho}$-dependence of the couplings defined at $\htilde = 0$.

\subsection*{Scaling solution for the standard model}

\indent We first solve the scaling equation for the region $\tilde{\rho} \gg 1$ where the fluctuations of the metric and possible particles with masses near the Planck mass can be neglected.
The particle content is given by the standard model plus the cosmon.
For this local scaling solution we set "initial conditions" at a value of $\tilde{\rho}$ which corresponds to $k=m_Z$.
At this scale we can employ the measured values of the gauge and Yukawa couplings as well as the quartic scalar coupling.
The only free initial value is then the cosmon-Higgs coupling $\lambda_m$.
We neglect all Yukawa couplings except the one for the top quark and use one-loop $\beta$-functions for the gauge and Yukawa couplings.

\indent For an expansion of $u(\tilde{\rho},\tilde{h})$ in powers of $\tilde{h}$ at $\tilde{h}=0$ one finds in our truncation a closed system of non-linear differential equations for the scaling solution that may be solved numerically.
We are mainly interested in the $\tilde{\rho}$-dependence of $\lambda_m(\tilde{\rho}) = \lambda_m(\tilde{\rho},h=0)$ and consider for this purpose the scaling equation for the quantity
\begin{equation}
    \label{eq:MA1}
    \sigma_m(\tilde{\rho}) = \tilde{\rho} \lambda_m(\tilde{\rho},h=0)\,,
\end{equation}
which reads
\begin{align}
    \label{eq:MA2}
    \tilde{\rho} \partial_{\tilde{\rho}} \sigma_m &= \sigma_m + \tilde{\rho}^2 \partial_{\tilde{\rho}} \lambda_m = \sigma_m - \frac{1}{2} \tilde{\rho} \beta_m \nn\\
						  &= \sigma_m - \frac{1}{128\pi^2} \bigg\{ \sum_i{}^\prime \hat{n}_i \beta_i - 12 \tilde{\rho} \partial_{\tilde{\rho}} \lambda_h (1 + \partial_{\tilde{h}} u)^{-2} \nn\\
						  &\hspace{73pt}+ 24 \lambda_h \sigma_m (1+ \partial_{\tilde{h}} u)^{-3} \bigg\}\,.
\end{align}
Here the sum $\sum_i{}^\prime$ extends over all particles except for the scalars.
For these particles one has $\tilde{m}_i^2 = 0$ for $h = 0$,
\begin{equation}
    \label{eq:MA3}
    \tilde{m}_i^2 = \alpha_i h \tilde{\rho}\,,\quad
    \tilde{\rho} \partial_{\tilde{\rho}} \alpha_i = -\frac{1}{2} \beta_i\,.
\end{equation}

\indent The last term in eq.~\eqref{eq:MA2} involves the anomalous dimension $A_m$, such that we may write
\begin{align}
    \label{eq:MA4}
    \tilde{\rho} \partial_{\tilde{\rho}} \sigma_m =&\; \left(1 - \frac{A_m}{2}\right)\sigma_m \\
    &- \frac{1}{128\pi^2} \left\{\sum_i{}^\prime \hat{n}_i \beta_i + 6 \beta_h(1+\mu_h)^{-2}\right\}\,,\nn
\end{align}
where, 
\begin{gather}
    \label{eq:MA5}
    \mu_h(\tilde{\rho}) = \partial_{\tilde{h}}u(\tilde{\rho},\tilde{h}=0)\,,\quad
    \lambda_h(\tilde{\rho}) = \partial_{\tilde{h}}^2 u(\tilde{\rho},\tilde{h}=0)\,,\nn\\
    \beta_h = -2\tilde{\rho} \partial_{\tilde{\rho}} \lambda_h(\tilde{\rho})\,.
\end{gather}
For the anomalous dimension of the cosmon-Higgs coupling $A_m$ we include the part of the anomalous dimension $\eta_H$ of the Higgs field. 
This replaces $A_m \to A_m + \eta_H$, resulting in
\begin{equation}
    \label{eq:MA6}
    A_m = \frac{3\lambda_h}{8\pi^2(1+\mu_h)^3} + \eta_H\,,
\end{equation}
with
\begin{equation}
    \label{eq:MA7A}
    \eta_H = \frac{3}{8\pi^2} \left(y_t^2 - \frac{3(5g_2^2 + g_1^2)}{20(1+\mu_h)}\right)\,.
\end{equation}

\indent For the $\beta$-functions all particles except the scalars are massless for $h=0$, implying in our approximation
\begin{align}
    \label{eq:MA8}
    \beta_t &= - 2\tilde{\rho} \partial_{\tilde{\rho}} y_t^2 \nn\\
	    &= \frac{1}{16\pi^2} \bigg\{\frac{3y_t^4}{1+\mu_h} -\left(16g_3^2  + \frac{4}{5} g_1^2\right)y_t^2\bigg\}+\eta_H y^2_t\,,\nn\\
    \beta_3 &= -2 \tilde{\rho} \partial_{\tilde{\rho}} g_3^2 = -\frac{7g_3^4}{8\pi^2}\,,\nn\\
    \beta_2 &= -2 \tilde{\rho} \partial_{\tilde{\rho}} g_2^2 = -\frac{1}{12\pi^2}\left(5-\frac{1}{4(1+\mu_h)^3}\right)g_2^4\,,\nn\\
    \beta_1 &= -2 \tilde{\rho} \partial_{\tilde{\rho}} g_1^2 = \frac{1}{2\pi^2}\left(1 +\frac{1}{40(1+\mu_h)^3}\right)g_1^4\,.
\end{align}
For $\beta_h$ we add the contribution from the scalar field anomalous dimension $\eta_H$,
\begin{align}
    \label{eq:MA9}
    \beta_h =&\; \frac{3}{4\pi^2} \left\{\frac{\lambda_h^2}{(1+\mu_h)^3} - y_t^4 + \frac{3}{16}g_2^4 + \frac{3}{40}g_2^2 g_1^2 + \frac{9}{400}g_1^4\right\} \nn\\
&+ 2\eta_H \lambda_h\,.
\end{align}
For some of the expressions the dependence on $\mu_h$ is somewhat more complicated than a simple power $(1+\mu_h)^{-\gamma}$.
In this case our simplified formula expresses only the leading power for the suppression for large $\mu_h$.
In any case, we will see that the effect of non-zero $\mu_h$ is minor.
The mass term $\mu_h$ is related to $\lambda_m$ by
\begin{equation}
    \label{eq:MM1}
    \lambda_m = \partial_{\tilde{\rho}} \mu_h\,,
\end{equation}
implying the scaling equation
\begin{equation}
    \label{eq:MM2}
    \tilde{\rho} \partial_{\tilde{\rho}} \mu_h = \sigma_m\,.
\end{equation}

\indent The system of the scaling equations \eqref{eq:MA2}, \eqref{eq:MA8}, \eqref{eq:MA9}, \eqref{eq:MM2} for the seven couplings $\mu_h$, $\sigma_m$, $\lambda_h$, $y_t^2$, $g_3^2$, $g_2^2$, $g_1^2$ is closed in our approximation.
It can be solved numerically.
The flow variable $x = \ln \tilde{\rho}$ takes the value zero for $k^2 = \chi^2/2$, which corresponds to $k=M_p/\sqrt{2}$ in the normalization of $\chi$ with $\xi=1$.
We may set initial conditions at $x_F$ corresponding to $k=m_Z=91.2\;\mathrm{GeV}$,
\begin{equation}
    \label{eq:MM3}
    x_F = \ln \left(\frac{M_p^2}{2m_Z^2}\right) = 74.9\,.
\end{equation}
For the gauge couplings $g_i^2(x_F)$ we insert the observed values normalized at a renormalization scale equal to $M_Z$.
For $\lambda_h$ we use for this purpose the measured mass of the Higgs boson $m_H^2 = 2\lambda_h \varphi_0^2$.
The ratio $\lambda_h /y_t^2$ is predicted for the scaling solution from the condition $\lambda_h(x=0)\approx 0$, since $\lambda_h$ is an irrelevant coupling.
We adapt $y_t^2(x_F)$ in order to realize this condition.
This results in a ``one-loop value'' for the top quark mass slightly smaller than the observed top mass.
(For a precise determination of the predicted ratio $m_t/m_H$ one has to proceed beyond a one-loop approximation.)
We take $\sigma_m(x_F)$ as a free input parameter and approximate,
\begin{equation}
    \label{eq:MM4}
    \mu(x_F) = \sigma_m(x_F) + \delta\mu(x_F)\,.
\end{equation}
We choose $\delta\mu$ in the vicinity of
\begin{equation}
    \label{eq:MM4B}
    \mu_{\text{cr}} = -\frac{3}{32\pi^2} \left(\lambda_h + \frac{3}{4} g_2^2 + \frac{3}{20}g_1^2 - 2y_t^2\right)\,.
\end{equation}
The numerical results for $x<x_F$ are rather insensitive to the precise choice of $\delta\mu (x_F)$.

\begin{figure}[ht]
    \centering
    \includegraphics[width=0.9\linewidth]{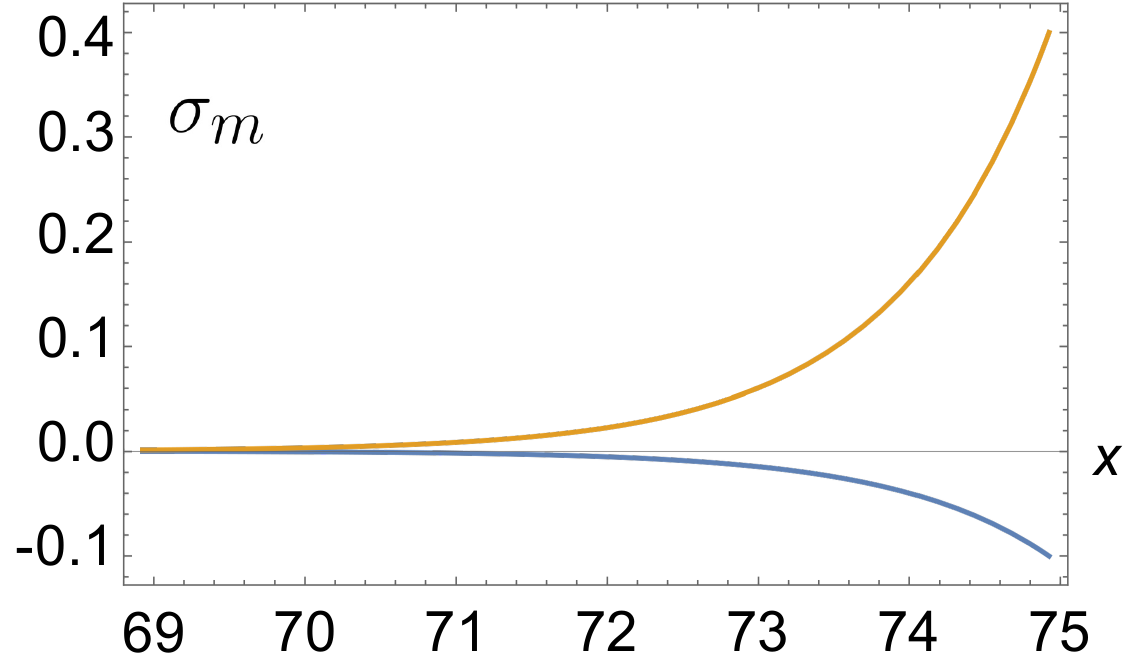}
    \caption{Infrared flow of cosmon-Higgs coupling. We plot $\sigma_m = \tilde{\rho} \lambda_m$ as a function of $x=\ln \tilde{\rho}$. The two curves correspond to initial values $\sigma_m(x_F) = -0.1$ (blue) and $\sigma_m(x_F) = 0.4$ (orange).}
    \label{fig:1}
\end{figure}

\indent In Fig.~\ref{fig:1} we plot $\sigma_m(x)$ for two different boundary values at $x_F$, namely $\sigma_m(x_F) = -0.1$ (blue) and $\sigma_m(x_F) = 0.4$ (orange).
Both values approach quickly the common critical coupling as $x$ decreases.
Continuing the flow for smaller $x$ both curves follow with high precision the trajectory given by the critical cosmon-Higgs coupling $\lambda_m^{\cri}$ according to eq.~\eqref{eq:263HA}.

\begin{figure}[ht]
    \centering
    \includegraphics[width=0.9\linewidth]{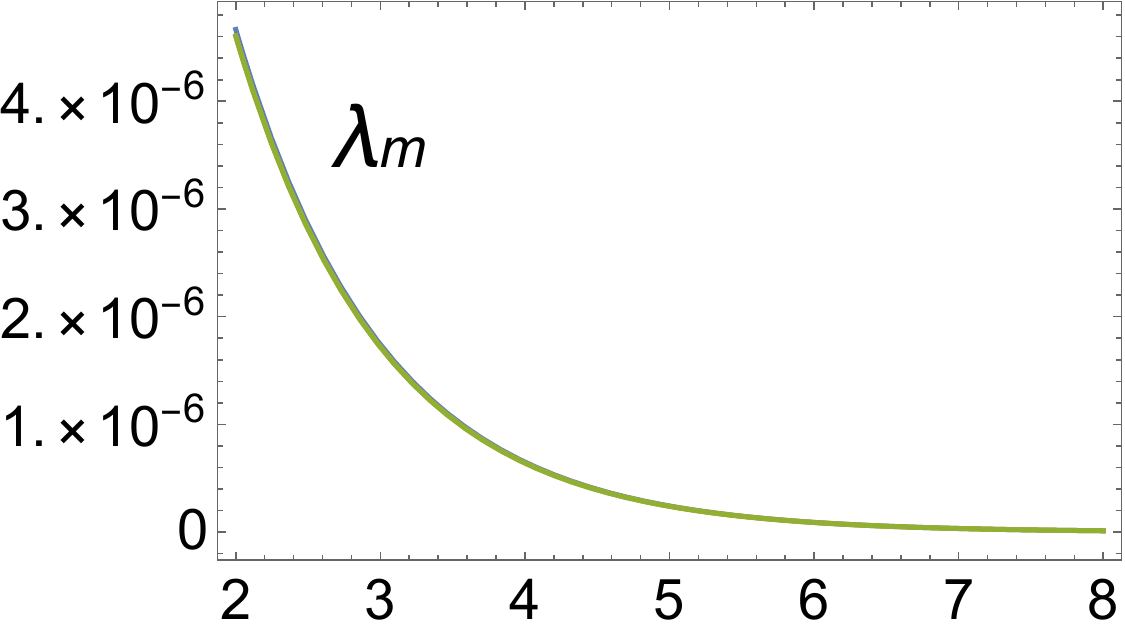}
    \caption{Flow of cosmon-Higgs coupling $\lambda_m$ in the Planck region as a function of $x = \ln \tilde{\rho}$. Two of the three curves correspond to the initial values $\sigma_m(x_F)=-0.1$ and $\sigma_m(x_F)=0.4$ as for Fig.~\ref{fig:1}. The third curve is on approximation to $\lambda_m^{\cri}$. The three curves almost coincide.}
    \label{fig:2}
\end{figure}

\indent In Fig.~\ref{fig:2} we show the corresponding values of $\lambda_m(x)$ for a range of $x$ corresponding to $k$ near the Planck mass.
One observes the increase of $\lambda_m(\tilde{\rho})$ proportional $\tilde{\rho}^{-1}$, as predicted by $\lambda_m^{\cri}(\tilde{\rho})$.
Near the Planck scale at $\tilde{\rho} \approx 1 $ the cosmon-Higgs coupling reaches a value around $10^{-5}$.
We also plot the critical trajectory $\lambda_m^{\cri}(x)$. 
We take here the approximation
\begin{equation}
    \label{eq:MA7}
    \lambda_m^{\cri} = \frac{1}{2} \tilde{\beta}_m \left(1-\frac{A_m}{2}\right)^{-1}\,,
\end{equation}
where
\begin{equation}
    \label{eq:MA8A}
    \tilde{\beta}_m = \frac{1}{64 \pi^2 \tilde{\rho}} \left(\sum_i{}^\prime \hat{n}_i \beta_i + \frac{6 \beta_h}{(1+\mu_h)^2}\right)\,.
\end{equation}
Here $\tilde{\beta}_m$ subtracts from $\beta_m$ the contribution $\sim\sigma_m$, which is incorporated in $A_m$
\begin{equation}
    \label{eq:MA6A}
    \tilde{\rho} \partial_{\tilde{\rho}} \lambda_m = -\frac{1}{2} \tilde{\beta}_m - \frac{1}{2} A_m \lambda_m\,.
\end{equation}
The three curves almost coincide.
On the level of precision visible in Fig.~\ref{fig:2} the difference between the improved value of $\lambda_m^{\cri}(\tilde{\rho})$ which includes in eq.~\eqref{eq:263HA} the contribution from $d_m \neq 0$ and the approximation \eqref{eq:MA7} is not visible.

\indent For the initial value $\lambda_m(x_F) = -0.1$ one finds $\lambda_m(0) = 3.3359\times 10^{-5}$, while $\lambda_m(x_F) = 0.4$ yields $\lambda_m(0) = 3.3311\times10^{-5}$.
The difference between the two initial values of $\lambda_m$ at $x_F$ is of the order $10^{-32}$.
It has increased to around $5\cdot10^{-8}$ at $x=1$.
More in detail, the difference remains almost constant $\approx 10^{-32}$ for a certain range of $x$ near $x_F$, and subsequently increases almost $\sim \tilde{\rho}^{-1}$, with almost constant difference in $\sigma_m$.
This behavior is precisely what one expects from the mixing effect in eq.~\eqref{eq:CLA9}.
Also the difference to our approximation for $\lambda_m^{\cri}$ is for $x=0$ in the range of a few times $10^{-7}$ or $10^{-8}$, respectively.
At this level one cannot decide which one of the two boundary values is preferred.
For this question one has to include the gravitational fluctuations.
A continuation to $\tilde{\rho} \to 0$ or $x \to -\infty$ will decide for which boundary value a finite value of $\lambda_m$ is reached for $\tilde{\rho} \to 0$, and for which other boundary values $\lambda_m(\tilde{\rho} \to 0)$ diverges.
The corresponding tuning of $\delta_m(x_F)$ yields the prediction for the Fermi scale.

\subsection*{Ultraviolet fixed point}
\indent We assume that for $\tilde{\rho} \to 0$ all couplings approach an ultraviolet fixed point, as appropriate for an ultraviolet complete theory.
For the gauge couplings we assume for simplicity a common fixed point at $g_*^2$, as realized for grand unified theories.
We realize this by choosing $D_F$ in eq.~\eqref{eq:UV7} as 
\begin{equation}
    \label{eq:UVR1}
    D_{F,i} = \bar{\beta}_{\text{UV}} - \bar{\beta}_i\,,\quad
    \beta_i = \bar{\beta}_i g_i^4\,,
\end{equation}
The constant $\bar{\beta}_{\text{UV}} = - n_{\text{UV}}/(8\pi^2)$ reflects the running in a GUT-model.
For $c_F$ we take $c_F = 1/\tilde{\rho}_\GUT$, where $\tilde{\rho}_{\text{GUT}} = \exp(x_{\text{GUT}})$, with $x_{\text{GUT}} \approx 11$ corresponding to $k=10^{16}\;\mathrm{GeV}$.
For $\beta_2$ and $\beta_1$ we take slightly larger values of $x_{\text{GUT}}$ such that for $\tilde{\rho} \ll \tilde{\rho}_{\text{GUT}}$ the three couplings $g_i$ coincide.
For the gravity induced anomalous dimension we add to $\beta_i$ a term
\begin{align}
    \label{eq:UVR2}
    \beta_i^{\gra} &= - B_g g_i^2\,,\nn\\
    B_g &= \frac{5}{288\pi^2 w}\left(\frac{4}{1-v} - \frac{3}{(1-v)^2}\right)\,,\nn\\
    w &= w_0 + \tilde{\rho}\,, \quad\quad v = \frac{u_0}{w}\,.
\end{align}
Here we employ the normalization $\xi=1$, $u_0 = u(\tilde{\rho} \to 0)$, and $w_0$ determines the ratio $M_p^2/k^2$ for $k\to \infty$.
For $\beta_t$ we add a contribution from fluctuations of the metric and heavy particles
\begin{equation}
    \label{eq:UVR3}
    \Delta \beta_t = \frac{y_t^2}{16\pi^2} \left\{\frac{a_y y_t^2 - a_g g_3^2}{1+c_F\tilde{\rho}} + \frac{2 b_t}{3 w(1-v)^2}\right\}\,.
\end{equation}
Finally, we add in the scalar sector $A_{\text{grav}}$ to $A_m$.

\begin{figure}[ht]
    \centering
    \includegraphics[width=0.9\linewidth]{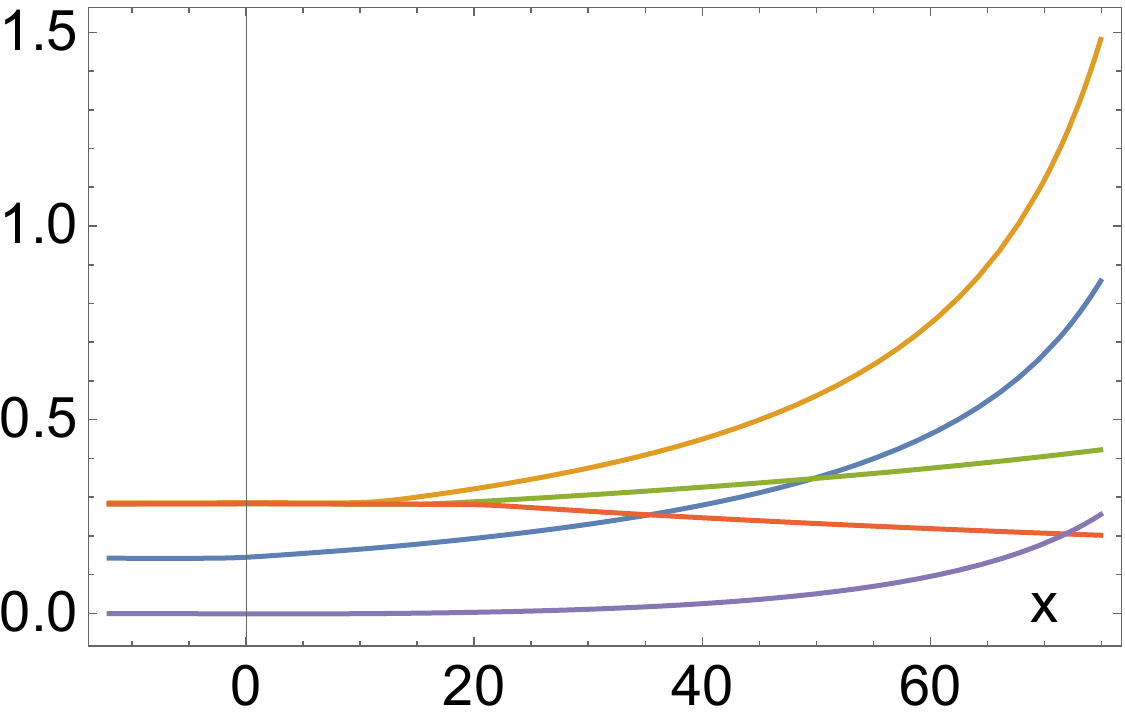}
    \caption{Field-dependent couplings of the standard model. As a function of $x= \ln \tilde{\rho}$ we plot the squared top-quark Yukawa coupling $y^2_t$ (blue), the squared gauge couplings $g^2_3$ (orange), $g^2_2$ (green) and $g^2_1$ (red), as well as the quartic scalar coupling $\lambda_h$ (violet).}
    \label{fig:3}
\end{figure}

\indent It is not our aim here to construct a realistic ultraviolet model.
We rather want to demonstrate the predictive power of the scaling solution for the Fermi scale.
For this purpose we choose the parameters for the $\beta$-functions of gauge and Yukawa couplings rather arbitrarily.
We take a setting for which the gauge couplings have in the presence of gravity an infrared stable fixed point $g_*^2$, which is realized for $B_g > 0$, $B_F < 0$.
With our choice $v_0 = v(\tilde{\rho}=0) = 0.2$ one realizes $B_g > 0$, and we take positive $\bar{\beta}_\text{UV}$ implying negative $B_F$ for $\tilde{\rho} = 0$.
The fixed point occurs for $g_*^2 = B_g/\bar{\beta}_{\text{UV}}$.
The running away from this fixed point sets in only once the gravitational fluctuations decouple for $\tilde{\rho} > w_0$.
This type of fixed point predicts the values of the gauge couplings.
In order to be compatible with observation $g_*^2$ needs to occur at the right value.
This can be achieved approximately with a mild tuning of parameters.
Following the evolution from large $\tilde{\rho}$ towards $\tilde{\rho} = 0$ the $\beta$-functions for the gauge couplings, and similarly for the Yukawa couplings should vanish for $\tilde{\rho} \to 0$ if the assumed values of the couplings at the Fermi scale coincide with the predicted values.
For our mild tuning these $\beta$-functions are of the order $10^{-4}$ for $\tilde{\rho} \to 0$, which is sufficient for our purpose.
We show the $\tilde{\rho}$-dependence of the gauge couplings, top-Yukawa coupling and quartic Higgs coupling in Fig.~\ref{fig:3}.

\begin{figure}[ht]
    \centering
    \includegraphics[width=0.9\linewidth]{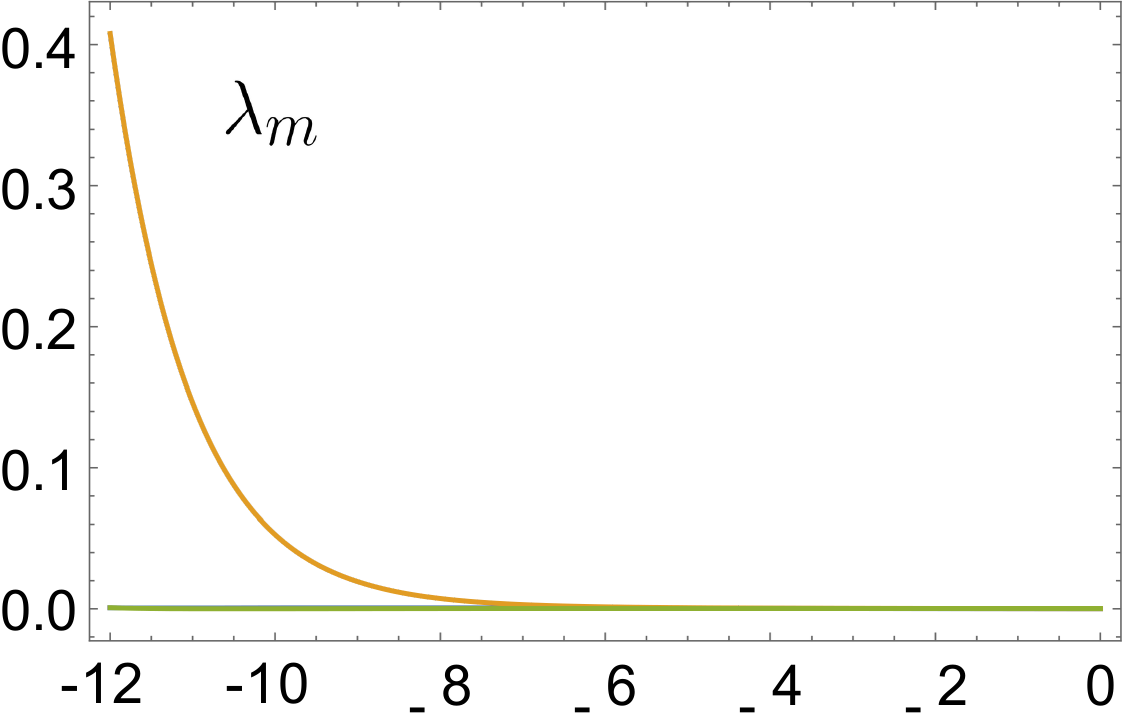}
    \caption{Ultraviolet flow of cosmon-Higgs coupling. We plot $\lambda_m$ as a function of $x = \ln \tilde{\rho}$, for two initial values $\sigma_m(x_F)=-0.1$ (green) and $\sigma_m(x_F)=0.4$ (orange). The curve for $\sigma_m(x_F)=-0.1$ coincides with $\lambda_m^{\cri}(\tilde{\rho})$.
    This figure should be interpreted as a demonstration of predictivity rather than an actual prediction of the values of the cosmon-Higgs coupling at the Fermi scale.}
    \label{fig:4}
\end{figure}

\indent The extrapolation of $\lambda_m$ towards $\tilde{\rho} = 0$ becomes sensitive to the boundary value of $\delta_m$ at $x_F$.
In Fig.~\ref{fig:4} we plot $\lambda_m(x)$ in the UV-range $x<0$ for the same boundary values as for Figs.~\ref{fig:1},\ref{fig:2}.
The green curve for $\delta_m(x_F) = -0.1$ stays very close to zero and coincides with $\lambda_m^{\cri}$.
On the other hand, for the orange curve with boundary value $\delta_m(x_F) = 0.4$ one observes a substantial increase of $\lambda_m$ for $x<-8$.
For this boundary value a finite value $\lambda_m(\tilde{\rho}=0)$ is not realized.
On this level of accuracy the boundary value $\delta_m(x_F) = -0.1$ is compatible with an analytic scaling solution, while $\delta_m(x_F) = 0.4$ is not.

\indent The plot in Fig.~\ref{fig:4} should be seen as demonstration of the "predictibility in principle" of the Fermi scale.
It should not be interpreted as a "prediction in practice" which precise value of $h_0$ is selected.
The latter may depend on the choice of the ultraviolet physics.
The dependence of $\bar{\lambda}_m$ on the assumptions for fixed points, gravitational contributions and contributions of heavy particles with mass $\sim \chi$ is not investigated here.
We have verified numerically, however, that for slight shifts of the parameters used here an extrapolation of the observed value of $h_0$ to a regular scaling solution for $x\rightarrow0$ remains possible.
There is no obstacle for realising a realistic value of the Fermi scale.
The prediction for the value of the Fermi scale becomes a matter of the ultraviolet model.

\indent Taking things together, this first numerical investigation should be seen as a confirmation of our analytic results. 
In particular, the predictivity for the Fermi scale can be seen clearly in Fig.~\ref{fig:4}.
A next step towards practical predictions for certain classes of short distance models should consider a variation of the parameters characterizing the short distance model, and an investigation of the dependence on assumptions for the effective low energy theory.
Presumably the variation of the short distance model has the largest effect and may be done first. 
We plan to perform such an analysis in the future.
The complication arises from the fact that we consider a system of several couplings which are all predicted for a given short distance model.
For given values of the observed couplings and given parameters of the short distance model a violation of analyticity can arise from non-analyticity in the flow of each coupling. 
Analytic scaling solutions will correspond to a relation between infrared couplings and short distance model parameters in a multi-dimensional coupling constant space. 
For this reason the next numerical step needs already a dedicated effort.

\section{Conclusions}
\label{sec:XI}
\indent This work investigates the question if the ratio of Fermi scale over Planck mass $\varphi_0/M_p$ can be predicted by quantum gravity. 
Our basic assumption is the existence of an ultraviolet complete quantum field theory for the metric and particle fields, being realized by an ultraviolet fixed point for the renormalisation flow of couplings.
We focus on a setting where the largest intrinsic mass scale characterizing the flow away from the ultraviolet fixed point is much smaller than the Fermi scale.
In this case the observed couplings of the standard model are given by the scaling solution of the functional flow equation.

\indent Our main result states that in this case the ratio $\varphi_0/M_p$ can be predicted.
There is no free parameter that can be adjusted in order to obtain an arbitrary value of $\varphi_0/M_p$.
The predicted value depends, however, on the details of the short distance physics for momenta around and above the Planck scale.
For a given short distance model the Fermi scale is predicted - this is "predictivity in principle".
"Predictivity in practice" needs the selection of the model and the computational capacity to extract the predicted value of $\varphi_0/M_p$.
In other words, the proposal of a given short distance model can be falsified if the predicted value of $\varphi_0/M_p$ does not coincide with observation.
The predictivity is lost if the largest
intrinsic mass scale is above the Fermi scale.
In this case the Fermi scale becomes a relevant parameter of the renormalization flow whose value can be chosen freely.
The discussion of this paper can therefore also be seen as a test if fundamental scale invariance or a largest intrinsic mass scale below the Fermi scale are viable concepts.
So far our answer is positive.

\indent A theory of quantum gravity becomes "ultraviolet complete" if an ultraviolet fixed point for the flow of couplings exists. In this case it can be extrapolated to infinitely high momentum or infinitesimal length scales without further change of the couplings. Within functional renormalisation an ultraviolet fixed point does not only concern a finite number of couplings. 
Infinitely many couplings, as generated for example, by a Taylor expansion of functions as the scalar potential, need to assume fixed point values. An ultraviolet fixed point is associated to a scaling solution for which dimensionless functions of dimensionless scaling fields do not depend on the renormalisation scale.

\indent A condition for predictivity is the existence of a unique scaling solution, or perhaps a discrete set of scaling solutions. In contrast, if a continuous family of scaling solutions would exist, there will be free parameters specifying the individual members of this family. If the Fermi scale depends on these free parameters it cannot be predicted. A first key question therefore asks if the scaling solutions form a discrete set or a continuous family.

\indent Scaling solutions for the dimensionless scalar potential or similar quantities have to obey a "scaling equation". 
These are first order non-linear differential equations for functions of fields. 
A "local scaling solution" for some local region in field space typically has free integration constants or initial values. Local scaling solutions form continuous families. The restrictions leading to discreteness arise from the necessity to extend the local scaling solutions to the whole field space. The scaling potential for the Higgs field has to be valid for a zero field value and extend to arbitrarily large field values. The compatibility with boundary conditions for small and large fields may be given only for a finite set of solutions, somewhat analogous to the bound states in quantum mechanics.

\indent Infinitely many couplings may be defined by derivatives of the scalar potential with respect to renormalized fields. For example, a quartic coupling for the Higgs scalar can be defined as the fourth derivative of the potential with respect to the Higgs field. The standard setting for the treatment of irrelevant parameters in critical phenomena is the assumption that all couplings take finite values at the ultraviolet fixed point. This means that the scaling solution for the potential is analytic at the point defined by zero field values.

\indent For an analytic scaling solution we find that no continuous family of scaling solutions exists. Scaling solutions are unique or form a discrete set. For an analytic scaling solution the ratio $\varphi_0/M_p$ does not depend on a continuous parameter. This has far reaching consequences. For example, if a scaling solution would describe the critical surface of a second order quantum electroweak phase transition, no neighbouring scaling solutions with arbitrarily small $\varphi_0/M_p$ are allowed. In this case the scaling solution would imply $\varphi_0=0$ in the limit where the infrared cutoff scale $k$ vanishes. This could explain why the Fermi scale is tiny as compared to the Planck mass. 
The predicted value would turn out, however, to be too small. 

\indent Our second question asks if a tiny ratio $\varphi_0/M_p$ is predicted by the scaling solution.
The ratio $\varphi_0/M_p$ is mainly set by the infrared value of the dimensionless cosmon-Higgs coupling $\lambda_m$.
We find that $\lambda_m$ is determined by the short-distance physics for momenta around or above the field-dependent Planck mass $\chi$.
The $\beta$-function $\beta_m$ dictating the dependence of $\lambda_m$ on the dimensionless ratio $\tilde{\rho} = \chi^2 / (2k^2)$ vanishes $\sim \tilde{\rho}^{-1}$ for $\tilde{\rho} \gg 1$.
This holds provided that only the particles of the standard model contribute  in this range of $k$.
Within the effective standard model for $k\ll \chi$ we find no approximately constant contribution to $\beta_m$.
Such a constant value would lead to a  logarithmic dependence of $\lambda_m$ on $\tilde{\rho}$ which would provide for a simple possible solution for the gauge hierarchy.
An effective theory with light particles beyond the standard model would be needed for such a "low energy solution" of the gauge hierarchy problem.

\indent If the theory below the Planck mass, or below some grand unified scale in its vicinity, is given by the standard model, one is left with a "high energy solution" for the gauge hierarchy $\varphi_0/M_p$.
Typically, the value of $\varphi_0/M_p$ is determined in this case by properties of the ultraviolet fixed point.
We find that $\varphi_0/M_p$ is generically a small parameter $\lesssim 10 ^{-6}$.
For obtaining the observed value of $10^{-30}$ additional properties are needed.
This may involve, in particular, enhanced symmetries realised by the scaling solution.
A natural candidate is the "particle scale symmetry" \cite{CWQS} associated to a second order quantum electroweak phase transition.
We have discussed the formal conditions for the realisation of this symmetry, but have not attempted to prove that a scaling solution with these properties exists independently of details of the short distnace model.

\indent Another interesting ultraviolet symmetry is given if the potential $u$ at $\tilde{\rho} = 0$ becomes independent of $\htilde$.
This realizes shift symmetry in $H$.
A consequence is the vanishing of $\mu_h$ and $\lambda_h$ for $\tilde{\rho} = 0$.
This entails additional relations between parameters of the standard model.
In particular, the boundary condition $\lambda_h (\tilde{\rho} = 0) = 0$ induces a relation between the top quark Yukawa coupling and the gauge coupling.
The corresponding prediction for the mass of the top quark seems compatible with observation.

\indent The particle scale symmetry associated to a second order quantum electroweak phase transition predicts $\varphi_0/M_p =0$.
The zero temperature or quantum electroweak transition is, however, not precisely a second order phase transition. While it shows the typical features of a second order phase transition for scales sufficiently above the confinement scale of QCD, the non-perturbative spontaneous chiral symmetry breaking in QCD turns it to a crossover around the confinement scale.
The particular "particle scale symmetry" associated to an exact second order phase transition is violated by the running of gauge and Yukawa couplings.
In the limit of constant couplings one finds a second order quantum phase transition with a plausible prediction $\varphi_0 = 0$.
Strong QCD effects induce an approximate bound $\varphi_0 \gtrsim \Lambda_\QCD$.
The question arises if running couplings could be responsible for a value $\varphi_0/\Lambda_\QCD$ of a few times $10^2$.
In other words: how strong is the deviation from a second order quantum phase transition?
We find that the non-perturbative QCD-effects set a lower bound of around $50\,\text{MeV}$ for the Fermi scale, which is not sufficient for an explanation of its observed value.

\indent A first numerical investigation of a setting with an ultraviolet fixed point at nonvanishing gauge and Yukawa couplings suggests that a realistic Fermi scale may indeed be possible.
The result is far from being conclusive since a systematic investigation how $\varphi_0/M_p$ depends on assumptions for the ultraviolet fixed point needs to be performed.
Nevertheless, it is intriguing that already a first investigation of a model for an UV-fixed point has found a tiny ratio $\varphi_0/M_p$ without any particular tuning of parameters.
This may, of course, also be accidental.
We leave a systematic investigation of the dependence of $\varphi_0/\Lambda_\QCD$ on properties of the UV-fixed point for future research.

\indent The ratio $\varphi_0/M_p$ predicted for the scaling solution for a given UV-model is not necessarily the end of the story.
In a general functional renormalisation setting the couplings flow away from the scaling solution due to the presence of relevant parameters.
If the Fermi scale is associated to a relevant parameter it cannot be predicted in a general setting.
If for a given short distance model the predicted value of the Fermi scale conflicts with observations, the largest intrinsic mass scale has to be above the Fermi scale.
The second condition for predictivity of the ratio $\varphi_0/M_p$ is therefore that the largest intrinsic mass scale is much below the Fermi scale.

\indent On the other hand the setting of fundamental scale invariance postulates that no deviation from the scaling solution occurs.
Fundamental physics is described by the exact scaling solution of functional renormalisation.
In this case $\varphi_0/M_p$ is predicted for a given model.
It is highly non-trivial that this prediction coincides with observation.
If one can find acceptable short distance models which yield the correct prediction for $\varphi_0/M_p$ this could be interpreted as a strong hint towards fundamental scale invariance.
If not, fundamental scale invariance may be excluded.
These statements extend to settings where the largest intrinsic mass scale generated by the flow away from the scaling solution is far below the Fermi scale.

\indent Fundamental scale invariance offers a dynamical mechanism for a vanishing cosmological constant and dynamical dark energy. 
Its predictivity for the ratio of the Fermi scale over the Planck mass may have excluded this possibility. 
With the present investigation this has not happened.
The option of fundamental scale invariance provides a powerful tool for restricting possible short distance models.

\indent Finally, a given short distance model for physics around and beyond the Planck scale may not be valid to infinitely short distances.
It may be itself an intermediate effective theory which could be replaced far beyond the Planck scale.
In this case the sharp prediction for $\varphi_0/M_p$ would be weakened to an infrared interval.
The allowed interval for $\varphi_0/M_p$ would result from the requirement that $\lambda_m$ does not grow to huge values for $\tilde{\rho}$ within the range of validity of the intermediate model for quantum gravity.
The difference between the observed value of $\varphi_0/M_p$ and the precise prediction of the short-distance model would indicate the scale $\tilde{\rho}$ where the intermediate model of quantum gravity has to be replaced.

\indent In summary, the present work has laid the ground for a systematic investigation of the ratio $\varphi_0/M_p$ for the scaling solution of quantum gravity.
The next step will have to investigate how this ratio, or the ratio $\varphi_0/\Lambda_\QCD$, depends on the properties of the microscopic model.

\section*{Acknowledgment}
The author would like to thank J. Pawlowski and M. Yamada for discussions.

\appendix
\section{Notations}
\label{app:notations}

\begingroup
\setlength{\parskip}{0.5\baselineskip}

\indent For the convenience of the reader we summarize here our notation for scalar fields and the scalar potential.

\subsection*{Fields}

$\chi$: cosmon field, $\rho=\frac{\chi^2}{2}$, $\tilde\rho=\frac{\rho}{k^2}$.

$H$: Higgs-doublet field, $\tilde h=\frac{H^\dagger H}{k^2}$, $h=\frac{\tilde h}{\tilde\rho}=\frac{2H^\dagger H}{\chi^2}$.

\subsection*{Potential}

$U$: scalar potential, $u=\frac{U}{k^4}=u(\tilde\rho,\tilde h)=u(\tilde\rho,h)$.

For partial derivatives of $u$ it matters which fields are kept fixed.
If not indicated explicitly, these are the fields in the argument, e.g.
$\partial_{\tilde h}u(\tilde\rho,\tilde h) =\left.\frac{\partial u}{\partial\tilde h}\right|_{\tilde\rho, \htilde}$.

\subsection*{Potential derivatives}

$\mu_h(\tilde\rho,\tilde h)=\partial_{\tilde h}u(\tilde\rho,\tilde h)$,

$\lambda_m(\tilde\rho,\tilde h)=\partial_{\tilde\rho}\partial_{\tilde h}u(\tilde\rho,\tilde h)=\partial_{\tilde\rho}\mu_h(\tilde\rho,\tilde h)$,

$\lambda_\chi(\tilde\rho,\tilde h)=\partial_{\tilde\rho}^2u(\tilde\rho,\tilde h)$,

$\lambda_h(\tilde\rho,\tilde h)=\partial_{\tilde h}^2u(\tilde\rho,\tilde h) =\partial_{\tilde h}\mu_h(\tilde\rho,\tilde h)$.

$\mu_h(\tilde\rho)=\mu_h(\tilde\rho,0)$,

$\lambda_m(\tilde\rho)=\lambda_m(\tilde\rho,0)$,

$\sigma_m(\tilde\rho)=\tilde\rho\lambda_m(\tilde\rho)$,

$\lambda_h(\tilde\rho)=\lambda_h(\tilde\rho,0)$,

$\lambda_\chi(\tilde\rho)=\lambda_\chi(\tilde\rho,0)$.

$\tilde \nu(\tilde\rho,h)=\partial_hu(\tilde\rho,h)=\tilde\rho\,\mu_h(\tilde\rho,h)$,

$\tilde \nu(\tilde\rho)=\tilde \nu(\tilde\rho,0)$,

$\hat\lambda_h (\tilde\rho, h)=\partial^2_h u(\tilde\rho, h)$.

\subsection*{Location of partial minimum}

$\tilde h_0(\tilde\rho)$: location of partial minimum, 

$\left.\partial_{\tilde h}\,(\tilde\rho,\tilde h)u \right|_{\tilde h_0(\tilde\rho)}=0$,

$\varphi_0^2(\tilde\rho) = k^2 \tilde h_0 (\tilde\rho)$, $\,\varphi_0 = \varphi_0 (\tilde\rho \to \infty)$: Fermi scale,

$h_0(\tilde\rho)=\tilde\rho \tilde h_0 (\tilde\rho)$,
$\left.\partial_h u(\tilde\rho,h)\right|_{h_0(\tilde\rho)}=0$,

$h_0=h_0(\tilde\rho\to\infty)=\frac{\varphi_0^2}{\chi^2} =\frac{2\xi\varphi_0^2}{M_p^2}$.

\subsection*{Quartic couplings}

$\lambda_{h_0}(\tilde\rho) =\lambda_h(\tilde\rho,\tilde h_0(\tilde\rho)) =\left.\frac{\partial^2U}{\partial(H^\dagger H)^2} \right|_{H^\dagger H=\varphi_0^2(\tilde\rho)}$,

$\bar\lambda_h=\lambda_{h_0}(\tilde\rho\to\infty)$, 

$\bar\lambda_m=\lambda_m(\tilde\rho\to\infty)$, 

$\bar\lambda_\chi=\lambda_\chi(\tilde\rho\to\infty)$.

\endgroup

\section{Strong interactions}
\label{app:A}

\indent In this appendix we investigate the non-perturbative effects of the strong interactions. 
They are responsible for turning the second order quantum electroweak phase transition into a crossover. 
The dominant effect for electroweak symmetry breaking is the generation of effective four-quark interactions by the functional flow.
By flowing bosonization they can be absorbed into modifications of the running of the Yukawa coupling and Higgs mass term, leading finally to electroweak symmetry breaking.
We focus on the top-quark.
Including the other quarks will not change the result qualitatively.

\subsection*{Top-antitop composite}

\indent Let us introduce the quark bilinear for the top quark

\begin{equation}
    \label{eq:CT1}
    \tilde{H}_t = \bar{t}_R\begin{pmatrix}
        t_L\\b_L
    \end{pmatrix}\,, 
\end{equation}
where a summation over colours is implied. In terms of $\tilde{H}_t$ the top quark Yukawa coupling term can be written in the form 

\begin{equation}
    \label{eq:CT2}
    \mathscr{L}_{Y,t} = y_t H^\dagger\tilde{H}_t + h.c. \,.
\end{equation}
The quark bilinear $\tilde{H_t}$ has the same transformation properties as the Higgs doublet $H$ with respect to the gauge symmetries of the standard model. A four-quark interaction takes the form

\begin{equation}
    \label{eq:CT3}
    \mathscr{L}_{4,t} = -\frac{1}{2}\bar{\lambda}_t\tilde{H}_t^\dagger\tilde{H}_t\,.
\end{equation}
The functional renormalisation flow generates this interaction through box diagrams involving the exchange of gauge bosons and scalars. 
Omitting mass threshold functions one finds \cite{Gies_2002, Gies_2004} for fixed fields $\chi$, $H$

\begin{equation}
    \label{eq:CT4}
    \begin{aligned}
        &k\partial_k \bar{\lambda}_t = - \frac{D}{16\pi^2k^2},\\
        &D = \frac{80}{3}g^4_3 - 5y^4_t\,.
    \end{aligned}
\end{equation}

\subsection*{Flowing bosonisation}

\indent Flowing bosonisation \cite{Gies_2002}, absorbs this term by a $k$-dependent field redefinition of the Higgs doublet

\begin{equation}
    \label{eq:CT5}
    H_k = H + \alpha_k\tilde{H}_t\,.
\end{equation}
Evaluating the flow equation at fixed $H_k$ instead of fixed $H$ one has 

\begin{equation}
\label{eq:CT6}
\begin{aligned}
    &k\partial_k\Gamma_k|_{H_k} = k\partial_k \Gamma_k|_H + \Delta_k\,,\\
    &\Delta_k = -\int_x \left(  \frac{\partial\Gamma_k}{\partial H_k} \partial_t H_k + \partial_tH^\dagger_k \frac{\partial\Gamma_k}{\partial H^\dagger_k} \right) \, ,
\end{aligned}
\end{equation}
with

\begin{equation}
    \label{eq:CT7}
    \partial_t H_k = k\partial_k H_k|_{H,\tilde{H}_t} = k\partial_k \alpha_k\tilde{H}_t\,,
\end{equation}
The first term on the r.h.s of eq.~\eqref{eq:CT6} arises from the $k$-dependence of the IR-cutoff and amounts to the usual flow equation. The second term reflects the $k$-dependence of the field variables. We employ a truncation where $H$ is replaced by $H_k$.

\indent Neglecting derivatives of $H_k$ our truncation amounts to

\begin{equation}
    \label{eq:CT8}
    \begin{aligned}
        \frac{\partial \Gamma_k}{\partial H_k} &= \frac{\partial U}{\partial (H_k^\dagger H_k)}H_k^\dagger + y_t\tilde{H}_t^\dagger \,,\\
        \frac{\partial \Gamma_k}{\partial H_k^\dagger} &= \frac{\partial U}{\partial (H_k^\dagger H_k)} H_k + y_t \tilde{H}_t \, ,
    \end{aligned}
\end{equation}
and one finds the correction term

\begin{equation}
    \label{eq:CT9}
    \Delta_k = - \int_x k\partial_k \alpha_k \left\{  \frac{\partial U}{\partial(H_k^\dagger H_k)}H_k^\dagger \tilde{H}_t + y_t \tilde{H}_t^\dagger \tilde{H}_t + h.c.  \right\} \, .
\end{equation}
In the following we omit the label $k$ for $H_k$. With the choice

\begin{equation}
    \label{eq:CT10}
    -y_t k \partial_k \alpha_k = \frac{1}{4}k\partial_k \bar{\lambda}_t = -\frac{D}{64\pi^2k^2}\, ,
\end{equation}
the contribution of the field transformation for the four-quark interaction cancels the term \eqref{eq:CT4} arising from $k\partial_k \Gamma_k|_H$. We can therefore work with $\bar{\lambda}_t=0$. 

\indent The first part of the correction term reads then

\begin{equation}
    \label{eq:CT11}
    \tilde{\Delta}_k  = -\int_x \frac{D}{64\pi^2 k^2 y_t} \frac{\partial U}{\partial(H^\dagger H)} \left(H^\dagger\tilde{H}_t + h.c.\right)\, .
\end{equation}
This yields a correction to the flow of the Yukawa coupling 

\begin{equation}
    \label{eq:CT12}
    \Delta(k\partial_ky_t) =-\frac{D}{64\pi^2k^2y_t} \frac{\partial U}{\partial(H^\dagger H)}\, .
\end{equation}
The expression $\partial \Gamma_k/\partial H_k$ in eq.~\eqref{eq:CT8} depends on momentum due to the kinetic term for the Higgs doublet.
This generates a momentum dependence of the Yukawa coupling.
We define $y_t$ as the Yukawa coupling at zero momentum and neglect the momentum dependence.

\indent Accordingly, the flow of $y^2_t$ is modified to

\begin{equation}
    \label{eq:CT13}
    k\partial_k y^2_t = \beta_{t,H} - \frac{D\partial_{\tilde{h}}u}{32\pi^2}=\beta_{t,H} - \frac{D\mu_h}{32\pi^2}\, ,
\end{equation}
with $\beta_{t,H}$ the part without the field transformation. In our setting $y^2_t$, $\mu_h$ and $D$ are functions of the scalar fields $\chi$ and $H$. 
We switch again to dimensionless scaling fields.
The four fermion vertex $\bar{\lambda}_t$ depends on momenta.
A dominant momentum dependence can be accounted for by  a further rescaling of the renormalised Higgs field.
This adds to the anomalous field dimension $\eta_H$ a term $\eta_{tc}$.
The field-flow of the Yukawa couplings receives then an additional contribution $\sim \eta_{tc} y^2_t$.
We obtain the scaling equation (omitting mass thresholds)

\begin{align}
\label{eq:CT14}
\tilde{\rho}\,\partial_{\tilde{\rho}} y_t^2
  = -\frac{1}{32\pi^2} \Bigg\{
      9 y_t^4
      &-\left(16 g_3^2 + \frac{9}{2} g_2^2 + \frac{17}{10} g_1^2\right) y_t^2 \nonumber\\
  &
      - \frac{1}{2} D\mu_h
    \Bigg\}-\frac{1}{2}\eta_{tc}y^2_t \, .
\end{align}
As long as $|\mu_h|$ remains small the incorporation of the four-quark interaction by a $k$-dependent field transformation remains a small correction for the field-dependence of the Yukawa coupling. For the scaling solution corresponding to a second order phase transition one has in lowest order $\mu_h = 3y^2_t/(16\pi^2)+\cdots$, which is indeed a small quantity. The correction $\sim D \mu_h$ is of the order of a two-loop contribution.

\subsection*{QCD-induced IR-fixed point}

\indent For larger values of $|\mu_h|$ the term $\sim D \mu_h$ can no longer be neglected. For understanding its effect it is useful to consider the evolution of the ratio 

\begin{equation}
    \label{eq:CT15}
    \begin{aligned}
    &\tilde{\epsilon}_t = \frac{\mu_h}{y^2_t}\, , \\
    \tilde{\rho} \partial_{\tilde{\rho}} &\tilde{\epsilon}_t = \frac{1}{y^2_t}\left( \tilde{\rho}\partial_{\tilde{\rho}}\mu_h -\tilde{\epsilon}_t \tilde{\rho} \partial_{\tilde{\rho}} y^2_t   \right)\, .
    \end{aligned}
\end{equation}
We use eqs.~\eqref{eq:NYO1}~\eqref{eq:NYB},
\begin{equation}
    \label{eq:CT15A}
    \tilde{\rho} \partial_{\tilde{\rho}} \mu_h = \mu_h + \beta_\nu - \frac{1}{2} \left( \eta_H + \eta_{ct} \right)\mu_h \,,
\end{equation}
with $\beta_\nu$ given by eq.~\eqref{eq:118C}.
The top-composite induced anomalous dimension $\eta_{ct}$ cancels out in the field dependence of $\tilde{\epsilon}_t$.
Omitting mass threshold and anomalous dimensions, one finds

\begin{equation}
    \label{eq:CT15A}
    \tilde{\rho}\partial_{\tilde{\rho}} \tilde{\epsilon}_t = - \frac{D}{64\pi^2} \tilde{\epsilon}^2_t + E \tilde{\epsilon}_t -\frac{F}{64\pi^2}\, , 
\end{equation}
with 

\begin{equation}
    \label{eq:CT16}
    \begin{aligned}
    E &= 1 + \frac{1}{32\pi^2} \left[ 3y^2_t -\left(16g^2_3  + \frac{4}{5}g^2_1  \right) \right] \, , \\
    F &= 12 - \left( \frac{9}{2}\frac{g^2_2}{y^2_t}  +\frac{9}{10}\frac{g^2_1}{y^2_t} +\frac{2\lambda_h}{y^2_t}  \right)\, .
    \end{aligned}
\end{equation}
A qualitative understanding can be gained by taking the gauge and Yukawa couplings constant.
A more quantitative analysis can be done numerically.

\indent If one neglects the $\sim D$ and takes the gauge and Yukawa couplings approximately constant, one finds a fixed point

\begin{equation}
    \label{eq:CT17}
    \tilde{\epsilon}_{t,\text{UV}} = \frac{F}{64\pi^2 E}\, .
\end{equation}
This fixed point is IR-unstable. It corresponds to the scaling solution for $\mu_h$ for constant gauge couplings, 

\begin{equation}
    \label{eq:CT18}
    \mu_h = \frac{3}{16\pi^2} \left[ y^2_t - \left(  \frac{3}{8}g^2_2 + \frac{3}{40}g^2_1 +\frac{1}{6}\lambda_h \right) \right]\, ,
\end{equation}
which denotes the critical surface for a second order electroweak phase transition.

\indent For $D\neq0$ a second IR-stable fixed point is present in the limit of constant threshold functions and constant gauge and Yukawa couplings. 
The r.h.s~of eq.~\eqref{eq:CT15A} vanishes for the two possible fixed point solutions

\begin{equation}
    \label{eq:CT19}
    \epsilon_{t*} = \frac{32\pi^2}{D}\left( E \pm \sqrt{E^2-DF/(32\pi^2)^2}  \right)\, . 
\end{equation}
These fixed points exist as long as $DF<(32\pi^2 E)^2$. For $|DF|\ll(32\pi^2)^2$ the fixed point with the relative minus sign can be identified with the IR-unstable fixed point \eqref{eq:CT17}. The second fixed point occurs approximately for 

\begin{equation}
    \label{eq:CT20}
    \tilde{\epsilon}_{t,\text{IR}} = \frac{64\pi^2E}{D}\, .
\end{equation}
It is approached for increasing $\tilde{\rho}$. For $D>0$, $E \approx 1$ this fixed point is located at positive $\mu_h$, 

\begin{equation}
    \label{eq:CT21}
    \mu_H \approx \frac{64\pi^2 y^2_t}{D}\, .
\end{equation}

\indent For $D<0$ one has $\mu_h<0$, which would correspond to a minimum of $u$ at non-zero $\tilde{h}_0$,

\begin{equation}
    \label{eq:CT22}
    \tilde{h}_0 \approx \frac{64 \pi^2 y^2_t}{\lambda_h |D|}\,.
\end{equation}
The fixed point with negative $D$ is not realized at the minimum of $u$, however. Negative $D$ is only possible if the Yukawa coupling dominates in eq.~\eqref{eq:CT4}. For the large value $\tilde{h}_0$ in eq.~\eqref{eq:CT22} both the top quark mass and the mass of the radial scalar mode are large. Mass thresholds with powers of $(1+y^2_t\tilde{h})^{-1}$ suppress the contribution $\sim D$ in eq.~\eqref{eq:CT15A}. For negative $\tilde{\epsilon}_t$ one should adapt the definition of $\tilde{\epsilon}_t$ for an evaluation at the minimum $\tilde{h}_0$ of $u$, i.e. $\tilde{\epsilon}_t = -\lambda_h \tilde{h}_0/y^2_t$. This implies $\tilde{m}^2_t = - y^4_t\tilde{\epsilon}_t/\lambda_h$, such that the suppression factors $(1+\tilde{m}^2_t)^{-1}=(1-y^4_t\tilde{\epsilon}_t/\lambda_h)^{-1}$ annihilate the increase $\sim \tilde{\epsilon}^2_t$ of the term $\sim D$ in eq.~\eqref{eq:CT15A}. There is no second fixed point for negative $\tilde{\epsilon}_t$.

\indent Let us concentrate on the case of positive $D$, which is realised if the gauge interactions dominate $D$ in eq.~\eqref{eq:CT4}. This is indeed the case in the vicinity of the fixed point \eqref{eq:CT20}, \eqref{eq:CT21}. The contribution to $D$ involving the Yukawa coupling arises from box diagrams with exchange of a Higgs scalar. It is suppressed by threshold functions involving factors $(1+\mu_h)^{-1}$ with large $\mu_h$ according to eq.~\eqref{eq:CT21}.

\indent In the vicinity of the fixed point \eqref{eq:CT20} we can therefore omit in eq.~\eqref{eq:CT4} the contributions $\sim y^4_t$, resulting in $D>0$. For smaller values of $\mu_h$ the contributions  $\sim y^4_t $ are present for $D$, but the overall size of the term $\sim D$ is small. With reasonable accuracy we may only include the contributions of the gauge couplings in eq.~\eqref{eq:CT4}.

\subsection*{Electroweak crossover}

\indent If for some $\tilde{\rho}$ the scaling solution deviates from the UV-fixed point $\tilde{\epsilon}_{t,\text{UV}}$ with $\tilde{\epsilon}_t>\tilde{\epsilon}_{t,\text{UV}}$, the value of $\tilde{\epsilon}_{t}(\tilde{\rho})$ will increase until it reaches the IR-fixed point $\tilde{\epsilon}_{t,\text{IR}}$. 
The trajectory describes a crossover between the two fixed points. For large $\tilde{\rho}$ one ends in an IR-stable scaling solution with a positive mass term of the Higgs double $m^2_H \sim k^2$. 
This contrasts with a flow towards the symmetric phase for which $m^2_H$ ceases to decrease due to the decoupling of fluctuations because of the threshold functions.

\indent As $\tilde{\rho}$ increases the approximation of weak and slowly varying gauge couplings remains no longer valid. Due to the large couplings of QCD the two fixed points approach each other and finally disappear. This happens at a critical scale $\tilde{\rho}_\chi$ for which $g_3^2$ is large enough such that the square root in eq.~\eqref{eq:CT19} vanishes, i.e. for 

\begin{equation}
    \label{eq:CT23}
    D = (32\pi^2)^2E^2/F \,.
\end{equation}
At this value of $\tilde{\rho}$ one has 

\begin{equation}
    \label{eq:CT24}
    \tilde{\epsilon}_t = \sqrt{\frac{F}{D}} = \frac{F}{32\pi^2E}\,, \quad \mu_h = \sqrt{\frac{F}{D}}y^2_t =\frac{Fy^2_t}{32\pi^2E}\,.
\end{equation}
For $\tilde{\rho}$ increasing further beyond $\tilde{\rho}_\chi$ the r.h.s.~of eq.~\eqref{eq:CT15A} turns negative and $\tilde{\epsilon}_t$ reaches rapidly zero. 
From there on a nontrivial minimum of $u$ at $h_0(\tilde{\rho})>0$ develops, and finally the $\tilde{\rho}$-dependence of $h_0$ stops due to threshold functions. One ends in the phase with spontaneous breaking of the electroweak symmetry. The final value of $h_0$ sets the Fermi scale $\varphi_0$. The value $\tilde{\rho}_\chi$ for which the fixed point for $\tilde{\epsilon}_t$ disappears corresponds to a renormalisation scale $k_\chi$. This scale sets the order of magnitude for the Fermi scale. In turn, $k_\chi$ is the scale for which $g^2_3$ reaches the critical value  for which $D$ obeys eq.~\eqref{eq:CT23}.

\indent For an estimate of $\tilde{\rho}_\chi$ we employ eq.~\eqref{eq:YU3},

\begin{equation}
    \label{eq:CT25}
    E = 1 -\frac{g_3^2}{2\pi^2}\left(  1-\frac{9}{16R_t}  \right)R_g \,,
\end{equation}
with $R_t\approx 1.9$ and 

\begin{equation}
    \label{eq:CT26}
    R_g = 1+\frac{9g_2^2}{32g_3^2} + \frac{17g_1^2}{160g^2_3} \approx 1.1
\end{equation}
close to one. 
We observe that $E$ approaches zero as $g^2_3$ increases. 
The numerical values are taken at the $Z$-boson mass, with $R_t$ approaching $4.5$ and $R_g$ approaching $1$ for smaller $k$. 
The scale $k_\chi$ is above the scale where $E$ reaches zero.
We may estimate the size of $F$ by inserting the measured particle masses

\begin{equation}
    \label{eq:CT27}
    F = 12-\frac{6m^2_W}{m^2_t} - \frac{3m^2_Z}{m^2_t} - \frac{m^2_H}{m^2_t}\,.
\end{equation}
Evaluated at $k=m_Z$ one finds $F\approx9.3$, while it approaches $12$ for decreasing $k$. 

\indent The leading contribution to $D$ arises from a box diagram with exchange of two gluons and reads \cite{Gies_2002}

\begin{equation}
    \label{eq:CT28}
    D_3 = \frac{80}{3}g^4_3t_{D3}(\tilde{m}^2_t)\, ,
\end{equation}
with threshold function

\begin{equation}
    \label{eq:CT29}
    t_{D3}(\tilde{m}^2_t)= \left(1+\tilde{m}^2_t\right)^{-2} \left(  \frac{1}{3} + \frac{2}{3(1+\tilde{m}^2_t)} \right)\,.
\end{equation}
Keeping only this contribution the scale $k_\chi$ corresponds to $(g^2_3=4\pi\alpha_s)$

\begin{equation}
    \label{eq:CT30}
    \alpha_s^2 = \frac{12\pi^2E^2}{5F} = \frac{12\pi^2}{5F} \left[ 1-\frac{2\alpha_s}{\pi}\left( 1-\frac{9}{16R_t}\right)R_g \right]^2 \, .
\end{equation}
In the approximation of constant $F$ and $B_t = (1-\frac{9}{16R_t})R_g$ this results in the quadratic equation

\begin{equation}
    \label{eq:CT31}
    \left( 1-\frac{48B_t^2}{5F} \right)\alpha_s^2 + \frac{48\pi B_t}{5F}\alpha_s -\frac{12\pi^2}{5F} = 0\,.
\end{equation}
The solution yields $\alpha_s(k_\chi)$ which is reached for $k_\chi$ somewhat above the confinement scale.

\indent We have solved eqs.~\eqref{eq:CT13}, \eqref{eq:CT14} numerically, approximating $\eta_{ct}=0$.
One finds indeed a lower bound for $k_\chi$ where $\mu_h$ reaches zero.
In units where $\chi$ is the present Planck mass this is given by $k_\chi\approx 45 \text{MeV}$, slightly larger than the scale where $g^2_3$ diverges.

\subsection*{Correction to mass of of Higgs boson and top quark}
\indent The four-quark interaction $\sim \bar{\lambda}_t$ in eq.~\eqref{eq:CT3} is not taken into account for a perturbative computation of the flow of the Yukawa coupling and quartic scalar coupling. 
The operator mixing between the Higgs field $H$ and top-antitop composite $\tilde{H}_t$ matters, however, for the precise value of the mass of the Higgs boson and its Yukawa coupling to the top quark. 
Flowing bosonisation is an efficient way to account for this operator mixing. The additional term $\sim D$ in the flow equation for the top quark Yukawa coupling will affect the prediction for the ratio of Higgs boson mass and top quark mass $m_H/m_t$. 
It concerns directly the quantum gravity prediction for this ratio, which is given in our setting by the ratio $\lambda_h/y^2_t = m^2_H/(2m^2_t)$. 
The effect on this ratio is small, typically at the percent level or below. 
It decreases this ratio due to a slightly faster increase of $y^2_t$ towards the infrared.
For a given measured mass of the Higgs boson this could lead to an increase of the top quark mass on the percent level. This may be sufficient to improve compatibility with "vacuum stability" or the quantum gravity prediction for the Higgs boson mass \cite{SHAW}. 
The latter may be turned into a prediction of the top quark mass since the Higgs boson mass is measured precisely.

\indent Since the effect is small, we can use an expansion linear in the correction. For this purpose we consider the flow of the ratio

\begin{equation}
    \label{eq:HT1}
    S_t = \frac{\lambda_h}{y^2_t}\,, \quad k\partial_k S_t = \beta_{S,p} + \delta\beta_S\,,
\end{equation}
where $\beta_{S,p}$ is the perturbative result in some loop order (all computed perturbative parts are induced here).
In perturbation theory the ratio $S_t$ approaches towards the IR a partial fixed point \cite{CWFP, SWI}.
The top-antitop bilinear induces a correction

\begin{equation}
    \label{eq:HT2}
    \delta \beta_S = \frac{D\mu_h\lambda_h}{32\pi^2y^4_t} = \frac{(80g^4_3-15y^4_t)\mu_h\lambda_h}{96\pi^2y^4_t}\, .
\end{equation}
with

\begin{equation}
    \label{eq:HT3}
    S_t = S_{t,p} + \delta_S\, , \quad k\partial_k S_{t,p} =\beta_{S,p}\,,
\end{equation}
the change of $S_t$ due to the inclusion of the term $\sim D$ obeys 

\begin{equation}
    \label{eq:HT4}
    k\partial_k \delta_S = \delta\beta_S\, .
\end{equation}

\indent These formulae should only be seen as a rough guess on the possible size of this correction.
For a reliable computation the momentum structure of the four-quark vertex matters.
It seems preferable to introduce a separate field for the top-antitop bilinear, and to compute its potential as well as its interaction with the Higgs doublet.

\section{Threshold functions for field-dependent couplings}
\label{app:B}

\indent The threshold functions defined by eq.~\eqref{eq:G2} constitute the formal solution for the scaling form of the effective potential.
They encode the global form of $u$ for arbitrary values of $\tilde{\rho}$ and $\htilde$.
Their main content is the decoupling of particles with masses larger than $k$.
We will mainly be concerned with the behavior of $u$ for very small values of $h$ for which an expansion in powers of $h$ is valid.
While the necessary information is contained in the exact solution for the threshold functions we will employ more direct methods following flow equations for the $h$-derivatives as functions of $\tilde{\rho}$.
The threshold functions remain, nevertheless, an important ingredient for the understanding of properties of $u$ for a wider field range.
In the presence of field-dependent gauge and Yukawa couplings explicit analytical solutions become difficult.
The qualitative features and important qualitative details remain computable and will be discussed in this appendix.

\indent Eq.~\eqref{eq:NYA} is a differential equation whose solution involves integration constants.
We choose these integration constants such that $t_u^{(i)}(\tilde{\rho})$ vanishes for $\tilde{\rho} \to \infty$.
With this definition the solution of eq.~\eqref{eq:G2} take the form
\begin{align}
    \label{eq:TR1}
    t_u^{(i)}(\tilde\rho, h) &= 2\tilde{\rho}^2 \int_{\tilde{\rho}}^{\infty} \frac{\mathrm{d}\rho'}{\rho^{\prime 3} (1+\tilde{m}_i^2(\rho',h))} \nn\\
			     &= 1 - 2\tilde{\rho}^2 \int_{\tilde{\rho}}^{\infty} \frac{\mathrm{d} \rho'\; \tilde{m}_i^2(\rho',h)}{\rho^{\prime 3} \left(1+\tilde{m}_i^2(\rho',h)\right)}\,.
\end{align}
For the $h$-dependence at fixed $\tilde{\rho}$ one finds
\begin{equation}
    \label{eq:TR2}
    \partial_h t_u^{(i)}(\tilde{\rho},h) = -2\tilde{\rho}^2 \int_{\tilde\rho}^{\infty} \mathrm{d} \rho' \; \frac{\partial_h \tilde{m}_i^2}{\rho^{\prime 3}(1+\tilde{m}_i^2)^2}\,.
\end{equation}
We may parameterize
\begin{equation}
    \label{eq:TR3}
    \tilde{m}_i^2 (\tilde\rho, h) = \alpha_i (\tilde\rho,h) h \tilde{\rho}
\end{equation}
resulting in 
\begin{equation}
    \label{eq:TR4}
    \partial_h t_u^{(i)} = -2\tilde{\rho}^2 \int_{\tilde{\rho}}^{\infty} \mathrm{d} \rho'\; \frac{\alpha_i + h \partial_h \alpha_i}{\rho^{\prime 2}(1+\alpha_i h \rho')^2}\,.
\end{equation}

\indent For fixed $H\neq 0$ and $\chi\neq 0$ the particle masses $m_i$ take for $k\to 0$ values that no longer depend on $k$.
This implies $\tilde{m}_i^2 \sim k^{-2}$, and therefore for $\tilde{\rho} \to \infty$ the behavior $\tilde{m}_i^2(\tilde{\rho},h) \sim \tilde{\rho}$, with $\alpha_i(\tilde{\rho}\to \infty, h)\to \bar{\alpha}_i(h)$.
In consequence, one finds for $\tilde{\rho} \to \infty$
\begin{align}
    \label{eq:TR5}
    t_u^{(i)}(\tilde{\rho},h) &= 2 \tilde{\rho}^2 \int_{\tilde{\rho}}^{\infty} \frac{\mathrm{d}\rho'}{\rho^{\prime 3}(1+\bar{\alpha}_i(h) h\rho')} = \frac{2}{3\bar{\alpha}_i(h) h \tilde{\rho}} \nn\\
			      &= \frac{2}{3\tilde{m}_i^2(\tilde{\rho},h)}\,.
\end{align}
This generalizes eq.~\eqref{eq:FS31} to the case of running dimensionless couplings.
In the limit $\tilde{\rho} \to \infty$ also the $h$-dependence of $t_u^{(i)}$ vanishes for $h > 0$,
\begin{align}
    \label{eq:TR6}
    \partial_h t_u^{(i)}(\tilde{\rho},h) &= - \frac{2(\bar{\alpha}_i + h \partial_h \bar{\alpha}_i)}{3 \bar{\alpha}_i^2 h^2 \tilde{\rho}} \nn\\
					 &= - \frac{2}{3\tilde{m}_i^2} \left(\frac{1}{h} + \partial_h \ln \bar{\alpha}_i\right)\,.
\end{align}

\indent Since the threshold functions vanish for $\tilde{\rho} \to \infty$, the effective potential $u$ for $\tilde{\rho} \to \infty$ is determined by the boundary term $L(h)\tilde{\rho}^2$.
As for the case of field-independent gauge and Yukawa couplings, $L(h)$ can be associated to the limiting behavior of $u$ for $\tilde{\rho} \to \infty$.
If for a given UV-range of $\tilde{\rho}$ the form of $u_{\text{UV}}(\tilde{\rho})$ is given, one can determine $L(h)$ by comparison with eq.~\eqref{eq:G1} 
\begin{equation}
    \label{eq:TR7}
    L(h) \tilde{\rho}^2 = u_{\text{UV}} (\tilde{\rho}) - \frac{1}{128\pi^2}\sum_i \hat{n}_i t_u^{(i)} (\tilde{m}_i^2)\,.
\end{equation}
Here $\tilde{\rho}$ may typically be chosen in a range where the gravitational fluctuations have decoupled, such that eq.~\eqref{eq:G1} is valid, and small enough such that perturbation theory is valid.
In this way the threshold functions describe the matching between the ultraviolet part of $u$ for small $\tilde{\rho}$ and the infrared part for large $\tilde{\rho}$.
(The ultraviolet part involves here still values of $k$ smaller than the Planck scale since the effects of the metric fluctuations are not included.)

\indent For a determination of $L(h)$ one needs the behavior of the threshold functions $t_u^{(i)}$ for sufficiently small $\tilde{\rho}$.
Let us focus on the $h$-dependence according to eq.~\eqref{eq:TR4}.
For $\tilde{m}_i^2 \ll 1$ the mass thresholds involving $h$ do not yet play a role.
In this range the running dimensionless couplings, and therefore $\alpha_i$, are functions of $\tilde{\rho}$ independent of $h$.

\indent We may divide the integral \eqref{eq:TR4} into two regions, separated by a transition at
\begin{equation}
    \label{eq:TR8}
    \tilde{\rho}_{\text{tr}} = \frac{c_{\text{tr}}}{h}\,,\quad
    c_{\text{tr}} \ll 1\,,
\end{equation}
such that
\begin{align}
    \label{eq:TR9}
    \int_{\tilde{\rho}}^{\infty} \mathrm{d}\rho' &\frac{\alpha_i + h \partial_h \alpha_i}{\rho^{\prime 3}(1+\alpha_i h \rho')^2}
    = \Delta_1 + \Delta_2\,,\nn\\
    \Delta_1 &= \int_{\tilde{\rho}}^{\tilde{\rho}_{\text{tr}}} \frac{\mathrm{d}\rho' \alpha_i(\rho')}{\rho^{\prime 2}}\,, \nn\\
    \Delta_2 &= \int_{\tilde{\rho}_{\text{tr}}}^{\infty} \mathrm{d}\rho'\; \frac{\alpha_i + h \partial_h \alpha_i}{\rho^{\prime 2}(1+\alpha_i h \rho')^2}\,.
\end{align}
Only the first part $\Delta_1$ depends on $\tilde{\rho}$, while the second part $\Delta_2$ gives a $\tilde{\rho}$-independent contribution.
The ultraviolet-infrared connection \eqref{eq:TR7} can be realized by a proper matching of these regions.

\indent For the first integral we employ (for $\alpha(\rho') = \alpha_i(\rho',h)$)
\begin{equation}
    \label{eq:TR10}
    \frac{\alpha(\rho')}{\rho^{\prime 2}}\;\mathrm{d}\rho' = - \mathrm{d} \left(\frac{f(\rho')}{\rho'}\right)\,,
\end{equation}
where $f(\rho')$ obeys
\begin{equation}
    \label{eq:TR11}
    f \left(1-\frac{\partial \ln f}{\partial \ln \rho'}\right) = \alpha(\rho')\,.
\end{equation}
We may solve eq.~\eqref{eq:TR11} iteratively
\begin{align}
    \label{eq:TR12}
    f_0(\rho') &= \alpha(\rho')\,, \nn\\
    f_1(\rho') &= \frac{\alpha(\rho')}{1 - \frac{\partial \ln \alpha(\rho')}{\partial \ln \rho'}}\,, \nn\\
    f_n(\rho') &= \frac{\alpha(\rho')}{1 - \frac{\partial\ln f_{n-1}(\rho')}{\partial \ln \rho'}}\,.
\end{align}
With perturbative $\beta$ functions (at fixed $\chi$, $H$) given by $k \partial_k \alpha_i = \beta_i$ one has for the scaling solution
\begin{equation}
    \label{eq:TR13}
    \frac{\partial \alpha_i}{\partial\ln\tilde{\rho}} = - \frac{1}{2} \beta_i\,.
\end{equation}
The $\beta$-functions are higher order in the couplings, with typical values $\beta_i/\alpha_i \ll 1$.
We may therefore stop at $f_1(\rho')$, approximating
\begin{equation}
    \label{eq:TR14}
    \Delta_1 = \frac{\alpha_i(\tilde{\rho})}{\tilde{\rho}\big(1-(\beta_i/\alpha_i)(\tilde\rho)\big)} - \frac{\alpha_i(\tilde{\rho}_{\text{tr}})h}{c_{\text{tr}}\big(1-(\beta_i/\alpha_i)(\tilde{\rho}_{\text{tr}})\big)}\,.
\end{equation}

\indent For the second region we divide $\Delta_2$ into three parts
\begin{align}
    \label{eq:TR15}
    \Delta_2 &= \Delta_3 + \Delta_4 + \Delta_5\,,\nn\\
    \Delta_3 &= \int_{\tilde{\rho}_{\text{tr}}}^{\infty} \frac{\mathrm{d} \rho'\; \alpha_i(\rho')}{\rho^{\prime 2}}\,,\quad
    \Delta_4 = \int_{\tilde{\rho}_{\text{tr}}}^{\infty} \mathrm{d} \rho'\; \frac{h \partial_h \alpha_i}{\rho^{\prime 2}(1 + \alpha_i h \rho')^2}\,, \nn\\
    \Delta_5 &= -h\int_{\tilde{\rho}_{\text{tr}}}^{\infty} \mathrm{d} \rho'\; \frac{2\alpha_i^2 + \alpha_i^3 h \rho'}{\rho'(1 + \alpha_i h \rho')^2}\,.
\end{align}
The part $\Delta_3$ cancels the second term of eq.~\eqref{eq:TR14}, such that a dependence on the chosen boundary $\tilde{\rho}_{\text{tr}}$ only can arise from $\Delta_4 + \Delta_5$.
The integral $\Delta_5$ is essentially a logarithm and we may take the approximation of constant $\alpha_i$
\begin{align}
    \label{eq:TR16}
    \Delta_5 &= - h \alpha_i^2 \int_{\tilde{\rho}_{\text{tr}}}^{\infty} \mathrm{d}\rho' \left(\frac{2}{\rho'(1+\alpha_i h \rho')} - \frac{\alpha_i h}{(1 + \alpha_i h \rho')^2}\right) \nn\\
	       &= h \alpha_i^2 \left(2\ln \left(\frac{\alpha_i c_{\text{tr}}}{1 + \alpha_i c_{\text{tr}}}\right) + \frac{1}{1 + \alpha_i c_{\text{tr}}}\right)
\end{align}
With $\Delta_5 \sim h$ the logarithmic contribution from $\Delta_5$ reflects the running of the quartic scalar coupling $\lambda_h$.
A similar logarithmic term arises from $\Delta_1$ if we expand in linear order in $h$, canceling the dependence of $\Delta_5$ on $c_{\text{tr}}$.
For our purpose we may approximate the $h$-dependence of the threshold function for sufficiently small $\tilde{\rho}$ by
\begin{equation}
    \label{eq:TR17}
    \partial_h t_u^{(i)}(\tilde{\rho}, h) \approx - \frac{2\alpha_i(\tilde{\rho})\tilde{\rho}}{1 - (\beta_i / \alpha_i) (\tilde{\rho})} - 2\tilde{\rho}^2 \Delta_4\,,
\end{equation}
with corrections linear in $h$.
This maps $\partial_{h} L(h)$ to the ultraviolet physics by eq.~\eqref{eq:TR7}.

\section{Threshold functions for low energy effective theory}
\label{app:C}

\indent In this appendix we provide further details on the threshold functions. 
We formulate them here as functions of $\htilde$ and $h$ as adapted to an effective theory for the standard model degrees of freedom.

\indent The threshold functions for the potential $t_u^{(i)}(\tilde{h},h)$ obey the defining equation
\begin{equation}
    \label{eq:TH1}
    \hat{h} \partial_{\hat{h}} t^{(i)}_{u\;|h} = 2 \left(t_u^{(i)} - \frac{1}{1 + \tilde{m}_i^2}\right)\,,
\end{equation}
with boundary condition
\begin{equation}
    \label{eq:TH2}
    t_u^{(i)} (\tilde{h} \to \infty, h) = 0\,.
\end{equation}
With
\begin{equation}
    \label{eq:TH3}
    \tilde{h} \partial_{\tilde{h}} \left(\frac{t_u^{(i)}}{\tilde{h}^2}\right) = - \frac{2}{(1 + \tilde{m}_i^2) \tilde{h}^2}\,,
\end{equation}
the solution takes the form
\begin{align}
    \label{eq:TH4}
    t_u^{(i)} (\tilde{h},h) &= 2\tilde{h}^2 \int_{\tilde{h}}^{\infty} \frac{\mathrm{d} h'}{h^{\prime 3} (1+\tilde{m}_i^2(h',h))} \nn\\
			    &= 1 - 2\tilde{h}^2 \int_{\tilde h}^{\infty} \frac{\mathrm{d} h' \; \tilde{m}_i^2(h',h)}{h^{\prime 3}(1+\tilde{m}_i^2(h',h))}\,.
\end{align}
If $\tilde{m}_i^2(\tilde{h},h)$ increases for $\tilde{h} \to \infty$ with some positive power of $\tilde{h}$ the boundary condition \eqref{eq:TH2} is obeyed.
Parameterizing
\begin{equation}
    \label{eq:TH5}
    \tilde{m}_i^2(\tilde{h},h) = \alpha_i (\tilde{h},h)\tilde{h}\,,
\end{equation}
yields
\begin{equation}
    \label{eq:TH6}
    t_u^{(i)}(\tilde{h},h) = 1 - 2\tilde{h}^2 \int_{\tilde{h}}^{\infty} \frac{\mathrm{d}h'\; \alpha_i(h',h)}{h^{\prime 2}(1+\alpha_i(h',h)h')}\,.
\end{equation}

\indent The dependence of $t_u^{(i)}$ on $h$ (at fixed $\tilde{h}$) arises from the $h$-dependence of $\alpha_i$
\begin{equation}
    \label{eq:TH7}
    \partial_h t_u^{(i)} (\tilde{h},h) = - \frac{2 \tilde{h}^2}{h} \int_{\tilde h}^{\infty} \frac{\mathrm{d} h'\; h \partial_h \alpha_i(h',h)}{h^{\prime 2}(1+\alpha_i h')^2}\,.
\end{equation}
The integral is dominated by the region near the lower boundary.
For $\tilde{h} \ll 1$ the coefficients $\alpha_i(\tilde{\rho},h)$ depend mainly on $\tilde{\rho}$ since the mass thresholds play not yet a role for the running dimensionless couplings.
Using
\begin{equation}
    \label{eq:TH8}
    h \partial_{h\;|\tilde{h}} = h \partial_{h\;|\tilde{\rho}} - \tilde{\rho} \partial_{\tilde{\rho}\; |h}\,,
\end{equation}
we can write equivalently, with $h' = h \rho'$,
\begin{equation}
    \label{eq:TH9}
    \partial_h t^{(i)}_{u\;|\tilde{h}} = 2 \tilde{\rho}^2 \int_{\tilde{\rho}}^{\infty} \frac{\mathrm{d} \rho'\; (\rho'\partial_{\rho'} - h \partial_h) \alpha_i (\rho',h)}{\rho^{\prime 2}(1+\alpha_i h \rho')^2}\,.
\end{equation}

\indent Let us consider the threshold function for the top quark in the limit of vanishing gauge couplings, with $\alpha_t = y_t^2$.
For fixed $H$ and $\chi$ the flow of the field-dependent top Yukawa coupling $y_t^2(k,H,\chi)$ is given by
\begin{equation}
    \label{eq:TH10}
    k \partial_k y_t^2 = \frac{b_t y_t^4}{1 + y_t^2 \tilde{h}}\,,\quad
    b_t = \frac{9}{16\pi^2}\,.
\end{equation}
For the scaling solution this implies
\begin{equation}
    \label{eq:TH11}
    \tilde{h} \partial_{\tilde{h}} y^2_{t\;|h} = -\frac{1}{2} b_t y_t^4(1+y_t^2\tilde{h})^{-1}\,.
\end{equation}
The flow of $y_t^2$ is logarithmic for $y_t^2 \tilde{h} \ll 1$, and essentially stops for $y_t^2 \tilde{h} \gg 1$.
We may therefore approximate $(1+y_t^2 \tilde{h})^{-1}$ by $(1+\bar{y}_t^2 \tilde{h})^{-1}$, with $\bar{y}_t^2$ a suitable average, typically close to the value at the threshold
\begin{equation}
    \label{eq:TH12}
    \bar{y}_t^2 \approx y_t^2(\tilde{h})\,.
\end{equation}
The solution in this approximation reads
\begin{equation}
    \label{eq:TH13}
    y_t^{-2}(\tilde h) = y_t^{-2}(\tilde{h}_{\text{in}}) + \frac{b_t}{2} \ln \left(\frac{\tilde{h} (1 + \bar{y}_t^2 \tilde{h}_{\text{in}})}{\tilde{h}_{\text{in}} (1+\bar{y}_t^2 \tilde{h})}\right)\,.
\end{equation}
For $\bar{y}_t^2 \tilde{h} \gg 1$ the flow of $y_t^2(\tilde{h})$ indeed stops, according to 
\begin{equation}
    \label{eq:TH14}
    y_t^{-2}(\tilde{h}) = y_t^{-2}(\tilde{h}_{\text{in}}) + \frac{b_t}{2} \ln \frac{1 + \bar{y}_t^2 \tilde{h}_{\text{in}}}{\bar{y}_t^2 \tilde{h}_{\text{in}}} - \frac{b_t}{2 \bar{y}_t^2 \tilde{h}}\,.
\end{equation}

\indent The dependence of $y_t^2(\tilde{h},h)$ on $h$ at fixed $\tilde h$ only arises from the $h$-dependence of the initial value $y_t^2(\tilde{h}_{\text{in}})$,
\begin{equation}
    \label{eq:TH15}
    h \partial_h \alpha_t = h \partial_h y_t^2 = -y_t^4 h \partial_h y_t^{-2}(\tilde{h},h) = -y_t^4 h \partial_h y_t^{-2}(\tilde{h}_{\text{in}})\,.
\end{equation}
The $h$-dependence of $y_t^2$ at fixed $\tilde{h}_{\text{in}}$ is due to the fact that the ultraviolet behavior is parametrized by fixing $y_t^2$ at sufficiently small $\tilde{\rho}_0$ independently of $h$,
\begin{equation}
    \label{eq:TH16}
    y_t^2 (\tilde{\rho}_0,h) = \bar{y}_{t,0}^2\,.
\end{equation}
A given $\tilde{\rho}_0$ is connected to a value $\bar{h} = h \tilde{\rho}_0$ by a relation which depends on $h$.
Keeping $\tilde{h}_{\text{in}}$ fixed the condition \eqref{eq:TH16} results therefore in different $y_t^2(\tilde{h}_{\text{in}})$.
Equivalently, we can keep a fixed initial value $\bar{y}_{t,0}^2$ which is realized for different $\tilde{h}_{\text{in}} = h \tilde{\rho}_0$.
From
\begin{equation}
    \label{eq:TH17}
    y_t^{-2}(\tilde{h}) = \bar{y}_{t,0}^{-2} - \frac{b_t}{2} \ln \left(\frac{h \tilde{\rho}_0}{1 + \bar{y}_t^2 h \tilde{\rho}_0}\right) + \frac{b_t}{2} \ln \left(\frac{\tilde{h}}{1 + \bar{y}_t^2 \tilde{h}}\right)\,,
\end{equation}
one infers
\begin{equation}
    \label{eq:TH18}
    h \partial_h y_t^2(\tilde{h}, h) = \frac{b_t}{2} y_t^4(\tilde{h}, h)\,,
\end{equation}
where we use $\tilde{\rho}_0$ in a range where $\bar{y}_t^2 h \tilde{\rho}_0 \ll 1$.

\indent One obtains the $h$-dependence of the top-quark threshold function as
\begin{equation}
    \label{eq:TH19}
    h \partial_h t_u^{(t)}(\tilde{h}, h) = - b_t \tilde{h}^2 \int_{\tilde{h}}^{\infty} \frac{\mathrm{d} h'\; y_t^4(h',h)}{h^{\prime 2}(1+y_t^2(h',h)h')^2}\,.
\end{equation}
This integral is dominated by the range of $h'$ close to the lower boundary at $\tilde{h}$.
We may approximate it by
\begin{equation}
    \label{eq:TH20}
h \partial_h t_u^{(t)} (\tilde{h},h) = - b_t \tilde{h} y_t^4(\tilde{h}) J(\tilde h)\,,
\end{equation}
with
\begin{align}
    \label{eq:TH21}
    J(\tilde h) &= \tilde{h} \int_{\tilde h}^{\infty} \frac{\mathrm{d} h'}{h^{\prime 2}(1+\bar{y}_t^2 h')^2} \nn\\
		&= 1 + \bar{y}_t^2 \tilde{h} \left(2\ln \left(\frac{\bar{y}_t^2 \tilde{h}}{1 + \bar{y}_t^2 \tilde{h}}\right) + \frac{1}{1 + \bar{y}_t^2 \tilde{h}}\right)\,.
\end{align}
We conclude 
\begin{equation}
    \label{eq:TH22}
    t_u^{(t)}(\tilde h, h) = t_u^{(t)}(\tilde h, h_0) - b_t y_t^4(\tilde h) \tilde{h} J(\tilde{h}) \ln \left(\frac{h}{h_0}\right)\,.
\end{equation}
Where $y_t^2(\tilde{h})$ is given by eq.~\eqref{eq:TH17}.
For $\bar{y}_t^2 \tilde{h} \gg 1$ one finds
\begin{equation}
    \label{eq:TH23}
    J(\tilde{h} \gg \bar{y}_t^2) = \frac{1}{3\bar{y}_t^4 \tilde{h}^2}\,,
\end{equation}
such that the $h$-dependence of $t_u^{(t)}$ vanishes for $\tilde{h} \to \infty$.

\indent If we only include the contribution of the top quark Yukawa coupling, the $\chi$-dependent part in the effective potential takes the form
\begin{align}
    \label{eq:TH24}
    &\Delta U =\nn\\
    &\frac{1}{4} L(h) \chi^4 + \frac{3b_t}{32\pi^2}y_t^4 \! \left(\!\frac{H^\dagger H}{k^2}\!\right)\! J \!\left(\!\frac{H^\dagger H}{k^2}\!\right) \ln\!\left(\!\frac{2H^\dagger \!H}{\chi^2}\!\right)\! H^\dagger \! H k^2\!.
\end{align}
This $\chi$-dependence does not modify the second order character of the phase transition.
For $L(h) = \bar{\lambda}_h h^2/2$ there is no $\chi$-dependence of the boundary term, and one finds for $k^2 \ll H^\dagger H$
\begin{equation}
    \label{eq:TH25}
    \Delta U = \frac{b_t k^6}{32\pi^2 H^\dagger H} \ln \left(\frac{2H^\dagger H}{\chi^2}\right)\,.
\end{equation}

\section{Dependence of $\nu$ on $h$}
\label{app:D}

\indent The parameter $\nu$ describes the $h$-derivative of the potential with the boundary term subtracted.
For the standard model the dependence of $\nu$ on $h$ is rather complex.
It will be discussed in this appendix.
In particular we investigate if for $\bar{\lambda}_m=0$ a partial minimum of $u$ with respect to variation of $h$ is possible.
The answer is negative for the observed values of Yukawa and gauge couplings.

\indent In the approximation \eqref{eq:NY2} the $h$-dependence of $\nu$ at finite $\tilde{\rho}$ is non-trivial, even if we approximate the dimensionless couplings by constants.
Let us first consider only the contributions of top-quarks, $W$- and $Z$-gauge bosons and the Higgs scalar,
\begin{align}
    \label{eq:NY6}
    \nu =&\; \frac{A_\nu \tilde{\rho}}{64\pi^2}\,,\nn\\
    A_\nu =&\; 12 y_t^2 s_u^{(t)} (y_t^2 h \tilde{\rho}) - 3 g_2^2 s_u^{(W)} \left(\frac{1}{2}g_2^2 h \tilde{\rho}\right) \nn\\
	&- \!\left(\frac{3g_2^2}{2} + \frac{9 g_1^2}{10}\right)\! s_u^{(Z)} \!\left(\!\left( \frac{g_2^2}{2} + \frac{3g_1^2}{10}\right)\! h\tilde{\rho}\right)\! \nn\\
	&-3 \lambda_h s_u^{(H)} \left(2\lambda_h h \tilde{\rho} + \frac{\tilde{\nu}}{\tilde{\rho}}\right)\,.
\end{align}
For $h \ll (y_t^2 \tilde{\rho})^{-1}$ one finds $\nu > 0$, the sign determined by the sign of the combination
\begin{equation}
    \label{eq:NY7}
    \Delta_+ = \frac{1}{\bar{\varphi}_0} \left(12\bar{m}_t^2 - 6\bar{m}_W^2 - 3\bar{m}_Z^2 - 3\bar{m}_H^2\right) > 0\,.
\end{equation}
Here we have expressed the Yukawa and gauge couplings by the observed values of the particle masses divided by the observed Fermi scale $\bar{\varphi}_0$.

\indent In contrast, for $h\gg (y_t^2\tilde{\rho})^{-1}$ the sign of $\nu$ turns negative, according to
\begin{equation}
    \label{eq:NY8}
    \Delta_- = \bar{\varphi}_0^2 \left(\frac{4}{\bar{m}_t^2} - \frac{2}{\bar{m}_W^2} - \frac{1}{\bar{m}_Z^2} - \frac{1}{\bar{m}_H^2}\right) < 0\,.
\end{equation}
One concludes that $u-L(h) \tilde{\rho}^2$ has a partial maximum at
\begin{equation}
    \label{eq:NY9}
    h_{\text{max}} = \frac{c_{\text{max}}}{y_t^2 \tilde{\rho}}\,.
\end{equation}
For a given finite $\tilde{\rho}$ one has
\begin{equation}
    \label{eq:NY10}
    \lim_{h \to \infty} \nu(\tilde{\rho},h) = 0\,,
\end{equation}
since all $\tilde{m}_i^2$ grow much larger than one as $h$ increases to very large values.
One concludes that $u(\tilde{\rho}, h) - L(h)\tilde{\rho}$ has also a partial minimum at $h_{\text{min}} = \hat{h}_0(\tilde{\rho})$.
This minimum typically occurs once the fluctuations of the bottom and charm quark and the tau-lepton become dominant and turn $\nu$ again to positive values.
The location of this minimum is roughly given by
\begin{equation}
    \label{eq:NY11}
    12 \tilde{m}_b^2 + 12 \tilde{m}_c^2 + 4 \tilde{m}_\tau^2 = \frac{2}{\tilde{m}_W^2} + \frac{1}{\tilde{m}_Z^2} + \frac{2}{\tilde{m}_H^2} - \frac{4}{\tilde{m}_t^2}\,,
\end{equation}
or
\begin{equation}
    \label{eq:NY12}
    \hat{h}_0(\tilde{\rho}) = \sqrt{-\Delta_- / \Delta_b} \tilde{\rho}^{-1}\,,
\end{equation}
with
\begin{equation}
    \label{eq:NY13}
    \Delta_b = \frac{12 \bar{m}_b^2 + 12\bar{m}_c^2 + 4\bar{m}_\tau^2}{\bar{\varphi}_0^2}\,.
\end{equation}
In this approximation $\hat{h}_0$ decreases $\sim\tilde{\rho}^{-1}$, according to constant $\tilde{h}_0 = \hat{h}_0 \tilde{\rho}$.

\indent A zero of $\nu(\tilde{\rho},h)$ at $\hat{h}_0(\tilde{\rho})$ corresponds to an extremum of $u-L(h)\tilde{\rho}^2$, not of $u$.
If $L(h)$ can be expanded for small $h$, a zero of $\tilde{\nu}(\tilde{\rho},h)$ requires
\begin{equation}
    \label{eq:NY14}
    \nu(\tilde{\rho},h_0) + (\bar{\lambda}_m + \bar{\lambda}_h h_0) \tilde{\rho}^2 = 0\,.
\end{equation}
For $\bar{\lambda}_m = 0$ this requires for a non-vanishing $h_0(\tilde{\rho})$ the condition
\begin{equation}
    \label{eq:NY15}
    h_0 = - \frac{\nu(\tilde{\rho}, h_0)}{\bar{\lambda}_h \tilde{\rho}^2}\,.
\end{equation}
With eq.~\eqref{eq:NY6} this condition reads
\begin{equation}
    \label{eq:NY16}
    \tilde{h}_0 = h_0 \tilde{\rho} = - \frac{A_\nu (\tilde{h}_0)}{64\pi^2\bar{\lambda}_h}\,,
\end{equation}
or
\begin{equation}
    \label{eq:NY17}
    \tilde{m}^2_{t\;|0} = y_t^2 \tilde{h}_0 = -\frac{A_\nu}{64\pi^2}\frac{\bar{m}_t^2}{\bar{m}_H^2}\,.
\end{equation}
In the range of negative $A_\nu$ this quantity does not exceed a few.
With $\tilde{m}_{t,0}^2 \lesssim 10^{-2}$ according to eq.~\eqref{eq:NY17} this leads to a contradiction with the condition $\tilde{m}_{t,0}^2 \gtrsim 1$ necessary for negative $A_\nu$.
We conclude that for $\bar{\lambda}_m=0$ a non-trivial partial minimum of $u(\tilde{\rho},h)$ is not possible for the observed values of the Yukawa and gauge couplings.
A partial minimum at $h_0(\tilde{\rho}) > 0$ requires $\bar{\lambda}_m < 0$.
In this case $h_0(\tilde{\rho} \to \infty)$ remains larger than zero.

\indent The situation would be different for a smaller value of the top Yukawa coupling.
In this case $A_\nu$ may be negative for all values $\tilde{h}_0 \lesssim 1$.
Eq.~\eqref{eq:NY16} would have for large enough $k$ a solution for non-zero $\tilde{h}_0(\tilde{\rho})$ even for $\bar{\lambda}_m = 0$.
This would also be the case if one adds to the standard model bosons with couplings to the Higgs scalar overwhelming the ones of the top quark.

\section{Polynomial expansions of effective potential in ratio of Fermi over Planck scale}
\label{app:E}

\indent In this appendix we translate a polynomial expansion of $u$ in powers of $h$ to a corresponding dependence of the Higgs potential $U$ on $H$.
We consider both expansions around $H=0$ and around a minimum at non-zero $H_0$.
This relates the parameters discussed for the scaling form of $u$ to observable quantities as the Fermi scale and the mass of the Higgs boson.

\indent If the first two derivatives of $u(\tilde{\rho},h)$ in $h$ exist at $h=0$ the expansion of $u(\tilde{\rho},h)$ in $h$ at fixed $\tilde{\rho}$ yields
\begin{align}
    \label{eq:118P}
    u &= u_0 + \tilde{\nu} h + \frac{1}{2} \partial_h \tilde{\nu} h^2 + \dots\nn\\
      &= u_0 + \left(h \partial_h + \frac{1}{2} h^2 \partial_h^2\right) L_{|h=0} \tilde{\rho}^2 - \tilde{\beta}_\nu \tilde{h} - \frac{1}{2} h \partial_h \tilde{\beta}_\nu \tilde{h}\,,
\end{align}
where the $h$-derivatives are taken at fixed $\tilde{\rho}$ and evaluated for $h\to 0$.
Employing $\tilde{\beta}_\nu(\tilde{h},h)$ from eqs.~\eqref{eq:118L}, \eqref{eq:118M}, with $h\partial_h \tilde{\beta}_{\nu\;|\tilde{\rho}} = h\partial_h \tilde{\beta}_{\nu\;|\tilde{h}} + \tilde{h}\partial_{\tilde{h}} \tilde{\beta}_{\nu\;|h}$, and turning to dimensionful quantities leads for the scaling solution for the effective potential
\begin{align}
    \label{eq:118Q}
    U &= u_0(\tilde{\rho}) k^4 + \frac{1}{2} \bar{\lambda}_m \chi^2 H^\dagger H + \frac{1}{2} \bar{\lambda}_h (H^\dagger H)^2 \nn\\
       &\;\;- \tilde{\beta}_\nu (\tilde{h})\!\left(\!1+\frac{1}{2} \frac{\partial\ln\tilde{\beta}_\nu}{\partial\ln\tilde{h}}\!\right)\! k^2 H^\dagger\! H - \partial_h \tilde{\beta}_\nu(\tilde{h}) \frac{k^2}{\chi^2}(H^\dagger H)^2.
\end{align}
Here $\tilde{\beta}_\nu(\tilde{h}) = \tilde{\beta}_\nu(\tilde{h},h=0)$, $\partial_h \tilde{\beta}_\nu(\tilde{h}) = \partial_h \tilde{\beta}_\nu (\tilde{h},h=0)$, with $\tilde{h} = H^\dagger H / k^2$ inserted in eqs.~\eqref{eq:118M}, \eqref{eq:118N}.
For $\bar{\lambda}_m = 0$ the only $\chi$-dependence of $U$ results from $u_0$ and the contribution to the term quartic in $H$ which is suppressed by $k^2/\chi^2$.
For all $k > 0$ the potential for the critical surface involves a term quadratic in $H^\dagger H$ proportional to $k^2$, with additional logarithms of $H^\dagger H / k^2$ and $\rho/k^2$ from $\tilde{\beta}_\nu$.

\indent As an alternative second case we rather expand around some nonzero small value $\bar{h}_0$. This is needed if $h \partial_h \tilde{\beta}_\nu$ differs from zero for $h\to 0$, typically because of a logarithmic dependence of $\alpha_i(\tilde{h},h)$ on $h$.
This expansion reads
\begin{align}
    \label{eq:118S}
    u =&\; u_0(\tilde{\rho},\bar{h}_0) + \tilde{\nu}(\tilde{\rho},\bar{h}_0) (h-\bar{h}_0) \nn\\
       & + \frac{1}{2}\partial_h \tilde{\nu}(\tilde{\rho},\bar{h}_0)(h-\bar{h}_0)^2 + \dots \nn\\
    =&\; u_0 + \left[(h-\bar{h}_0) \partial_h + \frac{1}{2}(h - \bar{h}_0)^2 \partial_{h^2}\right] L_{\;|h=\bar{h}_0} \tilde{\rho}^2 \nn\\
     &- \left[\tilde{\beta}_\nu(\tilde{h},\bar{h}_0) + \frac{1}{2} (h \partial_h + \tilde{h} \partial_{\tilde{h}}) \tilde{\beta}_\nu (\tilde{h},h_0) \left(1-\frac{\bar{h}_0}{h}\right)\right] \nn\\
     &\quad\quad \times(\tilde{h}-\bar{h}_0\tilde{\rho})\,.
\end{align}
For example, we may choose $\bar{h}_0$ such that
\begin{equation}
    \label{eq:118T}
    \partial_h L(\bar{h}_0) = 0\,,\quad \partial_h^2 L(\bar{h}_0) = \bar{\lambda}_{h,0}\,.
\end{equation}
The corresponding scaling from of the effective potential reads
\begin{align}
    \label{eq:118U}
    U =&\; u_0 k^4 + \frac{1}{2}\left[\bar{\lambda}_{h,0} - \frac{1}{\tilde{\rho}} \partial_h \tilde{\beta}_\nu(\tilde{\rho},\bar{h}_0)\right] \left(H^\dagger H - \frac{1}{2}\bar{h}_0 \chi^2\right)^2 \nn\\
    &- \tilde{\beta}_\nu(\tilde{\rho},\bar{h}_0)k^2 \left(H^\dagger H - \frac{1}{2} \bar{h}_0 \chi^2\right)\,.
\end{align}
For $k \to 0$ the terms involving $\tilde{\beta}_\nu$ vanish and we can associate $\bar{h}_0$ with the ratio $2 \varphi_0^2/\chi^2$ and $\bar{\lambda}_{h,0} = \bar{\lambda}_h$.
For $k > 0$ the minimum of $U$ moves away from $H^\dagger H = \bar{h}_0 \chi^2 / 2$, according to
\begin{align}
    \label{eq:118V}
    \frac{\partial U}{\partial(H^\dagger H)} =&\; \left[\bar{\lambda}_h - \frac{1}{\tilde{\rho}} \partial_h \tilde{\beta}_\nu (\tilde{\rho}, \bar{h}_0)\right] \left(H^\dagger H - \frac{1}{2}\bar{h}_0 \chi^2\right) \nn\\
    &- \tilde{\beta}_\nu(\tilde{\rho},\bar{h}_0) k^2\,.
\end{align}

\section{Flow in SSB regime}
\label{app:G}

\indent In this appendix we discuss the case of a partial minimum of the effective potential $u$ at a non-zero value $\tilde{h}_0 (\tilde{\rho})$.
We use the scaling equation for the dependence of $\tilde{h}_0$ on $\tilde{\rho}_0$ for a "stability discussion" for the behavior of small deviations from a given scaling solution.

\indent Let us assume a non-zero partial minimum of $u$ at $\tilde{h}_0 (\tilde{\rho})$.
For $u(\tilde{\rho}, \htilde)$ it obeys the condition
\begin{equation}
    \label{eq:SB1}
    \partial_{\htilde} u (\tilde{\rho}, \tilde{h}_0 (\tilde{\rho}) )=0\,.
\end{equation}
This holds for all $\tilde\rho$ as long as $\tilde{h}_0(\tilde{\rho})$ remains positive.
Taking a logarithmic $\tilde{\rho}$-derivative yields the scaling equation for $\tilde{h}_0(\tilde{\rho})$
\begin{equation}
    \label{eq:SB2}
    \tilde{\rho} \partial_{\tilde{\rho}} \partial_{\htilde} (\tilde{\rho}, \htilde_0 (\tilde{\rho})) + \partial_{\htilde}^2u\tilde{\rho} \partial_{\tilde{\rho}} \htilde_0(\tilde{\rho})=0\,.
\end{equation}
(Here the $\tilde{\rho}$-derivatives are taken at fixed $\htilde$, and the $\htilde$-derivatives at fixed $\tilde{\rho}$.)
We recognise the couplings
\begin{equation}
    \label{eq:SB3}
    \lambda_m (\tilde{\rho}) = \partial_{\tilde{\rho}} \partial_{\htilde} u|_{\htilde_0}\,, \quad \lambda_h = \partial_{\htilde}^2 u|_{\htilde_0}\,. 
\end{equation}
They are now evaluated at $\htilde_0 (\tilde{\rho}) > 0$ instead of $\htilde = 0$.
The flow of $\htilde_0 (\tilde{\rho})$ is given by
\begin{align}
    \label{eq:SB4}
    \tilde{\rho} \partial_{\tilde{\rho}} \htilde_0 &=- \frac{\tilde{\rho} \lambda_m}{\lambda_h} = -\frac{\sigma_m}{\lambda_h} = \zeta\,, \nn \\
    \sigma_m &= \tilde{\rho} \lambda_m\,, \quad \zeta= -\frac{\sigma_m}{\lambda_h} \,.
\end{align}

\indent We need the flow of $\zeta$,
\begin{align}
    \label{eq:SB5}
    \tilde{\rho} \partial_{\tilde{\rho}} \zeta &= -\frac{1}{\lambda_h} \tilde{\rho}\partial_{\tilde{\rho}}\sigma_m + \frac{\sigma_m}{\lambda_h^2}\tilde{\rho} \partial_{\tilde{\rho}} \lambda_h \nn \\
    &= \zeta + \frac{\tilde{\rho} \beta_m}{2\lambda_h} + \frac{\zeta \beta_h}{2\lambda_h}\,.
\end{align}
For $\beta_m$ we use eqs.~\eqref{eq:262A}\eqref{eq:262B}
\begin{equation}
    \label{eq:SB6}
    \tilde{\rho}\beta_m = \tilde{\rho}\tilde{\beta}_m + A_m \sigma_m \,,
\end{equation}
with
\begin{align}
    \label{eq:SB7}
    \tilde{\rho}\tilde{\beta}_m = &- \frac{1}{64\pi^2} \Bigg\{ \frac{12\beta_t}{(1 + y^2_t \htilde_0)^2} -3\beta_h \left( 1 + \frac{1}{(1+2\lambda_h \htilde_0)^2} \right) \nn \\
    &- \frac{3\beta_2}{(1+g^2_2 \htilde_0/2)^2} - \frac{3\beta_2 /2 + 9\beta_1 /10}{((1 + g^2_2/2 + 3 g^2_1/10)\htilde_0)^2}\Bigg\} \,.
\end{align}
Here we have neglected a contribution $\sim \tilde{\rho} \partial_{\tilde{\rho}} \htilde_0$ due to the fact that $\beta_m$ should be evaluated at a $\tilde{\rho}$-dependent value of $\htilde_0$.
One obtains
\begin{align}
    \label{eq:SB8}
    \tilde{\rho} \partial_{\tilde{\rho}} \zeta = \gamma_\zeta (\htilde_0) \zeta + \eta_\zeta (\htilde_0)\, &, \nn \\
    \gamma_\zeta = 1 - \frac{A_m}{2} + \frac{\beta_h}{2\lambda_h} \, &, \quad \eta_\zeta = \frac{\tilde{\rho}\tilde{\beta}_m}{2\lambda_h}\,.
\end{align}
Due to the value of $\gamma_\zeta$ close to one the flow of $\zeta$ is unstable towards the IR of increasing $\tilde{\rho}$.

\indent This induces an instability in the flow of $\tilde h_0$
With $x = \ln \tilde{\rho}$ we can write
\begin{equation}
    \label{eq:SB9}
    \partial_x^2 \htilde_0 = \partial_x \zeta = \gamma_\zeta(\htilde_0) \partial_x \htilde_0 + \eta_\zeta (\htilde_0)\,.
\end{equation}
Let us neglect the flow of the couplings $y^2_t$ etc., as well as the dependence of $\gamma_\zeta$ on $\htilde_0$.
Let us further suppose that $\eta_\zeta$ has a zero for some $\hat{h}_0$. 
Linearizing in $\delta \htilde = \htilde_0 -\hat{h}_0$,
\begin{equation}
    \label{eq:SB10}
    \eta_\zeta(\hat{h}_0) = \eta^\prime_\zeta \delta \htilde \,,
\end{equation}
yields
\begin{equation}
    \label{eq:SB11}
    \partial_x^2 \delta \htilde = \gamma_\zeta \partial_x \delta \htilde - \eta^\prime_\zeta \delta\htilde\,.
\end{equation}
The solution $\delta \htilde = \delta \htilde (x_0) \exp\{ \omega(x-x_0)\}$ involves the solutions of the quadratic equation
\begin{equation}
    \label{eq:SB12}
    \omega^2 - \gamma_\zeta \omega - \eta^\prime_\zeta = 0\,,
\end{equation}
given by
\begin{equation}
    \label{eq:SB13}
    \omega_{\pm} = \frac{\gamma_\zeta}{2} \left( 1 \pm \sqrt{1+ \dfrac{4\eta^\prime}{\gamma_\zeta^2}} \right)\,,
\end{equation}
with approximate values
\begin{equation}
    \label{eq:SB14}
    \omega_+ = \gamma_\zeta + \frac{\eta^\prime}{\gamma_\zeta}\,, \quad \omega_- = - \frac{\eta^\prime}{\gamma_\zeta}\,.
\end{equation}
The leading contribution for $\delta \htilde$ for increasing $\tilde{\rho}$ reads
\begin{equation}
    \label{eq:SB15}
    \delta \htilde(\tilde{\rho}) = \delta \htilde(\tilde{\rho}_0) \left( \dfrac{\tilde{\rho}}{\tilde{\rho}_0} \right)^{\omega_+}\,, 
\end{equation}
while for decreasing $\tilde{\rho}$ the exponent $\omega_-$ dominates.
The issue of stability is similar to the case where the partial minimum occurs at $\tilde{h}_0 = 0$.

\nocite{*}
\bibliography{refs}

\begin{thebibliography}{119}%
\makeatletter
\providecommand \@ifxundefined [1]{%
 \@ifx{#1\undefined}
}%
\providecommand \@ifnum [1]{%
 \ifnum #1\expandafter \@firstoftwo
 \else \expandafter \@secondoftwo
 \fi
}%
\providecommand \@ifx [1]{%
 \ifx #1\expandafter \@firstoftwo
 \else \expandafter \@secondoftwo
 \fi
}%
\providecommand \natexlab [1]{#1}%
\providecommand \enquote  [1]{``#1''}%
\providecommand \bibnamefont  [1]{#1}%
\providecommand \bibfnamefont [1]{#1}%
\providecommand \citenamefont [1]{#1}%
\providecommand \href@noop [0]{\@secondoftwo}%
\providecommand \href [0]{\begingroup \@sanitize@url \@href}%
\providecommand \@href[1]{\@@startlink{#1}\@@href}%
\providecommand \@@href[1]{\endgroup#1\@@endlink}%
\providecommand \@sanitize@url [0]{\catcode `\\12\catcode `\$12\catcode `\&12\catcode `\#12\catcode `\^12\catcode `\_12\catcode `\%12\relax}%
\providecommand \@@startlink[1]{}%
\providecommand \@@endlink[0]{}%
\providecommand \url  [0]{\begingroup\@sanitize@url \@url }%
\providecommand \@url [1]{\endgroup\@href {#1}{\urlprefix }}%
\providecommand \urlprefix  [0]{URL }%
\providecommand \Eprint [0]{\href }%
\providecommand \doibase [0]{https://doi.org/}%
\providecommand \selectlanguage [0]{\@gobble}%
\providecommand \bibinfo  [0]{\@secondoftwo}%
\providecommand \bibfield  [0]{\@secondoftwo}%
\providecommand \translation [1]{[#1]}%
\providecommand \BibitemOpen [0]{}%
\providecommand \bibitemStop [0]{}%
\providecommand \bibitemNoStop [0]{.\EOS\space}%
\providecommand \EOS [0]{\spacefactor3000\relax}%
\providecommand \BibitemShut  [1]{\csname bibitem#1\endcsname}%
\let\auto@bib@innerbib\@empty
\bibitem [{\citenamefont {Gildener}(1976)}]{GIL}%
  \BibitemOpen
  \bibfield  {author} {\bibinfo {author} {\bibfnamefont {E.}~\bibnamefont {Gildener}},\ }\bibfield  {title} {\bibinfo {title} {{Gauge Symmetry Hierarchies}},\ }\href {https://doi.org/10.1103/PhysRevD.14.1667} {\bibfield  {journal} {\bibinfo  {journal} {Phys. Rev. D}\ }\textbf {\bibinfo {volume} {14}},\ \bibinfo {pages} {1667} (\bibinfo {year} {1976})}\BibitemShut {NoStop}%
\bibitem [{\citenamefont {Weinberg}(1979)}]{SWGH}%
  \BibitemOpen
  \bibfield  {author} {\bibinfo {author} {\bibfnamefont {S.}~\bibnamefont {Weinberg}},\ }\bibfield  {title} {\bibinfo {title} {{Gauge Hierarchies}},\ }\href {https://doi.org/10.1016/0370-2693(79)90248-X} {\bibfield  {journal} {\bibinfo  {journal} {Phys. Lett. B}\ }\textbf {\bibinfo {volume} {82}},\ \bibinfo {pages} {387} (\bibinfo {year} {1979})}\BibitemShut {NoStop}%
\bibitem [{\citenamefont {Wetterich}(1984)}]{FTP}%
  \BibitemOpen
  \bibfield  {author} {\bibinfo {author} {\bibfnamefont {C.}~\bibnamefont {Wetterich}},\ }\bibfield  {title} {\bibinfo {title} {{Fine Tuning Problem and the Renormalization Group}},\ }\href {https://doi.org/10.1016/0370-2693(84)90923-7} {\bibfield  {journal} {\bibinfo  {journal} {Phys. Lett. B}\ }\textbf {\bibinfo {volume} {140}},\ \bibinfo {pages} {215} (\bibinfo {year} {1984})}\BibitemShut {NoStop}%
\bibitem [{\citenamefont {Bardeen}(1995)}]{WBAR}%
  \BibitemOpen
  \bibfield  {author} {\bibinfo {author} {\bibfnamefont {W.~A.}\ \bibnamefont {Bardeen}},\ }\bibfield  {title} {\bibinfo {title} {{On naturalness in the standard model}},\ }in\ \href@noop {} {\emph {\bibinfo {booktitle} {{Ontake Summer Institute on Particle Physics}}}}\ (\bibinfo {year} {1995})\BibitemShut {NoStop}%
\bibitem [{\citenamefont {Hempfling}(1996)}]{HEM}%
  \BibitemOpen
  \bibfield  {author} {\bibinfo {author} {\bibfnamefont {R.}~\bibnamefont {Hempfling}},\ }\bibfield  {title} {\bibinfo {title} {The next-to-minimal coleman-weinberg model},\ }\href {https://api.semanticscholar.org/CorpusID:14434956} {\bibfield  {journal} {\bibinfo  {journal} {Physics Letters B}\ }\textbf {\bibinfo {volume} {379}},\ \bibinfo {pages} {153} (\bibinfo {year} {1996})}\BibitemShut {NoStop}%
\bibitem [{\citenamefont {{Peccei}}\ \emph {et~al.}(1987)\citenamefont {{Peccei}}, \citenamefont {{Sol{\`a}}},\ and\ \citenamefont {{Wetterich}}}]{PSW}%
  \BibitemOpen
  \bibfield  {author} {\bibinfo {author} {\bibfnamefont {R.~D.}\ \bibnamefont {{Peccei}}}, \bibinfo {author} {\bibfnamefont {J.}~\bibnamefont {{Sol{\`a}}}},\ and\ \bibinfo {author} {\bibfnamefont {C.}~\bibnamefont {{Wetterich}}},\ }\bibfield  {title} {\bibinfo {title} {{Adjusting the cosmological constant dynamically: Cosmons and a new force weaker than gravity}},\ }\href {https://doi.org/10.1016/0370-2693(87)91191-9} {\bibfield  {journal} {\bibinfo  {journal} {Physics Letters B}\ }\textbf {\bibinfo {volume} {195}},\ \bibinfo {pages} {183} (\bibinfo {year} {1987})}\BibitemShut {NoStop}%
\bibitem [{\citenamefont {{Meissner}}\ and\ \citenamefont {{Nicolai}}(2007)}]{MENI}%
  \BibitemOpen
  \bibfield  {author} {\bibinfo {author} {\bibfnamefont {K.~A.}\ \bibnamefont {{Meissner}}}\ and\ \bibinfo {author} {\bibfnamefont {H.}~\bibnamefont {{Nicolai}}},\ }\bibfield  {title} {\bibinfo {title} {{Conformal symmetry and the Standard Model}},\ }\href {https://doi.org/10.1016/j.physletb.2007.03.023} {\bibfield  {journal} {\bibinfo  {journal} {Physics Letters B}\ }\textbf {\bibinfo {volume} {648}},\ \bibinfo {pages} {312} (\bibinfo {year} {2007})},\ \Eprint {https://arxiv.org/abs/hep-th/0612165} {arXiv:hep-th/0612165 [hep-th]} \BibitemShut {NoStop}%
\bibitem [{\citenamefont {{Meissner}}\ and\ \citenamefont {{Nicolai}}(2008)}]{MENI2}%
  \BibitemOpen
  \bibfield  {author} {\bibinfo {author} {\bibfnamefont {K.~A.}\ \bibnamefont {{Meissner}}}\ and\ \bibinfo {author} {\bibfnamefont {H.}~\bibnamefont {{Nicolai}}},\ }\bibfield  {title} {\bibinfo {title} {{Effective action, conformal anomaly and the issue of quadratic divergences}},\ }\href {https://doi.org/10.1016/j.physletb.2007.12.035} {\bibfield  {journal} {\bibinfo  {journal} {Physics Letters B}\ }\textbf {\bibinfo {volume} {660}},\ \bibinfo {pages} {260} (\bibinfo {year} {2008})},\ \Eprint {https://arxiv.org/abs/0710.2840} {arXiv:0710.2840 [hep-th]} \BibitemShut {NoStop}%
\bibitem [{\citenamefont {{Foot}}\ \emph {et~al.}(2007)\citenamefont {{Foot}}, \citenamefont {{Kobakhidze}},\ and\ \citenamefont {{Volkas}}}]{FKV}%
  \BibitemOpen
  \bibfield  {author} {\bibinfo {author} {\bibfnamefont {R.}~\bibnamefont {{Foot}}}, \bibinfo {author} {\bibfnamefont {A.}~\bibnamefont {{Kobakhidze}}},\ and\ \bibinfo {author} {\bibfnamefont {R.~R.}\ \bibnamefont {{Volkas}}},\ }\bibfield  {title} {\bibinfo {title} {{Electroweak Higgs as a pseudo-Goldstone boson of broken scale invariance}},\ }\href {https://doi.org/10.1016/j.physletb.2007.06.084} {\bibfield  {journal} {\bibinfo  {journal} {Physics Letters B}\ }\textbf {\bibinfo {volume} {655}},\ \bibinfo {pages} {156} (\bibinfo {year} {2007})},\ \Eprint {https://arxiv.org/abs/0704.1165} {arXiv:0704.1165 [hep-ph]} \BibitemShut {NoStop}%
\bibitem [{\citenamefont {{Shaposhnikov}}\ and\ \citenamefont {{Zenh{\"a}usern}}(2009)}]{SHZEN}%
  \BibitemOpen
  \bibfield  {author} {\bibinfo {author} {\bibfnamefont {M.}~\bibnamefont {{Shaposhnikov}}}\ and\ \bibinfo {author} {\bibfnamefont {D.}~\bibnamefont {{Zenh{\"a}usern}}},\ }\bibfield  {title} {\bibinfo {title} {{Quantum scale invariance, cosmological constant and hierarchy problem}},\ }\href {https://doi.org/10.1016/j.physletb.2008.11.041} {\bibfield  {journal} {\bibinfo  {journal} {Physics Letters B}\ }\textbf {\bibinfo {volume} {671}},\ \bibinfo {pages} {162} (\bibinfo {year} {2009})},\ \Eprint {https://arxiv.org/abs/0809.3406} {arXiv:0809.3406 [hep-th]} \BibitemShut {NoStop}%
\bibitem [{\citenamefont {Wetterich}(2012)}]{CWWTL}%
  \BibitemOpen
  \bibfield  {author} {\bibinfo {author} {\bibfnamefont {C.}~\bibnamefont {Wetterich}},\ }\bibfield  {title} {\bibinfo {title} {{Where to look for solving the gauge hierarchy problem?}},\ }\href {https://doi.org/10.1016/j.physletb.2012.11.020} {\bibfield  {journal} {\bibinfo  {journal} {Phys. Lett. B}\ }\textbf {\bibinfo {volume} {718}},\ \bibinfo {pages} {573} (\bibinfo {year} {2012})},\ \Eprint {https://arxiv.org/abs/1112.2910} {arXiv:1112.2910 [hep-ph]} \BibitemShut {NoStop}%
\bibitem [{\citenamefont {{Shaposhnikov}}\ and\ \citenamefont {{Shkerin}}(2018)}]{SHSH}%
  \BibitemOpen
  \bibfield  {author} {\bibinfo {author} {\bibfnamefont {M.}~\bibnamefont {{Shaposhnikov}}}\ and\ \bibinfo {author} {\bibfnamefont {A.}~\bibnamefont {{Shkerin}}},\ }\bibfield  {title} {\bibinfo {title} {{Gravity, scale invariance and the hierarchy problem}},\ }\href {https://doi.org/10.1007/JHEP10(2018)024} {\bibfield  {journal} {\bibinfo  {journal} {Journal of High Energy Physics}\ }\textbf {\bibinfo {volume} {2018}},\ \bibinfo {eid} {24} (\bibinfo {year} {2018})},\ \Eprint {https://arxiv.org/abs/1804.06376} {arXiv:1804.06376 [hep-th]} \BibitemShut {NoStop}%
\bibitem [{\citenamefont {{Karananas}}\ \emph {et~al.}(2024)\citenamefont {{Karananas}}, \citenamefont {{Shaposhnikov}},\ and\ \citenamefont {{Zell}}}]{KSZ}%
  \BibitemOpen
  \bibfield  {author} {\bibinfo {author} {\bibfnamefont {G.~K.}\ \bibnamefont {{Karananas}}}, \bibinfo {author} {\bibfnamefont {M.}~\bibnamefont {{Shaposhnikov}}},\ and\ \bibinfo {author} {\bibfnamefont {S.}~\bibnamefont {{Zell}}},\ }\bibfield  {title} {\bibinfo {title} {{Weyl-invariant Einstein-Cartan gravity: unifying the strong CP and hierarchy puzzles}},\ }\href {https://doi.org/10.1007/JHEP11(2024)146} {\bibfield  {journal} {\bibinfo  {journal} {Journal of High Energy Physics}\ }\textbf {\bibinfo {volume} {2024}},\ \bibinfo {eid} {146} (\bibinfo {year} {2024})},\ \Eprint {https://arxiv.org/abs/2406.11956} {arXiv:2406.11956 [hep-th]} \BibitemShut {NoStop}%
\bibitem [{\citenamefont {{de Boer}}\ \emph {et~al.}(2025)\citenamefont {{de Boer}}, \citenamefont {{Lindner}},\ and\ \citenamefont {{Trautner}}}]{BOLI}%
  \BibitemOpen
  \bibfield  {author} {\bibinfo {author} {\bibfnamefont {T.}~\bibnamefont {{de Boer}}}, \bibinfo {author} {\bibfnamefont {M.}~\bibnamefont {{Lindner}}},\ and\ \bibinfo {author} {\bibfnamefont {A.}~\bibnamefont {{Trautner}}},\ }\bibfield  {title} {\bibinfo {title} {{Hidden Sector Custodial Naturalness}},\ }\href {https://doi.org/10.48550/arXiv.2507.22980} {\bibfield  {journal} {\bibinfo  {journal} {arXiv e-prints}\ ,\ \bibinfo {eid} {arXiv:2507.22980}} (\bibinfo {year} {2025})},\ \Eprint {https://arxiv.org/abs/2507.22980} {arXiv:2507.22980 [hep-ph]} \BibitemShut {NoStop}%
\bibitem [{\citenamefont {Wetterich}(1993)}]{EEE}%
  \BibitemOpen
  \bibfield  {author} {\bibinfo {author} {\bibfnamefont {C.}~\bibnamefont {Wetterich}},\ }\bibfield  {title} {\bibinfo {title} {{Exact evolution equation for the effective potential}},\ }\href {https://doi.org/10.1016/0370-2693(93)90726-X} {\bibfield  {journal} {\bibinfo  {journal} {Phys. Lett. B}\ }\textbf {\bibinfo {volume} {301}},\ \bibinfo {pages} {90} (\bibinfo {year} {1993})},\ \Eprint {https://arxiv.org/abs/1710.05815} {arXiv:1710.05815 [hep-th]} \BibitemShut {NoStop}%
\bibitem [{\citenamefont {Reuter}\ and\ \citenamefont {Wetterich}(1994)}]{REUW}%
  \BibitemOpen
  \bibfield  {author} {\bibinfo {author} {\bibfnamefont {M.}~\bibnamefont {Reuter}}\ and\ \bibinfo {author} {\bibfnamefont {C.}~\bibnamefont {Wetterich}},\ }\bibfield  {title} {\bibinfo {title} {{Effective average action for gauge theories and exact evolution equations}},\ }\href@noop {} {\bibfield  {journal} {\bibinfo  {journal} {Nucl. Phys. B}\ }\textbf {\bibinfo {volume} {417}},\ \bibinfo {pages} {181} (\bibinfo {year} {1994})}\BibitemShut {NoStop}%
\bibitem [{\citenamefont {Tetradis}\ and\ \citenamefont {Wetterich}(1994)}]{TETW}%
  \BibitemOpen
  \bibfield  {author} {\bibinfo {author} {\bibfnamefont {N.}~\bibnamefont {Tetradis}}\ and\ \bibinfo {author} {\bibfnamefont {C.}~\bibnamefont {Wetterich}},\ }\bibfield  {title} {\bibinfo {title} {Critical exponents from the effective average action},\ }\href {https://doi.org/10.1016/0550-3213(94)90446-4} {\bibfield  {journal} {\bibinfo  {journal} {Nuclear Physics B}\ }\textbf {\bibinfo {volume} {422}},\ \bibinfo {pages} {541–592} (\bibinfo {year} {1994})}\BibitemShut {NoStop}%
\bibitem [{\citenamefont {Dupuis}\ \emph {et~al.}(2021)\citenamefont {Dupuis}, \citenamefont {Canet}, \citenamefont {Eichhorn}, \citenamefont {Metzner}, \citenamefont {Pawlowski}, \citenamefont {Tissier},\ and\ \citenamefont {Wschebor}}]{RGREV}%
  \BibitemOpen
  \bibfield  {author} {\bibinfo {author} {\bibfnamefont {N.}~\bibnamefont {Dupuis}}, \bibinfo {author} {\bibfnamefont {L.}~\bibnamefont {Canet}}, \bibinfo {author} {\bibfnamefont {A.}~\bibnamefont {Eichhorn}}, \bibinfo {author} {\bibfnamefont {W.}~\bibnamefont {Metzner}}, \bibinfo {author} {\bibfnamefont {J.}~\bibnamefont {Pawlowski}}, \bibinfo {author} {\bibfnamefont {M.}~\bibnamefont {Tissier}},\ and\ \bibinfo {author} {\bibfnamefont {N.}~\bibnamefont {Wschebor}},\ }\bibfield  {title} {\bibinfo {title} {The nonperturbative functional renormalization group and its applications},\ }\href@noop {} {\bibfield  {journal} {\bibinfo  {journal} {Physics Reports}\ }\textbf {\bibinfo {volume} {910}},\ \bibinfo {pages} {1–114} (\bibinfo {year} {2021})},\ \Eprint {https://arxiv.org/abs/2006.04853} {arXiv:2006.04853 [cond-mat.stat-mech]} \BibitemShut {NoStop}%
\bibitem [{\citenamefont {{Ellwanger}}(1994)}]{ELL}%
  \BibitemOpen
  \bibfield  {author} {\bibinfo {author} {\bibfnamefont {U.}~\bibnamefont {{Ellwanger}}},\ }\bibfield  {title} {\bibinfo {title} {{Flow equations for N point functions and bound states}},\ }\href {https://doi.org/10.1007/BF01555911} {\bibfield  {journal} {\bibinfo  {journal} {Zeitschrift fur Physik C Particles and Fields}\ }\textbf {\bibinfo {volume} {62}},\ \bibinfo {pages} {503} (\bibinfo {year} {1994})},\ \Eprint {https://arxiv.org/abs/hep-ph/9308260} {arXiv:hep-ph/9308260 [hep-ph]} \BibitemShut {NoStop}%
\bibitem [{\citenamefont {{Morris}}(1994)}]{MOR}%
  \BibitemOpen
  \bibfield  {author} {\bibinfo {author} {\bibfnamefont {T.~R.}\ \bibnamefont {{Morris}}},\ }\bibfield  {title} {\bibinfo {title} {{The Exact Renormalization Group and Approximate Solutions}},\ }\href {https://doi.org/10.1142/S0217751X94000972} {\bibfield  {journal} {\bibinfo  {journal} {International Journal of Modern Physics A}\ }\textbf {\bibinfo {volume} {9}},\ \bibinfo {pages} {2411} (\bibinfo {year} {1994})},\ \Eprint {https://arxiv.org/abs/hep-ph/9308265} {arXiv:hep-ph/9308265 [hep-ph]} \BibitemShut {NoStop}%
\bibitem [{\citenamefont {Wetterich}(1990)}]{CWQR}%
  \BibitemOpen
  \bibfield  {author} {\bibinfo {author} {\bibfnamefont {C.}~\bibnamefont {Wetterich}},\ }\bibfield  {title} {\bibinfo {title} {{Quadratic Renormalization of the Average Potential and the Naturalness of Quadratic Mass Relations for the Top Quark}},\ }\href {https://doi.org/10.1007/BF01614706} {\bibfield  {journal} {\bibinfo  {journal} {Z. Phys. C}\ }\textbf {\bibinfo {volume} {48}},\ \bibinfo {pages} {693} (\bibinfo {year} {1990})}\BibitemShut {NoStop}%
\bibitem [{\citenamefont {Gies}\ \emph {et~al.}(2014)\citenamefont {Gies}, \citenamefont {Gneiting},\ and\ \citenamefont {Sondenheimer}}]{GGS}%
  \BibitemOpen
  \bibfield  {author} {\bibinfo {author} {\bibfnamefont {H.}~\bibnamefont {Gies}}, \bibinfo {author} {\bibfnamefont {C.}~\bibnamefont {Gneiting}},\ and\ \bibinfo {author} {\bibfnamefont {R.}~\bibnamefont {Sondenheimer}},\ }\bibfield  {title} {\bibinfo {title} {{Higgs Mass Bounds from Renormalization Flow for a simple Yukawa model}},\ }\href {https://doi.org/10.1103/PhysRevD.89.045012} {\bibfield  {journal} {\bibinfo  {journal} {Phys. Rev. D}\ }\textbf {\bibinfo {volume} {89}},\ \bibinfo {pages} {045012} (\bibinfo {year} {2014})},\ \Eprint {https://arxiv.org/abs/1308.5075} {arXiv:1308.5075 [hep-ph]} \BibitemShut {NoStop}%
\bibitem [{\citenamefont {Gies}\ and\ \citenamefont {Sondenheimer}(2015)}]{GISO}%
  \BibitemOpen
  \bibfield  {author} {\bibinfo {author} {\bibfnamefont {H.}~\bibnamefont {Gies}}\ and\ \bibinfo {author} {\bibfnamefont {R.}~\bibnamefont {Sondenheimer}},\ }\bibfield  {title} {\bibinfo {title} {{Higgs Mass Bounds from Renormalization Flow for a Higgs-top-bottom model}},\ }\href {https://doi.org/10.1140/epjc/s10052-015-3284-1} {\bibfield  {journal} {\bibinfo  {journal} {Eur. Phys. J. C}\ }\textbf {\bibinfo {volume} {75}},\ \bibinfo {pages} {68} (\bibinfo {year} {2015})},\ \Eprint {https://arxiv.org/abs/1407.8124} {arXiv:1407.8124 [hep-ph]} \BibitemShut {NoStop}%
\bibitem [{\citenamefont {{Krajewski}}\ and\ \citenamefont {{Lalak}}(2015{\natexlab{a}})}]{KRALA1}%
  \BibitemOpen
  \bibfield  {author} {\bibinfo {author} {\bibfnamefont {T.}~\bibnamefont {{Krajewski}}}\ and\ \bibinfo {author} {\bibfnamefont {Z.}~\bibnamefont {{Lalak}}},\ }\bibfield  {title} {\bibinfo {title} {{Fine-tuning and vacuum stability in the Wilsonian effective action}},\ }\href {https://doi.org/10.1103/PhysRevD.92.075009} {\bibfield  {journal} {\bibinfo  {journal} {\prd}\ }\textbf {\bibinfo {volume} {92}},\ \bibinfo {eid} {075009} (\bibinfo {year} {2015}{\natexlab{a}})},\ \Eprint {https://arxiv.org/abs/1411.6435} {arXiv:1411.6435 [hep-ph]} \BibitemShut {NoStop}%
\bibitem [{\citenamefont {{Krajewski}}\ and\ \citenamefont {{Lalak}}(2015{\natexlab{b}})}]{KRALA2}%
  \BibitemOpen
  \bibfield  {author} {\bibinfo {author} {\bibfnamefont {T.}~\bibnamefont {{Krajewski}}}\ and\ \bibinfo {author} {\bibfnamefont {Z.}~\bibnamefont {{Lalak}}},\ }\bibfield  {title} {\bibinfo {title} {{Naturalness of effective theories in Wilsonian approach}},\ }\href {https://doi.org/10.48550/arXiv.1510.02154} {\bibfield  {journal} {\bibinfo  {journal} {arXiv e-prints}\ ,\ \bibinfo {eid} {arXiv:1510.02154}} (\bibinfo {year} {2015}{\natexlab{b}})},\ \Eprint {https://arxiv.org/abs/1510.02154} {arXiv:1510.02154 [hep-ph]} \BibitemShut {NoStop}%
\bibitem [{\citenamefont {Eichhorn}\ \emph {et~al.}(2015)\citenamefont {Eichhorn}, \citenamefont {Gies}, \citenamefont {Jaeckel}, \citenamefont {Plehn}, \citenamefont {Scherer},\ and\ \citenamefont {Sondenheimer}}]{EGJP}%
  \BibitemOpen
  \bibfield  {author} {\bibinfo {author} {\bibfnamefont {A.}~\bibnamefont {Eichhorn}}, \bibinfo {author} {\bibfnamefont {H.}~\bibnamefont {Gies}}, \bibinfo {author} {\bibfnamefont {J.}~\bibnamefont {Jaeckel}}, \bibinfo {author} {\bibfnamefont {T.}~\bibnamefont {Plehn}}, \bibinfo {author} {\bibfnamefont {M.~M.}\ \bibnamefont {Scherer}},\ and\ \bibinfo {author} {\bibfnamefont {R.}~\bibnamefont {Sondenheimer}},\ }\bibfield  {title} {\bibinfo {title} {{The Higgs Mass and the Scale of New Physics}},\ }\href {https://doi.org/10.1007/JHEP04(2015)022} {\bibfield  {journal} {\bibinfo  {journal} {JHEP}\ }\textbf {\bibinfo {volume} {04}},\ \bibinfo {pages} {022}},\ \Eprint {https://arxiv.org/abs/1501.02812} {arXiv:1501.02812 [hep-ph]} \BibitemShut {NoStop}%
\bibitem [{\citenamefont {Gies}\ and\ \citenamefont {Sondenheimer}(2018)}]{GISO2}%
  \BibitemOpen
  \bibfield  {author} {\bibinfo {author} {\bibfnamefont {H.}~\bibnamefont {Gies}}\ and\ \bibinfo {author} {\bibfnamefont {R.}~\bibnamefont {Sondenheimer}},\ }\bibfield  {title} {\bibinfo {title} {{Renormalization Group Flow of the Higgs Potential}},\ }\href {https://doi.org/10.1098/rsta.2017.0120} {\bibfield  {journal} {\bibinfo  {journal} {Phil. Trans. Roy. Soc. Lond. A}\ }\textbf {\bibinfo {volume} {376}},\ \bibinfo {pages} {20170120} (\bibinfo {year} {2018})},\ \Eprint {https://arxiv.org/abs/1708.04305} {arXiv:1708.04305 [hep-ph]} \BibitemShut {NoStop}%
\bibitem [{\citenamefont {Reichert}\ \emph {et~al.}(2018)\citenamefont {Reichert}, \citenamefont {Eichhorn}, \citenamefont {Gies}, \citenamefont {Pawlowski}, \citenamefont {Plehn},\ and\ \citenamefont {Scherer}}]{REG}%
  \BibitemOpen
  \bibfield  {author} {\bibinfo {author} {\bibfnamefont {M.}~\bibnamefont {Reichert}}, \bibinfo {author} {\bibfnamefont {A.}~\bibnamefont {Eichhorn}}, \bibinfo {author} {\bibfnamefont {H.}~\bibnamefont {Gies}}, \bibinfo {author} {\bibfnamefont {J.~M.}\ \bibnamefont {Pawlowski}}, \bibinfo {author} {\bibfnamefont {T.}~\bibnamefont {Plehn}},\ and\ \bibinfo {author} {\bibfnamefont {M.~M.}\ \bibnamefont {Scherer}},\ }\bibfield  {title} {\bibinfo {title} {{Probing baryogenesis through the Higgs boson self-coupling}},\ }\href {https://doi.org/10.1103/PhysRevD.97.075008} {\bibfield  {journal} {\bibinfo  {journal} {Phys. Rev. D}\ }\textbf {\bibinfo {volume} {97}},\ \bibinfo {pages} {075008} (\bibinfo {year} {2018})},\ \Eprint {https://arxiv.org/abs/1711.00019} {arXiv:1711.00019 [hep-ph]} \BibitemShut {NoStop}%
\bibitem [{\citenamefont {Sondenheimer}(2019)}]{SON}%
  \BibitemOpen
  \bibfield  {author} {\bibinfo {author} {\bibfnamefont {R.}~\bibnamefont {Sondenheimer}},\ }\bibfield  {title} {\bibinfo {title} {{Nonpolynomial Higgs interactions and vacuum stability}},\ }\href {https://doi.org/10.1140/epjc/s10052-018-6507-4} {\bibfield  {journal} {\bibinfo  {journal} {Eur. Phys. J. C}\ }\textbf {\bibinfo {volume} {79}},\ \bibinfo {pages} {10} (\bibinfo {year} {2019})},\ \Eprint {https://arxiv.org/abs/1711.00065} {arXiv:1711.00065 [hep-ph]} \BibitemShut {NoStop}%
\bibitem [{\citenamefont {Held}\ and\ \citenamefont {Sondenheimer}(2019)}]{HESO}%
  \BibitemOpen
  \bibfield  {author} {\bibinfo {author} {\bibfnamefont {A.}~\bibnamefont {Held}}\ and\ \bibinfo {author} {\bibfnamefont {R.}~\bibnamefont {Sondenheimer}},\ }\bibfield  {title} {\bibinfo {title} {{Higgs stability-bound and fermionic dark matter}},\ }\href {https://doi.org/10.1007/JHEP02(2019)166} {\bibfield  {journal} {\bibinfo  {journal} {JHEP}\ }\textbf {\bibinfo {volume} {02}},\ \bibinfo {pages} {166}},\ \Eprint {https://arxiv.org/abs/1811.07898} {arXiv:1811.07898 [hep-ph]} \BibitemShut {NoStop}%
\bibitem [{\citenamefont {{Pastor-Guti{\'e}rrez}}\ \emph {et~al.}(2023)\citenamefont {{Pastor-Guti{\'e}rrez}}, \citenamefont {{Pawlowski}},\ and\ \citenamefont {{Reichert}}}]{PASPA}%
  \BibitemOpen
  \bibfield  {author} {\bibinfo {author} {\bibfnamefont {{\'A}.}~\bibnamefont {{Pastor-Guti{\'e}rrez}}}, \bibinfo {author} {\bibfnamefont {J.~M.}\ \bibnamefont {{Pawlowski}}},\ and\ \bibinfo {author} {\bibfnamefont {M.}~\bibnamefont {{Reichert}}},\ }\bibfield  {title} {\bibinfo {title} {{The Asymptotically Safe Standard Model: From quantum gravity to dynamical chiral symmetry breaking}},\ }\href {https://doi.org/10.21468/SciPostPhys.15.3.105} {\bibfield  {journal} {\bibinfo  {journal} {SciPost Physics}\ }\textbf {\bibinfo {volume} {15}},\ \bibinfo {eid} {105} (\bibinfo {year} {2023})},\ \Eprint {https://arxiv.org/abs/2207.09817} {arXiv:2207.09817 [hep-th]} \BibitemShut {NoStop}%
\bibitem [{\citenamefont {{Gies}}\ \emph {et~al.}(2025)\citenamefont {{Gies}}, \citenamefont {{Schmieden}},\ and\ \citenamefont {{Zambelli}}}]{GISZ}%
  \BibitemOpen
  \bibfield  {author} {\bibinfo {author} {\bibfnamefont {H.}~\bibnamefont {{Gies}}}, \bibinfo {author} {\bibfnamefont {R.}~\bibnamefont {{Schmieden}}},\ and\ \bibinfo {author} {\bibfnamefont {L.}~\bibnamefont {{Zambelli}}},\ }\bibfield  {title} {\bibinfo {title} {{Interplay of chiral transitions in the standard model}},\ }\href {https://doi.org/10.1140/epjc/s10052-024-13689-3} {\bibfield  {journal} {\bibinfo  {journal} {European Physical Journal C}\ }\textbf {\bibinfo {volume} {85}},\ \bibinfo {eid} {56} (\bibinfo {year} {2025})},\ \Eprint {https://arxiv.org/abs/2306.05943} {arXiv:2306.05943 [hep-th]} \BibitemShut {NoStop}%
\bibitem [{\citenamefont {Garc{\'e}s}\ \emph {et~al.}(2025)\citenamefont {Garc{\'e}s}, \citenamefont {Goertz}, \citenamefont {Lindner},\ and\ \citenamefont {Pastor-Guti{\'e}rrez}}]{GGL}%
  \BibitemOpen
  \bibfield  {author} {\bibinfo {author} {\bibfnamefont {J.~P.}\ \bibnamefont {Garc{\'e}s}}, \bibinfo {author} {\bibfnamefont {F.}~\bibnamefont {Goertz}}, \bibinfo {author} {\bibfnamefont {M.}~\bibnamefont {Lindner}},\ and\ \bibinfo {author} {\bibfnamefont {{\'A}.}~\bibnamefont {Pastor-Guti{\'e}rrez}},\ }\href@noop {} {\bibinfo {title} {{The quantum criticality of the Standard Model and the hierarchy problem}}} (\bibinfo {year} {2025}),\ \Eprint {https://arxiv.org/abs/2506.15919} {arXiv:2506.15919 [hep-ph]} \BibitemShut {NoStop}%
\bibitem [{\citenamefont {Weinberg}(1980)}]{WEI}%
  \BibitemOpen
  \bibfield  {author} {\bibinfo {author} {\bibfnamefont {S.}~\bibnamefont {Weinberg}},\ }\bibfield  {title} {\bibinfo {title} {{Ultraviolet divergences in quantum theories of gravitation}},\ }in\ \href@noop {} {\emph {\bibinfo {booktitle} {{General Relativity: an Einstein Centenary Survey}}}}\ (\bibinfo  {publisher} {Cambridge University Press},\ \bibinfo {year} {1980})\ p.\ \bibinfo {pages} {790}\BibitemShut {NoStop}%
\bibitem [{\citenamefont {Reuter}(1998)}]{REU}%
  \BibitemOpen
  \bibfield  {author} {\bibinfo {author} {\bibfnamefont {M.}~\bibnamefont {Reuter}},\ }\bibfield  {title} {\bibinfo {title} {{Nonperturbative evolution equation for quantum gravity}},\ }\href {http://dx.doi.org/10.1103/PhysRevD.57.971} {\bibfield  {journal} {\bibinfo  {journal} {Phys. Rev. D}\ }\textbf {\bibinfo {volume} {57}},\ \bibinfo {pages} {971} (\bibinfo {year} {1998})},\ \Eprint {https://arxiv.org/abs/hep-th/9605030} {arXiv:hep-th/9605030 [hep-th]} \BibitemShut {NoStop}%
\bibitem [{\citenamefont {Stelle}(1977)}]{STE}%
  \BibitemOpen
  \bibfield  {author} {\bibinfo {author} {\bibfnamefont {K.~S.}\ \bibnamefont {Stelle}},\ }\bibfield  {title} {\bibinfo {title} {{Renormalization of higher-derivative quantum gravity}},\ }\href@noop {} {\bibfield  {journal} {\bibinfo  {journal} {Phys. Rev. D}\ }\textbf {\bibinfo {volume} {16}},\ \bibinfo {pages} {953} (\bibinfo {year} {1977})}\BibitemShut {NoStop}%
\bibitem [{\citenamefont {Fradkin}\ and\ \citenamefont {Tseytlin}(1982)}]{FRT}%
  \BibitemOpen
  \bibfield  {author} {\bibinfo {author} {\bibfnamefont {E.~S.}\ \bibnamefont {Fradkin}}\ and\ \bibinfo {author} {\bibfnamefont {A.~A.}\ \bibnamefont {Tseytlin}},\ }\bibfield  {title} {\bibinfo {title} {{Renormalizable asymptotically free quantum theory of gravity}},\ }\href@noop {} {\bibfield  {journal} {\bibinfo  {journal} {Nucl. Phys. B}\ }\textbf {\bibinfo {volume} {201}},\ \bibinfo {pages} {469} (\bibinfo {year} {1982})}\BibitemShut {NoStop}%
\bibitem [{\citenamefont {Avramidy}\ and\ \citenamefont {Barvinsky}(1985)}]{AVBAR}%
  \BibitemOpen
  \bibfield  {author} {\bibinfo {author} {\bibfnamefont {I.~G.}\ \bibnamefont {Avramidy}}\ and\ \bibinfo {author} {\bibfnamefont {A.~O.}\ \bibnamefont {Barvinsky}},\ }\bibfield  {title} {\bibinfo {title} {{Asymptotic freedom in higher-derivative quantum gravity}},\ }\href@noop {} {\bibfield  {journal} {\bibinfo  {journal} {Phys. Lett. B}\ }\textbf {\bibinfo {volume} {159}},\ \bibinfo {pages} {269} (\bibinfo {year} {1985})}\BibitemShut {NoStop}%
\bibitem [{\citenamefont {Sen}\ \emph {et~al.}(2022)\citenamefont {Sen}, \citenamefont {Wetterich},\ and\ \citenamefont {Yamada}}]{SWY}%
  \BibitemOpen
  \bibfield  {author} {\bibinfo {author} {\bibfnamefont {S.}~\bibnamefont {Sen}}, \bibinfo {author} {\bibfnamefont {C.}~\bibnamefont {Wetterich}},\ and\ \bibinfo {author} {\bibfnamefont {M.}~\bibnamefont {Yamada}},\ }\bibfield  {title} {\bibinfo {title} {{Asymptotic freedom and safety in quantum gravity}},\ }\href@noop {} {\bibfield  {journal} {\bibinfo  {journal} {JHEP}\ }\textbf {\bibinfo {volume} {03}},\ \bibinfo {pages} {130}},\ \Eprint {https://arxiv.org/abs/2111.04696} {arXiv:2111.04696 [hep-th]} \BibitemShut {NoStop}%
\bibitem [{\citenamefont {Shaposhnikov}\ and\ \citenamefont {Wetterich}(2010)}]{SHAW}%
  \BibitemOpen
  \bibfield  {author} {\bibinfo {author} {\bibfnamefont {M.}~\bibnamefont {Shaposhnikov}}\ and\ \bibinfo {author} {\bibfnamefont {C.}~\bibnamefont {Wetterich}},\ }\bibfield  {title} {\bibinfo {title} {{Asymptotic safety of gravity and the Higgs boson mass}},\ }\href {https://doi.org/10.1016/j.physletb.2009.12.022} {\bibfield  {journal} {\bibinfo  {journal} {Phys. Lett. B}\ }\textbf {\bibinfo {volume} {683}},\ \bibinfo {pages} {196} (\bibinfo {year} {2010})},\ \Eprint {https://arxiv.org/abs/0912.0208} {arXiv:0912.0208 [hep-th]} \BibitemShut {NoStop}%
\bibitem [{\citenamefont {Eichhorn}\ \emph {et~al.}(2018)\citenamefont {Eichhorn}, \citenamefont {Hamada}, \citenamefont {Lumma},\ and\ \citenamefont {Yamada}}]{EHLY}%
  \BibitemOpen
  \bibfield  {author} {\bibinfo {author} {\bibfnamefont {A.}~\bibnamefont {Eichhorn}}, \bibinfo {author} {\bibfnamefont {Y.}~\bibnamefont {Hamada}}, \bibinfo {author} {\bibfnamefont {J.}~\bibnamefont {Lumma}},\ and\ \bibinfo {author} {\bibfnamefont {M.}~\bibnamefont {Yamada}},\ }\bibfield  {title} {\bibinfo {title} {Quantum gravity fluctuations flatten the planck-scale higgs potential},\ }\href@noop {} {\bibfield  {journal} {\bibinfo  {journal} {Physical Review D}\ }\textbf {\bibinfo {volume} {97}} (\bibinfo {year} {2018})},\ \Eprint {https://arxiv.org/abs/1712.00319} {arXiv:1712.00319 [hep-th]} \BibitemShut {NoStop}%
\bibitem [{\citenamefont {Pawlowski}\ \emph {et~al.}(2019)\citenamefont {Pawlowski}, \citenamefont {Reichert}, \citenamefont {Wetterich},\ and\ \citenamefont {Yamada}}]{PRWY}%
  \BibitemOpen
  \bibfield  {author} {\bibinfo {author} {\bibfnamefont {J.~M.}\ \bibnamefont {Pawlowski}}, \bibinfo {author} {\bibfnamefont {M.}~\bibnamefont {Reichert}}, \bibinfo {author} {\bibfnamefont {C.}~\bibnamefont {Wetterich}},\ and\ \bibinfo {author} {\bibfnamefont {M.}~\bibnamefont {Yamada}},\ }\bibfield  {title} {\bibinfo {title} {Higgs scalar potential in asymptotically safe quantum gravity},\ }\href@noop {} {\bibfield  {journal} {\bibinfo  {journal} {Physical Review D}\ }\textbf {\bibinfo {volume} {99}} (\bibinfo {year} {2019})},\ \Eprint {https://arxiv.org/abs/1811.11706} {arXiv:1811.11706 [hep-th]} \BibitemShut {NoStop}%
\bibitem [{\citenamefont {Wetterich}(2019)}]{CWQS}%
  \BibitemOpen
  \bibfield  {author} {\bibinfo {author} {\bibfnamefont {C.}~\bibnamefont {Wetterich}},\ }\href@noop {} {\bibinfo {title} {Quantum scale symmetry}} (\bibinfo {year} {2019}),\ \Eprint {https://arxiv.org/abs/1901.04741} {arXiv:1901.04741 [hep-th]} \BibitemShut {NoStop}%
\bibitem [{\citenamefont {Wetterich}\ and\ \citenamefont {Yamada}(2019)}]{CWMY}%
  \BibitemOpen
  \bibfield  {author} {\bibinfo {author} {\bibfnamefont {C.}~\bibnamefont {Wetterich}}\ and\ \bibinfo {author} {\bibfnamefont {M.}~\bibnamefont {Yamada}},\ }\bibfield  {title} {\bibinfo {title} {Variable planck mass from the gauge invariant flow equation},\ }\href@noop {} {\bibfield  {journal} {\bibinfo  {journal} {Physical Review D}\ }\textbf {\bibinfo {volume} {100}} (\bibinfo {year} {2019})},\ \Eprint {https://arxiv.org/abs/1906.01721} {arXiv:1906.01721 [hep-th]} \BibitemShut {NoStop}%
\bibitem [{\citenamefont {Wetterich}(2020)}]{CWEP}%
  \BibitemOpen
  \bibfield  {author} {\bibinfo {author} {\bibfnamefont {C.}~\bibnamefont {Wetterich}},\ }\bibfield  {title} {\bibinfo {title} {Effective scalar potential in asymptotically safe quantum gravity},\ }\href@noop {} {\bibfield  {journal} {\bibinfo  {journal} {Universe}\ }\textbf {\bibinfo {volume} {7}},\ \bibinfo {pages} {45} (\bibinfo {year} {2020})},\ \Eprint {https://arxiv.org/abs/1911.06100} {arXiv:1911.06100 [hep-th]} \BibitemShut {NoStop}%
\bibitem [{\citenamefont {Wetterich}(2021)}]{CWFSI}%
  \BibitemOpen
  \bibfield  {author} {\bibinfo {author} {\bibfnamefont {C.}~\bibnamefont {Wetterich}},\ }\bibfield  {title} {\bibinfo {title} {{Fundamental Scale Invariance}},\ }\href@noop {} {\bibfield  {journal} {\bibinfo  {journal} {Nuclear Physics B}\ }\textbf {\bibinfo {volume} {964}},\ \bibinfo {pages} {115326} (\bibinfo {year} {2021})},\ \Eprint {https://arxiv.org/abs/2007.08805} {arXiv:2007.08805 [hep-th]} \BibitemShut {NoStop}%
\bibitem [{\citenamefont {Henz}\ \emph {et~al.}(2017)\citenamefont {Henz}, \citenamefont {Pawlowski},\ and\ \citenamefont {Wetterich}}]{DQG1}%
  \BibitemOpen
  \bibfield  {author} {\bibinfo {author} {\bibfnamefont {T.}~\bibnamefont {Henz}}, \bibinfo {author} {\bibfnamefont {J.}~\bibnamefont {Pawlowski}},\ and\ \bibinfo {author} {\bibfnamefont {C.}~\bibnamefont {Wetterich}},\ }\bibfield  {title} {\bibinfo {title} {Scaling solutions for dilaton quantum gravity},\ }\href@noop {} {\bibfield  {journal} {\bibinfo  {journal} {Physics Letters B}\ }\textbf {\bibinfo {volume} {769}},\ \bibinfo {pages} {105} (\bibinfo {year} {2017})}\BibitemShut {NoStop}%
\bibitem [{\citenamefont {Henz}\ \emph {et~al.}(2013)\citenamefont {Henz}, \citenamefont {Pawlowski}, \citenamefont {Rodigast},\ and\ \citenamefont {Wetterich}}]{DQG2}%
  \BibitemOpen
  \bibfield  {author} {\bibinfo {author} {\bibfnamefont {T.}~\bibnamefont {Henz}}, \bibinfo {author} {\bibfnamefont {J.}~\bibnamefont {Pawlowski}}, \bibinfo {author} {\bibfnamefont {A.}~\bibnamefont {Rodigast}},\ and\ \bibinfo {author} {\bibfnamefont {C.}~\bibnamefont {Wetterich}},\ }\bibfield  {title} {\bibinfo {title} {Dilaton quantum gravity},\ }\href@noop {} {\bibfield  {journal} {\bibinfo  {journal} {Physics Letters B}\ }\textbf {\bibinfo {volume} {727}},\ \bibinfo {pages} {298–302} (\bibinfo {year} {2013})},\ \Eprint {https://arxiv.org/abs/1304.7743} {arXiv:1304.7743} \BibitemShut {NoStop}%
\bibitem [{\citenamefont {Maitiniyazi}\ \emph {et~al.}(2025)\citenamefont {Maitiniyazi}, \citenamefont {Wetterich},\ and\ \citenamefont {Yamada}}]{A2last}%
  \BibitemOpen
  \bibfield  {author} {\bibinfo {author} {\bibfnamefont {Y.}~\bibnamefont {Maitiniyazi}}, \bibinfo {author} {\bibfnamefont {C.}~\bibnamefont {Wetterich}},\ and\ \bibinfo {author} {\bibfnamefont {M.}~\bibnamefont {Yamada}},\ }\href {https://arxiv.org/abs/2512.14009} {\bibinfo {title} {Scaling solutions for gauge invariant flow equations in dilaton quantum gravity}} (\bibinfo {year} {2025}),\ \Eprint {https://arxiv.org/abs/2512.14009} {arXiv:2512.14009 [hep-th]} \BibitemShut {NoStop}%
\bibitem [{\citenamefont {Dou}\ and\ \citenamefont {Percacci}(1998)}]{DP}%
  \BibitemOpen
  \bibfield  {author} {\bibinfo {author} {\bibfnamefont {D.}~\bibnamefont {Dou}}\ and\ \bibinfo {author} {\bibfnamefont {R.}~\bibnamefont {Percacci}},\ }\bibfield  {title} {\bibinfo {title} {The running gravitational couplings},\ }\href@noop {} {\bibfield  {journal} {\bibinfo  {journal} {Classical and Quantum Gravity}\ }\textbf {\bibinfo {volume} {15}},\ \bibinfo {pages} {3449–3468} (\bibinfo {year} {1998})},\ \Eprint {https://arxiv.org/abs/hep-th/9707239} {arXiv:hep-th/9707239} \BibitemShut {NoStop}%
\bibitem [{\citenamefont {{Souma}}(1999)}]{WSOU}%
  \BibitemOpen
  \bibfield  {author} {\bibinfo {author} {\bibfnamefont {W.}~\bibnamefont {{Souma}}},\ }\bibfield  {title} {\bibinfo {title} {{Non-Trivial Ultraviolet Fixed Point in Quantum Gravity}},\ }\href {https://doi.org/10.1143/PTP.102.181} {\bibfield  {journal} {\bibinfo  {journal} {Progress of Theoretical Physics}\ }\textbf {\bibinfo {volume} {102}},\ \bibinfo {pages} {181} (\bibinfo {year} {1999})},\ \Eprint {https://arxiv.org/abs/hep-th/9907027} {arXiv:hep-th/9907027 [hep-th]} \BibitemShut {NoStop}%
\bibitem [{\citenamefont {{Lauscher}}\ and\ \citenamefont {{Reuter}}(2001)}]{LAUREU}%
  \BibitemOpen
  \bibfield  {author} {\bibinfo {author} {\bibfnamefont {O.}~\bibnamefont {{Lauscher}}}\ and\ \bibinfo {author} {\bibfnamefont {M.}~\bibnamefont {{Reuter}}},\ }\bibfield  {title} {\bibinfo {title} {{Ultraviolet fixed point and generalized flow equation of quantum gravity}},\ }\href {https://doi.org/10.1103/PhysRevD.65.025013} {\bibfield  {journal} {\bibinfo  {journal} {\prd}\ }\textbf {\bibinfo {volume} {65}},\ \bibinfo {eid} {025013} (\bibinfo {year} {2001})},\ \Eprint {https://arxiv.org/abs/hep-th/0108040} {arXiv:hep-th/0108040 [hep-th]} \BibitemShut {NoStop}%
\bibitem [{\citenamefont {{Reuter}}\ and\ \citenamefont {{Saueressig}}(2002)}]{SAUREU}%
  \BibitemOpen
  \bibfield  {author} {\bibinfo {author} {\bibfnamefont {M.}~\bibnamefont {{Reuter}}}\ and\ \bibinfo {author} {\bibfnamefont {F.}~\bibnamefont {{Saueressig}}},\ }\bibfield  {title} {\bibinfo {title} {{Renormalization group flow of quantum gravity in the Einstein-Hilbert truncation}},\ }\href {https://doi.org/10.1103/PhysRevD.65.065016} {\bibfield  {journal} {\bibinfo  {journal} {\prd}\ }\textbf {\bibinfo {volume} {65}},\ \bibinfo {eid} {065016} (\bibinfo {year} {2002})},\ \Eprint {https://arxiv.org/abs/hep-th/0110054} {arXiv:hep-th/0110054 [hep-th]} \BibitemShut {NoStop}%
\bibitem [{\citenamefont {Narain}\ and\ \citenamefont {Percacci}(2010)}]{NP}%
  \BibitemOpen
  \bibfield  {author} {\bibinfo {author} {\bibfnamefont {G.}~\bibnamefont {Narain}}\ and\ \bibinfo {author} {\bibfnamefont {R.}~\bibnamefont {Percacci}},\ }\bibfield  {title} {\bibinfo {title} {Renormalization group flow in scalar-tensor theories: I},\ }\href@noop {} {\bibfield  {journal} {\bibinfo  {journal} {Classical and Quantum Gravity}\ }\textbf {\bibinfo {volume} {27}},\ \bibinfo {pages} {075001} (\bibinfo {year} {2010})},\ \Eprint {https://arxiv.org/abs/0911.0386} {arXiv:0911.0386 [hep-th]} \BibitemShut {NoStop}%
\bibitem [{\citenamefont {Percacci}\ and\ \citenamefont {Vacca}(2015)}]{PV}%
  \BibitemOpen
  \bibfield  {author} {\bibinfo {author} {\bibfnamefont {R.}~\bibnamefont {Percacci}}\ and\ \bibinfo {author} {\bibfnamefont {G.~P.}\ \bibnamefont {Vacca}},\ }\bibfield  {title} {\bibinfo {title} {Search of scaling solutions in scalar–tensor gravity},\ }\href@noop {} {\bibfield  {journal} {\bibinfo  {journal} {The European Physical Journal C}\ }\textbf {\bibinfo {volume} {75}} (\bibinfo {year} {2015})},\ \Eprint {https://arxiv.org/abs/1501.00888} {arXiv:1501.00888 [hep-th]} \BibitemShut {NoStop}%
\bibitem [{\citenamefont {Donà}\ \emph {et~al.}(2016)\citenamefont {Donà}, \citenamefont {Eichhorn}, \citenamefont {Labus},\ and\ \citenamefont {Percacci}}]{DELP}%
  \BibitemOpen
  \bibfield  {author} {\bibinfo {author} {\bibfnamefont {P.}~\bibnamefont {Donà}}, \bibinfo {author} {\bibfnamefont {A.}~\bibnamefont {Eichhorn}}, \bibinfo {author} {\bibfnamefont {P.}~\bibnamefont {Labus}},\ and\ \bibinfo {author} {\bibfnamefont {R.}~\bibnamefont {Percacci}},\ }\bibfield  {title} {\bibinfo {title} {Asymptotic safety in an interacting system of gravity and scalar matter},\ }\href@noop {} {\bibfield  {journal} {\bibinfo  {journal} {Physical Review D}\ }\textbf {\bibinfo {volume} {93}} (\bibinfo {year} {2016})},\ \Eprint {https://arxiv.org/abs/1512.01589} {arXiv:1512.01589 [hep-th]} \BibitemShut {NoStop}%
\bibitem [{\citenamefont {Eichhorn}\ and\ \citenamefont {Pauly}(2021)}]{EP}%
  \BibitemOpen
  \bibfield  {author} {\bibinfo {author} {\bibfnamefont {A.}~\bibnamefont {Eichhorn}}\ and\ \bibinfo {author} {\bibfnamefont {M.}~\bibnamefont {Pauly}},\ }\bibfield  {title} {\bibinfo {title} {Constraining power of asymptotic safety for scalar fields},\ }\href@noop {} {\bibfield  {journal} {\bibinfo  {journal} {Physical Review D}\ }\textbf {\bibinfo {volume} {103}} (\bibinfo {year} {2021})},\ \Eprint {https://arxiv.org/abs/2009.13543} {arXiv:2009.13543 [hep-th]} \BibitemShut {NoStop}%
\bibitem [{\citenamefont {Laporte}\ \emph {et~al.}(2021)\citenamefont {Laporte}, \citenamefont {Pereira}, \citenamefont {Saueressig},\ and\ \citenamefont {Wang}}]{LPF}%
  \BibitemOpen
  \bibfield  {author} {\bibinfo {author} {\bibfnamefont {C.}~\bibnamefont {Laporte}}, \bibinfo {author} {\bibfnamefont {A.~D.}\ \bibnamefont {Pereira}}, \bibinfo {author} {\bibfnamefont {F.}~\bibnamefont {Saueressig}},\ and\ \bibinfo {author} {\bibfnamefont {J.}~\bibnamefont {Wang}},\ }\bibfield  {title} {\bibinfo {title} {Scalar-tensor theories within asymptotic safety},\ }\href@noop {} {\bibfield  {journal} {\bibinfo  {journal} {Journal of High Energy Physics}\ }\textbf {\bibinfo {volume} {2021}} (\bibinfo {year} {2021})},\ \Eprint {https://arxiv.org/abs/2110.09566} {arXiv:2110.09566 [hep-th]} \BibitemShut {NoStop}%
\bibitem [{\citenamefont {{Wetterich}}\ and\ \citenamefont {{Yamada}}(2017)}]{CWMYGH}%
  \BibitemOpen
  \bibfield  {author} {\bibinfo {author} {\bibfnamefont {C.}~\bibnamefont {{Wetterich}}}\ and\ \bibinfo {author} {\bibfnamefont {M.}~\bibnamefont {{Yamada}}},\ }\bibfield  {title} {\bibinfo {title} {{Gauge hierarchy problem in asymptotically safe gravity - The resurgence mechanism}},\ }\href {https://doi.org/10.1016/j.physletb.2017.04.049} {\bibfield  {journal} {\bibinfo  {journal} {Physics Letters B}\ }\textbf {\bibinfo {volume} {770}},\ \bibinfo {pages} {268} (\bibinfo {year} {2017})},\ \Eprint {https://arxiv.org/abs/1612.03069} {arXiv:1612.03069 [hep-th]} \BibitemShut {NoStop}%
\bibitem [{\citenamefont {Bornholdt}\ and\ \citenamefont {Wetterich}(1992)}]{BOW}%
  \BibitemOpen
  \bibfield  {author} {\bibinfo {author} {\bibfnamefont {S.}~\bibnamefont {Bornholdt}}\ and\ \bibinfo {author} {\bibfnamefont {C.}~\bibnamefont {Wetterich}},\ }\bibfield  {title} {\bibinfo {title} {{Selforganizing criticality, large anomalous mass dimension and the gauge hierarchy problem}},\ }\href {https://doi.org/10.1016/0370-2693(92)90659-R} {\bibfield  {journal} {\bibinfo  {journal} {Phys. Lett. B}\ }\textbf {\bibinfo {volume} {282}},\ \bibinfo {pages} {399} (\bibinfo {year} {1992})}\BibitemShut {NoStop}%
\bibitem [{\citenamefont {Gies}\ and\ \citenamefont {Scherer}(2010)}]{GIS}%
  \BibitemOpen
  \bibfield  {author} {\bibinfo {author} {\bibfnamefont {H.}~\bibnamefont {Gies}}\ and\ \bibinfo {author} {\bibfnamefont {M.~M.}\ \bibnamefont {Scherer}},\ }\bibfield  {title} {\bibinfo {title} {{Asymptotic safety of simple Yukawa systems}},\ }\href {https://doi.org/10.1140/epjc/s10052-010-1256-z} {\bibfield  {journal} {\bibinfo  {journal} {Eur. Phys. J. C}\ }\textbf {\bibinfo {volume} {66}},\ \bibinfo {pages} {387} (\bibinfo {year} {2010})},\ \Eprint {https://arxiv.org/abs/0901.2459} {arXiv:0901.2459 [hep-th]} \BibitemShut {NoStop}%
\bibitem [{\citenamefont {Gies}\ \emph {et~al.}(2010)\citenamefont {Gies}, \citenamefont {Rechenberger},\ and\ \citenamefont {Scherer}}]{GIRS}%
  \BibitemOpen
  \bibfield  {author} {\bibinfo {author} {\bibfnamefont {H.}~\bibnamefont {Gies}}, \bibinfo {author} {\bibfnamefont {S.}~\bibnamefont {Rechenberger}},\ and\ \bibinfo {author} {\bibfnamefont {M.~M.}\ \bibnamefont {Scherer}},\ }\bibfield  {title} {\bibinfo {title} {{Towards an Asymptotic-Safety Scenario for Chiral Yukawa Systems}},\ }\href {https://doi.org/10.1140/epjc/s10052-010-1257-y} {\bibfield  {journal} {\bibinfo  {journal} {Eur. Phys. J. C}\ }\textbf {\bibinfo {volume} {66}},\ \bibinfo {pages} {403} (\bibinfo {year} {2010})},\ \Eprint {https://arxiv.org/abs/0907.0327} {arXiv:0907.0327 [hep-th]} \BibitemShut {NoStop}%
\bibitem [{\citenamefont {Gies}\ and\ \citenamefont {Picciau}(2025)}]{GIPI}%
  \BibitemOpen
  \bibfield  {author} {\bibinfo {author} {\bibfnamefont {H.}~\bibnamefont {Gies}}\ and\ \bibinfo {author} {\bibfnamefont {M.}~\bibnamefont {Picciau}},\ }\href@noop {} {\bibinfo {title} {{Self-organized criticality in a relativistic Yukawa theory with Luttinger fermions}}} (\bibinfo {year} {2025}),\ \Eprint {https://arxiv.org/abs/2506.13441} {arXiv:2506.13441 [hep-th]} \BibitemShut {NoStop}%
\bibitem [{\citenamefont {Wetterich}(1988{\natexlab{a}})}]{CWQ}%
  \BibitemOpen
  \bibfield  {author} {\bibinfo {author} {\bibfnamefont {C.}~\bibnamefont {Wetterich}},\ }\bibfield  {title} {\bibinfo {title} {Cosmology and the fate of dilatation symmetry},\ }\href@noop {} {\bibfield  {journal} {\bibinfo  {journal} {Nuclear Physics B}\ }\textbf {\bibinfo {volume} {302}},\ \bibinfo {pages} {668–696} (\bibinfo {year} {1988}{\natexlab{a}})},\ \Eprint {https://arxiv.org/abs/1711.03844} {arXiv:1711.03844} \BibitemShut {NoStop}%
\bibitem [{\citenamefont {Wetterich}(2022{\natexlab{a}})}]{CWIQ}%
  \BibitemOpen
  \bibfield  {author} {\bibinfo {author} {\bibfnamefont {C.}~\bibnamefont {Wetterich}},\ }\bibfield  {title} {\bibinfo {title} {{The Quantum Gravity Connection between Inflation and Quintessence}},\ }\href@noop {} {\bibfield  {journal} {\bibinfo  {journal} {Galaxies}\ }\textbf {\bibinfo {volume} {10}},\ \bibinfo {pages} {50} (\bibinfo {year} {2022}{\natexlab{a}})},\ \Eprint {https://arxiv.org/abs/2201.12213} {arXiv:2201.12213 [astro-ph.CO]} \BibitemShut {NoStop}%
\bibitem [{\citenamefont {Wetterich}(2023)}]{CWQGSS}%
  \BibitemOpen
  \bibfield  {author} {\bibinfo {author} {\bibfnamefont {C.}~\bibnamefont {Wetterich}},\ }\href@noop {} {\bibinfo {title} {Quantum gravity and scale symmetry in cosmology}} (\bibinfo {year} {2023}),\ \Eprint {https://arxiv.org/abs/2211.03596} {arXiv:2211.03596 [gr-qc]} \BibitemShut {NoStop}%
\bibitem [{\citenamefont {Wetterich}(2024{\natexlab{a}})}]{DEQG}%
  \BibitemOpen
  \bibfield  {author} {\bibinfo {author} {\bibfnamefont {C.}~\bibnamefont {Wetterich}},\ }\href@noop {} {\bibinfo {title} {{Dark energy evolution from quantum gravity}}} (\bibinfo {year} {2024}{\natexlab{a}}),\ \Eprint {https://arxiv.org/abs/2407.03465} {arXiv:2407.03465 [gr-qc]} \BibitemShut {NoStop}%
\bibitem [{\citenamefont {Wetterich}(1988{\natexlab{b}})}]{CWTVC}%
  \BibitemOpen
  \bibfield  {author} {\bibinfo {author} {\bibfnamefont {C.}~\bibnamefont {Wetterich}},\ }\bibfield  {title} {\bibinfo {title} {{Cosmologies with variable Newton's “constant”}},\ }\href {https://doi.org/https://doi.org/10.1016/0550-3213(88)90192-7} {\bibfield  {journal} {\bibinfo  {journal} {Nuclear Physics B}\ }\textbf {\bibinfo {volume} {302}},\ \bibinfo {pages} {645} (\bibinfo {year} {1988}{\natexlab{b}})}\BibitemShut {NoStop}%
\bibitem [{\citenamefont {{Wetterich}}(2003)}]{primeiro}%
  \BibitemOpen
  \bibfield  {author} {\bibinfo {author} {\bibfnamefont {C.}~\bibnamefont {{Wetterich}}},\ }\bibfield  {title} {\bibinfo {title} {{Probing quintessence with time variation of couplings}},\ }\href {https://doi.org/10.1088/1475-7516/2003/10/002} {\bibfield  {journal} {\bibinfo  {journal} {JCAP}\ }\textbf {\bibinfo {volume} {2003}}\bibfield  {number} {\bibinfo  {number} { (10)},\ \bibinfo {eid} {002}},\ }\Eprint {https://arxiv.org/abs/hep-ph/0203266} {arXiv:hep-ph/0203266 [hep-ph]} \BibitemShut {NoStop}%
\bibitem [{\citenamefont {{M{\"u}ller}}\ \emph {et~al.}(2004)\citenamefont {{M{\"u}ller}}, \citenamefont {{Sch{\"a}fer}},\ and\ \citenamefont {{Wetterich}}}]{segundo}%
  \BibitemOpen
  \bibfield  {author} {\bibinfo {author} {\bibfnamefont {C.~M.}\ \bibnamefont {{M{\"u}ller}}}, \bibinfo {author} {\bibfnamefont {G.}~\bibnamefont {{Sch{\"a}fer}}},\ and\ \bibinfo {author} {\bibfnamefont {C.}~\bibnamefont {{Wetterich}}},\ }\bibfield  {title} {\bibinfo {title} {{Nucleosynthesis and the variation of fundamental couplings}},\ }\href {https://doi.org/10.1103/PhysRevD.70.083504} {\bibfield  {journal} {\bibinfo  {journal} {\prd}\ }\textbf {\bibinfo {volume} {70}},\ \bibinfo {eid} {083504} (\bibinfo {year} {2004})},\ \Eprint {https://arxiv.org/abs/astro-ph/0405373} {arXiv:astro-ph/0405373 [astro-ph]} \BibitemShut {NoStop}%
\bibitem [{\citenamefont {{Dent}}\ \emph {et~al.}(2007)\citenamefont {{Dent}}, \citenamefont {{Stern}},\ and\ \citenamefont {{Wetterich}}}]{terceiro}%
  \BibitemOpen
  \bibfield  {author} {\bibinfo {author} {\bibfnamefont {T.}~\bibnamefont {{Dent}}}, \bibinfo {author} {\bibfnamefont {S.}~\bibnamefont {{Stern}}},\ and\ \bibinfo {author} {\bibfnamefont {C.}~\bibnamefont {{Wetterich}}},\ }\bibfield  {title} {\bibinfo {title} {{Primordial nucleosynthesis as a probe of fundamental physics parameters}},\ }\href {https://doi.org/10.1103/PhysRevD.76.063513} {\bibfield  {journal} {\bibinfo  {journal} {\prd}\ }\textbf {\bibinfo {volume} {76}},\ \bibinfo {eid} {063513} (\bibinfo {year} {2007})},\ \Eprint {https://arxiv.org/abs/0705.0696} {arXiv:0705.0696 [astro-ph]} \BibitemShut {NoStop}%
\bibitem [{\citenamefont {{Meyer}}\ and\ \citenamefont {{Mei{\ss}ner}}(2024)}]{quarto}%
  \BibitemOpen
  \bibfield  {author} {\bibinfo {author} {\bibfnamefont {H.}~\bibnamefont {{Meyer}}}\ and\ \bibinfo {author} {\bibfnamefont {U.-G.}\ \bibnamefont {{Mei{\ss}ner}}},\ }\bibfield  {title} {\bibinfo {title} {{Improved constraints on the variation of the weak scale from Big Bang nucleosynthesis}},\ }\href {https://doi.org/10.1007/JHEP06(2024)074} {\bibfield  {journal} {\bibinfo  {journal} {Journal of High Energy Physics}\ }\textbf {\bibinfo {volume} {2024}},\ \bibinfo {eid} {74} (\bibinfo {year} {2024})},\ \Eprint {https://arxiv.org/abs/2403.09325} {arXiv:2403.09325 [hep-ph]} \BibitemShut {NoStop}%
\bibitem [{\citenamefont {{Uzan}}(2025)}]{quinto}%
  \BibitemOpen
  \bibfield  {author} {\bibinfo {author} {\bibfnamefont {J.-P.}\ \bibnamefont {{Uzan}}},\ }\bibfield  {title} {\bibinfo {title} {{Fundamental constants: from measurement to the universe, a window on gravitation and cosmology}},\ }\href {https://doi.org/10.1007/s41114-025-00059-y} {\bibfield  {journal} {\bibinfo  {journal} {Living Reviews in Relativity}\ }\textbf {\bibinfo {volume} {28}},\ \bibinfo {eid} {6} (\bibinfo {year} {2025})},\ \Eprint {https://arxiv.org/abs/2410.07281} {arXiv:2410.07281 [astro-ph.CO]} \BibitemShut {NoStop}%
\bibitem [{\citenamefont {Shaposhnikov}\ and\ \citenamefont {Zenhäusern}(2009{\natexlab{a}})}]{SHAZEN1}%
  \BibitemOpen
  \bibfield  {author} {\bibinfo {author} {\bibfnamefont {M.}~\bibnamefont {Shaposhnikov}}\ and\ \bibinfo {author} {\bibfnamefont {D.}~\bibnamefont {Zenhäusern}},\ }\bibfield  {title} {\bibinfo {title} {Scale invariance, unimodular gravity and dark energy},\ }\href@noop {} {\bibfield  {journal} {\bibinfo  {journal} {Physics Letters B}\ }\textbf {\bibinfo {volume} {671}},\ \bibinfo {pages} {187–192} (\bibinfo {year} {2009}{\natexlab{a}})},\ \Eprint {https://arxiv.org/abs/0809.3395} {arXiv:0809.3395 [hep-th]} \BibitemShut {NoStop}%
\bibitem [{\citenamefont {Shaposhnikov}\ and\ \citenamefont {Zenhäusern}(2009{\natexlab{b}})}]{SHAZEN2}%
  \BibitemOpen
  \bibfield  {author} {\bibinfo {author} {\bibfnamefont {M.}~\bibnamefont {Shaposhnikov}}\ and\ \bibinfo {author} {\bibfnamefont {D.}~\bibnamefont {Zenhäusern}},\ }\bibfield  {title} {\bibinfo {title} {Quantum scale invariance, cosmological constant and hierarchy problem},\ }\href@noop {} {\bibfield  {journal} {\bibinfo  {journal} {Physics Letters B}\ }\textbf {\bibinfo {volume} {671}},\ \bibinfo {pages} {162} (\bibinfo {year} {2009}{\natexlab{b}})},\ \Eprint {https://arxiv.org/abs/0809.3406} {arXiv:0809.3406 [hep-th]} \BibitemShut {NoStop}%
\bibitem [{\citenamefont {Shaposhnikov}\ and\ \citenamefont {Tkachev}(2009)}]{SHT}%
  \BibitemOpen
  \bibfield  {author} {\bibinfo {author} {\bibfnamefont {M.}~\bibnamefont {Shaposhnikov}}\ and\ \bibinfo {author} {\bibfnamefont {I.}~\bibnamefont {Tkachev}},\ }\bibfield  {title} {\bibinfo {title} {Quantum scale invariance on the lattice},\ }\href@noop {} {\bibfield  {journal} {\bibinfo  {journal} {Physics Letters B}\ }\textbf {\bibinfo {volume} {675}},\ \bibinfo {pages} {403} (\bibinfo {year} {2009})},\ \Eprint {https://arxiv.org/abs/0811.1967} {arXiv:0811.1967 [hep-th]} \BibitemShut {NoStop}%
\bibitem [{\citenamefont {Blas}\ \emph {et~al.}(2011)\citenamefont {Blas}, \citenamefont {Shaposhnikov},\ and\ \citenamefont {Zenhäusern}}]{BSZ}%
  \BibitemOpen
  \bibfield  {author} {\bibinfo {author} {\bibfnamefont {D.}~\bibnamefont {Blas}}, \bibinfo {author} {\bibfnamefont {M.}~\bibnamefont {Shaposhnikov}},\ and\ \bibinfo {author} {\bibfnamefont {D.}~\bibnamefont {Zenhäusern}},\ }\bibfield  {title} {\bibinfo {title} {Scale-invariant alternatives to general relativity},\ }\href@noop {} {\bibfield  {journal} {\bibinfo  {journal} {Physical Review D}\ }\textbf {\bibinfo {volume} {84}} (\bibinfo {year} {2011})},\ \Eprint {https://arxiv.org/abs/1104.1392} {arXiv:1104.1392 [hep-th]} \BibitemShut {NoStop}%
\bibitem [{\citenamefont {Karananas}\ and\ \citenamefont {Shaposhnikov}(2016)}]{KASH}%
  \BibitemOpen
  \bibfield  {author} {\bibinfo {author} {\bibfnamefont {G.~K.}\ \bibnamefont {Karananas}}\ and\ \bibinfo {author} {\bibfnamefont {M.}~\bibnamefont {Shaposhnikov}},\ }\bibfield  {title} {\bibinfo {title} {Scale-invariant alternatives to general relativity. {II}. dilaton properties},\ }\href@noop {} {\bibfield  {journal} {\bibinfo  {journal} {Physical Review D}\ }\textbf {\bibinfo {volume} {93}} (\bibinfo {year} {2016})},\ \Eprint {https://arxiv.org/abs/1603.01274} {arXiv:1603.01274 [hep-th]} \BibitemShut {NoStop}%
\bibitem [{\citenamefont {García-Bellido}\ \emph {et~al.}(2011)\citenamefont {García-Bellido}, \citenamefont {Rubio}, \citenamefont {Shaposhnikov},\ and\ \citenamefont {Zenhäusern}}]{GBRSZ}%
  \BibitemOpen
  \bibfield  {author} {\bibinfo {author} {\bibfnamefont {J.}~\bibnamefont {García-Bellido}}, \bibinfo {author} {\bibfnamefont {J.}~\bibnamefont {Rubio}}, \bibinfo {author} {\bibfnamefont {M.}~\bibnamefont {Shaposhnikov}},\ and\ \bibinfo {author} {\bibfnamefont {D.}~\bibnamefont {Zenhäusern}},\ }\bibfield  {title} {\bibinfo {title} {Higgs-dilaton cosmology: From the early to the late universe},\ }\href@noop {} {\bibfield  {journal} {\bibinfo  {journal} {Physical Review D}\ }\textbf {\bibinfo {volume} {84}} (\bibinfo {year} {2011})},\ \Eprint {https://arxiv.org/abs/1107.2163} {arXiv:1107.2163 [hep-ph]} \BibitemShut {NoStop}%
\bibitem [{\citenamefont {Ferreira}\ \emph {et~al.}(2017)\citenamefont {Ferreira}, \citenamefont {Hill},\ and\ \citenamefont {Ross}}]{FHR}%
  \BibitemOpen
  \bibfield  {author} {\bibinfo {author} {\bibfnamefont {P.~G.}\ \bibnamefont {Ferreira}}, \bibinfo {author} {\bibfnamefont {C.~T.}\ \bibnamefont {Hill}},\ and\ \bibinfo {author} {\bibfnamefont {G.~G.}\ \bibnamefont {Ross}},\ }\bibfield  {title} {\bibinfo {title} {Weyl current, scale-invariant inflation, and planck scale generation},\ }\href@noop {} {\bibfield  {journal} {\bibinfo  {journal} {Physical Review D}\ }\textbf {\bibinfo {volume} {95}} (\bibinfo {year} {2017})},\ \Eprint {https://arxiv.org/abs/1610.09243} {arXiv:1610.09243 [hep-th]} \BibitemShut {NoStop}%
\bibitem [{\citenamefont {Casas}\ \emph {et~al.}(2018)\citenamefont {Casas}, \citenamefont {Pauly},\ and\ \citenamefont {Rubio}}]{CPRU}%
  \BibitemOpen
  \bibfield  {author} {\bibinfo {author} {\bibfnamefont {S.}~\bibnamefont {Casas}}, \bibinfo {author} {\bibfnamefont {M.}~\bibnamefont {Pauly}},\ and\ \bibinfo {author} {\bibfnamefont {J.}~\bibnamefont {Rubio}},\ }\bibfield  {title} {\bibinfo {title} {Higgs-dilaton cosmology: An inflation{\textendash}dark-energy connection and forecasts for future galaxy surveys},\ }\href@noop {} {\bibfield  {journal} {\bibinfo  {journal} {Physical Review D}\ }\textbf {\bibinfo {volume} {97}} (\bibinfo {year} {2018})},\ \Eprint {https://arxiv.org/abs/1712.04956} {arXiv:1712.04956 [astro-ph.CO]} \BibitemShut {NoStop}%
\bibitem [{\citenamefont {Ferreira}\ \emph {et~al.}(2018)\citenamefont {Ferreira}, \citenamefont {Hill}, \citenamefont {Noller},\ and\ \citenamefont {Ross}}]{FHNR}%
  \BibitemOpen
  \bibfield  {author} {\bibinfo {author} {\bibfnamefont {P.~G.}\ \bibnamefont {Ferreira}}, \bibinfo {author} {\bibfnamefont {C.~T.}\ \bibnamefont {Hill}}, \bibinfo {author} {\bibfnamefont {J.}~\bibnamefont {Noller}},\ and\ \bibinfo {author} {\bibfnamefont {G.~G.}\ \bibnamefont {Ross}},\ }\bibfield  {title} {\bibinfo {title} {Inflation in a scale-invariant universe},\ }\href@noop {} {\bibfield  {journal} {\bibinfo  {journal} {Physical Review D}\ }\textbf {\bibinfo {volume} {97}} (\bibinfo {year} {2018})},\ \Eprint {https://arxiv.org/abs/1802.06069} {arXiv:1802.06069 [astro-ph.CO]} \BibitemShut {NoStop}%
\bibitem [{\citenamefont {Rubio}(2019)}]{RUB}%
  \BibitemOpen
  \bibfield  {author} {\bibinfo {author} {\bibfnamefont {J.}~\bibnamefont {Rubio}},\ }\bibfield  {title} {\bibinfo {title} {Higgs inflation},\ }\href@noop {} {\bibfield  {journal} {\bibinfo  {journal} {Frontiers in Astronomy and Space Sciences}\ }\textbf {\bibinfo {volume} {5}} (\bibinfo {year} {2019})},\ \Eprint {https://arxiv.org/abs/1807.02376} {arXiv:1807.02376 [hep-ph]} \BibitemShut {NoStop}%
\bibitem [{\citenamefont {Casas}\ \emph {et~al.}(2019)\citenamefont {Casas}, \citenamefont {Karananas}, \citenamefont {Pauly},\ and\ \citenamefont {Rubio}}]{CKPR}%
  \BibitemOpen
  \bibfield  {author} {\bibinfo {author} {\bibfnamefont {S.}~\bibnamefont {Casas}}, \bibinfo {author} {\bibfnamefont {G.~K.}\ \bibnamefont {Karananas}}, \bibinfo {author} {\bibfnamefont {M.}~\bibnamefont {Pauly}},\ and\ \bibinfo {author} {\bibfnamefont {J.}~\bibnamefont {Rubio}},\ }\bibfield  {title} {\bibinfo {title} {Scale-invariant alternatives to general relativity. {III}. the inflation-dark energy connection},\ }\href@noop {} {\bibfield  {journal} {\bibinfo  {journal} {Physical Review D}\ }\textbf {\bibinfo {volume} {99}} (\bibinfo {year} {2019})},\ \Eprint {https://arxiv.org/abs/1811.05984} {arXiv:1811.05984 [astro-ph.CO]} \BibitemShut {NoStop}%
\bibitem [{\citenamefont {Fujii}(1982)}]{FUJ}%
  \BibitemOpen
  \bibfield  {author} {\bibinfo {author} {\bibfnamefont {Y.}~\bibnamefont {Fujii}},\ }\bibfield  {title} {\bibinfo {title} {{Origin of the Gravitational Constant and Particle Masses in Scale Invariant Scalar - Tensor Theory}},\ }\href {https://doi.org/10.1103/PhysRevD.26.2580} {\bibfield  {journal} {\bibinfo  {journal} {Phys. Rev. D}\ }\textbf {\bibinfo {volume} {26}},\ \bibinfo {pages} {2580} (\bibinfo {year} {1982})}\BibitemShut {NoStop}%
\bibitem [{\citenamefont {{Ellwanger}}\ and\ \citenamefont {{Wetterich}}(1994)}]{ELLW}%
  \BibitemOpen
  \bibfield  {author} {\bibinfo {author} {\bibfnamefont {U.}~\bibnamefont {{Ellwanger}}}\ and\ \bibinfo {author} {\bibfnamefont {C.}~\bibnamefont {{Wetterich}}},\ }\bibfield  {title} {\bibinfo {title} {{Evolution equations for the quark-meson transition}},\ }\href {https://doi.org/10.1016/0550-3213(94)90568-1} {\bibfield  {journal} {\bibinfo  {journal} {Nuclear Physics B}\ }\textbf {\bibinfo {volume} {423}},\ \bibinfo {pages} {137} (\bibinfo {year} {1994})},\ \Eprint {https://arxiv.org/abs/hep-ph/9402221} {arXiv:hep-ph/9402221 [hep-ph]} \BibitemShut {NoStop}%
\bibitem [{\citenamefont {Bornholdt}\ and\ \citenamefont {Wetterich}(1993)}]{BOW2}%
  \BibitemOpen
  \bibfield  {author} {\bibinfo {author} {\bibfnamefont {S.}~\bibnamefont {Bornholdt}}\ and\ \bibinfo {author} {\bibfnamefont {C.}~\bibnamefont {Wetterich}},\ }\bibfield  {title} {\bibinfo {title} {{Average action for models with fermions}},\ }\href {https://doi.org/10.1007/BF01553018} {\bibfield  {journal} {\bibinfo  {journal} {Z. Phys. C}\ }\textbf {\bibinfo {volume} {58}},\ \bibinfo {pages} {585} (\bibinfo {year} {1993})}\BibitemShut {NoStop}%
\bibitem [{\citenamefont {Wetterich}(2018)}]{CWGIF}%
  \BibitemOpen
  \bibfield  {author} {\bibinfo {author} {\bibfnamefont {C.}~\bibnamefont {Wetterich}},\ }\bibfield  {title} {\bibinfo {title} {Gauge invariant flow equation},\ }\href@noop {} {\bibfield  {journal} {\bibinfo  {journal} {Nuclear Physics B}\ }\textbf {\bibinfo {volume} {931}},\ \bibinfo {pages} {262} (\bibinfo {year} {2018})},\ \Eprint {https://arxiv.org/abs/1607.02989} {arXiv:1607.02989 [hep-th]} \BibitemShut {NoStop}%
\bibitem [{\citenamefont {Wetterich}(2025)}]{SFE}%
  \BibitemOpen
  \bibfield  {author} {\bibinfo {author} {\bibfnamefont {C.}~\bibnamefont {Wetterich}},\ }\bibfield  {title} {\bibinfo {title} {{Simplified functional flow equation}},\ }\href {https://doi.org/10.1016/j.physletb.2025.139435} {\bibfield  {journal} {\bibinfo  {journal} {Phys. Lett. B}\ }\textbf {\bibinfo {volume} {864}},\ \bibinfo {pages} {139435} (\bibinfo {year} {2025})},\ \Eprint {https://arxiv.org/abs/2403.17523} {arXiv:2403.17523 [hep-th]} \BibitemShut {NoStop}%
\bibitem [{\citenamefont {{Papenbrock}}\ and\ \citenamefont {{Wetterich}}(1995)}]{PAPW}%
  \BibitemOpen
  \bibfield  {author} {\bibinfo {author} {\bibfnamefont {T.}~\bibnamefont {{Papenbrock}}}\ and\ \bibinfo {author} {\bibfnamefont {C.}~\bibnamefont {{Wetterich}}},\ }\bibfield  {title} {\bibinfo {title} {{Two-loop results from improved one loop computations}},\ }\href {https://doi.org/10.1007/BF01556140} {\bibfield  {journal} {\bibinfo  {journal} {Zeitschrift fur Physik C Particles and Fields}\ }\textbf {\bibinfo {volume} {65}},\ \bibinfo {pages} {519} (\bibinfo {year} {1995})},\ \Eprint {https://arxiv.org/abs/hep-th/9403164} {arXiv:hep-th/9403164 [hep-th]} \BibitemShut {NoStop}%
\bibitem [{\citenamefont {Litim}(2001)}]{LIT}%
  \BibitemOpen
  \bibfield  {author} {\bibinfo {author} {\bibfnamefont {D.~F.}\ \bibnamefont {Litim}},\ }\bibfield  {title} {\bibinfo {title} {Optimized renormalization group flows},\ }\href@noop {} {\bibfield  {journal} {\bibinfo  {journal} {Physical Review D}\ }\textbf {\bibinfo {volume} {64}} (\bibinfo {year} {2001})},\ \Eprint {https://arxiv.org/abs/hep-th/0103195} {arXiv:hep-th/0103195} \BibitemShut {NoStop}%
\bibitem [{\citenamefont {Wetterich}(2024{\natexlab{b}})}]{CWFT}%
  \BibitemOpen
  \bibfield  {author} {\bibinfo {author} {\bibfnamefont {C.}~\bibnamefont {Wetterich}},\ }\href {https://arxiv.org/abs/2402.04679} {\bibinfo {title} {Field transformations in functional integral, effective action and functional flow equations}} (\bibinfo {year} {2024}{\natexlab{b}}),\ \Eprint {https://arxiv.org/abs/2402.04679} {arXiv:2402.04679 [hep-th]} \BibitemShut {NoStop}%
\bibitem [{\citenamefont {Ratra}\ and\ \citenamefont {Peebles}(1988)}]{RP2}%
  \BibitemOpen
  \bibfield  {author} {\bibinfo {author} {\bibfnamefont {B.}~\bibnamefont {Ratra}}\ and\ \bibinfo {author} {\bibfnamefont {P.~J.~E.}\ \bibnamefont {Peebles}},\ }\bibfield  {title} {\bibinfo {title} {Cosmological consequences of a rolling homogeneous scalar field},\ }\href@noop {} {\bibfield  {journal} {\bibinfo  {journal} {Phys. Rev. D}\ }\textbf {\bibinfo {volume} {37}},\ \bibinfo {pages} {3406} (\bibinfo {year} {1988})}\BibitemShut {NoStop}%
\bibitem [{\citenamefont {Wetterich}(1995)}]{CWCMAV}%
  \BibitemOpen
  \bibfield  {author} {\bibinfo {author} {\bibfnamefont {C.}~\bibnamefont {Wetterich}},\ }\bibfield  {title} {\bibinfo {title} {{The Cosmon model for an asymptotically vanishing time dependent cosmological 'constant'}},\ }\href@noop {} {\bibfield  {journal} {\bibinfo  {journal} {Astron. Astrophys.}\ }\textbf {\bibinfo {volume} {301}},\ \bibinfo {pages} {321} (\bibinfo {year} {1995})},\ \Eprint {https://arxiv.org/abs/9408025} {arXiv:9408025 [hep-th]} \BibitemShut {NoStop}%
\bibitem [{\citenamefont {Frieman}\ \emph {et~al.}(1995)\citenamefont {Frieman}, \citenamefont {Hill}, \citenamefont {Stebbins},\ and\ \citenamefont {Waga}}]{FTSW}%
  \BibitemOpen
  \bibfield  {author} {\bibinfo {author} {\bibfnamefont {J.~A.}\ \bibnamefont {Frieman}}, \bibinfo {author} {\bibfnamefont {C.~T.}\ \bibnamefont {Hill}}, \bibinfo {author} {\bibfnamefont {A.}~\bibnamefont {Stebbins}},\ and\ \bibinfo {author} {\bibfnamefont {I.}~\bibnamefont {Waga}},\ }\bibfield  {title} {\bibinfo {title} {Cosmology with ultralight pseudo nambu-goldstone bosons},\ }\href@noop {} {\bibfield  {journal} {\bibinfo  {journal} {Physical Review Letters}\ }\textbf {\bibinfo {volume} {75}},\ \bibinfo {pages} {2077–2080} (\bibinfo {year} {1995})},\ \Eprint {https://arxiv.org/abs/astro-ph/9505060} {arXiv:astro-ph/9505060} \BibitemShut {NoStop}%
\bibitem [{\citenamefont {Ferreira}\ and\ \citenamefont {Joyce}(1997)}]{FEJO}%
  \BibitemOpen
  \bibfield  {author} {\bibinfo {author} {\bibfnamefont {P.~G.}\ \bibnamefont {Ferreira}}\ and\ \bibinfo {author} {\bibfnamefont {M.}~\bibnamefont {Joyce}},\ }\bibfield  {title} {\bibinfo {title} {Structure formation with a self-tuning scalar field},\ }\href@noop {} {\bibfield  {journal} {\bibinfo  {journal} {Phys. Rev. Lett.}\ }\textbf {\bibinfo {volume} {79}},\ \bibinfo {pages} {4740} (\bibinfo {year} {1997})}\BibitemShut {NoStop}%
\bibitem [{\citenamefont {Viana}\ and\ \citenamefont {Liddle}(1998)}]{VL}%
  \BibitemOpen
  \bibfield  {author} {\bibinfo {author} {\bibfnamefont {P.~T.~P.}\ \bibnamefont {Viana}}\ and\ \bibinfo {author} {\bibfnamefont {A.~R.}\ \bibnamefont {Liddle}},\ }\bibfield  {title} {\bibinfo {title} {Perturbation evolution in cosmologies with a decaying cosmological constant},\ }\href@noop {} {\bibfield  {journal} {\bibinfo  {journal} {Phys. Rev. D}\ }\textbf {\bibinfo {volume} {57}},\ \bibinfo {pages} {674} (\bibinfo {year} {1998})}\BibitemShut {NoStop}%
\bibitem [{\citenamefont {Copeland}\ \emph {et~al.}(1998)\citenamefont {Copeland}, \citenamefont {Liddle},\ and\ \citenamefont {Wands}}]{CLW}%
  \BibitemOpen
  \bibfield  {author} {\bibinfo {author} {\bibfnamefont {E.~J.}\ \bibnamefont {Copeland}}, \bibinfo {author} {\bibfnamefont {A.~R.}\ \bibnamefont {Liddle}},\ and\ \bibinfo {author} {\bibfnamefont {D.}~\bibnamefont {Wands}},\ }\bibfield  {title} {\bibinfo {title} {Exponential potentials and cosmological scaling solutions},\ }\href@noop {} {\bibfield  {journal} {\bibinfo  {journal} {Phys. Rev. D}\ }\textbf {\bibinfo {volume} {57}},\ \bibinfo {pages} {4686} (\bibinfo {year} {1998})}\BibitemShut {NoStop}%
\bibitem [{\citenamefont {Caldwell}\ \emph {et~al.}(1998)\citenamefont {Caldwell}, \citenamefont {Dave},\ and\ \citenamefont {Steinhardt}}]{CDS}%
  \BibitemOpen
  \bibfield  {author} {\bibinfo {author} {\bibfnamefont {R.~R.}\ \bibnamefont {Caldwell}}, \bibinfo {author} {\bibfnamefont {R.}~\bibnamefont {Dave}},\ and\ \bibinfo {author} {\bibfnamefont {P.~J.}\ \bibnamefont {Steinhardt}},\ }\bibfield  {title} {\bibinfo {title} {Cosmological imprint of an energy component with general equation of state},\ }\href@noop {} {\bibfield  {journal} {\bibinfo  {journal} {Physical Review Letters}\ }\textbf {\bibinfo {volume} {80}},\ \bibinfo {pages} {1582–1585} (\bibinfo {year} {1998})},\ \Eprint {https://arxiv.org/abs/astro-ph/9708069} {arXiv:astro-ph/9708069} \BibitemShut {NoStop}%
\bibitem [{\citenamefont {Amendola}(1999)}]{LA1}%
  \BibitemOpen
  \bibfield  {author} {\bibinfo {author} {\bibfnamefont {L.}~\bibnamefont {Amendola}},\ }\bibfield  {title} {\bibinfo {title} {Scaling solutions in general nonminimal coupling theories},\ }\href@noop {} {\bibfield  {journal} {\bibinfo  {journal} {Physical Review D}\ }\textbf {\bibinfo {volume} {60}} (\bibinfo {year} {1999})},\ \Eprint {https://arxiv.org/abs/astro-ph/9904120} {arXiv:astro-ph/9904120} \BibitemShut {NoStop}%
\bibitem [{\citenamefont {Amendola}(2000)}]{LACQ}%
  \BibitemOpen
  \bibfield  {author} {\bibinfo {author} {\bibfnamefont {L.}~\bibnamefont {Amendola}},\ }\bibfield  {title} {\bibinfo {title} {Coupled quintessence},\ }\href@noop {} {\bibfield  {journal} {\bibinfo  {journal} {Physical Review D}\ }\textbf {\bibinfo {volume} {62}} (\bibinfo {year} {2000})},\ \Eprint {https://arxiv.org/abs/9908023} {arXiv:9908023 [astro-ph]} \BibitemShut {NoStop}%
\bibitem [{\citenamefont {Linder}(2007)}]{LIN}%
  \BibitemOpen
  \bibfield  {author} {\bibinfo {author} {\bibfnamefont {E.~V.}\ \bibnamefont {Linder}},\ }\bibfield  {title} {\bibinfo {title} {The dynamics of quintessence, the quintessence of dynamics},\ }\href@noop {} {\bibfield  {journal} {\bibinfo  {journal} {General Relativity and Gravitation}\ }\textbf {\bibinfo {volume} {40}},\ \bibinfo {pages} {329–356} (\bibinfo {year} {2007})},\ \Eprint {https://arxiv.org/abs/0704.2064} {arXiv:0704.2064 [astro-ph]} \BibitemShut {NoStop}%
\bibitem [{\citenamefont {Wetterich}(2014)}]{CWVG}%
  \BibitemOpen
  \bibfield  {author} {\bibinfo {author} {\bibfnamefont {C.}~\bibnamefont {Wetterich}},\ }\bibfield  {title} {\bibinfo {title} {Variable gravity universe},\ }\href@noop {} {\bibfield  {journal} {\bibinfo  {journal} {Phys. Rev. D}\ }\textbf {\bibinfo {volume} {89}},\ \bibinfo {pages} {024005} (\bibinfo {year} {2014})},\ \Eprint {https://arxiv.org/abs/1308.1019} {arXiv:1308.1019 [astro-ph.CO]} \BibitemShut {NoStop}%
\bibitem [{\citenamefont {Wetterich}(2007)}]{CWGN}%
  \BibitemOpen
  \bibfield  {author} {\bibinfo {author} {\bibfnamefont {C.}~\bibnamefont {Wetterich}},\ }\bibfield  {title} {\bibinfo {title} {Growing neutrinos and cosmological selection},\ }\href@noop {} {\bibfield  {journal} {\bibinfo  {journal} {Physics Letters B}\ }\textbf {\bibinfo {volume} {655}},\ \bibinfo {pages} {201–208} (\bibinfo {year} {2007})},\ \Eprint {https://arxiv.org/abs/0706.4427} {arXiv:0706.4427 [hep-ph]} \BibitemShut {NoStop}%
\bibitem [{\citenamefont {Amendola}\ \emph {et~al.}(2008)\citenamefont {Amendola}, \citenamefont {Baldi},\ and\ \citenamefont {Wetterich}}]{ABW}%
  \BibitemOpen
  \bibfield  {author} {\bibinfo {author} {\bibfnamefont {L.}~\bibnamefont {Amendola}}, \bibinfo {author} {\bibfnamefont {M.}~\bibnamefont {Baldi}},\ and\ \bibinfo {author} {\bibfnamefont {C.}~\bibnamefont {Wetterich}},\ }\bibfield  {title} {\bibinfo {title} {Quintessence cosmologies with a growing matter component},\ }\href@noop {} {\bibfield  {journal} {\bibinfo  {journal} {Physical Review D}\ }\textbf {\bibinfo {volume} {78}} (\bibinfo {year} {2008})},\ \Eprint {https://arxiv.org/abs/0706.3064} {arXiv:0706.3064 [astro-ph]} \BibitemShut {NoStop}%
\bibitem [{\citenamefont {Wetterich}(2017)}]{CWGFC}%
  \BibitemOpen
  \bibfield  {author} {\bibinfo {author} {\bibfnamefont {C.}~\bibnamefont {Wetterich}},\ }\bibfield  {title} {\bibinfo {title} {Graviton fluctuations erase the cosmological constant},\ }\href@noop {} {\bibfield  {journal} {\bibinfo  {journal} {Physics Letters B}\ }\textbf {\bibinfo {volume} {773}},\ \bibinfo {pages} {6–19} (\bibinfo {year} {2017})},\ \Eprint {https://arxiv.org/abs/1704.08040} {arXiv:1704.08040 [gr-qc]} \BibitemShut {NoStop}%
\bibitem [{\citenamefont {Wetterich}(2022{\natexlab{b}})}]{CWSSFGC}%
  \BibitemOpen
  \bibfield  {author} {\bibinfo {author} {\bibfnamefont {C.}~\bibnamefont {Wetterich}},\ }\bibfield  {title} {\bibinfo {title} {{Scaling solution for field-dependent gauge couplings in quantum gravity}},\ }\href {https://doi.org/10.1016/j.nuclphysb.2022.116017} {\bibfield  {journal} {\bibinfo  {journal} {Nucl. Phys. B}\ }\textbf {\bibinfo {volume} {985}},\ \bibinfo {pages} {116017} (\bibinfo {year} {2022}{\natexlab{b}})},\ \Eprint {https://arxiv.org/abs/2205.07029} {arXiv:2205.07029 [hep-th]} \BibitemShut {NoStop}%
\bibitem [{\citenamefont {{Gies}}\ and\ \citenamefont {{Wetterich}}(2002)}]{Gies_2002}%
  \BibitemOpen
  \bibfield  {author} {\bibinfo {author} {\bibfnamefont {H.}~\bibnamefont {{Gies}}}\ and\ \bibinfo {author} {\bibfnamefont {C.}~\bibnamefont {{Wetterich}}},\ }\bibfield  {title} {\bibinfo {title} {{Renormalization flow of bound states}},\ }\href {https://doi.org/10.1103/PhysRevD.65.065001} {\bibfield  {journal} {\bibinfo  {journal} {\prd}\ }\textbf {\bibinfo {volume} {65}},\ \bibinfo {eid} {065001} (\bibinfo {year} {2002})},\ \Eprint {https://arxiv.org/abs/hep-th/0107221} {arXiv:hep-th/0107221 [hep-th]} \BibitemShut {NoStop}%
\bibitem [{\citenamefont {{Gies}}\ and\ \citenamefont {{Wetterich}}(2004)}]{Gies_2004}%
  \BibitemOpen
  \bibfield  {author} {\bibinfo {author} {\bibfnamefont {H.}~\bibnamefont {{Gies}}}\ and\ \bibinfo {author} {\bibfnamefont {C.}~\bibnamefont {{Wetterich}}},\ }\bibfield  {title} {\bibinfo {title} {{Universality of spontaneous chiral symmetry breaking in gauge theories}},\ }\href {https://doi.org/10.1103/PhysRevD.69.025001} {\bibfield  {journal} {\bibinfo  {journal} {\prd}\ }\textbf {\bibinfo {volume} {69}},\ \bibinfo {eid} {025001} (\bibinfo {year} {2004})},\ \Eprint {https://arxiv.org/abs/hep-th/0209183} {arXiv:hep-th/0209183 [hep-th]} \BibitemShut {NoStop}%
\bibitem [{\citenamefont {{Wetterich}}(2001)}]{CWSBC}%
  \BibitemOpen
  \bibfield  {author} {\bibinfo {author} {\bibfnamefont {C.}~\bibnamefont {{Wetterich}}},\ }\bibfield  {title} {\bibinfo {title} {{Spontaneously broken color}},\ }\href {https://doi.org/10.1103/PhysRevD.64.036003} {\bibfield  {journal} {\bibinfo  {journal} {\prd}\ }\textbf {\bibinfo {volume} {64}},\ \bibinfo {eid} {036003} (\bibinfo {year} {2001})},\ \Eprint {https://arxiv.org/abs/hep-ph/0008150} {arXiv:hep-ph/0008150 [hep-ph]} \BibitemShut {NoStop}%
\bibitem [{\citenamefont {{Wetterich}}(2004)}]{CWHP}%
  \BibitemOpen
  \bibfield  {author} {\bibinfo {author} {\bibfnamefont {C.}~\bibnamefont {{Wetterich}}},\ }\bibfield  {title} {\bibinfo {title} {{Higgs picture of the QCD-vacuum}},\ }in\ \href {https://doi.org/10.1063/1.1843594} {\emph {\bibinfo {booktitle} {IX Hadron Physics and VII Relativistic Aspects of Nuclear Physics: A Joint Meeting on QCD and Qcp}}},\ \bibinfo {series} {American Institute of Physics Conference Series}, Vol.\ \bibinfo {volume} {739},\ \bibinfo {editor} {edited by\ \bibinfo {editor} {\bibfnamefont {M.~E.}\ \bibnamefont {{Bracco}}}, \bibinfo {editor} {\bibfnamefont {M.}~\bibnamefont {{Chiapparini}}}, \bibinfo {editor} {\bibfnamefont {E.}~\bibnamefont {{Ferreira}}},\ and\ \bibinfo {editor} {\bibfnamefont {T.}~\bibnamefont {{Kodama}}}}\ (\bibinfo  {publisher} {AIP},\ \bibinfo {year} {2004})\ pp.\ \bibinfo {pages} {123--159},\ \Eprint {https://arxiv.org/abs/hep-ph/0410057} {arXiv:hep-ph/0410057 [hep-lat]} \BibitemShut {NoStop}%
\bibitem [{\citenamefont {{Daum}}\ \emph {et~al.}(2010)\citenamefont {{Daum}}, \citenamefont {{Harst}},\ and\ \citenamefont {{Reuter}}}]{DHR}%
  \BibitemOpen
  \bibfield  {author} {\bibinfo {author} {\bibfnamefont {J.-E.}\ \bibnamefont {{Daum}}}, \bibinfo {author} {\bibfnamefont {U.}~\bibnamefont {{Harst}}},\ and\ \bibinfo {author} {\bibfnamefont {M.}~\bibnamefont {{Reuter}}},\ }\bibfield  {title} {\bibinfo {title} {{Running gauge coupling in asymptotically safe quantum gravity}},\ }\href {https://doi.org/10.1007/JHEP01(2010)084} {\bibfield  {journal} {\bibinfo  {journal} {Journal of High Energy Physics}\ }\textbf {\bibinfo {volume} {2010}},\ \bibinfo {eid} {84} (\bibinfo {year} {2010})},\ \Eprint {https://arxiv.org/abs/0910.4938} {arXiv:0910.4938 [hep-th]} \BibitemShut {NoStop}%
\bibitem [{\citenamefont {{Folkerts}}\ \emph {et~al.}(2012)\citenamefont {{Folkerts}}, \citenamefont {{Litim}},\ and\ \citenamefont {{Pawlowski}}}]{FDP}%
  \BibitemOpen
  \bibfield  {author} {\bibinfo {author} {\bibfnamefont {S.}~\bibnamefont {{Folkerts}}}, \bibinfo {author} {\bibfnamefont {D.~F.}\ \bibnamefont {{Litim}}},\ and\ \bibinfo {author} {\bibfnamefont {J.~M.}\ \bibnamefont {{Pawlowski}}},\ }\bibfield  {title} {\bibinfo {title} {{Asymptotic freedom of Yang-Mills theory with gravity}},\ }\href {https://doi.org/10.1016/j.physletb.2012.02.002} {\bibfield  {journal} {\bibinfo  {journal} {Physics Letters B}\ }\textbf {\bibinfo {volume} {709}},\ \bibinfo {pages} {234} (\bibinfo {year} {2012})},\ \Eprint {https://arxiv.org/abs/1101.5552} {arXiv:1101.5552 [hep-th]} \BibitemShut {NoStop}%
\bibitem [{\citenamefont {Harst}\ and\ \citenamefont {Reuter}(2011)}]{HARE}%
  \BibitemOpen
  \bibfield  {author} {\bibinfo {author} {\bibfnamefont {U.}~\bibnamefont {Harst}}\ and\ \bibinfo {author} {\bibfnamefont {M.}~\bibnamefont {Reuter}},\ }\bibfield  {title} {\bibinfo {title} {{QED coupled to QEG}},\ }\href {https://doi.org/10.1007/JHEP05(2011)119} {\bibfield  {journal} {\bibinfo  {journal} {JHEP}\ }\textbf {\bibinfo {volume} {05}},\ \bibinfo {pages} {119}},\ \Eprint {https://arxiv.org/abs/1101.6007} {arXiv:1101.6007 [hep-th]} \BibitemShut {NoStop}%
\bibitem [{\citenamefont {{Christiansen}}\ and\ \citenamefont {{Eichhorn}}(2017)}]{CHE}%
  \BibitemOpen
  \bibfield  {author} {\bibinfo {author} {\bibfnamefont {N.}~\bibnamefont {{Christiansen}}}\ and\ \bibinfo {author} {\bibfnamefont {A.}~\bibnamefont {{Eichhorn}}},\ }\bibfield  {title} {\bibinfo {title} {{An asymptotically safe solution to the U(1) triviality problem}},\ }\href {https://doi.org/10.1016/j.physletb.2017.04.047} {\bibfield  {journal} {\bibinfo  {journal} {Physics Letters B}\ }\textbf {\bibinfo {volume} {770}},\ \bibinfo {pages} {154} (\bibinfo {year} {2017})},\ \Eprint {https://arxiv.org/abs/1702.07724} {arXiv:1702.07724 [hep-th]} \BibitemShut {NoStop}%
\bibitem [{\citenamefont {{Christiansen}}\ \emph {et~al.}(2018)\citenamefont {{Christiansen}}, \citenamefont {{Litim}}, \citenamefont {{Pawlowski}},\ and\ \citenamefont {{Reichert}}}]{CLPR}%
  \BibitemOpen
  \bibfield  {author} {\bibinfo {author} {\bibfnamefont {N.}~\bibnamefont {{Christiansen}}}, \bibinfo {author} {\bibfnamefont {D.~F.}\ \bibnamefont {{Litim}}}, \bibinfo {author} {\bibfnamefont {J.~M.}\ \bibnamefont {{Pawlowski}}},\ and\ \bibinfo {author} {\bibfnamefont {M.}~\bibnamefont {{Reichert}}},\ }\bibfield  {title} {\bibinfo {title} {{Asymptotic safety of gravity with matter}},\ }\href {https://doi.org/10.1103/PhysRevD.97.106012} {\bibfield  {journal} {\bibinfo  {journal} {\prd}\ }\textbf {\bibinfo {volume} {97}},\ \bibinfo {eid} {106012} (\bibinfo {year} {2018})},\ \Eprint {https://arxiv.org/abs/1710.04669} {arXiv:1710.04669 [hep-th]} \BibitemShut {NoStop}%
\bibitem [{\citenamefont {{Eichhorn}}\ and\ \citenamefont {{Versteegen}}(2018)}]{EVER}%
  \BibitemOpen
  \bibfield  {author} {\bibinfo {author} {\bibfnamefont {A.}~\bibnamefont {{Eichhorn}}}\ and\ \bibinfo {author} {\bibfnamefont {F.}~\bibnamefont {{Versteegen}}},\ }\bibfield  {title} {\bibinfo {title} {{Upper bound on the Abelian gauge coupling from asymptotic safety}},\ }\href {https://doi.org/10.1007/JHEP01(2018)030} {\bibfield  {journal} {\bibinfo  {journal} {Journal of High Energy Physics}\ }\textbf {\bibinfo {volume} {2018}},\ \bibinfo {eid} {30} (\bibinfo {year} {2018})},\ \Eprint {https://arxiv.org/abs/1709.07252} {arXiv:1709.07252 [hep-th]} \BibitemShut {NoStop}%
\bibitem [{\citenamefont {Wetterich}(1981)}]{CWFP}%
  \BibitemOpen
  \bibfield  {author} {\bibinfo {author} {\bibfnamefont {C.}~\bibnamefont {Wetterich}},\ }\bibfield  {title} {\bibinfo {title} {{Gauge Hierarchy due to strong interactions?}},\ }\href {https://doi.org/10.1016/0370-2693(81)90124-6} {\bibfield  {journal} {\bibinfo  {journal} {Phys. Lett. B}\ }\textbf {\bibinfo {volume} {104}},\ \bibinfo {pages} {269} (\bibinfo {year} {1981})}\BibitemShut {NoStop}%
\bibitem [{\citenamefont {Schrempp}\ and\ \citenamefont {Wimmer}(1996)}]{SWI}%
  \BibitemOpen
  \bibfield  {author} {\bibinfo {author} {\bibfnamefont {B.}~\bibnamefont {Schrempp}}\ and\ \bibinfo {author} {\bibfnamefont {M.}~\bibnamefont {Wimmer}},\ }\bibfield  {title} {\bibinfo {title} {{Top quark and Higgs boson masses: Interplay between infrared and ultraviolet physics}},\ }\href {https://doi.org/10.1016/0146-6410(96)00059-2} {\bibfield  {journal} {\bibinfo  {journal} {Prog. Part. Nucl. Phys.}\ }\textbf {\bibinfo {volume} {37}},\ \bibinfo {pages} {1} (\bibinfo {year} {1996})},\ \Eprint {https://arxiv.org/abs/hep-ph/9606386} {arXiv:hep-ph/9606386} \BibitemShut {NoStop}%
\end{thebibliography}%

\end{document}